\documentclass[12pt]{article}
\usepackage{graphicx}

\usepackage{amssymb}
\usepackage{amsmath}
\usepackage{amscd}
\usepackage{latexsym}

\numberwithin{equation}{section}

\newcommand{\be}{\begin{equation}}
\newcommand{\ee}{\end{equation}}
\newcommand{\Dlt}{\Delta}
\newcommand{\dlt}{\delta}
\newcommand{\prt}{\partial}
\newcommand{\br}{{\bf r}}
\newcommand{\bk}{{\bf k}}

\newcommand{\ba}{{\bf a}}
\newcommand{\bn}{{\bf n}}
\newcommand{\bp}{{\bf p}}
\newcommand{\bg}{{\bf g}}
\newcommand{\bq}{{\bf q}}
\newcommand{\bP}{{\bf P}}
\newcommand{\bM}{{\bf M}}
\newcommand{\bS}{{\bf S}}
\newcommand{\bv}{{\bf v}}
\newcommand{\bt}{\beta}
\newcommand{\vp}{\varphi}
\newcommand{\ep}{\varepsilon}
\newcommand{\al}{\alpha}
\newcommand{\ra}{\rightarrow}
\newcommand{\sgm}{\sigma}

\newcommand{\gm}{\gamma}
\newcommand{\om}{\omega}
\newcommand{\Om}{\Omega}
\newcommand{\Gm}{\Gamma}
\newcommand{\dgr}{\dagger}
\newcommand{\lbd}{\lambda}
\newcommand{\Lbd}{\Lambda}
\newcommand{\cF}{{\cal F}}
\newcommand{\cH}{{\cal H}}
\newcommand{\cD}{{\cal D}}

\begin{document}

\begin{center}

{\Large {\bf Cold Bosons in Optical Lattices} \\ [5mm]

V.I. Yukalov} \\ [3mm]

{\it
Bogolubov Laboratory of Theoretical Physics, \\
Joint Institute for Nuclear Research, Dubna 141980, Russia}

\end{center}

\vskip 3cm

\begin{abstract}

Basic properties of cold Bose atoms in optical lattices are reviewed. The
main principles of correct self-consistent description of arbitrary systems
with Bose-Einstein condensate are formulated. Theoretical methods for
describing regular periodic lattices are presented. A special attention is
paid to the discussion of Bose-atom properties in the frame of the boson
Hubbard model. Optical lattices with arbitrary strong disorder, induced by
random potentials, are treated. Possible applications of cold atoms in
optical lattices are discussed, with an emphasis of their usefulness for
quantum information processing and quantum computing. An important feature
of the present review article, distinguishing it from other review works,
is that theoretical fundamentals here are not just mentioned in brief, but
are thoroughly explained. This makes it easy for the reader to follow the
principal points without the immediate necessity of resorting to numerous
publications in the field.

\end{abstract}

\vskip 7cm

{\bf E-mail:} yukalov@theor.jinr.ru

\newpage

{\bf{\Large Contents}}

\vskip 5mm

{\parindent=0pt
{\bf 1. Tools for Manipulating Atoms} }

\vskip 2mm

{\bf 1.1}. Cold Atoms

\vskip 2mm

{\bf 1.2}. Control Parameters

\vskip 2mm

{\bf 1.3}. Atomic Fractions

\vskip 2mm

{\bf 1.4}. System Classification

\vskip 2mm

{\bf 1.5}. Cold Molecules

\vskip 5mm

{\parindent=0pt
{\bf 2. Systems with Bose-Einstein Condensate} }

\vskip 2mm

{\bf 2.1}. Bose-Einstein Condensation

\vskip 2mm

{\bf 2.2}. Penrose-Onsager Scheme

\vskip 2mm

{\bf 2.3}. Order Indices

\vskip 2mm

{\bf 2.4}. Representative Ensembles

\vskip 2mm

{\bf 2.5}. Field Operators

\vskip 2mm

{\bf 2.6}. Gauge Symmetry

\vskip 2mm

{\bf 2.7}. Bogolubov Shift

\vskip 2mm

{\bf 2.8}. Self-Consistent Approach

\vskip 2mm

{\bf 2.9}. Condensate Existence

\vskip 2mm

{\bf 2.10}. Superfluid Fraction

\vskip 2mm

{\bf 2.11}. Equations of Motion

\vskip 2mm

{\bf 2.12}. Uniform System

\vskip 2mm

{\bf 2.13}. Anomalous Averages

\vskip 2mm

{\bf 2.14}. Particle Fluctuations

\vskip 2mm

{\bf 2.15}. Fragmented Condensates

\vskip 2mm

{\bf 2.16}. Multicomponent Condensates

\vskip 2mm

{\bf 2.17}. Model Condensates

\vskip 5mm

{\parindent=0pt
{\bf 3. Regular Optical Lattices} }

\vskip 2mm

{\bf 3.1}. Optical Lattices

\vskip 2mm

{\bf 3.2}. Periodic Structures

\vskip 2mm

{\bf 3.3}. Condensate in Lattices

\vskip 2mm

{\bf 3.4}. Operator of Momentum

\vskip 2mm

{\bf 3.5}. Tight-Binding Approximation

\vskip 2mm

{\bf 3.6}. Superfluidity in Lattices

\vskip 2mm

{\bf 3.7}. Transverse Confinement

\vskip 2mm

{\bf 3.8}. Bloch Spectrum

\vskip 2mm

{\bf 3.9}. Spectrum Parameters

\vskip 2mm

{\bf 3.10}. Elementary Excitations

\vskip 2mm

{\bf 3.11}. Wave Stability

\vskip 2mm

{\bf 3.12}. Moving Lattices

\vskip 2mm

{\bf 3.13}. Soliton Formation

\vskip 2mm

{\bf 3.14}. Transverse Resonances

\vskip 2mm

{\bf 3.15}. Lagrange Variation

\vskip 5mm

{\parindent=0pt
{\bf 4. Boson Hubbard Model} }

\vskip 2mm

{\bf 4.1}. Wannier Representation

\vskip 2mm

{\bf 4.2}. Grand Hamiltonian

\vskip 2mm

{\bf 4.3}. Bose-Condensed System

\vskip 2mm

{\bf 4.4}. Thermodynamic Characteristics

\vskip 2mm

{\bf 4.5}. Superfluid Fraction

\vskip 2mm

{\bf 4.6}. Single-Site Approximation

\vskip 2mm

{\bf 4.7}. Localized State

\vskip 5mm

{\parindent=0pt
{\bf 5. Phase States and Transitions} }

\vskip 2mm

{\bf 5.1}. Existence of Pure Phases

\vskip 2mm

{\bf 5.2}. Hard-Core Lattice Gas

\vskip 2mm

{\bf 5.3}. Effective Interaction Parameter

\vskip 2mm

{\bf 5.4}. Gutzwiller Single-Site Approximation

\vskip 2mm

{\bf 5.5}. Dynamical Mean-Field Approximation

\vskip 2mm

{\bf 5.6}. Small-System Numerical Diagonalization

\vskip 2mm

{\bf 5.7}. Density-Matrix Renormalization Group

\vskip 2mm

{\bf 5.8}. Strong-Coupling Perturbation Theory

\vskip 2mm

{\bf 5.9}. Monte Carlo Simulations

\vskip 2mm

{\bf 5.10}. Order of Phase Transition

\vskip 2mm

{\bf 5.11}. Experiments on Superfluid-Insulator Transition

\vskip 2mm

{\bf 5.12}. Layered Superfluid-Insulator Structure

\vskip 2mm

{\bf 5.13}. Models with Neighbor Interactions

\vskip 2mm

{\bf 5.14}. Quasiperiodic Optical Lattices

\vskip 2mm

{\bf 5.15}. Rotating Optical Lattices

\vskip 5mm

{\parindent=0pt
{\bf 6. Optical Lattices with Disorder} }

\vskip 2mm

{\bf 6.1}. Random Potentials

\vskip 2mm

{\bf 6.2}. Uniform Limit

\vskip 2mm

{\bf 6.3}. Disordered Lattice

\vskip 2mm

{\bf 6.4}. Disordered Superfluid

\vskip 2mm

{\bf 6.5}. Phase Diagram

\vskip 5mm

{\parindent=0pt
{\bf 7. Nonstandard Lattice Models} }

\vskip 2mm

{\bf 7.1}. Coexisting States

\vskip 2mm

{\bf 7.2}. Vibrational Excitations

\vskip 2mm

{\bf 7.3}. Phonon-Induced Interactions

\vskip 5mm

{\parindent=0pt
{\bf 8. Double-Well Optical Lattices} }

\vskip 2mm

{\bf 8.1}. Effective Hamiltonians

\vskip 2mm

{\bf 8.2}. Phase Transitions

\vskip 2mm

{\bf 8.3}. Collective Excitations

\vskip 2mm

{\bf 8.4}. Nonequilibrium States

\vskip 2mm

{\bf 8.5}. Heterophase Lattices

\vskip 5mm

{\parindent=0pt
{\bf 9. Tools for Quantum Computing} }

\vskip 2mm

{\bf 9.1}. Entanglement Production

\vskip 2mm

{\bf 9.2}. Topological Modes

\vskip 2mm

{\bf 9.3}. Coherent States

\vskip 2mm

{\bf 9.4}. Coherent-Mode Register

\vskip 2mm

{\bf 9.5}. Double-Well Register

\vskip 5mm

{\parindent=0pt
{\bf 10. Brief Concluding Remarks}

\vskip 5mm

{\bf References} }

\newpage

\section{Tools for Manipulating Atoms}

\subsection{Cold Atoms}

Physics of cold trapped atoms has become nowadays a very fastly developing
field of research, both theoretical and experimental. Magnetic,
magneto-optical, and all optical traps are employed for trapping atoms.
Several atomic species have been cooled down to low temperatures, when
their quantum degeneracy could be observed. The Bose-Einstein condensation
(BEC) of trapped atoms was experimentally realized [1--3]. At the present
time, BEC has been achieved for 12 atomic species: $^1$H, $^4$He, $^7$Li,
$^{23}$Na, $^{39}$K, $^{41}$K, $^{52}$Cr, $^{85}$Rb, $^{87}$Rb, $^{133}$Cs,
$^{170}$Yb, and $^{174}$Yb. Quantum degeneracy in trapped Fermi gases was
achieved for $^{40}$K and $^6$Li atoms [4--6]. Now there exist several books
[7,8] and review articles treating the Bose-Einstein condensation of Bose
atoms [9--14] and the quantum properties of ultracold Fermi gases [15].

An important development has been the realization of optical lattices,
formed by interfering laser beams producing a standing wave. Cold atoms
can be trapped for a long time in the minima of the created periodic
potential [16]. There are several surveys considering the properties of
cold atoms in optical lattices, e.g., [17--20].

In the present review paper, the emphasis is made on the theory of cold
Bose atoms in periodic potentials. Such potentials are usually formed by
optical lattices, though recently magnetic lattices have also been realized
[21].

There are two principal features making this review paper distinct from all
other review articles. First, the basic theoretical points are thoroughly
explained here, but not just mentioned in brief. This should allow the reader
to better understand the theoretical fundamentals and to easily follow the
logic of the used mathematical methods. The material of this paper can serve
as a reference source for researchers in the field. Second, this paper covers
the most recent theoretical results that have not yet been described in other
review articles.

\subsection{Control Parameters}

Optical lattices with cold atoms provide an extraordinary possibility of
creating systems with a wide variety of properties, which can be manipulated
in several ways. First of all, the lattice parameters themselves can be
varied in a wide range. In experiments, optical lattices can be formed,
having different spacing, depth, and filling factors. The latter can be
either integer or fractional and can vary between one and $10^4$ atoms per
lattice site [22,23]. The number of lattice sites can also be different. One
- two - and three - dimensional lattices can be formed. The lattices can be
periodic and quasiperiodic. Different atomic species, or their mixtures, can
be loaded in the lattice. The strength of interatomic interactions can be
regulated in a very wide range by employing the Feshbach resonance techniques
[10,24]. Varying temperature and/or lattice depth, it is possible to induce
phase transitions between localized and delocalized states of atoms, as
well as between the normal and superfluid phases.

Lattice properties can also be regulated by imposing additional external
potentials. In particular, random external fields can be used, producing
disordered lattices. By means of alternating external fields, one can
manipulate the motion of atomic clouds. Employing special resonant
alternating fields makes it possible to create an unusual state of matter,
the nonground-state Bose-Einstein condensates.

These rich potentialities of manipulating cold atoms in optical lattices
make this object of high importance for various applications. But the latter
can become practicable only being based on effective and correct theoretical
investigations.

\subsection{Atomic Fractions}

The total number of atoms $N$, loaded into a lattice, can consist of several
parts. An important part of a Bose system is that one forming Bose-Einstein
condensate (BEC) of $N_0$ atoms, which characterizes the coherent portion of
atoms. As a rule the arising BEC leads to the appearance of superfluidity
involving $N_{sup}$ atoms. There is no simple relation between $N_0$ and
$N_{sup}$ and even it is not compulsory that they be present simultaneously.
The physical origins of $N_0$ and $N_{sup}$ are different. The appearance
of BEC manifests the existence of coherence in the system. Superfluidity
demonstrates the presence of nontrivial response to a velocity boost, which
is caused by strong atomic correlations.

Atoms in a lattice can also be distinguished by the region of their motion.
Atoms can be localized in their lattice sites or can be delocalized and
moving through the whole sample. The localized atoms are associated with
the solid state of matter, possessing small compressibility and a gap in
the single-particle spectrum. The number of atoms, forming a solid, will be
denoted by $N_{sol}$. Delocalized atoms are typical of the liquid or gaseous
state of matter, with gapless single-particle spectra.

Since the total number of atoms $N$, as well as the atomic numbers $N_0$,
$N_{sup}$, and $N_{sol}$, can be very large, it is more appropriate to deal
with the related atomic fractions, which are:

the {\it condensate fraction}
$$
n_0 \equiv \frac{N_0}{N} \; ,
$$

{\it superfluid fraction}
$$
n_{sup} \equiv \frac{N_{sup}}{N} \; ,
$$

and the {\it solid fraction}
$$
n_{sol} \equiv \frac{N_{sol}}{N} \; .
$$

Similarly, the number of uncondensed atoms $N_1$ defines the normal fraction
$n_1\equiv N_1/N$. And one can define the fraction of atoms in fluid phase.
But the fractions $n_0$, $n_{sup}$, and $n_{sol}$ are the main for the
classification of the major system features.

\subsection{System Classification}

The atomic fractions $n_0$, $n_{sup}$, and $n_{sol}$ characterize the basic
properties of systems formed by Bose atoms in optical lattices. Thus, the
existence of BEC means the presence of coherence, because of which such a
system can be termed coherent,
$$
n_0 > 0 \qquad (coherent) \; .
$$
Vice versa, the absence of BEC permits to call the system incoherent
$$
n_0 = 0 \qquad (incoherent) \; .
$$
In the same way, the system is superfluid, when there is the superfluid
fraction,
$$
n_{sup} > 0 \qquad (superfluid) \; .
$$
The absence of the superfluid fraction implies that the system is not
superfluid, that is, normal,
$$
n_{sup} = 0 \qquad (normal) \; .
$$
The presence of the solid fraction gives to the system rigidity typical of
solids,
$$
n_{sol} > 0 \qquad (solid) \; .
$$
While, if there is no solid fraction, the system is either liquid or gaseous,
generally speaking, fluid,
$$
n_{sol} = 0 \qquad (fluid) \; .
$$

This terminology allows us to suggest the following classification of
admissible systems, depending on the presence or absence of the related
atomic fractions.

\vskip 2mm

(1) {\it Incoherent normal fluid}:
$$
n_0 = 0 \; , \qquad n_{sup} = 0 \; , \qquad n_{sol} = 0 \; .
$$
Ubiquitous examples are classical liquids and gases.

\vskip 2mm

(2) {\it Coherent normal fluid}:
$$
n_0 > 0 \; , \qquad n_{sup} = 0 \; , \qquad n_{sol} = 0 \; .
$$
This case looks a bit exotic, though the situation, when there is BEC but
there is no superfluidity can be attributed to what one calls Bose glass,
the state that may develop in the presence of disorder.

\vskip 2mm

(3) {\it Incoherent superfluid}:
$$
n_0 = 0 \; , \qquad n_{sup} > 0 \; , \qquad n_{sol} = 0 \; .
$$
The known examples are two-dimensional superfluid films without BEC.

\vskip 2mm

(4) {\it Coherent superfluid}:
$$
n_0 > 0 \; , \qquad n_{sup} > 0 \; , \qquad n_{sol} = 0 \; .
$$
This is superfluid $^4$He.

\vskip 2mm

(5) {\it Incoherent normal solid}:
$$
n_0 = 0 \; , \qquad n_{sup} = 0 \; , \qquad n_{sol} > 0 \; .
$$
The majority of solids are exactly of this type.

\vskip 2mm

(6) {\it Coherent normal solid}:
$$
n_0 > 0 \; , \qquad n_{sup} = 0 \; , \qquad n_{sol} > 0 \; .
$$
This type of solids can also be attributed to the so-called Bose glass.

\vskip 2mm

(7) {\it Incoherent superfluid solid}:
$$
n_0 = 0 \; , \qquad n_{sup} > 0 \; , \qquad n_{sol} > 0 \; .
$$
The possibility of such solids is currently under discussion.

\vskip 2mm

(8) {\it Coherent superfluid solid}:
$$
n_0 > 0 \; , \qquad n_{sup} > 0 \; , \qquad n_{sol} > 0 \; .
$$
This state looks admissible in optical lattices.

Thus, there can exist 8 classes of systems, depending on the presence or
absence of the fractions $n_0$, $n_{sup}$, and $n_{sol}$. These different
states can be achieved by appropriately adjusting the system parameters.

\subsection{Cold Molecules}

Bose-Einstein condensate can, in principle, be created in different Bose
systems. As is mentioned in subsection 1.1, at the present time, BEC has
been achieved in 12 atomic species. The latest of them was $^{170}$Yb [25].
In addition, there exist Bose molecules formed of either Bose or Fermi atoms
[10,15,24,26,27]. In systems, composed of Bose molecules, BEC can also arise.
Thus, BEC was produced in molecular systems, where the molecules were formed
by Bose atoms ($^{23}$Na$_2$, $^{85}$Rb$_2$, $^{87}$Rb$_2$, $^{133}$Cs$_2$)
as well as by Fermi atoms ($^6$Li$_2$, $^{40}$K$_2$). Among other systems,
that could exhibit BEC, it is possible to mention boson quark clusters and
hadronic molecules [28,29]. Pion condensation in nuclear matter could be one
more example [30--34], though in this case the condensate itself possesses a
periodic structure.

The theory, presented in the following sections, is applicable to Bose
systems of arbitrary nature, whether the constituents are atoms or molecules,
or some kind of bosonic clusters. The sole thing is that these constituents
are treated as Bose particles, characterized by their masses and interactions.
Also, the main attention is paid to particles without internal degrees of
freedom. For instance, spins are assumed to be frozen, so that particles can
be treated as spinless. The consideration of particles with spin degrees of
freedom requires a separate paper.

Theoretical methods are general for describing any type of bosons, whether
the latter are atoms or molecules. However, it is important to keep in mind
that the possibility of creating molecules provides the way of enriching the
system properties. Molecules can also be loaded in optical lattices [35--39].

\section{Systems with Bose-Einstein Condensate}

\subsection{Bose-Einstein Condensation}

Lattices can be periodic, quasiperiodic or even random, representing
different external potentials making the system nonuniform. It is worth
starting the consideration by formulating the general criteria characterizing
the occurrence of BEC in nonuniform systems.

Generally, BEC is the occupation of a single, or several, quantum states by
a large number of identical particles. For simplicity, we shall be talking
about a single quantum state. The generalization to several macroscopically
occupied quantum states is straightforward and will be done in Sec. 2.15.

Historically, BEC was described by Einstein for ideal uniform Bose gas. The
history and related historical references can be found in Ref. [40]. The
quantum states of a uniform gas are characterized by the momentum $\bk$. Here
and in what follows, we use the system of units, where the Planck constant
and the Boltzmann constant are set to one, $\hbar\equiv 1$, $k_B\equiv 1$.

In the general case of nonuniform systems, quantum states are labelled by
a multi-index $k$, whose concrete representation depends on the considered
problem. There exists the state occupation number $n_k$ showing the number
of particles in a quantum state $k$. Suppose that among all quantum states
there occurs a single state $k_0$, for which the occupation number
\be
\label{2.1}
N_0 \equiv n_{k_0}
\ee
is large. Here "large" means not merely that $N_0$ is much larger than one,
but that it is comparable to the total number of particles $N$, such that
$N_0\propto N$. Then we can say that there occurs BEC into the state $k_0$.

One says that the condensate state is macroscopically occupied. To make this
phrase mathematically accurate, one resorts to the notion of the {\it
thermodynamic limit}
\be
\label{2.2}
N \; \ra \; \infty \; , \qquad V \; \ra \; \infty \; , \qquad
\frac{N}{V} \; \ra \; const \; > \; 0 \; ,
\ee
where $N$ is the total number of particles in the system of volume $V$. The
state $k_0$ is termed {\it macroscopically occupied}, when
\be
\label{2.3}
\lim_{N\ra\infty} \; \frac{N_0}{V} \; > \; 0 \; ,
\ee
where the thermodynamic limit (\ref{2.2}) is implied. Condition (\ref{2.3})
is, actually, the {\it Einstein criterion} of BEC.

For trapped atoms, the system volume $V$ may be not well defined. Then the
thermodynamic limit can be specified in a different way [41]. If the system
contains $N$ trapped atoms, for which extensive observable quantities are
defined, then the following limit can be considered. Let $A_N$ be an extensive
observable quantity, then the {\it effective thermodynamic limit} is
\be
\label{2.4}
N\; \ra \; \infty \; , \qquad \frac{A_N}{N} \; \ra \; const \; .
\ee
For instance, taking for the observable quantity the internal energy $E_N$ of
$N$ particles, we have [41] limit (\ref{2.4}) as
\be
\label{2.5}
N \; \ra \; \infty \; , \qquad \frac{E_N}{N} \; \ra \; const \; .
\ee
In what follows, writing $N\ra\infty$, we shall assume one of the forms of
thermodynamic limit.

Confined systems contain finite numbers of atoms $N$, though the latter is
large. All finite systems with $N\gg 1$ can be treated by the standard methods
of statistical mechanics. For finite systems, thermodynamic limits (\ref{2.2})
or (\ref{2.4}) and (\ref{2.5}) are interpreted as {\it thermodynamic tests},
not merely allowing for the simplification of calculations, but, which is the
most important, making it possible to check the correctness of theories.

\subsection{Penrose-Onsager Scheme}

The Einstein criterion of BEC (\ref{2.3}) is easily applicable to uniform
ideal gases. But for interacting systems, especially for nonuniform cases,
to make criterion (\ref{2.3}) useful requires, first, to specify how the
quantum state occupation numbers $n_k$ could be found. Penrose and Onsager
[42] suggested the following scheme.

Assume that the single-particle density matrix $\rho(\br,\br')$ of the
considered system is known. This matrix is a function of the real-space
variables $\br$ and, generally, of time $t$. The latter does not enter
$\rho(\br,\br')$ for equilibrium systems, but for nonequilibrium systems,
$\rho(\br,\br',t)$ depends on time. In what follows, we shall omit, for the
sake of brevity, the time dependence, where this is not important. However,
we may keep in mind that the time variable can always be included, when the
consideration concerns nonequilibrium cases.

If the density matrix $\rho(\br,\br')$ is known, then one could solve the
eigenvalue problem
\be
\label{2.6}
\int \rho (\br,\br') \vp_k(\br') \; d\br' = n_k \vp_k(\br) \; ,
\ee
where the integration is over the whole volume specifying the system. The
eigenfunctions $\vp_k(\br)$ are called the {\it natural orbitals} [43]. The
family $\{\vp_k(\br)\}$ forms a complete orthonormal basis, for which
$$
\int \vp_k^*(\br) \vp_p(\br) \; d\br = \dlt_{kp} \; .
$$
Since the single-particle density matrix is normalized to the total number
of particles
$$
N = \int \rho(\br,\br) \; d\br \; ,
$$
the eigenvalues
$$
n_k = \int \vp_k^*(\br) \rho(\br,\br') \vp_k(\br') \; d\br d\br'
$$
have the meaning of the occupation numbers of quantum states labelled by a
multi-index $k$.

In terms of the {\it natural orbitals}, the density matrix enjoys the {\it
diagonal expansion}
\be
\label{2.7}
\rho(\br,\br') = \sum_k n_k \vp_k(\br) \vp_k^*(\br') \; .
\ee
Suppose that the maximal of the eigenvalues $n_k$ corresponds to a quantum
state $k_0$, for which we may write
\be
\label{2.8}
N_0 \equiv \sup_k \; n_k = n_{k_0} \; .
\ee
Separating the state $k_0$ from sum (\ref{2.7}) gives
\be
\label{2.9}
\rho(\br,\br') = N_0 \vp_0(\br) \vp_0^*(\br') \; + \;
\sum_{k\neq k_0} n_k \vp_k(\br) \vp_k^*(\br') \; ,
\ee
where $\vp_0(\br)\equiv\vp_{k_0}(\br)$. Then the total number of particles
can be written as the sum
\be
\label{2.10}
N = N_0 + N_1 \; , \qquad N_1 \equiv \sum_{k\neq k_0} n_k
\ee
of the number of particles $N_0$ in the state $k_0$ and the number of all
other particles $N_1$.

One says that there happens BEC into the state $k_0$, if the latter is
macroscopically occupied, such that the largest eigenvalue (\ref{2.8})
satisfies condition (\ref{2.3}) in the sense of one of the thermodynamic
limits (\ref{2.2}) or (\ref{2.4}). Thus, we return to the Einstein criterion
(\ref{2.3}). The novelty in the Penrose-Onsager scheme is the interpretation
of the number of condensed particles (\ref{2.1}) as the largest eigenvalue
(\ref{2.8}) of the single-particle density matrix.

The Penrose-Onsager interpretation of the Einstein criterion for BEC is very
general, being applicable to arbitrary statistical systems, including confined
systems of trapped atoms. This is contrary to the concept of the off-diagonal
long-range order [44], introducing the number of condensed particles $N_0$
through the limiting relation
$$
\frac{N_0}{V} \; = \; \lim_{|\br-\br'|\ra\infty} \;
\rho(\br,\br') \; .
$$
This concept has a meaning solely for uniform infinite systems, while for
trapped atoms it is not applicable [9], always strictly giving $N_0=0$.

\subsection{Order Indices}

The Penrose-Onsager scheme can be generalized by introducing the notion
of order indices. The latter can be formulated for arbitrary operators [45].
Let $\hat A$ be an operator possessing a norm $||\hat A||$ and a trace
${\rm Tr}\hat A$. Then the {\it operator order index} of $\hat A$ is defined
[45] as
$$
\om(\hat A) \equiv
\frac{\log||\hat A||}{\log|{\rm Tr}\hat A|} \; .
$$
Here the logarithm can be taken to any convenient base, for instance, it can
be the natural logarithm ${\rm ln}$.

The reduced density matrices can be treated as matrices with respect to their
real-space variables [43]. Thus, the single-particle density matrix defines
the first-order density matrix $\hat\rho_1\equiv[\rho(\br,\br')]$. Then the
order index of $\hat\rho_1$ is
\be
\label{2.11}
\om(\hat\rho_1) =
\frac{\log||\hat\rho_1||}{\log{\rm Tr}\hat\rho_1} \; .
\ee
Similarly, one can introduce the order indices of higher-order density
matrices [46--49]. For the order index (\ref{2.11}), since
$$
||\hat\rho_1|| = N_0 \; , \qquad {\rm Tr}\hat\rho_1 = N \; ,
$$
we obtain
\be
\label{2.12}
\om(\hat\rho_1) = \frac{\log N_0}{\log N} \; .
\ee
This index is not larger than one, $\om(\hat\rho_1)\leq 1$, because
$N_0\leq N$.

The order indices are convenient for classifying different types of order that
can arise in the system. For Bose systems, there can be three possibilities
for the order indices (\ref{2.11}) or (\ref{2.12}). When
\be
\label{2.13}
\om(\hat\rho_1) \leq 0 \qquad (no \; order)\; ,
\ee
there is no ordering in the system, or at the most, a kind of short-range
order may appear. In the interval
\be
\label{2.19}
0 \; < \; \om(\hat\rho_1) \; < \; 1 \qquad (mid-range) \; ,
\ee
the order index demonstrates the amount of mid-range order. There is no true
BEC in this case, but there exists some ordering that can be associated with
{\it quasicondensate}. The true BEC corresponds to the index
\be
\label{2.20}
\om(\hat\rho_1) = 1 \qquad (long-range) \; ,
\ee
which happens in thermodynamic limit $N\ra\infty$. In the finite uniform
systems, as well as in low-dimensional uniform systems, such as one- and
two-dimensional systems, there can be no true BEC, but there can arise
quasicondensate. In confined systems, BEC can happen in low-dimensional
systems, depending on the type of the trapping potential [41]. The BEC
criterion (\ref{2.3}) is equivalent to condition (\ref{2.20}) occurring in
thermodynamic limit,
$$
\om(\hat\rho_1) \; \ra \; 1 \qquad (N\; \ra\; \infty) \; .
$$

The notion of order indices is applicable to arbitrary statistical
systems, whether finite or infinite, uniform or nonuniform, equilibrium
or nonequilibrium. The order indices retain their meaning, when the order
parameters cannot be defined [45--49].

\subsection{Representative Ensembles}

The BEC criteria of the previous sections signals the appearance of BEC.
But these criteria assume the knowledge of the density matrix supposed to
be found beforehand. Such criteria do not prescribe the way of solving the
problem.

The very first step in considering any statistical system is the choice of a
statistical ensemble to be used. The {\it statistical ensemble} is a triplet
$\{\cF,\hat\rho,\prt t\}$, in which $\cF$ is the space of microstates,
$\hat\rho=\hat\rho(0)$ is the initial form of the statistical operator,
and $\prt t$ signifies the evolution law for the considered system. With
the given $\cF$ and $\hat\rho$, one can find the statistical average
\be
\label{2.21}
<\hat A(t)>\; \equiv \; {\rm Tr}_\cF \; \hat\rho\; \hat A(t)
\ee
for an operator $\hat A(t)$. The prescribed evolution law makes it possible
to define the temporal evolution of average (\ref{2.21}),
\be
\label{2.22}
\frac{\prt}{\prt t} < \hat A(t)>\; = \; {\rm Tr}_\cF \; \hat\rho\;
\frac{\prt\hat A(t)}{\prt t} \; .
\ee
The set of all operators $\hat A(t)$, corresponding to observable quantities,
forms the algebra of observables ${\cal O}\equiv\{\hat A(t)\}$. The collection
of the statistical averages (\ref{2.21}) for all operators from the algebra of
observables ${\cal O}$ is termed the {\it statistical state} $<{\cal O}>$.

When defining a statistical ensemble, it is necessary that it would correctly
represent the studied statistical system. This means that all conditions and
constraints, uniquely defining the system, must be taken into account when
constructing the statistical operator $\hat\rho$ and formulating the evolution
law. Such an ensemble is called representative.

In the case of equilibrium systems, one usually tells that a statistical
ensemble is defined by a Gibbs statistical operator, either canonical or
grand canonical. One often calls this the "Gibbs prescription". In many
situations, this is sufficient. However in general such a point of view is
a strong trivialization of the Gibbs ideas. Gibbs did write [50] that just
prescribing a distribution, whether canonical or grand canonical, may be not
sufficient, but the description must be complimented by all those constraints
and conditions that make the statistical systems uniquely defined. Thus, the
idea of representative statistical ensembles is actually due to Gibbs [50].
The term "representative ensembles" was employed by ter Haar [51,52], who
discussed the necessity of correctly representing statistical systems. Such
ensembles, equipped with additional conditions, are also called conditional
[53]. The general theory of equilibrium and quasiequilibrium representative
ensembles was described in the review article [54] and book [55]. Representative
ensembles for Bose systems with broken gauge symmetry were covered in detail in
Refs. [56,57].

To specify the state of microstates, it is necessary to fix the system
variables. Suppose we choose as the variables the field operators $\psi(\br)$
and $\psi^\dgr(\br)$, with the Bose commutation relations
$$
\left [ \psi(\br),\; \psi^\dgr(\br') \right ] =
\dlt(\br -\br') \; ,
$$
other relations being zero. The creation operator $\psi^\dgr(\br)$ generates
the Fock space $\cF(\psi)$, which is the space of microstates [55]. Then the
statistical state is given by the averages (\ref{2.21}), with the trace over
$\cF(\psi)$.

In order to define the statistical operator $\hat\rho$, we need to specify
the conditions making the statistical ensemble representative. One evident
condition is the normalization of the statistical operator,
\be
\label{2.23}
< \hat 1_\cF > \; = \; 1 \; ,
\ee
where $\hat 1_\cF$ is the unity operator in $\cF(\psi)$. The Hamiltonian
energy operator $\hat H[\psi]$, which is a functional of $\psi$ and
$\psi^\dgr$, defines the internal energy
\be
\label{2.24}
 < \hat H[\psi] > \; = \; E \; ,
\ee
which is another statistical condition. The total number of particles $N$ is
given by the average
\be
\label{2.25}
< \hat N[\psi] > \; = \; N \; ,
\ee
of the number-of-particle operator
\be
\label{2.26}
\hat N[\psi] \equiv \int \psi^\dgr(\br) \psi(\br) \; d\br \; .
\ee
Similarly, there can exist other condition operators $\hat C_i[\psi]$, with
$i=1,2,\ldots$, whose averages define additional {\it statistical conditions}
\be
\label{2.27}
< \hat C_i[\psi] > \; = \; C_i \; .
\ee

The statistical operator $\hat\rho$ of an equilibrium system is defined as
the minimizer of the information functional [55]
$$
I[\hat\rho] = {\rm Tr} \hat\rho \; \ln\hat\rho + \lbd_0 \left (
{\rm Tr}\hat\rho - 1 \right ) +
$$
\be
\label{2.28}
+ \bt \left ( {\rm Tr} \hat\rho \hat H[\psi] - E \right ) -
\bt\mu \left ( {\rm Tr} \hat\rho \hat N[\psi] - N \right ) +
\bt \sum_i \nu_i \left ( {\rm Tr} \hat\rho \hat C_i[\psi] -
C_i \right ) \; ,
\ee
in which $\lbd_0$, $\bt$, $\bt\mu$, and $\bt\nu_i$ are the appropriate
Lagrange multipliers. Minimizing (\ref{2.28}) gives
\be
\label{2.29}
\hat\rho =
\frac{\exp(-\bt H[\psi])}{{\rm Tr}_{\cF(\psi)}\exp(-\bt H[\psi])} \; ,
\ee
where the trace is over $\cF(\psi)$ and
\be
\label{2.30}
H[\psi] \equiv \hat H[\psi] - \mu \hat N[\psi] +
\sum_i \nu_i \hat C_i [\psi]
\ee
is the {\it grand Hamiltonian}. The Lagrange multiplier $\bt=1/T$ is the
inverse temperature and $\mu$ is called the chemical potential.

After this, one can explicitly define what actually is the single-particle
density matrix, which till now has appeared as an abstract notion. This
density matrix is
\be
\label{2.31}
\rho(\br,\br') \; \equiv \; < \psi^\dgr(\br') \psi(\br) > \; .
\ee

The evolution equations for the field variables $\psi(\br)$ are obtained as
follows [57]. By introducing the temporal energy operator
\be
\label{2.32}
\hat E[\psi] \equiv \int \psi^\dgr(\br,t) i \;
\frac{\prt}{\prt t}\; \psi(\br,t) \; d\br \; ,
\ee
we define the {\it action functional}
\be
\label{2.33}
\Gm[\psi] \equiv \int \left ( \hat E[\psi] - H[\psi]
\right )\; dt \; .
\ee
The evolution equations for $\psi(\br,t)$ and $\psi^\dgr(\br,t)$ are given by
the extremization of the action functional,
\be
\label{2.34}
\frac{\dlt\Gm[\psi]}{\dlt\psi^\dgr(\br,t)} \; = \; 0 \; ,
\ee
and by the Hermitian conjugation of the latter variational equation. In view
of the action functional (\ref{2.33}), Eq. (\ref{2.34}) yields
\be
\label{2.35}
i\; \frac{\prt}{\prt t} \; \psi(\br,t) =
\frac{\dlt H[\psi]}{\dlt\psi^\dgr(\br,t)} \; .
\ee
This equation is equivalent [55] to the Heisenberg equation
$$
i\; \frac{\prt}{\prt t} \; \psi(\br,t) = \left [\psi(\br,t), \;
H[\psi] \right ] \; .
$$
The initial condition for the evolution equation is $\psi(\br,0)=\psi(\br)$.
The evolution is governed by the same grand Hamiltonian (\ref{2.30}) as that
characterizing the statistical operator (\ref{2.29}).

In the case of a nonequilibrium statistical system, additional conditions
(\ref{2.27}) should include the information on the initial values $<\hat
A(0)>$ for the considered operators $\hat A(t)$.

The procedure, described above, determines the standard way of characterizing
a representative statistical ensemble. Here, it is the triplet of the Fock
space of microstates $\cF(\psi)$, the statistical operator (\ref{2.29}), and
of the evolution equations (\ref{2.34}) or (\ref{2.35}). For an equilibrium
system, the grand Hamiltonian (\ref{2.30}) may not need the last term with
conditional operators.

\subsection{Field Operators}

After a statistical ensemble has been constructed, we may pose the question
whether BEC occurs in the system. Then we remember that BEC implies the
macroscopic occupation of a single quantum state. Quantum states, labelled
by a multi-index $k$, are associated with an orthonormal basis
$\{\vp_k(\br)\}$. Expanding the field operator over this basis, we have
\be
\label{2.36}
\psi(\br) = \sum_k a_k \vp_k(\br) \; ,
\ee
where the operators $a_k$ obey the commutation relations
$$
\left [ a_k, \; a_p^\dgr \right ] = \dlt_{kp} \; , \qquad
[a_k, \; a_p] = 0 \; .
$$
With expansion (\ref{2.36}), the density matrix (\ref{2.31}) takes the form
\be
\label{2.37}
\rho(\br,\br') = \sum_{kp} < a_k^\dgr a_p> \vp_p(\br)
\vp_k^*(\br') \; .
\ee

For a while, there is arbitrariness in choosing a basis in expansion
(\ref{2.36}). However, BEC is a physical phenomenon and can occur not
for an arbitrary chosen quantum state, but for a state naturally related
to the considered physical system. This means that the expansion basis
$\{\vp_k(\br)\}$ is not arbitrary, but is to be formed by natural orbitals.
In terms of the natural orbitals, the density matrix (\ref{2.37}) has to
enjoy the diagonal expansion [43], which implies the {\it quantum-number
conservation condition}
\be
\label{2.38}
< a_k^\dgr a_p> \; = \; \dlt_{kp} n_k \; ,
\ee
where
\be
\label{2.39}
n_k \; \equiv \; < a_k^\dgr a_k >
\ee
is the occupation number.

If BEC is associated with a quantum state $k_0$, then the field operator
(\ref{2.36}) can be separated into two parts,
\be
\label{2.40}
\psi(\br) = \psi_0(\br) + \psi_1(\br) \; ,
\ee
in which the first term is the operator of condensed particles,
\be
\label{2.41}
\psi_0(\br) \equiv a_0 \vp_0(\br) \; ,
\ee
where $a_0\equiv a_{k_0}$, and the second term is the operator of
uncondensed particles,
\be
\label{2.42}
\psi_1(\br) \equiv \sum_{k\neq k_0} a_k \vp_k(\br) \; .
\ee

From the {\it quantum-number conservation condition} (\ref{2.38}) it follows
that
\be
\label{2.43}
< \psi_0^\dgr(\br)\psi_1(\br') > \; = \; 0 \; ,
\ee
since
\be
\label{2.44}
< a_0^\dgr a_k > \; = \; 0 \qquad (k \neq k_0) \; .
\ee
And, because of the orthonormality of the basis $\{\vp_k(\br)\}$, we have the
{\it orthogonality condition}
\be
\label{2.45}
\int \psi_0^\dgr (\br) \psi_1(\br) \; d\br = 0 \; .
\ee
The density matrix (\ref{2.31}) takes the form
\be
\label{2.46}
\rho(\br,\br') \; = \; < \psi_0^\dgr(\br')\psi_0(\br) > +
< \psi_1^\dgr(\br')\psi_1(\br) > \; .
\ee

The number-of-particle operators are: for condensed particles,
\be
\label{2.47}
\hat N_0 [ \psi] \equiv \int \psi_0^\dgr(\br) \psi_0(\br)\; d\br
= a_0^\dgr a_0 \; ,
\ee
and for uncondensed particles
\be
\label{2.48}
\hat N_1 [\psi] \equiv \int \psi_1^\dgr(\br) \psi_1(\br)\; d\br =
\sum_{k\neq k_0} a_k^\dgr a_k \; .
\ee
In view of the orthogonality condition (\ref{2.45}), the number-of-particle
operator for the total number of particles is
\be
\label{2.49}
\hat N[\psi] = \hat N_0 [\psi] + \hat N_1 [\psi] \; .
\ee
The average number of particles in BEC is
\be
\label{2.50}
N_0 \; = \; < \hat N_0[\psi]> \; = \; < a_0^\dgr a_0> \; .
\ee
And the number of uncondensed particles is
\be
\label{2.51}
N_1 \; =\; < \hat N_1[\psi]> \; = \; \sum_{k\neq k_0} n_k \; .
\ee
The above equations are valid for any system, whether uniform or not uniform,
and for BEC of arbitrary nature, related to a quantum state $k_0$.

It is important to stress that operators (\ref{2.41}) and (\ref{2.42}) are not
separate independent operators, describing different particles, but $\psi_0(\br)$
and $\psi_1(\br)$ are simply two parts of one Bose field operator (\ref{2.40}).
This is evident from the commutation relations for $\psi_0(\br)$,
\be
\label{2.52}
\left [ \psi_0(\br), \; \psi_0^\dgr(\br') \right ] =
\vp_0(\br) \vp_0^*(\br') \; ,
\ee
and for $\psi_1(\br)$,
\be
\label{2.53}
\left [ \psi_1(\br), \; \psi_1^\dgr(\br') \right ] =
\sum_{k\neq k_0} \vp_k(\br) \vp_k^*(\br') \; ,
\ee
which show that neither $\psi_0(\br)$ nor $\psi_1(\br)$ characterize Bose
particles. There exists the sole field operator (\ref{2.40}) enjoying the
Bose commutation relations. This operator is defined on the Fock space
$\cF(\psi)$ generated by $\psi^\dgr$.

Using notation (\ref{2.47}), the BEC criterion (\ref{2.3}) can be written as
\be
\label{2.54}
\lim_{N\ra\infty} \; \frac{<\hat N_0[\psi]>}{N} \; > \; 0 \; ,
\ee
or, equivalently, as
\be
\label{2.55}
\lim_{N\ra\infty} \; \frac{<a_0^\dgr a_0>}{N} \; > \; 0 \; .
\ee

Calculating the averages, one employs the statistical ensemble with the grand
Hamiltonian (\ref{2.30}) containing the chemical potential $\mu$, which is
the Lagrange multiplier guaranteeing the normalization condition (\ref{2.25}).
There is here the sole normalization condition, since there exists only one
field operator $\psi(\br)$ describing Bose particles.

\subsection{Gauge Symmetry}

Phase transitions from a disordered phase to an ordered phase are usually
accompanied by some symmetry breaking [58]. BEC is associated with the
global $U(1)$ gauge symmetry breaking. The fundamental question is whether
the gauge symmetry breaking is necessary and sufficient for the occurrence
of BEC. In literature, one can meet controversial statements, some claiming
that BEC does not require any symmetry breaking. This, however, is not correct.
The {\it gauge symmetry breaking is necessary and sufficient for the occurrence
of BEC}.

The equivalence of BEC and gauge symmetry breaking has been discussed in
recent papers [59--62] and thoroughly explained in the review article [63].
Considering these phenomena, one should always keep in mind that, in finite
systems, there are neither rigorously defined phase transitions nor symmetry
breaking. Both of them can happen only in thermodynamic limit. So, the
existence or absence of these phenomena acquires a correct mathematical
meaning only under the thermodynamic limiting test, either in form (\ref{2.2})
or in forms (\ref{2.4}) and (\ref{2.5}). However, one often talks about BEC
or symmetry breaking even in the case of a finite, but large, statistical
system, with $N\gg 1$, keeping in mind that the properties of the system are
asymptotically close to those the system would possess in the thermodynamic
limit.

The $U(1)$ gauge transformation can be represented as the transformation
\be
\label{2.56}
\psi(\br) \; \longrightarrow \; \psi(\br) e^{i\al}
\ee
for the field operator, with $\al$ being a real number. The Hamiltonian
$H[\psi]$ is assumed to be invariant under the gauge transformation
(\ref{2.56}). The gauge symmetry of the system can be broken by the Bogolubov
method of infinitesimal sources [64,65], by defining
\be
\label{2.57}
H_\ep [\psi] \equiv H[\vp] + \ep \; \sqrt{\rho} \; \int
\left [ \psi_0^\dgr(\br) + \psi_0(\br) \right ] \; d\br \; ,
\ee
where $\ep$ is a real parameter and $\rho$ is the mean particle density. The
related statistical operator is
\be
\label{2.58}
\hat\rho_\ep \; \equiv \;
\frac{\exp(-\bt H_\ep[\psi])}{{\rm Tr}\exp(-\bt H_\ep[\psi])} \; ,
\ee
with the trace over $\cF(\psi)$. The operator averages are defined as
\be
\label{2.59}
< \hat A>_\ep \; \equiv \; {\rm Tr} \hat\rho_\ep \hat A \; .
\ee

According to the Bogolubov method of quasiaverages [64,65], one should, first,
take the thermodynamic limit, after which the limit $\ep\ra 0$ is to be taken.
There also exist some other methods of symmetry breaking [54,66]. In
particular, it would be admissible to make the parameter $\ep$ a function
$\ep_N$ depending on $N$, such that it would appropriately tend to zero
together with the thermodynamic limit [54,67,68]. But for the sake of clarity,
we shall use here the standard Bogolubov method [64,65] of quasiaverages.

One can say that there happens the {\it local spontaneous gauge-symmetry
breaking}, when
\be
\label{2.60}
\lim_{\ep\ra 0} \; \lim_{N\ra\infty} < \psi_0(\br)>_\ep \;
\neq \; 0 \; ,
\ee
at least for some $\br$. And the {\it global spontaneous gauge-symmetry
breaking} implies that
\be
\label{2.61}
\lim_{\ep\ra 0} \; \lim_{N\ra\infty} \; \frac{1}{N} \; \int
\left | <\psi_0(\br)>_\ep \right |^2 \; d\br \; > \; 0 \; .
\ee
Because of definition (\ref{2.41}), one has
\be
\label{2.62}
\int \left | <\psi_0(\br)>_\ep \right |^2 \; d\br =
| < a_0>_\ep|^2 \; .
\ee
Hence, inequality (\ref{2.61}) of spontaneous gauge-symmetry breaking becomes
\be
\label{2.63}
\lim_{\ep\ra 0} \; \lim_{N\ra\infty} \;
\frac{|<a_0>_\ep|^2}{N} \; > \; 0 \; .
\ee

By the Cauchy-Schwarz inequality,
$$
|<a_0>_\ep| \; \leq \; \sqrt{<a_0^\dgr a_0>_\ep }
$$
for any $\ep$ and $N$. Therefore
\be
\label{2.64}
\lim_{\ep\ra 0} \; \lim_{N\ra\infty} \;
\frac{|<a_0>_\ep|^2}{N} \; \leq \; \lim_{\ep\ra 0} \;
\lim_{N\ra\infty} \; \frac{<a^\dgr_0 a_0>_\ep}{N} \; .
\ee
This tells us, that the spontaneous gauge symmetry breaking (\ref{2.63})
results in BEC, in agreement with condition (\ref{2.55}).

Moreover, inequality (\ref{2.64}) can be made the equality. Recall that
Hamiltonian (\ref{2.57}) is a functional
$$
H_\ep[\psi] \equiv H_\ep [\psi_0, \; \psi_1]
$$
of $\psi_0(\br)$ and $\psi_1(\br)$. And let us introduce the statistical
operator
\be
\label{2.65}
\hat\rho_{\eta\ep} \equiv
\frac{\exp(-\bt H_\ep[\eta,\psi_1])}
{{\rm Tr}\exp(-\bt H_\ep[\eta,\psi_1])} \; ,
\ee
in which the operator $\psi_0(\br)$ is replaced by a function $\eta(\br)$.
The related average of an operator $\hat A$ is
\be
\label{2.66}
< \hat A>_{\eta\ep} \; \equiv \;
{\rm Tr}\hat\rho_{\eta\ep} \hat A \; .
\ee
Let us also define a class of correlation functions given by the form
\be
\label{2.67}
C_\ep(\psi_0,\psi_1) \; \equiv \; < \ldots \psi_0^\dgr
\ldots \psi_1^\dgr \ldots \psi_0 \ldots \psi_1>_\ep \; .
\ee
Replacing here all operators $\psi_0^\dgr$ and $\psi_0(\br)$ by functions
$\eta^*(\br)$ and $\eta(\br)$, we get the class of correlation functions
\be
\label{2.68}
C_\ep(\eta,\psi_1) \; \equiv \; < \ldots \eta^* \ldots
\psi_1^\dgr \ldots \eta \ldots \psi_1>_\ep \; .
\ee
It is assumed that the function $\eta(\br)$ is normalized to the same number
of condensed particles, as $\psi_0(\br)$, such that
\be
\label{2.69}
\int < \psi_0^\dgr(\br) \psi_0(\br) >_\ep d\br =
\int |\eta(\br)|^2 \; d\br = N_0 \; .
\ee
Then the following statement holds [65].

\vskip 2mm

{\bf Bogolubov theorem}. In thermodynamic limit, the correlation functions
(\ref{2.67}) and (\ref{2.68}), under normalization condition (\ref{2.69}),
coincide with each other,
\be
\label{2.70}
\lim_{N\ra\infty} C_\ep(\psi_0,\psi_1) =
\lim_{N\ra\infty} C_\ep(\eta,\psi_1) \; ,
\ee
for any real $\ep$. In particular,
\be
\label{2.71}
\lim_{\ep\ra 0} \; \lim_{N\ra\infty} C_\ep(\psi_0,\psi_1) =
\lim_{\ep\ra 0} \; \lim_{N\ra\infty} C_\ep(\eta,\psi_1) \; .
\ee

From the Bogolubov theorem it follows that
\be
\label{2.72}
\lim_{\ep\ra 0} \; \lim_{N\ra\infty}
< \psi_0(\br) >_\ep \; = \; \eta(\br) \; .
\ee
Also, we have
\be
\label{2.73}
\lim_{\ep\ra 0} \; \lim_{N\ra\infty} < \psi_0^\dgr(\br)
\psi_1(\br')  >_\ep \; = \; \eta^*(\br) \; \lim_{\ep\ra 0} \;
\lim_{N\ra\infty} < \psi_1(\br')  >_{\eta\ep} \; .
\ee
If $\eta(\br)$ is not identically zero, then the quantum-number conservation
condition (\ref{2.43}) acquires the form
\be
\label{2.74}
\lim_{\ep\ra 0} \; \lim_{N\ra\infty} <
\psi_1(\br)  >_{\eta\ep} \; = \; 0 \; .
\ee
Hence, for the field operator (\ref{2.40}) one gets
\be
\label{2.75}
\lim_{\ep\ra 0} \; \lim_{N\ra\infty} < \psi(\br) >_\ep \; =
\; \eta(\br) \; .
\ee
The condition (\ref{2.61}) of global spontaneous gauge-symmetry breaking
becomes
\be
\label{2.76}
\lim_{N\ra\infty} \; \frac{1}{N} \; \int
|\eta(\br)|^2 \; d\br \; > \; 0 \; .
\ee
This, according to normalization (\ref{2.69}), means the existence
of BEC.

To be more precise, we notice that
$$
\lim_{\ep\ra 0} \; \lim_{N\ra\infty} \;  \frac{1}{N} \; \int
| < \psi_0(\br) >_\ep |^2 \; d\br  =
$$
\be
\label{2.77}
= \; \lim_{\ep\ra 0} \; \lim_{N\ra\infty} \; \frac{|<a_0>_\ep|^2}{N} \; = \;
\lim_{N\ra\infty} \; \frac{1}{N} \; \int |\eta(\br)|^2 \; d\br \; .
\ee
At the same time, we find
\be
\label{2.78}
\lim_{\ep\ra 0} \; \lim_{N\ra\infty} \;
\frac{<\hat N_0[\psi]>_\ep}{N} \; = \;
\lim_{\ep\ra 0} \; \lim_{N\ra\infty} \;
\frac{<a_0^\dgr a_0>_\ep}{N} \; = \;  \lim_{N\ra\infty} \;
\frac{1}{N} \; \int |\eta(\br)|^2 \; d\br \; .
\ee
Comparing the latter equations, we obtain
\be
\label{2.79}
\lim_{\ep\ra 0} \; \lim_{N\ra\infty} \;
\frac{1}{N} \; \int | < \psi_0(\br) >_\ep |^2 \; d\br \;  = \;
\lim_{\ep\ra 0} \; \lim_{N\ra\infty} \;
\frac{<\hat N_0[\psi]>_\ep}{N} \; ,
\ee
or in another form,
\be
\label{2.80}
\lim_{\ep\ra 0} \; \lim_{N\ra\infty} \;
\frac{|<a_0>_\ep|^2}{N} \; = \; \lim_{\ep\ra 0} \;
\lim_{N\ra\infty} \; \frac{<a_0^\dgr a_0>_\ep}{N} \; .
\ee
Equations (\ref{2.79}) and (\ref{2.80}) demonstrate that the spontaneous gauge
symmetry breaking leads to the existence of BEC. This conclusion holds for any
equilibrium system, whether uniform or nonuniform. In the case of uniform
systems, Eq. (\ref{2.80}) was derived in Refs. [59--62].

For uniform systems, there also exist the following theorem [60--62], first,
proved by Ginibre [69]. One considers the thermodynamic potentials
\be
\label{2.81}
\Om_\ep \equiv - T \ln {\rm Tr}\left ( -
\bt H_\ep [\psi_0,\; \psi_1]\right )
\ee
and
\be
\label{2.82}
\Om_{\eta\ep} \equiv - T \ln {\rm Tr}\left ( -
\bt H_\ep [\eta,\; \psi_1]\right ) \; ,
\ee
where $\eta$ is the minimizer of Eq. (\ref{2.82}), such that
\be
\label{2.83}
\Om_{\eta\ep} = {\rm inf}_x \; \Om_{x\ep} \; .
\ee

\vskip 2mm

{\bf Ginibre theorem}. For the thermodynamic potentials (\ref{2.81}) and
(\ref{2.82}), under condition (\ref{2.83}), in the thermodynamic limit, one
has
\be
\label{2.84}
\lim_{N\ra\infty} \; \frac{\Om_\ep}{N} =
\lim_{N\ra\infty} \; \frac{\Om_{\eta\ep}}{N}
\ee
for any real $\ep$, including $\ep\ra 0$.

The Bogolubov and Ginibre theorems show that the {\it spontaneous gauge
symmetry breaking is a sufficient condition for the occurrence of BEC}.

The fact that the symmetry breaking is also a necessary condition for BEC
was, first, proved by Roepstorff [70] and recently this proof was generalized
by Lieb et al. [60,62]. For this purpose, one compares the average
$<a_0^\dgr a_0>$ for a uniform system without symmetry breaking and the
average $<a_0>_\ep$ in the presence of the gauge symmetry breaking.

\vskip 2mm

{\bf Roepstorff theorem}. In the thermodynamic limit,
\be
\label{2.85}
\lim_{N\ra\infty} \; \frac{<a_0^\dgr a_0>}{N} \; \leq \;
\lim_{\ep\ra 0} \; \lim_{N\ra\infty} \; \frac{|<a_0>_\ep|^2}{N} \; .
\ee

This theorem shows that the occurrence of BEC necessarily leads to the gauge
symmetry breaking. More details on the relation between BEC and gauge symmetry
breaking can be found in the review article [63].

Thus, the conclusion is:

\vskip 2mm

{\it The spontaneous gauge-symmetry breaking is the necessary and sufficient
condition for Bose-Einstein condensation}.

\subsection{Bogolubov Shift}

Describing a system with BEC, one can follow the procedure of the previous
sections, working with the field operator (\ref{2.40}) defined on the Fock
space $\cF(\psi)$. This operator can formally be partitioned into two terms.
However, neither of these terms represents particles, since the commutation
relations (\ref{2.52}) and (\ref{2.53}) are not of Bose type. Dealing with
the field operator (\ref{2.40}), one should accomplish calculations in finite
space, passing after this to the thermodynamic limit. Such an approach has
three weak points.

First, in practical calculations, it requires the use of perturbation theory
with respect to atomic interactions, as has been done by Belyaev [71]. Hence,
it is limited to weakly interacting Bose gases.

Second, the mentioned perturbation theory is singular, being plagued by
divergences. So that only the lowest orders of the perturbation theory are
meaningful.

Third, the commutation relations (\ref{2.52}) and (\ref{2.53}) are cumbersome
and not convenient in calculations.

But we know from the theorems of the previous section that in the
thermodynamic limit the operator term $\psi_0(\br)$ reduces to a function
$\eta(\br)$. Therefore, it is tempting to replace, from the very beginning,
the operator $\psi(\br)$ in Eq. (\ref{2.40}) by another operator
\be
\label{2.86}
\hat\psi(\br) \equiv \eta(\br) + \psi_1(\br) \; ,
\ee
in which the operator $\psi_0(\br)$ has been replaced by a function
$\eta(\br)$. The procedure of replacing $\psi(\br)$ by $\hat\psi(\br)$ is
called the {\it Bogolubov shift} [64,65,72,73].

From the requirement that $\hat\psi(\br)$ is a Bose operator, and because
$\eta(\br)$ is a nonoperator function, it follows that $\psi_1(\br)$ is a
Bose field operator, with the standard Bose commutation relations
\be
\label{2.87}
\left [ \psi_1(\br), \; \psi_1^\dgr(\br') \right ] =
\dlt(\br-\br') \; .
\ee
It is evident that to deal with these usual nice commutation relations is
much simpler than with the awkward relations (\ref{2.52}) and (\ref{2.53}).

In the shifted field operator (\ref{2.86}), the term $\eta(\br)$ is named
the {\it condensate wave function}, and the term $\psi_1(\br)$ is the operator
of uncondensed particles. To be correctly defined, these terms are assumed to
preserve the basic properties typical of $\psi_0(\br)$ and $\psi_1(\br)$.
Thus, the orthogonality condition (\ref{2.45}) now reads as
\be
\label{2.88}
\int \eta^*(\br) \psi_1(\br) \; d\br = 0 \; .
\ee
And the quantum-number conservation condition (\ref{2.43}), similarly to Eq.
(\ref{2.74}), now becomes
\be
\label{2.89}
< \psi_1(\br) > \; = \; 0 \; .
\ee
The number of condensed atoms (\ref{2.47}) can be represented as
\be
\label{2.90}
\hat N_0 = N_0 \hat 1_\cF \; ,
\ee
where, $\hat 1_\cF$ is the unity operator in the appropriate Fock space and
$N_0$, in agreement with normalization (\ref{2.69}), is
\be
\label{2.91}
N_0 = \int |\eta(\br)|^2 \; d\br \; .
\ee
By analogy with Eq. (\ref{2.50}), we have
\be
\label{2.92}
N_0 \; = \; < \hat N_0 > \; .
\ee
The number operator of noncondensed atoms (\ref{2.48}) is
$$
\hat N_1 = \int \psi_1^\dgr(\br) \psi_1(\br) \; d\br \; .
$$
So that the number (\ref{2.51}) of uncondensed atoms is
\be
\label{2.93}
N_1 \; = \; < \hat N_1 > \; .
\ee
Due to the orthogonality condition (\ref{2.88}), the operator of the total
number of particles is the sum
\be
\label{2.94}
\hat N \equiv \int \hat\psi(\br) \hat\psi(\br) \; d\br =
\hat N_0 + \hat N_1 \; .
\ee
Hence, the total number of particles is
$$
N \; = \; < \hat N > \; = \; N_0 + N_1 \; .
$$

The convenience of the Bogolubov shift (\ref{2.86}) is also in the fact that
it explicitly breaks the gauge symmetry since
\be
\label{2.95}
< \hat\psi(\br) > \; = \; \eta(\br) \; ,
\ee
which is necessary for correctly describing BEC. The latter equation makes
it possible to call the condensate wave function $\eta(\br)$ the system order
parameter.

The Bogolubov shift (\ref{2.86}) is the basis for the majority of calculations
for weakly nonideal Bose gases at low temperatures, when almost all
particles are in BEC [64,65,72,73]. Perturbation theory for asymptotically
weak interactions and low temperatures has been developed for uniform [74,75]
as well as for nonuniform [76,77] gases.

However, as soon as one tries to describe not asymptotically weak interactions
or higher temperatures, one encounters the Hohenberg-Martin dilemma [74].
Hohenberg and Martin [74] showed that the theory, based on the standard
grand canonical ensemble, where the gauge symmetry is broken by means of
the Bogolubov shift, is internally inconsistent. Depending on the way of
calculations, one gets either an unphysical gap in the spectrum of collective
excitations, or local conservation laws, together with general thermodynamic
relations, become invalid. Recall that the excitation spectrum, according to
the Hugenholtz-Pines theorem, must be gapless [65,78]. While conserving
approximations [79] usually give a gap in the spectrum [80--82].

The standard attempts to cure the problem are based on what Bogolubov [65]
named "the mismatch of approximations". One either arbitrarily adds some
phenomenological terms or removes, without justification, other terms. The
most popular trick is the omission of anomalous averages, as first, was
suggested by Shohno [83] and analysed by Reatto and Straley [84]. In recent
years, the Shohno trick [83] is often ascribed to Popov, although, as is
easy to infer from the Popov works [85--88], cited in this regard, he has
never suggested or used such an unjustified trick.

Of course, all phenomenological attempts, involving the mismatch of
approximations, as has already been mentioned by Bogolubov [65], cannot
cure the problem. Not self-consistent approaches render the system unstable,
spoil thermodynamic relations, and disrupt the Bose-Einstein condensation
phase transition from the second-order to the incorrect first-order transition
[89--92]. A detailed analysis of this problem has been done in Ref. [57].

\subsection{Self-Consistent Approach}

The origin of the Hohenberg-Martin dilemma is in the use of a
nonrepresentative ensemble for a Bose-condensed system with the gauge
symmetry broken by the Bogolubov shift (\ref{2.86}). Using a representative
ensemble [54,55] cures the problem [56,57] and makes it possible to develop
a fully self-consistent theory, free of paradoxes [93--95]. This theory is
conserving and gapless by construction, independently of the involved
approximation [93--98].

To determine a representative ensemble, one should start with the
specification of the space of microstates, which depends on the choice of the
accepted variables. When one works with the field operator $\psi(\br)$, as in
Secs. 2.4. and 2.5, the corresponding space of microstates is the Fock space
$\cF(\psi)$ generated by $\psi^\dgr(\br)$, as is explained in Ref. [55]. But,
as soon as the Bogolubov shift (\ref{2.86}) has been accomplished, the new
field operator $\hat\psi(\br)$ is defined on another space, which is the
Fock space $\cF(\psi_1)$ generated by $\psi_1^\dgr(\br)$, with the spaces
$\cF(\psi)$ and $\cF(\psi_1)$ being mutually orthogonal [99]. In the space
$\cF(\psi)$, there was the sole field variable $\psi(\br)$, while in the
space $\cF(\psi_1)$ there are now two variables, the condensate wave function
$\eta(\br)$ and the field operator of uncondensed particles $\psi_1(\br)$.
Respectively, instead of one normalization condition (\ref{2.25}), there are
two normalization conditions (\ref{2.92}) and (\ref{2.93}).

The conservation-number condition (\ref{2.89}) can be reduced to the standard
form of the statistical conditions (\ref{2.27}) by defining the Hermitian
operator
\be
\label{2.96}
\hat\Lbd \equiv \int \left [ \lbd(\br) \psi_1^\dgr(\br) +
\lbd^*(\br)\psi_1(\br) \right ]\; d\br \; ,
\ee
which can be called the {\it linear killer}. This is because the Lagrange
multiplier $\lbd(\br)$ has to be chosen so that
\be
\label{2.97}
< \hat\Lbd > \; = \; 0 \; ,
\ee
which requires the absence of the terms linear in $\psi_1(\br)$ in the related
grand Hamiltonian [57].

Statistical averages of operators from the algebra of observables
${\cal O}\equiv\{\hat A(t)\}$ are given as
\be
\label{2.98}
< \hat A(t) > \; \equiv \; {\rm Tr}_{\cF(\psi_1)}\; \hat\rho
\hat A(t) \; ,
\ee
with a statistical operator $\hat\rho\equiv\hat\rho(0)$. For instance, the
energy Hamiltonian
$$
\hat H = \hat H[\hat\psi] \equiv \hat H [\eta,\; \psi_1]
$$
defines the internal energy
\be
\label{2.99}
E\; = \; < \hat H> \; .
\ee
In what follows, we shall omit the notation of spaces, over which the trace
is taken, in order to avoid cumbersome expressions.

The statistical operator $\hat\rho$ is obtained from the minimization of
the information functional $I[\hat\rho]$ under the statistical conditions
(\ref{2.23}), (\ref{2.99}), (\ref{2.92}), (\ref{2.93}), and (\ref{2.97}).
The information functional is
$$
I[\hat\rho] = {\rm Tr}\hat\rho \; \ln\hat\rho +
\lbd_0 \left ( {\rm Tr}\hat\rho \; - \; 1 \right ) +
$$
\be
\label{2.100}
+ \bt \left ( {\rm Tr}\hat\rho \hat H \; - \; E \right ) -
\bt \mu_0 \left ( {\rm Tr}\hat\rho \hat N_0 \; - \; N_0 \right )
- \bt \mu_1 \left ( {\rm Tr}\hat\rho \hat N_1 \; - \; N_1 \right )
- \bt\; {\rm Tr}\hat\rho \hat\Lbd \; .
\ee
Its minimization yields the statistical operator
\be
\label{2.101}
\hat\rho = \frac{\exp(-\bt H)}{{\rm Tr}\exp(-\bt H)} \; ,
\ee
with the grand Hamiltonian
\be
\label{2.102}
H \equiv \hat H - \mu_0 \hat N_0 - \mu_1 \hat N_1 -
\hat\Lbd \; .
\ee
This Hamiltonian is, clearly, a functional $H=H[\eta,\psi_1]$ of $\eta(\br)$
and $\psi_1(\br)$.

The evolution laws are prescribed by extremizing the action functional, as
is described in Sec. 2.4. To this end, we define the temporal energy operator
\be
\label{2.103}
\hat E \equiv \int \hat\psi^\dgr(\br) i \;
\frac{\prt}{\prt t} \; \hat\psi(\br) \; d\br \; .
\ee
With the Bogolubov shift (\ref{2.86}), this takes the form
\be
\label{2.104}
\hat E = \int \left [ \eta^*(\br) i\;
\frac{\prt}{\prt t} \; \eta(\br) + \psi_1^\dgr(\br) i\;
\frac{\prt}{\prt t}\; \psi_1(\br) \right ] \; d\br \; ,
\ee
which shows that $\hat E=\hat E[\eta,\psi_1]$ is a functional of $\eta(\br)$
and $\psi_1(\br)$. The effective action is also a functional of these
variables,
\be
\label{2.105}
\Gm[\eta,\; \psi_1] \equiv \int \left ( \hat E \; - \; H
\right ) \; dt \; .
\ee
The evolution laws are given by the extremization of the action
functional (\ref{2.105}) with respect to the condensate wave function,
\be
\label{2.106}
\frac{\dlt\Gm[\eta,\psi_1]}{\dlt\eta^*(\br,t)} = 0 \; ,
\ee
and with respect to the field operator of uncondensed particles,
\be
\label{2.107}
\frac{\dlt\Gm[\eta,\psi_1]}{\dlt\psi_1^\dgr(\br,t)} = 0 \; .
\ee
These evolution equations, owing to the form of the action functional
(\ref{2.105}), are equivalent to the equation
\be
\label{2.108}
i\; \frac{\prt}{\prt t} \; \eta(\br,t) =
\frac{\dlt H[\eta,\psi_1]}{\dlt\eta^*(\br,t)}
\ee
for the condensate variable, and to the equation
\be
\label{2.109}
i\; \frac{\prt}{\prt t} \; \psi_1(\br,t) =
\frac{\dlt H[\eta,\psi_1]}{\dlt\psi_1^\dgr(\br,t)}
\ee
for the field variable of uncondensed particles.

Thus, the representative statistical ensemble for a Bose system with the
gauge symmetry breaking, induced by the Bogolubov shift (\ref{2.86}), is
the triplet $\{\cF(\psi_1),\hat\rho,\prt t\}$ formed by the Fock space
$\cF(\psi_1)$, generated by $\psi_1^\dgr(\br)$, the statistical operator
(\ref{2.101}), with the grand Hamiltonian (\ref{2.102}), and the evolution
laws (\ref{2.108}) and (\ref{2.109}).

In the case of an equilibrium system, we can introduce the grand thermodynamic
potential
\be
\label{2.110}
\Om = \; - T \ln{\rm Tr}\; e^{-\bt H} \; ,
\ee
defining all thermodynamics of the system. For example, the fraction of
condensed atoms is
\be
\label{2.111}
n_0 \equiv \frac{N_0}{N} = \; - \; \frac{1}{N} \;
\frac{\prt\Om}{\prt\mu_0} \; ,
\ee
and the fraction of uncondensed atoms is
\be
\label{2.112}
n_1 \equiv \frac{N_1}{N} = \; - \; \frac{1}{N} \;
\frac{\prt\Om}{\prt\mu_1} \; .
\ee
The equation for the condensate function is obtained from the statistical
averaging of Eq. (\ref{2.108}), under the condition that, for an equilibrium
system, $\eta(\br)$ does not depend on time. Then we get the equation
\be
\label{2.113}
\frac{\dlt\Om}{\dlt\eta^*(\br)} \; = \; <
\frac{\dlt H[\eta,\psi_1]}{\dlt\eta^*(\br)} > \; =
\; 0 \; ,
\ee
which is equivalent to the Bogolubov minimization of the thermodynamic
potential with respect to the condensate variable [64,65,72,73].

The free energy can be defined as
\be
\label{2.114}
F = \Om + \mu_0 N_0 + \mu_1 N_1 \; .
\ee
At the same time, keeping in mind the standard form of the free energy
\be
\label{2.115}
F = \Om + \mu N \; ,
\ee
we find the expression for the system chemical potential
\be
\label{2.116}
\mu = \mu_0 n_0 + \mu_1 n_1 \; .
\ee
The same form for the chemical potential (\ref{2.116}) can be derived from
the usual definition
\be
\label{2.117}
\mu \equiv \frac{\prt F}{\prt N} \; .
\ee
The right-hand side here can be written as
$$
\frac{\prt F}{\prt N} = \frac{\prt F}{\prt N_0} \;
\frac{\prt N_0}{\prt N} \; + \; \frac{\prt F}{\prt N_1} \;
\frac{\prt N_1}{\prt N} \; ,
$$
where
$$
\frac{\prt F}{\prt N_0} = \mu_0 \; , \qquad
\frac{\prt F}{\prt N_1} = \mu_1 \; .
$$
Assuming that $n_0$ and $n_1$ are fixed in the thermodynamic limit
$N\ra\infty$, from the relations
$$
N_0 = n_0 N \; , \qquad N_1 = n_1 N \; ,
$$
we have
$$
\frac{\prt N_0}{\prt N} =  n_0 \; , \qquad
\frac{\prt N_1}{\prt N} =  n_1 \; .
$$
Combining these derivatives in definition (\ref{2.117}), we get the same
expression (\ref{2.116}) for the chemical potential.

It is possible to show [57] that the dispersion of the number-of-particle
operator is given by
\be
\label{2.118}
\Dlt^2(\hat N) = T \; \frac{\prt N}{\prt\mu} \; ,
\ee
where the dispersion of a self-adjoint operator $\hat A$ is defined as
$$
\Dlt^2(\hat A) \; \equiv \; < \hat A^2> -
< \hat A>^2 \; .
$$

At the end, the free energy can be represented as a function $F=F(T,V,N)$ of
temperature $T$, volume $V$, and the particle number $N$, with the differential
\be
\label{2.119}
dF = - S\; dT - P\; dV + \mu \; dN \; ,
\ee
in which $S$ is entropy and $P$, pressure. And the grand potential
(\ref{2.110}) is a function $\Om=\Om(T,V,\mu)$, with the differential
\be
\label{2.120}
d\Om = - S\; dT - P \; dV - N \; d\mu \; .
\ee
All thermodynamics follows from the above expressions. A more detailed
discussion is given in Ref. [57].

\subsection{Condensate Existence}

The existence of BEC, as such, requires the validity of an important necessary
condition. Generally, the total number of atoms $N$ is the sum $\sum_k n_k$ of
the occupation numbers for quantum states labelled by a multi-index $k$. The
occurrence of BEC, by definition, means the microscopic occupation of a single
quantum state $k_0$, when $N_0\propto N$, in agreement with condition
(\ref{2.3}). Only in such a case, it is meaningful to separate out of the sum
$\sum_k n_k$ a single term, related to BEC, obtaining
\be
\label{2.121}
N = N_0 + \sum_{k\neq k_0} n_k \; .
\ee
Mathematically, the possibility of that separation necessarily implies that,
in thermodynamic limit, the distribution $n_k$ over quantum states diverges
when $k\ra k_0$. Hence, the {\it necessary condition for the BEC existence}
is
\be
\label{2.122}
\lim_{N\ra\infty}\; \lim_{k\ra k_0} \; \frac{1}{n_k} = 0 \; ,
\ee
where the thermodynamic limit is invoked in order to make the BEC rigorously
defined.

In the representative ensemble of the previous Sec. 2.8, the condensate wave
function can be written as
\be
\label{2.123}
\eta(\br) = \sqrt{N_0}\; \vp_0(\br) \; ,
\ee
while the operator of uncondensed atoms can be expanded over the natural
orbitals as
\be
\label{2.124}
\psi_1(\br) = \sum_{k\neq k_0} a_k \vp_k(\br) \; .
\ee
Hence, the occupation numbers are
\be
\label{2.125}
n_k \; \equiv \; < a_k^\dgr a_k > \; ,
\ee
with the statistical averaging defined in Eq. (\ref{2.98}).

If we turn to the terminology of Green functions, then the necessary
condition (\ref{2.122}) can be connected with the properties of the poles
of Green functions. Under the spontaneously broken gauge symmetry, the poles
of the first-order and second-order Green functions coincide, that is, the
single-particle spectrum coincides with the spectrum of collective excitations
$\ep_k$ [65,100]. For the latter, the {\it necessary condition of condensate
existence} (\ref{2.122}) translates into
\be
\label{2.126}
\lim_{k\ra k_0} \ep_k = 0 \; ,
\ee
with the condition $n_k\geq 0$ becoming the stability condition
\be
\label{2.127}
{\rm Re} \; \ep_k \geq 0 \; , \qquad {\rm Im}\; \ep_k \leq 0 \; .
\ee
Conditions (\ref{2.122}) or (\ref{2.126}) impose a constraint on the Lagrange
multiplier $\mu_1$, which has to be such that to make the spectrum $\ep_k$
gapless in the sense of the limit (\ref{2.126}).

To illustrate the properties (\ref{2.122}) and (\ref{2.126}), let us consider
a uniform system, when $k_0=0$. For a Bose-condensed uniform system, Bogolubov
[65] rigorously proved the inequalities for the occupation numbers
(\ref{2.125}) in the case of nonzero temperature,
\be
\label{2.128}
n_k \; \geq \; \frac{mn_0T}{2k^2} \; - \; \frac{1}{2} \; ,
\ee
and at zero temperature,
\be
\label{2.129}
n_k \; \geq \; \frac{mn_0\ep_k}{4k^2} \; - \; \frac{1}{2} \qquad
(T=0) \; ,
\ee
where $\ep_k$ is the real part of the spectrum of collective excitations.
These inequalities can be slightly improved [101], resulting, for finite
temperatures, in
\be
\label{2.130}
n_k \; \geq \; \frac{mn_0T}{k^2} \; - \; \frac{1}{2}
\ee
and for zero temperature, in
\be
\label{2.131}
n_k \; \geq \; \frac{mn_0\ep_k}{2k^2} \; - \; \frac{1}{2}
\qquad (T=0)  \; .
\ee
At zero temperature, one can use the Feynman relation [78,102,103]
\be
\label{2.132}
\ep_k = \frac{k^2}{2mS(\bk)} \qquad (T=0) \; ,
\ee
in which $S(\bk)$ is the structure factor. Then the Bogolubov inequality
(\ref{2.131}) takes the form
\be
\label{2.133}
n_k \; \geq \; \frac{n_0}{4S(\bk)} \; - \; \frac{1}{2}
\qquad (T=0) \; .
\ee
At zero temperature, the structure factor possesses the long-wave limit
[104] as
\be
\label{2.134}
S(\bk) \simeq \frac{k}{2mc} \qquad (T=0, \; k\ra 0) \; .
\ee
Hence, from the Feynman relation (\ref{2.132}), one has
\be
\label{2.135}
\ep_k \simeq ck \qquad (k\ra 0) \; .
\ee
The same long-wave limit exists for the spectrum of collective excitations
at finite temperatures [65]. That is, limit (\ref{2.126}) is valid for any
$T$. From the above inequalities, it follows that, for finite temperatures,
$$
n_k \; \geq \; \frac{mn_0T}{k^2} \qquad (k\ra 0) \; ,
$$
and for zero temperature,
$$
n_k \; \geq \; \frac{mn_0c}{2k} \qquad (T=0, \; k\ra 0) \; .
$$
Therefore, in any case, the BEC existence condition (\ref{2.122}) holds true.

Condition (\ref{2.122}) can also be generalized for nonequilibrium nonuniform
systems. But the latter should, at least, be locally equilibrium in order
that the meaning of thermodynamic phases be locally preserved. Then for
$\psi_1(\br,t)$ one has the same expansion (\ref{2.124}), but with $a_k(t)$
being a function of time. The occupation number (\ref{2.125}) becomes a
function of time, $n_k=n_k(t)$. Then condition (\ref{2.122}) defines the
Lagrange multiplier $\mu_1$ as a function of time $\mu_1(t)$. If one employs
the local-density approximation, then $\mu_1$ can also be a function of the
spatial variable.

\subsection{Superfluid Fraction}

Expressions, defining the superfluid fraction, can have different forms for
uniform and nonuniform systems. It is, therefore, important to recall the
most general definition of the superfluid fraction, which could be applied
to arbitrary systems, whether uniform or not. This general definition is
based on the calculation of the response to a velocity boost imposed on
the system.

Let $\hat H$ be the energy Hamiltonian of an immovable system, and let
$\hat H_v$ be the energy Hamiltonian of the system moving, as a whole, with
velocity $\bv$. The statistical operator, related to the moving system is
denoted as $\hat\rho_v$. The corresponding statistical average of an operator
$\hat A$ is
\be
\label{2.136}
< \hat A>_v \; \equiv \; {\rm Tr}\hat \rho_v \hat A \; .
\ee
The return to the immovable system is realized through the limit
\be
\label{2.137}
< \hat A > \; = \; \lim_{v\ra 0} < \hat A>_v \; .
\ee
The momentum operator of the total moving system can be represented as
\be
\label{2.138}
\hat\bP_v \equiv \frac{\prt\hat H_v}{\prt\bv} \; .
\ee
The momentum operator of the immovable system is
\be
\label{2.139}
\hat\bP = \lim_{v\ra 0} \hat\bP_v \; .
\ee
The superfluid fraction can be defined as a fraction of particles nontrivially
responding to the velocity boost,
\be
\label{2.140}
n_s \equiv \frac{1}{3mN} \; \lim_{v\ra 0} \;
\frac{\prt}{\prt\bv} \cdot < \hat\bP_v>_v \; .
\ee
This is the most general definition, valid for arbitrary systems [10,57].

For an equilibrium Bose system, with the grand Hamiltonian
\be
\label{2.141}
H_v \equiv \hat H_v - \mu_0 \hat N_0 - \mu_1 \hat N_1  -
\hat\Lbd \; ,
\ee
which differs from Eq. (\ref{2.102}) by the velocity boosted term $\hat H_v$,
the statistical operator is
\be
\label{2.142}
\hat\rho_v \equiv
\frac{\exp(-\bt H_v)}{{\rm Tr}\exp(-\bt H_v)} \; .
\ee
The differentiation in Eq. (\ref{2.140}) is accomplished according to the
rule of differentiation with respect to parameters [105], which gives
$$
\frac{\prt}{\prt\bv} \cdot < \hat\bP_v >_v \; = \;
< \frac{\prt}{\prt\bv} \cdot \hat\bP_v >_v - \;
\bt {\rm cov} \left ( \hat\bP_v, \frac{\prt H_v}{\prt\bv}
\right ) \; .
$$
Here the last term is the covariance defined as
$$
{\rm cov} \left ( \hat A,\hat B\right ) \equiv \frac{1}{2}
< \hat A \hat B + \hat B \hat A>_v -
< \hat A >_v < \hat B >_v \; .
$$
Using definition (\ref{2.138}), we obtain
\be
\label{2.143}
n_s = \frac{1}{3mN} \left [ \lim_{v\ra 0} <
\frac{\prt}{\prt\bv} \cdot \hat\bP_v > - \; \bt \Dlt^2(\hat\bP)
\right ] \; ,
\ee
where
$$
\Dlt^2(\hat\bP) \equiv \; < \hat\bP^2 > -  < \hat\bP >^2 \; .
$$

The same form (\ref{2.143}) can be derived from the definition [106,107] of
the superfluid fraction
\be
\label{2.144}
n_s \equiv \; \frac{1}{3mN} \; \lim_{v\ra 0} \;
\frac{\prt^2\Om_v}{\prt\bv^2} \; =\; \frac{1}{3mN} \; \lim_{v\ra 0} \;
\frac{\prt^2 F_v}{\prt\bv^2} \; ,
\ee
applicable to equilibrium systems. Here the grand potential for the moving
system is
\be
\label{2.145}
\Om_v \equiv - T\ln {\rm Tr} \exp ( - \bt H_v ) \; ,
\ee
and the free energy for that system is
\be
\label{2.146}
F_v \equiv - T \ln {\rm Tr} \exp ( - \bt \hat H_v ) \; .
\ee
The first derivative gives
\be
\label{2.147}
\frac{\prt F_v}{\prt\bv} = \frac{\prt\Om_v}{\prt\bv} \; = \;
< \bP_v>_v \; .
\ee
For an equilibrium system at rest, one has
\be
\label{2.148}
\Dlt^2(\hat\bP) \; = \; < \hat\bP^2> \qquad
\left ( < \hat\bP>\; = \; 0 \right ) \; .
\ee
Then, one gets
\be
\label{2.149}
n_s = \frac{1}{3mN} \left ( \lim_{v\ra 0}
< \frac{\prt}{\prt\bv} \cdot \hat\bP_v > - \;
\bt < \hat\bP^2> \right ) \; .
\ee
Taking into account Eq. (\ref{2.148}), we see that Eqs. (\ref{2.143}) and
(\ref{2.149}) coincide.

To specify the expression for the superfluid fraction, we can use the
definition for the operator of momentum
\be
\label{2.150}
\hat\bP_v \equiv \int \hat\psi_v^\dgr(\br,t) ( -i\nabla)
\hat\psi_v(\br,t) \; d\br
\ee
of the moving system. The field operator of the moving system can be
expressed through the field operator of the system at rest by means of the
Galilean transformation
\be
\label{2.151}
\hat\psi_v(\br,t) = \hat\psi(\br-\bv t,t)\; \exp\left \{ i
\left ( m\bv \cdot\br \; - \; \frac{mv^2}{2}\; t \right )
\right \} \; .
\ee
Consequently,
\be
\label{2.152}
\hat\bP_v = \int \hat\psi^\dgr(\br) \left ( -i\nabla + m\bv
\right ) \hat\psi(\br) \; d\br \; .
\ee
The energy Hamiltonian of the system at rest is $\hat H =\hat H[\hat\psi]$,
while that of the moving system is $\hat H_v =\hat H[\hat\psi_v]$.
Differentiating Eq. (\ref{2.152}) gives
$$
\frac{\prt}{\prt\bv} \cdot \hat\bP_v = 3m\hat N \; .
$$
We can define the {\it dissipated heat} of the considered quantum system as
\be
\label{2.153}
Q \equiv \frac{\Dlt^2(\hat\bP)}{2mN} =
\frac{<\hat\bP^2>}{2mN} \; ,
\ee
which is to be compared with the heat dissipated in a classical system,
\be
\label{2.154}
Q_0 \equiv \frac{3}{2}\; T \; .
\ee
Finally, the superfluid fraction (\ref{2.149}) reduces to the form
\be
\label{2.155}
n_s = 1 \; - \; \frac{Q}{Q_0} \; .
\ee
This formula is valid for arbitrary nonuniform equilibrium systems, including
periodic lattice potentials.

The superfluid fraction, as is known, is not directly related to the
condensate fraction. A straightforward example is the liquid $^4$He, which at
low temperature $T\ra 0$ is practically $100\%$ superfluid, while possessing
only about $10\%$ of BEC [108]. Superfluidity happens in two-dimensional
uniform systems, where BEC cannot exist. For trapped atoms, BEC can occur
in low-dimensional systems, depending on the type of the confining potential
[41], but again with no direct relation to the superfluid fraction
[109--114]. Quasi-low-dimensional atomic systems can be prepared by producing
a tight confinement in one or two directions [115]. It is only for ideal Bose
gases below $T_c$ [116--118], when the superfluid and condensate fractions
coincide [9]. Generally, superfluidity can exist without BEC, and vice versa.

\subsection{Equations of Motion}

All observable quantities are functionals of the condensate function
$\eta(\br)$ and the operator of uncondensed particles $\psi_1(\br)$.
These variables are defined by the evolution equations (\ref{2.108}) and
(\ref{2.109}). To derive these equations explicitly, we need to specify the
Hamiltonian $\hat H=\hat H[\eta,\psi_1]$.

Let us take the energy Hamiltonian in the standard form
$$
\hat H = \int \hat\psi(\br) \left ( -\; \frac{\nabla^2}{2m} +
U \right ) \hat\psi(\br) \; d\br \; +
$$
\be
\label{2.156}
+ \; \frac{1}{2} \;
\int \hat\psi^\dgr(\br) \hat\psi^\dgr(\br') \Phi(\br-\br')
\hat\psi(\br') \hat\psi(\br) \; d\br d\br' \; ,
\ee
in which $m$ is atomic mass, $U=U(\br)$ is an external potential, and
$\Phi(\br)=\Phi(-\br)$ is the binary atomic interaction potential. After
substituting here the Bogolubov shift (\ref{2.86}), Hamiltonian (\ref{2.156})
can be rewritten as the sum of five terms, depending on the number of factors
of $\psi_1$. The same concerns the grand Hamiltonian (\ref{2.102}), for which
we obtain
\be
\label{2.157}
H = \sum_{n=0}^4 H^{(n)} \; .
\ee
Here the zero-order term is
$$
H^{(0)} = \int \eta^*(\br) \left ( - \;
\frac{\nabla^2}{2m} + U - \mu_0 \right )
\eta(\br) \; d\br \; +
$$
\be
\label{2.158}
+ \; \frac{1}{2} \; \int \Phi(\br-\br')
|\eta(\br')|^2 |\eta(\br)|^2 \; d\br d\br' \; .
\ee
To satisfy the quantum-number conservation condition (\ref{2.89}), the
Hamiltonian should not contain the terms linear in $\psi_1$ [57]. For this
purpose, the Lagrange multiplier $\lbd(\br)$ in operator (\ref{2.96}) is to
be taken such that to cancel all linear terms, resulting in
\be
\label{2.159}
H^{(1)} = 0 \; .
\ee
For the second-order term, we have
$$
H^{(2)} = \int \psi_1^\dgr(\br) \left ( -\; \frac{\nabla^2}{2m} +
U - \mu_1 \right ) \psi_1(\br) d\br \; +
$$
$$
+ \int \Phi(\br-\br') \left [ |\eta(\br)|^2
\psi_1^\dgr(\br')\psi_1(\br') +
\eta^*(\br)\eta(\br')\psi_1^\dgr(\br')\psi_1(\br) + \right.
$$
\be
\label{2.160}
+ \left.
\frac{1}{2}\; \eta^*(\br)\eta^*(\br')\psi_1(\br')\psi_1(\br) +
\frac{1}{2}\; \eta(\br) \eta(\br') \psi_1^\dgr(\br')
\psi_1^\dgr(\br) \right ]
d\br d\br' \; .
\ee
The third-order term is
\be
\label{2.161}
H^{(3)} = \int \Phi(\br-\br') \left [
\eta^*(\br) \psi_1^\dgr(\br') \psi_1(\br') \psi_1(\br) +
\psi_1^\dgr(\br) \psi_1^\dgr(\br') \psi_1(\br') \eta(\br)
\right ] \; d\br d\br' \; .
\ee
And the fourth-order term is
\be
\label{2.162}
H^{(4)} = \frac{1}{2} \; \int \psi_1^\dgr(\br) \psi_1^\dgr(\br')
\Phi(\br-\br') \psi_1(\br') \psi_1(\br) \; d\br d\br' \; .
\ee

Inserting Hamiltonian (\ref{2.157}) into the evolution equation (\ref{2.108})
yields
$$
i\; \frac{\prt}{\prt t} \; \eta(\br,t) = \left ( -\;
\frac{\nabla^2}{2m} + U - \mu_0 \right ) \eta(\br,t) \; +
$$
\be
\label{2.163}
+ \; \int \Phi(\br-\br') \left [ |\eta(\br')|^2 \eta(\br) +
\hat X(\br,\br') \right ] \; d\br' \; ,
\ee
with the correlation operator
$$
\hat X(\br,\br') \equiv \psi_1^\dgr(\br') \psi_1(\br') \eta(\br)
+ \psi_1^\dgr(\br') \eta(\br') \psi_1(\br) +
\eta^*(\br') \psi_1(\br') \psi_1(\br) +
\psi_1^\dgr(\br') \psi_1(\br') \psi_1(\br) \; .
$$
And Eq. (\ref{2.109}), with Hamiltonian (\ref{2.157}), gives
$$
i\; \frac{\prt}{\prt t} \; \psi_1(\br,t) = \left ( -\;
\frac{\nabla^2}{2m} + U - \mu_1 \right ) \psi_1(\br,t) \; +
$$
\be
\label{2.164}
+ \int \Phi(\br-\br') \left [ |\eta(\br')|^2 \psi_1(\br) +
\eta^*(\br') \eta(\br) \psi_1(\br') +
\eta(\br') \eta(\br) \psi_1^\dgr(\br') + \hat X(\br,\br')
\right ] d\br' \; .
\ee

The equation for the condensate wave function is obtained by averaging Eq.
(\ref{2.163}). For this purpose, let us define the normal density matrix
\be
\label{2.165}
\rho_1(\br,\br') \; \equiv \; < \psi_1^\dgr(\br')
\psi_1(\br) > \; .
\ee
As soon as the gauge symmetry is broken, there arises the anomalous density
matrix
\be
\label{2.166}
\sgm_1(\br,\br') \; \equiv \; < \psi_1(\br') \psi_1(\br) > \; .
\ee
The density of BEC is
\be
\label{2.167}
\rho_0(\br) \equiv |\eta(\br)|^2 \; ,
\ee
and the density of uncondensed atoms is
\be
\label{2.168}
\rho_1(\br) \equiv \rho_1(\br,\br) \; = \; < \psi_1^\dgr(\br)
\psi_1(\br) > \; .
\ee
The diagonal part of Eq. (\ref{2.166}) is the anomalous average
\be
\label{2.169}
\sgm(\br) \equiv \sgm_1(\br,\br) \; = \;
< \psi_1(\br) \psi_1(\br) > \; .
\ee
The value $|\sgm(\br)|$ has the meaning of the density of pair-correlated
particles [99]. The total density of atoms
\be
\label{2.170}
\rho(\br) = \rho_0(\br) + \rho_1(\br)
\ee
is the sum of densities (\ref{2.167}) and (\ref{2.168}). Also, we need the
notation for the triple correlator
\be
\label{2.171}
\xi(\br,\br') \; \equiv \; < \psi_1^\dgr(\br') \psi_1(\br')
\psi_1(\br) > \; .
\ee
Using the above notation, we get
$$
< \hat X(\br,\br') > \; = \; \rho_1(\br') \eta(\br) +
\rho_1(\br,\br') \eta(\br') + \sgm_1(\br,\br') \eta^*(\br') +
\xi(\br,\br') \; .
$$
Finally, averaging Eq. (\ref{2.163}), we obtain
$$
i\; \frac{\prt}{\prt t} \; \eta(\br,t) = \left ( -\;
\frac{\nabla^2}{2m} + U - \mu_0 \right ) \eta(\br,t) \; +
$$
\be
\label{2.172}
+ \int \Phi(\br-\br') \left [ \rho(\br') \eta(\br) +
\rho_1(\br,\br') \eta(\br') + \sgm_1(\br,\br') \eta^*(\br') +
\xi(\br,\br') \right ] \; d\br' \; .
\ee
This is a general equation for the condensate wave function in the case of
an arbitrary Bose system. No approximation has been involved in deriving Eq.
(\ref{2.172}).

For an equilibrium system, we have
$$
\frac{\prt}{\prt t} \; \eta(\br) = 0 \qquad (equilibrium) \; .
$$
Then Eq. (\ref{2.172}) becomes the eigenproblem
$$
\left [ - \; \frac{\nabla^2}{2m} + U(\br) \right ] \eta(\br) \; +
$$
\be
\label{2.173}
+ \int \Phi(\br-\br') \left [ \rho(\br')\eta(\br) +
\rho_1(\br,\br') \eta(\br') + \sgm_1(\br,\br') \eta^*(\br') +
\xi(\br,\br') \right ] d\br' = \mu_0 \eta(\br)
\ee
defining $\eta(\br)$ and $\mu_0$.

The above equations are valid for any interaction potential $\Phi(\br)$, with
the sole restriction that it is integrable [55], so that
$$
\left | \int \Phi(\br) \; d\br \right | \; < \; \infty \; .
$$
For dilute gases, when the interaction radius is much shorter than the mean
interatomic distance [7--14], one uses the local interaction potential
\be
\label{2.174}
\Phi(\br) = \Phi_0 \dlt(\br) \qquad \left ( \Phi_0 \equiv 4\pi \;
\frac{a_s}{m} \right ) \; ,
\ee
in which $a_s$ is the scattering length. In that case, all equations simplify.
For example, the condensate-function equation (\ref{2.173}) reads as
$$
\left [ -\; \frac{\nabla^2}{2m} + U(\br) \right ] \eta(\br) +
$$
\be
\label{2.175}
+ \Phi_0 \left \{ [ \rho(\br) + \rho_1(\br) ] \eta(\br) +
\sgm_1(\br) \eta^*(\br) + \xi(\br,\br) \right \} =
\mu_0 \eta(\br) \; .
\ee
This equation is valid for any nonuniform equilibrium Bose system, which
can be treated as dilute.

Note that a dilute gas can, at the same time, be strongly interacting. Really,
the gas is dilute, when the interaction radius $r_0$ is much shorter than the
mean interatomic distance $a$, that is, $r_0\ll a$. Then the actual form of
the interaction potential is not important and this potential can be modelled
by the local expression (\ref{2.174}). The scattering length characterizes the
interaction strength, since
$$
\Phi_0 = \int \Phi(\br) \; d\br =  4\pi \; \frac{a_s}{m} \; .
$$
Nothing precludes the scattering length to be larger that the mean interatomic
distance. If $a_s>a$, then the average potential energy $\rho\Phi_0$ is larger
than the effective kinetic energy,
$$
\rho \Phi_0 \; > \; \frac{\rho^{2/3}}{2m} \qquad (a_s > a) \; ,
$$
which means that atoms strongly interact with each other. For a strongly
interacting system, atomic correlations can be rather important [10,119--122].
But for dilute gases, the influence of correlations can be taken into account
through defining an effective scattering length $a_s$.

\subsection{Uniform System}

Though this review article is devoted to nonuniform systems, it is
instructive to briefly touch the uniform case. First, this will illustrate
the self-consistency of the theory employing the representative statistical
ensemble [56,57,93--99]. Second, the theory for uniform systems can be used
for generalizing the approach to nonuniform systems by means of the
local-density approximation. Also, many formulas in the case of periodic
potentials have the structure very similar to that of expressions for the
uniform system.

In a uniform system, BEC occurs in the quantum state of zero momentum $k_0=0$.
Atomic densities do not depend on the spatial variable. Thus, the condensate
density
\be
\label{2.176}
\rho_0(\br) = \rho_0 \equiv \frac{N_0}{V} \; ,
\ee
the density of uncondensed atoms
\be
\label{2.177}
\rho_1(\br) = \rho_1 \equiv \frac{N_1}{V} \; ,
\ee
and the total density
\be
\label{2.178}
\rho(\br) = \rho =\rho_0 + \rho_1 \; ,
\ee
all are constants.

For the interaction potential $\Phi(\br)$ the Fourier transform
\be
\label{2.179}
\Phi_k = \int \Phi(\br) e^{-i\bk\cdot\br} \; d\br
\ee
is assumed to exist. Plane waves are the natural orbitals
$\vp_k(\br)= e^{i\bk\cdot\br}/\sqrt{V}$. Hence, it is convenient to use
everywhere the Fourier transforms.

The problem can be explicitly solved by using the Hartree-Fock-Bogolubov
(HFB) approximation for Bose-condensed systems [94--98]. Then the thermodynamic
potential (\ref{2.110}) becomes
\be
\label{2.180}
\Om = E_B + TV \int \ln \left ( 1 - e^{-\bt\ep_k}\right ) \;
\frac{d\bk}{(2\pi)^3} \; .
\ee
Here the first term is the nonoperator expression
\be
\label{2.181}
E_B = E_{HFB} + \frac{1}{2} \;
\sum_{k\neq 0} ( \ep_k - \om_k) \; ,
\ee
in which
$$
E_{HFB} = H^{(0)} \; - \; \frac{1}{2} \;
\rho_1^2 \Phi_0 V \; - \; \frac{1}{2V} \; \sum_{k\neq 0}
\Phi_{k+p} (n_k n_p + \sgm_k \sgm_p ) \; ,
$$
\be
\label{1.182}
H^{(0)} = - N_0 \left [ \frac{1}{2} (\rho + \rho_1) \Phi_0 +
\frac{1}{V} \; \sum_{p\neq 0} (n_p + \sgm_p) \Phi_p
\right ] \; .
\ee
The Bogolubov spectrum is
\be
\label{2.183}
\ep_k = \sqrt{\om_k^2 - \Dlt_k^2 } \; ,
\ee
where
\be
\label{2.184}
\om_k = \frac{k^2}{2m} + \rho_0 \Phi_k + \frac{1}{V} \;
\sum_{p\neq 0} \left ( n_p \Phi_{k+p} - n_p\Phi_p +
\sgm_p \Phi_p \right )
\ee
and
\be
\label{2.185}
\Dlt_k = \rho_0 \Phi_k + \frac{1}{V} \;
\sum_{p\neq 0} \sgm_p \Phi_{k+p} \; .
\ee

Equation (\ref{2.173}) gives
\be
\label{2.186}
\mu_0 = \rho \Phi_0 + \frac{1}{V} \;
\sum_{p\neq 0} (n_p + \sgm_p) \Phi_p \; .
\ee
And the condition of the BEC existence (\ref{2.122}) and (\ref{2.126}) define
\be
\label{2.187}
\mu_1 = \rho \Phi_0 + \frac{1}{V} \; \sum_{p\neq 0}
(n_p - \sgm_p) \Phi_p \; .
\ee
The distributions $n_p$ and $\sgm_p$ are the Fourier transforms of the
normal density matrix (\ref{2.165}) and of the anomalous matrix (\ref{2.166}),
respectively. For the former, we have
\be
\label{2.188}
n_k = \frac{\om_k}{2\ep_k} \; {\rm coth} \left (
\frac{\ep_k}{2T} \right ) \; - \; \frac{1}{2} \; ,
\ee
and for the latter, we find
\be
\label{2.189}
\sgm_k = -\; \frac{\Dlt_k}{2\ep_k} \; {\rm coth} \left (
\frac{\ep_k}{2T} \right ) \; .
\ee

The density of uncondensed atoms
\be
\label{2.190}
\rho_1 = \int n_k \; \frac{d\bk}{(2\pi)^3}
\ee
defines the condensate density $\rho_0=\rho-\rho_1$. Consequently, the
condensate fraction is
\be
\label{2.191}
n_0 =  1 \; - \; \frac{\rho_1}{\rho} \; .
\ee

The dissipated heat (\ref{2.153}) becomes
\be
\label{2.192}
Q = \frac{1}{\rho} \; \int \frac{k^2}{2m} \left ( n_k + n_k^2
- \sgm_k^2 \right )\; \frac{d\bk}{(2\pi)^3} \; ,
\ee
which, according to Eq. (\ref{2.155}), gives the superfluid fraction
\be
\label{2.193}
n_s =  1 \; - \; \frac{2Q}{3T} \; .
\ee

Again, these expressions are simplified in the case of dilute gases, with
the local interaction potential (\ref{2.174}). Then the Bogolubov spectrum
(\ref{2.183}) takes the standard form
\be
\label{2.194}
\ep_k =\sqrt{ (ck)^2 + \left ( \frac{k^2}{2m}\right )^2 } \; .
\ee
Notation (\ref{2.184}) gives
\be
\label{2.195}
\om_k = \frac{k^2}{2m} + mc^2 \; ,
\ee
while Eq. (\ref{2.185}) reduces to
\be
\label{2.196}
\Dlt_k =  (\rho_0 + \sgm_1) \Phi_0 \; .
\ee
The latter expression defines the sound velocity $c$ through the relation
\be
\label{2.197}
\Dlt_k \equiv \Dlt \equiv mc^2 \; .
\ee
And the anomalous average is
\be
\label{2.198}
\sgm_1 \equiv \int \sgm_k \; \frac{d\bk}{(2\pi)^3} \; .
\ee
For the Lagrange multipliers (\ref{2.186}) and (\ref{2.187}) we obtain
\be
\label{2.199}
\mu_0 = ( \rho +\rho_1 + \sgm_1 ) \Phi_0
\ee
and, respectively,
\be
\label{2.200}
\mu_1 = (\rho + \rho_1 -\sgm_1) \Phi_0 \; .
\ee
Clearly, $\mu_0\neq\mu_1$.

More details on the derivation of the above equations and on the
investigation of their properties can be found in the original papers
[57,94--98].

\subsection{Anomalous Averages}

Anomalous averages of the type (\ref{2.166}), (\ref{2.169}), and (\ref{2.198})
appear in all calculations for Bose systems with broken gauge symmetry. They
always exist together with BEC, since both of them, the anomalous averages
and the phenomenon of BEC, are caused by the same reason, by the spontaneous
breaking of symmetry [63]. And when the gauge symmetry is restored, both $n_0$
as well as $\sgm_1$ become zero. Therefore, $n_0$ and $\sgm_1$ are either
simultaneously nonzero, or simultaneously zero. It looks absolutely evident
that setting one of them zero, while keeping another nonzero would be
principally wrong. It is easy to check that the anomalous averages are often
of the same order as the normal averages [123], hence omitting the latter,
while keeping the former, is mathematically inappropriate. From these facts,
it is clear that neglecting the anomalous averages (as one often does) is
principally incorrect. It is also possible to check by direct calculations
that the omission of the anomalous averages makes all calculations not
self-consistent, dynamics not conserving, thermodynamics incorrect, disturbs
the phase transition order, and moreover, renders the system unstable
[57,123,124]. Therefore, it is absolutely compulsory to correctly keep account
of the anomalous averages.

Dealing with the interaction potentials of finite interaction radii, one
can use the results of the previous section, which requires to accomplish
numerical calculations. The situation simplifies for the local potential
(\ref{2.174}), when many calculations can be made analytically. The sole
thing, however, which does not make the life easier, is that the anomalous
average (\ref{2.198}), for the local potential (\ref{2.174}), becomes
divergent. So, a regularization method is needed.

In the case of the local interaction potential (\ref{2.174}), the anomalous
average (\ref{2.198}) reads as
\be
\label{2.201}
\sgm_1 = - \int \frac{\Dlt}{2\ep_k} \; {\rm coth} \left (
\frac{\ep_k}{2T} \right ) \frac{d\bk}{(2\pi)^3} \; .
\ee
The integral diverges for all $T<T_c$. The divergence is caused by the use of
the local potential (\ref{2.174}), resulting in $\Dlt_k$, in Eq. (\ref{2.197}),
containing no $k$-dependence. For interaction potentials, whose Fourier
transforms $\Phi_k$ depend on $k$, one should use $\Dlt_k$ from Eq.
(\ref{2.185}). Then, for $\Phi_k$ diminishing at large $k$ not slower than
$$
\Phi_k \; \leq \; \frac{const}{k^\al} \qquad
(\al>1, \; k\ra\infty) \; ,
$$
the integral in Eq. (\ref{2.201}) converges. However, then, instead of
simple Eqs. (\ref{2.194}) to (\ref{2.200}), one should return to much more
complicated equations, based on expressions (\ref{2.183}) to (\ref{2.189}).

Let us denote by
\be
\label{2.202}
\sgm_0 \equiv \lim_{T\ra 0} \sgm_1
\ee
the zero temperature limit of the anomalous average. Equation (\ref{2.201})
yields
\be
\label{2.203}
\sgm_0 = - \Dlt \int \frac{1}{2\ep_k} \;
\frac{d\bk}{(2\pi)^3} \; .
\ee
Substituting here the Bogolubov spectrum (\ref{2.194}), we meet the divergent
integral
\be
\label{2.204}
\int \frac{1}{2\ep_k} \; \frac{d\bk}{(2\pi)^3} =
\frac{m^2c}{\pi^2} \; \int_0^\infty
\frac{x\; dx}{\sqrt{1+x^2}} \; .
\ee
The integral in Eq. (\ref{2.204}) can be regularized by means of the
dimensional regularization [11,125], which gives
$$
\int_0^\infty \frac{x\; dx}{\sqrt{1+x^2}} \; \ra \; -1 \; .
$$
But, using this procedure, it is necessary to be cautious, keeping in mind
that the dimensional regularization has a well defined meaning only in the
limit of asymptotically weak interactions, when $\Phi_0\ra 0$. Consequently,
using this regularization presupposes that the value of $c$ in Eq.
(\ref{2.204}) has also to be taken in the same weak-coupling limit. In this
limit, Eqs. (\ref{2.196}) and (\ref{2.197}) give
\be
\label{2.205}
c \simeq c_B \; \sqrt{n_0} \qquad (\Phi_0\ra 0) \; ,
\ee
where
\be
\label{2.206}
c_B \equiv \sqrt{ \frac{\rho\Phi_0}{m} }
\ee
is the Bogolubov expression for the sound velocity. With this condition
$\Phi_0\ra 0$ in mind, from Eq. (\ref{2.204}), we get
\be
\label{2.207}
\int \frac{1}{2\ep_k} \; \frac{d\bk}{(2\pi)^3} \simeq -\;
\frac{m^2c_B}{\pi^2}\; \sqrt{n_0} \; .
\ee
In this way, in the weak-coupling limit, for the zero-temperature form of
the anomalous average (\ref{2.203}), we obtain
\be
\label{2.208}
\sgm_0 \simeq \frac{\Dlt m^2c_B}{\pi^2} \; \sqrt{n_0} \qquad
(\Phi_0\ra 0) \; .
\ee
The standard prescription in using the dimensional regularization is
to employ the latter in the region of its applicability, after which to
analytically continue the result to the whole region of parameters. Using
Eq. (\ref{2.197}), we finally obtain
\be
\label{2.209}
\sgm_0 = \frac{(mc)^2}{\pi^2} \; \sqrt{m\rho_0\Phi_0} \; .
\ee
One may notice that the procedure of the analytical continuation is not
uniquely defined. Fortunately, its different variants do not differ much
in the results, provided that the {\it restoration-symmetry condition}
$$
\sgm_0 \ra 0 \qquad (n_0 \ra 0 )
$$
explicitly holds [97,98]. The meaning of this condition is obvious. As has
been discussed above, the anomalous average and the condensate fraction either
are together nonzero or together zero. The BEC disappears as soon as the gauge
symmetry becomes restored, when $n_0\ra 0$, hence, simultaneously, it should
be that $\sgm_0\ra 0$.

To find the low-temperature behavior of $\sgm_1$, we may rewrite Eq.
(\ref{2.201}) as
\be
\label{2.210}
\sgm_1 = - \int \frac{\Dlt}{2\ep_k} \;
\frac{d\bk}{(2\pi)^3} \; - \; \int \frac{\Dlt}{2\ep_k} \left [
{\rm coth} \left ( \frac{\ep_k}{2T} \right ) \; - \; 1 \right ]
\frac{d\bk}{(2\pi)^3} \; .
\ee
The first term here, at low temperatures, can be replaced by form
(\ref{2.209}), which results in
\be
\label{2.211}
\sgm_1 =\simeq \sgm_0 \; - \; \frac{\sqrt{2}}{(2\pi)^2} \;
(mc)^3 \; \int_0^\infty \;
\frac{(\sqrt{1+x^2}-1)^{1/2}}{\sqrt{1+x^2}}
\left [ {\rm coth} \left ( \frac{mc^2}{2T} \; x \right ) \; -
\; 1 \right ] \; dx \; .
\ee
It is worth stressing that Eqs. (\ref{2.210}) and (\ref{2.211}) are not
identical. Equation (\ref{2.211}) is valid only for low temperatures, such
that
\be
\label{2.212}
\frac{2T}{mc^2} \; \ll \; 1 \; .
\ee
At these low temperatures, the main contribution to the integral in Eq.
(\ref{2.211}) comes from the region of small $x$. Then we can use the
expansion
$$
\frac{\sqrt{2}}{\sqrt{1+x^2}} \left ( \sqrt{1+x^2} \; - \; 1
\right )^{1/2} \simeq x - \; \frac{5}{8}\; x^3 \;  + \;
\frac{63}{128}\; x^5 \; - \; \frac{429}{1024}\; x^7
$$
and the integral
$$
\int_0^\infty x^{2n-1} \left [ {\rm coth}(px) -1 \right ] \; dx
= \frac{\pi^{2n}|B_{2n}|}{2n p^{2n} } \; ,
$$
where $B_n$ are the Bernoulli numbers. Let us introduce the notation
\be
\label{2.213}
\al \equiv \left ( \frac{\pi T}{2mc^2} \right )^2 \; ,
\ee
which is the squared ratio of the typical thermal energy $\pi T$ to the
characteristic kinetic energy
$$
\frac{k_0^2}{2m} =  2mc^2 \qquad (k_0 \equiv 2mc ) \; .
$$
From Eq. (\ref{2.211}), we find the low-temperature expansion
\be
\label{2.214}
\sgm_1 \simeq \sgm_0 \; - \; \frac{(mc)^3}{3\pi^2} \; \al
\left ( 1 - \al + 6\al^2 - \; \frac{429}{5}\; \al^3 \right )
\ee
for $\al\ra 0$. In the lowest order in $\al$, this gives
\be
\label{2.215}
\sgm_1 \simeq \frac{(mc)^3}{\pi^2} \left [
\frac{c_B}{c} \; \sqrt{n_0} \; - \; \frac{\pi^2}{12}
\left ( \frac{T}{mc^2} \right )^2 \right ] \qquad
(T\ra 0) \; .
\ee

Another asymptotic form of $\sgm_1$, which we can find, is its form at
$T\ra T_c$. The critical point $T_c$ is the temperature, where $n_0\ra 0$,
hence $\sgm_1\ra 0$. Respectively, from Eqs. (\ref{2.196}) and (\ref{2.197}),
it follows that $c\ra 0$, as $T\ra T_c$. Equation (\ref{2.201}), for any $T$,
can be identically rewritten as
\be
\label{2.216}
\sgm_1 = - \; \frac{\sqrt{2}}{(2\pi)^2} (mc)^3 \; \int_0^\infty \;
\frac{\left(\sqrt{1+x^2}-1\right )^{1/2}}{\sqrt{1+x^2}} \;
{\rm coth}\left ( \frac{mc^2}{2T}\; x\right ) \; dx \; .
\ee
When $c\ra 0$, we can use the asymptotic form
$$
{\rm coth}\left ( \frac{mc^2}{2T}\; x\right ) \simeq
\frac{2T}{mc^2 x} \qquad (c\ra 0) \; ,
$$
as a result of which, Eq. (\ref{2.216}) gives
\be
\label{2.217}
\sgm_1 \simeq  -\; \frac{m^2cT}{2\pi} \qquad (T\ra T_c) \; .
\ee

Thus, the correct anomalous average $\sgm_1$ should interpolate between
the low-temperature behavior (\ref{2.215}) and the critical asymptotic form
(\ref{2.217}). In order to better illustrate these asymptotic forms, it is
convenient to introduce the dimensionless anomalous average
\be
\label{2.218}
\sgm \equiv \frac{\sgm_1}{\rho} \; .
\ee
Also, let us define the dimensionless temperature
\be
\label{2.219}
t \equiv \frac{mT}{\rho^{2/3}}
\ee
and the dimensionless sound velocity
\be
\label{2.220}
s \equiv \frac{mc}{\rho^{1/3}} \; .
\ee
In this notation, the low-temperature expansion (\ref{2.214}) becomes
\be
\label{2.221}
\sgm \simeq \sgm_0 - \; \frac{s^3}{3\pi^2}\; \al
\left ( 1 - \al + 6\al^2 \right ) \; ,
\ee
when $t\ra 0$, with
\be
\label{2.222}
\al = \left ( \frac{\pi t}{2s^2} \right )^2 \; .
\ee
In the case of the local potential (\ref{2.174}), it is convenient to
introduce the {\it gas parameter}
\be
\label{2.223}
\gm \equiv \rho^{1/3} a_s \; .
\ee
Then the zero-temperature expression for the anomalous average (\ref{2.209})
reads as
\be
\label{2.224}
\sgm_0 = \frac{2s^2}{\pi^2} \; \sqrt{\pi\gm n_0} \; .
\ee

The critical limit (\ref{2.217}), in dimensionless units takes the form
\be
\label{2.225}
\sgm \simeq - \; \frac{st}{2\pi} \qquad (t\ra t_c) \; .
\ee
For the critical temperature, we obtain [57,97,98]
\be
\label{2.226}
t_c=3.312498 \; .
\ee
This coincides with the BEC temperature for the ideal Bose gas, as it should
be in the case of a mean-field picture [57].

It is important to use the correct form for the anomalous average in order
to get a self-consistent description of the system thermodynamics. At low
temperatures, outside of the critical region, expression (\ref{2.211}) can
be employed. But in the near vicinity of $T_c$, the correct behavior of the
anomalous average is prescribed by Eq. (\ref{2.217}). It is possible to check
by direct numerical calculations [98] that the asymptotic form (\ref{2.217})
guarantees the {\it second order} of the BEC phase transition for {\it any
value} of the gas parameter (\ref{2.223}). While, if one takes another
expression for the critical behavior of the anomalous average, one can get
a first-order transition, which would be incorrect. For instance, omitting
the anomalous average, as is done in the Shohno model [83], one always gets
the wrong first order of the BEC transition. It is worth stressing that the
BEC phase transition must be of second order for arbitrary interaction
strength [57].

To emphasize the second order of the BEC transition in the self-consistent
theory described above, let us present some asymptotic expansions in powers
of the relative temperature
$$
\tau \equiv \left | \frac{t-t_c}{t_c} \right | \ra 0 \; .
$$
Using the asymptotic expression (\ref{2.225}), we find the dimensionless
sound velocity (\ref{2.220}),
\be
\label{2.227}
s \simeq \frac{3\pi}{t_c} \; \tau + \frac{9\pi}{4t_c} \left (
1 \; - \; \frac{2\pi}{\gm t_c^2} \right ) \tau^2 \; ,
\ee
the condensate fraction (\ref{2.191}),
\be
\label{2.228}
n_0 \simeq \frac{3}{2}\; \tau \; - \; \frac{3}{8}\; \tau^2 \; ,
\ee
the anomalous average (\ref{2.218}),
\be
\label{2.229}
\sgm \simeq -\; \frac{3}{2}\; \tau + \frac{3}{8} \left ( 1 +
\frac{6\pi}{\gm t_c^2}\right ) \tau^2 \; ,
\ee
and the superfluid fraction (\ref{2.193}),
\be
\label{2.230}
n_s \simeq \frac{3}{2}\; \tau \; - \; \frac{3}{8} \left ( 1 +
\frac{132.413}{t_c^3} \right ) \tau^2 = \frac{3}{2}\; \tau -
1.741\; \tau^2 \; .
\ee
These expansions explicitly demonstrate the second order of the BEC
transition for arbitrary nonzero gas parameters $\gm>0$.

Let us also recall that liquid $^4$He is a strongly interacting
system exhibiting superfluid phase transition of {\it second
order} [108,126--130]. At low temperature, superfluid helium can be
characterized [130,131] by the gas parameter $\gm_{He}\cong0.6$. But
at high temperatures, more realistic potentials should be used. Such
potentials contain, as a rule, hard cores, which requires to take into
account short-range correlations [10,119--122]. The latter are often
described in the frame of the Jastrow approximation [132--136].

An important point is that the two-body scattering matrix [137] can be
shown to be directly related to the total anomalous average (\ref{2.185}).
The latter, for a general nonuniform system, can be represented as
\be
\label{2.231}
\Dlt_k = \int \Phi(\br-\br') < \vp_k^*(\br)\vp_{-k}(\br')
\hat\psi(\br') \hat\psi(\br) > d\br d\br' \; ,
\ee
which describes the scattering of two particles. For a uniform
system, the scattering process ends with the plane waves
$\vp_k(\br)=\exp(i\bk\cdot\br)/\sqrt{V}$. Then Eq. (\ref{2.231}) becomes
\be
\label{2.232}
\Dlt_k = \int \Phi(\br) e^{i\bk\cdot\br}
< \hat\psi(\br) \hat\psi(0)> d\br \; .
\ee
Using here the Bogolubov shift (\ref{2.86}) gives exactly Eq.
(\ref{2.185}).

On the other hand, the two-body scattering matrix can be defined as a
solution of a Lippman-Schwinger equation [138,139], which, in the limit
of weak interactions, results in the total anomalous average (\ref{2.185})
evaluated in the same weak-coupling limit [140--142].

For weak interactions, when $\rho_0\approx\rho$, the anomalous average
(\ref{2.185}), in view of Eq. (\ref{2.189}), can be rewritten as
\be
\label{2.233}
\Dlt_k \simeq \rho \widetilde\Phi_k \qquad (\rho_0 \ra \rho) \; ,
\ee
where the notation for an effective potential
\be
\label{2.234}
\widetilde\Phi_k = \Phi_k \; - \; \frac{1}{V} \;
\sum_p \widetilde\Phi_p \; \frac{\Phi_{k+p}}{2\ep_p} \;
{\rm coth} \left ( \frac{\ep_p}{2T} \right )
\ee
is introduced. As is clear, Eq. (\ref{2.234}) is nothing but a
particular form of the Lippman-Schwinger equation. For a symmetric
interaction potential, for which $\Phi_{-k}=\Phi_k$, the effective potential $\widetilde\Phi_{-k}=\widetilde\Phi_k$ is also symmetric.

Assuming that the potential $\Phi_k$ fastly diminishes as $k\ra\infty$,
with the maximum of $\Phi_k$ at the point $k=0$, and keeping in mind weak
interactions, we can invoke the following approximation:
\be
\label{2.235}
\sum_p \widetilde\Phi_p \; \frac{\Phi_{k+p}}{2\ep_p} \;
{\rm coth}\left ( \frac{\ep_k}{2T} \right ) \cong
\widetilde\Phi_k \sum_k \frac{\Phi_{k+p}}{2\ep_p} \left [
{\rm coth}\left ( \frac{\ep_k}{2T} \right ) -\; 1 \right ] \; .
\ee
Then Eq. (\ref{2.234}) is solved for the effective potential
\be
\label{2.236}
\widetilde\Phi_k = \frac{\Phi_k}{1+\frac{1}{V}\sum_p
\frac{\Phi_{k+p}}{2\ep_p}\left [{\rm coth} \left (
\frac{\ep_k}{2T}\right ) -1 \right ] } \; .
\ee
For the local potential (\ref{2.174}), the sum in the denominator of the
above expression can be represented as
\be
\label{2.237}
\frac{1}{V} \; \sum_p \; \frac{1}{2\ep_p} \left [
{\rm coth}\left ( \frac{\ep_p}{2T} - \; 1 \right ) \right ]
\equiv m^2 cJ \; ,
\ee
with the integral
\be
\label{2.238}
J = \frac{\sqrt{2}}{(2\pi)^2} \; \int_0^\infty \;
\frac{ (\sqrt{1+x^2}-1)^{1/2} } {\sqrt{1+x^2}} \;
\left [ {\rm coth} \left ( \frac{mc^2}{2T}\; x \right ) -\; 1
\right ] \; dx \; .
\ee

Thus, for the local potential (\ref{2.174}) since $\Phi_k=\Phi_0$, one gets
\be
\label{2.239}
\widetilde\Phi_0 = \frac{\Phi_0}{1+m^2 c J \Phi_0} \; .
\ee
Defining an {\it effective scattering length} $\widetilde a_s$ through the
notation
\be
\label{2.240}
\widetilde\Phi_0 \equiv 4\pi \; \frac{\widetilde a_s}{m} \; ,
\ee
we have
\be
\label{2.241}
\widetilde a_s = \frac{a_s}{1+4\pi a_s mc J} \; .
\ee
Similarly, one can introduce an {\it effective gas parameter}
\be
\label{2.242}
\widetilde\gm \equiv \rho^{1/3} \widetilde a_s \; ,
\ee
for which Eq. (\ref{2.241}) gives
\be
\label{2.243}
\widetilde\gm = \frac{\gm}{1+4\pi\gm s J} \; .
\ee

At low temperature, integral (\ref{2.238}) yields
$$
J \simeq \frac{t^2}{12s^4} \qquad (t\ra 0) \; .
$$
And the effective scattering length (\ref{2.241}) tends to the scattering
length $a_s$,
\be
\label{2.244}
\widetilde a_s \simeq \left ( 1 -\;
\frac{\pi\gm}{3s^3}\; t^2 \right ) a_s \qquad (t\ra 0) \; .
\ee

In the vicinity of the critical temperature, when $t\ra t_c$, the sound
velocity $s$ tends to zero, $s\ra 0$, according to Eq. (\ref{2.227}). Then
integral (\ref{2.238}) results in
$$
J \simeq \frac{t}{2\pi s^2} \qquad (t\ra t_c) \; ,
$$
hence Eq. (\ref{2.241}) gives
\be
\label{2.245}
\widetilde a_s \simeq \frac{s}{2\gm t} \; a_s \qquad (t\ra t_c) \; ,
\ee
which tends to zero.

The above analysis shows that the use of the two-body scattering matrix
is equivalent to the HFB approximation in the weak-coupling limit. However,
aiming at considering strong interactions, one is forced to return back to
the anomalous average (\ref{2.201}) expressed through a divergent integral.
The latter can be regularized involving some kind of an analytic
regularization, such as the dimensional regularization [11,125,143]. The
latter gives the zero-temperature anomalous average (\ref{2.209}). But near
the critical temperature the anomalous average behaves as in Eq. (\ref{2.217}).
The correct overall behavior of the anomalous average (\ref{2.201}) should
interpolate between the noncritical form (\ref{2.215}), valid outside of the
critical region, and the critical asymptotic expression (\ref{2.217}).

\subsection{Particle Fluctuations}

The problem of particle fluctuations in Bose-condensed systems has
attracted a great deal of attention provoking controversy in theoretical
literature. Many tens of papers have been published claiming the existence
of thermodynamically anomalous particle fluctuations in Bose-condensed systems
everywhere below the critical temperature. A detailed account of this trend,
with many citations, can be found in the recent survey [144]. However, the
occurrence of such thermodynamically anomalous particle fluctuations, as has
been explained in Refs. [10,93,145,146], contradicts the rigorous theoretical
relations as well as contravenes all known experiments. It is therefore
important to pay some more attention to this problem.

First of all, it is necessary to specify terminology. Observable quantities
are represented by self-adjoint operators. Fluctuations of an observable
quantity, associated with an operator $\hat A$, are characterized by the
operator dispersion
\be
\label{2.246}
\Dlt^2(\hat A) \; \equiv \; < \hat A^2 > - < \hat A>^2 \; .
\ee
Generally, in statistical mechanics, one distinguishes intensive and
extensive quantities [147]. Fluctuations of intensive quantities are
always finite, so that if $\hat A$ represents an intensive quantity,
then its dispersion (\ref{2.246}) is finite. Fluctuations of extensive
quantities are described by dispersions (\ref{2.246}) proportional to
the system volume or to the total number of particles $N$. Fluctuations
are termed {\it thermodynamically normal}, when
\be
\label{2.247}
0 \; \leq \; \frac{\Dlt^2(\hat A)}{N} \; < \; \infty
\ee
for any $N$, including the thermodynamic limit $N\ra\infty$. Condition
(\ref{2.247}) holds for any operators of observables, whether intensive
or extensive, which guarantees the system stability [10,93,146].

The number of particles in the system is represented by an operator $\hat N$.
So, particle fluctuations are characterized by the dispersion
\be
\label{2.248}
\Dlt^2(\hat N) \; \equiv \; < \hat N^2> - < \hat N >^2 \; .
\ee
Similarly to condition (\ref{2.247}), particle fluctuations are
thermodynamically normal, provided that
\be
\label{2.249}
0 \; \leq \; \frac{\Dlt^2(\hat N)}{N} \; < \; \infty
\ee
for any $N$, including $N\ra\infty$.

The fact why conditions (\ref{2.247}) or (\ref{2.249}) are to be valid for any
stable statistical system is that the reduced dispersions $\Dlt^2(\hat A)/N$
describe the system susceptibilities, which also are observable quantities.
More precisely, susceptibilities are intensive thermodynamic characteristics,
hence, they have to be finite for any stable statistical system, except, may
be, the points of phase transitions, where the system is, actually, unstable.
But the possible divergence of susceptibilities at phase-transition points
should not be confused with their thermodynamic divergence. At a
phase-transition point, a susceptibility could become divergent with respect
to some thermodynamic parameter, such as temperature, pressure, etc. However
it is never divergent with respect to the system volume or number of particles.

Particle fluctuations are directly related to the isothermal compressibility
\be
\label{2.250}
\kappa_T \equiv -\; \frac{1}{V} \left ( \frac{\prt P}{\prt V}
\right )^{-1}_T = \frac{\Dlt^2(\hat N)}{\rho TN} \; ,
\ee
where $P$ is pressure, and to the hydrodynamic sound velocity $s_T$, given
by the equation
\be
\label{2.251}
s_T \equiv \frac{1}{m} \left ( \frac{\prt P}{\prt\rho}
\right )_T = \frac{1}{m\rho\kappa_T} = \frac{NT}{m\Dlt^2(\hat N)} \; .
\ee
The structure factor
\be
\label{2.252}
S(\bk) = 1 + \rho \int \left [ g(\br) -1 \right ]
e^{-i\bk\cdot\br} \; d\br \; ,
\ee
in which $g(\br)$ is a pair correlation function [148], is also expressed
through the particle dispersion (\ref{2.248}), so that
\be
\label{2.253}
S(0) = \rho T\kappa_T = \frac{T}{ms_T^2} =
\frac{\Dlt^2(\hat N)}{N} \; .
\ee
As is evident, all these observable quantities, $\kappa_T$, $s_T$, and
$S(\bk)$, are finite then and only then, when the particle fluctuations
are normal, satisfying condition (\ref{2.249}).

Fluctuations of an observable, represented by an operator $\hat A$, are
called {\it thermodynamically anomalous}, when condition (\ref{2.247}) does
not hold, as a result of which
$$
\frac{\Dlt^2(\hat A)}{N} \; \ra \; \infty \qquad (N\ra\infty) \quad
(anomalous) \;.
$$
Clearly, if particle fluctuations would be thermodynamically anomalous,
then the isothermal compressibility (\ref{2.250}) would be infinite, sound
velocity (\ref{2.251}), zero, and the structure factor (\ref{2.253}) would
also be infinite, all that manifesting the system instability [10,93,146].

The thermodynamically normal properties of susceptibilities do not depend
on the used statistical ensemble, provided that the representative ensembles
are employed [54,56,57]. The microcanonical ensemble can be considered as a
projection of the canonical one, and the canonical ensemble, as a projection
of the grand canonical ensemble [149]. For any representative ensemble,
susceptibilities should be finite almost everywhere, except the points of 
phase transitions. For example, in the grand canonical ensemble, the 
compressibility can be found from the dispersion $\Dlt^2(\hat N)$, as in 
Eq. (\ref{2.250}). In the canonical ensemble, the total number of 
particles is fixed. But this does not mean that the compressibility here 
becomes zero. One simply has to use another formula for calculating the 
compressibility, which in the canonical ensemble can be calculated by 
means of the expression
\be
\label{2.254}
\kappa_T = \frac{1}{V} \left (
\frac{\prt^2 F}{\prt V^2}\right )^{-1}_{TN} \; ,
\ee
where $F$ is free energy. The compressibilities (\ref{2.250}) and
(\ref{2.254}) have to be the same, defining the same observable quantities as
the sound velocity (\ref{2.251}) or the structure factor (\ref{2.252}). This
concerns any statistical system, including Bose-condensed ones [150--152].
In all experiments, whether with cold trapped atoms or with superfluid helium,
all intensive quantities below $T_c$ are, of course, finite, including particle
fluctuations measured as $\Dlt^2(\hat N)/N$ (see Ref. [153]). Divergencies can
arise solely at the critical point
itself [154].

When one claims the occurrence of thermodynamically anomalous particle
fluctuations in Bose-condensed systems [144], one often tells that these
anomalous fluctuations may happen not for the total number of particles
but only separately for the number of condensed and uncondensed atoms. The
operator of the total number of atoms is the sum $\hat N=\hat N_0+\hat N_1$.
One assumes that the fluctuations of $\hat N$, given by the relative
dispersion $\Dlt^2(\hat N)/N$ could be normal, thus, not breaking the system
stability, while the fluctuations of $\hat N_0$ and $\hat N_1$ could be
anomalous. This assumption is, however, wrong [93,146].

Let us consider two operators $\hat A$ and $\hat B$, representing some
observable quantities. The dispersion of their sum
\be
\label{2.255}
\Dlt^2 (\hat A +\hat B) =\Dlt^2 (\hat A) + \Dlt^2(\hat B) + 2 {\rm cov}
\left ( \hat A,\hat B\right )
\ee
is expressed through the particle dispersions $\Dlt^2(\hat A)$ and
$\Dlt^2(\hat B)$ and the covariance
$$
{\rm cov}( \hat A + \hat B ) \equiv \frac{1}{2} < \hat A\hat B +
\hat B\hat A> - < \hat A> <\hat B> \; .
$$
The dispersions $\Dlt^2(\hat A)$ and $\Dlt^2(\hat B)$ are, by definition,
positive or, at least, non-negative, while the covariance can be of any
sign. However the covariance cannot compensate the partial dispersions, so
that the total dispersion (\ref{2.255}) is always governed by the largest
partial dispersion. This rigorously follows from the theorem below.

\vskip 2mm

{\bf Theorem} (Yukalov [93,146]) The dispersion of the sum of linearly
independent self-adjoint operators (\ref{2.255}) can be represented as
\be
\label{2.256}
\Dlt^2(\hat A + \hat B) = \left [ \sqrt{\Dlt^2(\hat A)} \; -
\; \sqrt{\Dlt^2(\hat B)} \right ]^2 \; + \;
c_{AB} \; \sqrt{\Dlt^2(\hat A)\Dlt^2(\hat B)} \; ,
\ee
where
$$
0 \; < \; c_{AB} \; <  \; 4 \; .
$$

From here it follows that fluctuations of the sum of two operators
$\hat A+\hat B$ is thermodynamically anomalous then and only then, when
at least one of the partial fluctuations of either $\hat A$ or $\hat B$
is anomalous, the anomaly of the total dispersion $\Dlt^2(\hat A+\hat B)$
being governed by the largest partial dispersion. Conversely, fluctuations
of the sum $\hat A+\hat B$ are thermodynamically normal if and only if all
partial fluctuations are thermodynamically normal.

\vskip 2mm

Applying this theorem to the sum $\hat N=\hat N_0+\hat N_1$, we see
that, if the total dispersion $\Dlt^2(\hat N)$ is thermodynamically normal,
which is compulsory for any stable system, then both partial dispersions,
$\Dlt^2(\hat N_0)$ as well as $\Dlt^2(\hat N_1)$, must be normal. Thus,
the normality of fluctuations of the total number of particles $\hat N$
necessarily requires the normality of fluctuations of both condensed as
well as uncondensed atoms.

A very widespread misconception is that the condensate fluctuations in
the grand canonical and canonical ensembles are different; in the grand
canonical ensemble the fluctuations are thermodynamically anomalous, such
that $\Dlt^2(\hat N_0)\sim N^2$, while in the canonical ensemble they are
normal. One even calls this "the grand canonical catastrophe". But there
is no any "catastrophe" here. The seeming paradox comes about only
because of the use of nonrepresentative ensembles [93]. The problem has
been explained long time ago by ter Haar [155]. The anomalous behavior
$\Dlt^2(\hat N_0)\sim N^2$ appears in the grand canonical ensemble
preserving the gauge symmetry, while in the canonical ensemble, the
gauge symmetry is effectively broken. However, if the gauge symmetry is
also broken in the grand canonical ensemble, then no anomalous behavior
of $\Dlt^2(\hat N_0)$ arises, this dispersion being the same in both
ensembles [93,155]. Recall that the gauge symmetry breaking is necessary
and sufficient for describing statistical systems with BEC [63].

Breaking the gauge symmetry by means of the Bogolubov shift (\ref{2.86}),
we pass to the Fock space $\cF(\psi_1)$, where the number operator of
condensed atoms is $\hat N_0 =N_0\hat  1_\cF$, as in Eq. (\ref{2.90}).
Then $\Dlt^2(\hat N_0)$ becomes identically zero. If one prefers to work
in the Fock space $\cF(\psi)$, as is explained in Sec. 2.6, then the field
operator $\psi_0(\br)$ becomes a function $\eta(\br)$ in the thermodynamic
limit. According to the Bogolubov theorem, it is easy to show [63] that
\be
\label{2.257}
\lim_{N\ra\infty} \; \frac{\Dlt^2(\hat N_0)}{N} \; \equiv \; 0
\ee
in the Fock space $\cF(\psi)$. The latter is orthogonal to the space
$\cF(\psi_1)$. The operator representations on these spaces are unitary
nonequivalent [99,156]. But in any case, the limit (\ref{2.257}) holds
true. Therefore, all particle fluctuations are, actually, caused by the
uncondensed atoms, since
\be
\label{2.258}
\frac{\Dlt^2(\hat N)}{N} \; \cong \; \frac{\Dlt^2(\hat N_1)}{N}
\ee
for large $N\ra\infty$.

In the Bogolubov approximation, as well as in the HFB approximation, for
a uniform system we have [10,145,146]
\be
\label{2.259}
\frac{\Dlt^2(\hat N_1)}{N} = \frac{T}{mc^2} \; ,
\ee
with the sound velocity $c$ in the corresponding approximation. All
these fluctuations, described by Eqs. (\ref{2.257}), (\ref{2.258}), and
(\ref{2.259}), are clearly thermodynamically normal.

It is worth mentioning that particle fluctuations in uniform systems with
BEC are normal for {\it interacting} systems. For an ideal Bose gas, they
are anomalous, which immediately follows from Eq. (\ref{2.259}), if one
sets there $c\ra 0$, that is, reducing interactions to zero. More precisely,
for an ideal uniform Bose-condensed gas, it is easy to find [93] that
$$
\frac{\Dlt^2(\hat N_1)}{N} \; \sim \; \left ( \frac{mT}{\pi}
\right )^2 \frac{N^{1/3}}{\rho^{4/3}} \qquad (ideal \; gas) \; ,
$$
which means anomalous fluctuations. This implies that the ideal uniform
gas with BEC cannot exist, being an unstable object [10,93,145,146].
Fortunately, the purely ideal gas certainly does not exist, being just
a cartoon of weakly interacting systems. Real atoms always interact with
each other, at least weakly. No matter how small the interaction, it does
stabilize uniform Bose-condensed systems. External fields, forming
trapping potentials, can also stabilize an ideal Bose gas, for instance,
if the trapping is realized by harmonic potentials [157] or power-law
(though not all) potentials [41].

Thus, any stable statistical system of interacting atoms, whether uniform
or not, must display thermodynamically normal particle fluctuations. At
the same time, as has been emphasized at the beginning of the present
section, many papers claim the occurrence of anomalous particle fluctuations
everywhere below $T_c$, as is summarized in Ref. [144]. These anomalous
fluctuations are claimed to be of the type $\Dlt^2(\hat N_1)\propto N^{4/3}$
for interacting Bose-condensed systems of any nature, whether uniform or
nonuniform trapped clouds. Moreover, since Bose-condensed systems are just
one particular example of a very general class of systems with a broken
continuous symmetry, the same type of fluctuations has to arise in all such
systems [10,146]. This class of systems is really rather wide. In addition
to cold trapped atoms, it includes superfluid $^4$He, with broken gauge
symmetry, isotropic magnets, with broken spin-rotational symmetry, and all
solids, with broken transitional and rotational symmetries. If such systems
would possess some divergent susceptibilities, this would mean that they
could not exist as a stable matter. There would be no superfluids, many
magnets, and no solids. It is evident that such an exotic conclusion would
be meaningless. - We perfectly know that all that matter does exist and no
one experiment has ever revealed any anomalous susceptibility that would
persistently be anomalous everywhere below the critical temperature
[158,159]. The same concerns theoretical investigations for exactly solvable
models [160] as well as correct calculations for other concrete models [58].

It is instructive to consider the magnetic susceptibility of isotropic
magnets. If the compressibility of Bose systems would be really divergent
as $N^{1/3}$, when $N\ra\infty$, then the magnetic susceptibility would
also display the same divergence as $N^{1/3}$.

Let us turn, first, to the isotropic ferromagnet described by the
Heisenberg model [161,162]. The sample magnetization is given by the vector
$\bM=\{ M_\al\}$, with $\al=x,y,z$, defined as the statistical average
$$
\bM \; = \; < \hat\bM> \; , \qquad \hat\bM = \mu_S N \hat\bS \; ,
$$
in which $\mu_S=g_S\mu_B$, $g_S$ is the gyromagnetic ratio for spin $S$,
$\mu_B$ is the Bohr magneton, and
$$
\bS \equiv \frac{1}{N} \; \sum_{j=1}^N \bS_j = \{ \hat S_\al \}
$$
is the reduced spin operator. The susceptibility tensor is defined as
the response of the magnetization to the variation of an external field
${\bf H}=\{ h_\al\}$, so that
\be
\label{2.260}
\chi_{\al\bt} \equiv \lim_{h\ra 0} \; \frac{1}{N} \;
\frac{\prt M_\al}{\prt h_\bt} = \frac{\mu_S^2}{T} \;
{\rm cov} (\hat S_\al,\hat S_\bt) \; .
\ee
The diagonal magnetic susceptibility
\be
\label{2.261}
\chi_{\al\al} = \frac{\mu_S^2}{T}\; \Dlt^2(\hat S_\al)
\ee
is expressed through the dispersion of the spin operator $\hat S_\al$. Let
$\bM$ be directed along the $z$-axis, $\bM=\{ M_z,0,0\}$, with
$$
M_z = N \mu_S S \sgm \; , \qquad \sgm \equiv \frac{<S_z>}{S} \; .
$$
Then it is convenient to distinguish the longitudinal susceptibility
\be
\label{2.262}
\chi_{||} \equiv \lim_{h\ra 0} \; \frac{1}{N} \;
\frac{\prt M_z}{\prt h_z} = \frac{\mu_S^2}{T} \; \Dlt^2(\hat S_z)
\ee
and the transverse susceptibility
\be
\label{2.263}
\chi_\perp \equiv \lim_{h\ra 0} \; \frac{1}{N} \;
\frac{\prt M_z}{\prt h_x} = \frac{\mu_S^2}{T} \;
{\rm cov}(\hat S_z,\hat S_x) \; .
\ee
In the mean-field approximation, the longitudinal susceptibility is
\be
\label{2.264}
\chi_{||} = \frac{S\mu_S^2(1-\sgm)}{2T-JS(1-\sgm)} \; ,
\ee
where, for spin one-half,
$$
\sgm = {\rm tanh}\; \frac{JS\sgm+\mu_S h_0}{2T} \; ,
$$
$J$ is an exchange integral $(J>0)$, and $h_0$ is an anisotropy field along
the $z$-axis. Expression (\ref{2.264}) is finite everywhere below $T_c$,
even when $h_0\ra 0$. This means, in view of Eq. (\ref{2.262}), that spin
fluctuations are normal.

For the transverse susceptibility (\ref{2.263}), one gets
\be
\label{2.265}
\chi_\perp = \frac{\mu_S S\sgm}{h_0} \; .
\ee
In all real magnets, there always exists an anisotropy field, caused
by one of many reasons, such as the natural magnetic lattice anisotropy,
spin-orbital interactions, demagnetizing shape factors, and so on,
including the Earth magnetic field. Therefore, in real life, the transverse
susceptibility (\ref{2.265}) is finite, no matter how small the anisotropy
field.

Even if we consider an ideal situation with no anisotropy, when
$\chi_\perp$ diverges with respect to $h_0\ra 0$, this divergence does
not make susceptibility (\ref{2.265}) thermodynamically anomalous. The
transverse susceptibility remains {\it thermodynamically normal}, since it
does not diverge with respect to $N\ra\infty$.

Recall that the mathematically correct order of the limits, according to
the Bogolubov method of quasi-averages [64,65], is in taking first the
thermodynamic limit $N\ra\infty$ and only after it, to consider the limit
$h_0\ra 0$. Since susceptibility (\ref{2.265}) does not diverge with respect
to $N\ra\infty$, it is thermodynamically normal.

Similarly, the divergence of the longitudinal susceptibility (\ref{2.262})
with respect to $T\ra T_c$, when
$$
\chi_{||} \; \propto \; \left | 1\; - \;
\frac{T}{T_c}\right |^{-\gm} \qquad (T\ra T_c) \; ,
$$
with $\gm>0$, does not imply thermodynamically anomalous behavior. For any
$T\neq T_c$, the susceptibility $\chi_{||}$ is finite under the limit $N
\ra\infty$. So, the above expression $\chi_{||}$ is also thermodynamically
normal. In addition, we should remember that, exactly at the critical
temperature $T_c$, the system is unstable.

In the same way, one can check that the susceptibilities of isotropic
antiferromagnets are also thermodynamically normal [161,163].

The appearance in some theoretical works of thermodynamically anomalous
susceptibilities in magnets or thermodynamically anomalous particle
fluctuations in Bose-condensed systems is due solely to calculational
mistakes. The standard such a mistake, as is explained in Refs.
[10,93,145,146], is as follows. One assumes the Bogolubov approximation
[72,73], which is a second-order accuracy approximation with respect to
the operators of uncondensed particles, or one accepts a hydrodynamic
approximation, which, actually, is mathematically equivalent to the
Bogolubov approximation, also being a second-order approximation with
respect to some field operators. The higher-order terms, higher than two,
are not well defined in the second-order theory. Forgetting this, one
calculates the fourth-order operator terms in the frame of a second-order
theory. This inconsistency results in the arising anomalous expressions,
which are just calculational artifacts, and thus have no physical meaning.
The detailed explanation can be found in Refs. [10,93,145,146].

The fact that condition (\ref{2.247}) has to be valid for any observable
can be simply understood in the following way. The average of an operator
$\hat A$, associated with an extensive observable quantity, is such that
$<\hat A>\propto N$. Hence, condition (\ref{2.247}) is equivalent to the
condition
$$
0 \; \leq \; \frac{\Dlt^2(\hat A)}{|<\hat A>|} \; < \; \infty \; ,
$$
meaning that the fluctuations of an observable cannot be infinitely larger
than the observable itself.

Thus, in any correct theory, all susceptibilities as well as fluctuations
of observables are always thermodynamically normal. Conditions (\ref{2.247}),
or (\ref{2.249}), are necessary for the stability of statistical systems.
There are no experimental observations that would display thermodynamically
anomalous susceptibilities in any system with broken continuous symmetry,
neither in trapped atoms, nor in liquid helium, nor in magnets, nor in solids.

\subsection{Fragmented Condensates}

The consideration of the previous sections, treating a Bose system with a
sole BEC, can be generalized to the case, when several condensates arise
in the system. There exist two principally different situations for the
appearance of multiple condensates, depending on how the latter are
distinguished. The distinction can be done according to two different kinds
of the quantum numbers labelling quantum states. One type of the indices
labels collective quantum states. Such has been the index (multi-index)
$k$ labelling natural orbitals $\vp_k(\br)$ in the previous sections. The
indices of another type are those characterizing the internal states of each
particle, because of which such indices can be termed internal or individual.
Examples of the individual indices are spin indices, hyperfine spin or isospin
indices, and like that. Briefly speaking, collective indices are associated
with collective quantum states of quasiparticles that are members of a
statistical system, while individual indices describe the internal states
of each separate particle. Respectively, there can be two types of multiple
condensates, depending on whether they are distinguished by collective or
internal quantum numbers.

In a single-component statistical system, the arising BEC corresponds to the
macroscopic occupation of one of the quantum states characterized by natural
orbitals, as has been described in Sections 2.1 and 2.2. One can assume that
not one but several occupation numbers, related to natural orbitals, become
macroscopic. Then the arising multiple condensates are distinguished by the
collective quantum index $k$, labelling the appropriate natural orbitals. This
case of multiple condensates is what one calls the condensate fragmentation.

The {\it fragmented condensate is a multiple condensate consisting of several
coexisting condensates distinguished by the collective quantum index labelling
the natural orbitals}.

Probably, the first example of the fragmented condensate was given by
Pollock [164], who considered the coexistence of two condensates, one
with the zero angular momentum and another with a nonzero angular momentum.
The term "fragmented condensates" was coined by Nozi\`eres and Saint James
[165], who considered the coexisting condensates with zero and nonzero
momenta. These condensates in equilibrium were shown [164,165] to be unstable.
This conclusion looks rather clear, since the appearance of a condensate with
either a nonzero momentum or nonzero angular momentum rises the system free
energy.

In order to remain stable, the fragmented condensate should consist of
degenerate parts, such that, though being described by different natural
orbitals, they contribute to the system the same energy. An example is the
fragmented quasicondensate whose parts possess the same modulus of momentum
but various arbitrary momentum directions [166--169]. Here, it is called
the quasicondensate, since the order index $\om(\hat\rho_1)$, defined in
Sec. 2.3., is $\om(\hat\rho_1)=1/3$, which corresponds to mid-range order,
as classified in Eq. (\ref{2.19}) (see details in Ref. [45]).

Fragmented condensates can be created in nonequilibrium systems. For example,
in the process of superradiant Rayleigh scattering [170--173] a single BEC
separates into several clouds with different momenta. Another example is a
Bose system subject to resonant external fields generating nonequilibrium
condensates corresponding to different coherent topological modes [174--196].

Fragmented condensates, as they are defined by Pollock [164] and Nozi\`eres
and Saint James [165], should not be confused with multicomponent condensates.
To avoid this confusion, let us give a precise mathematical definition of the
fragmented condensates.

Consider a single-component Bose system, whose natural orbitals $\vp_k(\br)$,
associated with the first-order density matrix, are labelled by a multi-index
$k$. Let the total set $\{ k\}$ of all collective indices $k$ contain a subset
$\{ k_\nu\}$, such that $\{ k_\nu\}\subset\{ k\}$. The enumeration of the members
$k_\nu$ of the subset $\{ k_\nu\}$ can be either continuous over a final
interval $0\leq\nu\leq\nu_{max}$ or can be discrete, when $\nu=0,1,2,\ldots$.

Expanding the field operator over natural orbitals, one has
\be
\label{2.266}
\psi(\br) = \sum_k a_k\vp_k(\br) = \sum_\nu \psi_{k_\nu}(\br) +
\psi_1(\br) \; .
\ee
Here the terms
\be
\label{2.267}
\psi_{k_\nu}(\br) \equiv a_{k_\nu} \vp_{k_\nu}(\br)
\ee
are assumed to be related to the expected fragmented condensates, while the
term
\be
\label{2.268}
\psi_1(\br) \equiv \sum_{k\neq k_\nu} a_k \vp_k(\br)
\ee
corresponds to uncondensed particles. The condition $k\neq k_\nu$ implies
that $k$ does not equal any $k_\nu$ from the set $\{ k_\nu\}$.

For $\vp_k(\br)$ to represent natural orbitals, it is necessary and sufficient
that the quantum-number conservation condition be valid,
\be
\label{2.269}
< a_k^\dgr a_p>\; = \; \dlt_{kp} n_k \; , \qquad
n_k\; \equiv \; < a_k^\dgr a_k > \; .
\ee
Then the density matrix
\be
\label{2.270}
\rho(\br,\br') \; \equiv \; < \psi^\dgr(\br') \psi(\br)>\; = \;
\sum_k n_k \vp_k(\br) \vp_k^*(\br')
\ee
is diagonal in the expansion over natural orbitals.

The terms (\ref{2.267}) correspond to condensates when
\be
\label{2.271}
\lim_{N\ra\infty} \; \frac{N_{k_\nu}}{N} \; > \; 0 \qquad
(N_{k_\nu} \equiv n_{k_\nu} ) \; .
\ee
If so, the density matrix (\ref{2.270}) can be separated into two parts,
\be
\label{2.272}
\rho(\br,\br') = \sum_\nu N_{k_\nu} \vp_{k_\nu}(\br) \vp_{k_\nu}^*(\br')
\; +  \; \sum_{k\neq k_\nu} n_k \vp_k(\br)\vp_k^*(\br') \; ,
\ee
where the first sum represents the fragmented condensate.

Realizing the Bogolubov shift, one has to replace the operator $\psi(\br)$
by the field operator
\be
\label{2.273}
\hat\psi(\br) = \sum_\nu \eta_\nu \; + \; \psi_1(\br) \; .
\ee
The grand Hamiltonian, in the case of the fragmented condensate, is
\be
\label{2.274}
H =\hat H[\hat\psi] \; - \; \sum_\nu \mu_{k_\nu} \hat N_{k_\nu} \; -
\; \mu_1 \hat N_1 \; - \; \hat\Lbd \; ,
\ee
where $\hat H[\hat\psi]$ is the Hamiltonian energy operator (\ref{2.156})
and the linear killer $\hat\Lbd=\hat\Lbd[\psi_1]$ is defined as in Eq.
(\ref{2.96}). The system chemical potential generalizes form (\ref{2.116})
to
\be
\label{2.275}
\mu = \sum_\nu \mu_{k_\nu} n_{k_\nu} \; + \; \mu_1 n_1 \; ,
\ee
where
$$
n_{k_\nu} \equiv \frac{N_{k_\nu}}{N} \; , \qquad
n_1 \equiv \frac{N_1}{N} \; .
$$

To stress it again, the fragmented condensate, by definition [164,165],
occurs, when several occupation numbers, associated with the {\it natural
orbitals}, become macroscopic. This usually happens for nonequilibrium systems.

\subsection{Multicomponent Condensates}

Contrary to fragmented condensates, which are rather rare and usually
are not equilibrium, the multicomponent condensates are ubiquitous and
may happen in any equilibrium system consisting of several kinds of Bose
particles, distinguished by different internal numbers. The simplest case,
studied long time ago [197--201], is the mixture of several components of
Bose particles with different masses and different interactions with each
other. Another example is the mixture of atoms with different spins or
hyperfine states [202]. And there are plenty of other examples [203].
Among the systems formed by composite bosons, one can remember the systems
of bipolarons composed of tightly bound electron pairs and of bound hole
pairs [204] and also the mixtures of multiquark bosonic clusters [28,29].

For a multicomponent system, one has several types of field operators
$\psi_\al(\br)$, labelled by an individual index $\al$, which can be either
discrete or continuous. Respectively, there exist several orthonormal bases
$\{\vp_{\al k}(\br)\}$, in which $k=k(\al)$ are the indices labelling
collective quantum states. Each component may possess BEC in a state
labelled by $k_0=k_0(\al)$. The related field operator can be expanded over
the given basis as
\be
\label{2.276}
\psi_\al(\br) = \sum_k a_{\al k} \vp_{\al k}(\br) =
\psi_{\al 0}(\br) + \psi_{\al 1}(\br) \; ,
\ee
where the first term
\be
\label{2.277}
\psi_{\al 0}(\br) \equiv a_{\al 0} \vp_{\al 0}(\br) \equiv
a_{\al k_0} \vp_{\al k_0} (\br)
\ee
is assumed to describe the BEC of the $\al$-component, and the second term
\be
\label{2.278}
\psi_{\al 1}(\br) \equiv \sum_{k\neq k_0}
a_{\al k} \vp_{\al k} (\br)
\ee
represents uncondensed particle of that component.

If $\vp_{\al k}(\br)$ are chosen as the natural orbitals, then the
quantum-number conservation condition must be valid,
\be
\label{2.279}
< a_{\al k}^\dgr a_{\bt p} > \; = \;
\dlt_{\al\bt} \dlt_{kp} n_{\al k} \; , \qquad
n_{\al k} \; \equiv \; < a_{\al k}^\dgr a_{\al k} > \; ,
\ee
similarly to condition (\ref{2.269}). This guarantees that the density matrix
\be
\label{2.280}
\rho_{\al\bt}(\br,\br') \; \equiv \;
< \psi_\bt^\dgr(\br') \psi_\al(\br) >
\ee
be diagonal, such that
\be
\label{2.281}
\rho_{\al\bt}(\br,\br') = \dlt_{\al\bt} \rho_\al(\br,\br') \; ,
\ee
and that it would possess the diagonal expansion
\be
\label{2.282}
\rho_\al(\br,\br') = \sum_k
n_{\al k} \vp_{\al k}(\br) \vp_{\al k}^*(\br') \; .
\ee

The $\al$-component enjoys condensation, when
\be
\label{2.283}
\lim_{N\ra\infty} \; \frac{N_{\al 0}}{N} \; > \; 0 \qquad
(N_{\al 0} \equiv n_{\al k_0} ) \; .
\ee
Then matrix (\ref{2.282}) can be separated into two parts as
\be
\label{2.284}
\rho_\al(\br,\br') = N_{\al 0}\vp_{\al 0}(\br)
\vp^*_{\al 0}(\br') \; + \; \sum_{k\neq k_0} n_{\al k} \vp_{\al k}(\br)
\vp^*_{\al k}(\br') \; .
\ee

The Bogolubov shift for a multicomponent system implies that each field operator $\psi_\al(\br)$ is to be replaced by
\be
\label{2.285}
\hat\psi_\al(\br) = \eta_\al(\br) + \psi_{\al 1}(\br) \; .
\ee
The corresponding grand Hamiltonian becomes
\be
\label{2.286}
H = \hat H \left [ \left \{ \hat\psi_\al \right\} \right ]
\; -\; \sum_\al \left ( \mu_{\al 0} \hat N_{\al 0} +
\mu_{\al 1} \hat N_{\al 1} + \hat\Lbd_\al \right ) \; ,
\ee
where $\hat\Lbd_\al=\hat\Lbd_\al[\psi_{\al 1}]$.

The total number of particles is
\be
\label{2.287}
N = \sum_\al N_\al \; , \qquad
N_\al = N_{\al 0} + N_{\al 1} \; .
\ee
If there are no mutual transformations between different components, so that
all $N_\al$ are fixed, then the system has so many chemical potentials
\be
\label{2.288}
\mu_\al = \mu_{\al 0} n_{\al 0} + \mu_{\al 1} n_{\al 1}
\qquad (N_\al = const)
\ee
how many components it has. But if mutual transformations between components
are allowed, so that only the total number of particles $N$ is fixed, then
all $\mu_\al=\mu$, and the system possesses the sole chemical potential
\be
\label{2.289}
\mu = \mu_{\al 0} n_{\al 0} + \mu_{\al 1} n_{\al 1}
\qquad (N = const) \; .
\ee
Here, the particle concentrations are
\be
\label{2.290}
n_{\al 0} \equiv \frac{N_{\al 0}}{N} \; , \qquad
n_{\al 1} \equiv \frac{N_{\al 1}}{N} \; .
\ee

Comparing Sections 2.15 and 2.16, we see that there are important
differences between fragmented and multicomponent condensates. Condensate
fragmentation occurs in a single-component system, whose several natural
orbitals are macroscopically occupied. Multicomponent condensation happens
in a multicomponent system, where several of the components acquire their
own condensates. Of course, more complicated situations may occur, when
multicomponent and fragmented condensates arise simultaneously.

In the case of multicomponent systems, a special care has to be taken with
respect to the stability of the considered multicomponent mixture. Depending
on the kind of the interactions in a multicomponent system, the latter can be
unstable with reference to the component stratification [198--201], when the
components spatially separate from each other, rendering the system to a set
of single-component parts.

\subsection{Model Condensates}

Simple models often are useful for quickly catching typical features of more
complicated realistic systems. However, one should keep in mind that BEC might
exist in a model, though could be absent in a realistic system that has been
mimicked by the model. Or the properties of BEC in a cartoon model could be
essentially distorted, as compared to the real case. As an example of such a
situation, let us consider a very popular two-level model with BEC.

Let us assume that $N$ atoms can occupy only two energy levels, one with
energy $E_1$ and another with energy $E_2$, such that $E_1<E_2$. The related
field operators, $a_1$ and $a_2$, satisfy the commutation relations
$$
\left [ a_1, a_1^\dgr \right ] = \left [ a_2, a_2^\dgr \right ] = 1 \; ,
\qquad
\left [ a_1, a_1\right ] = \left [ a_1, a_2 \right ] =
\left [ a_1, a_2^\dgr \right ] = \left [ a_2, a_2 \right ] = 0 \; .
$$
The condition that all $N$ atoms pertain to one of two levels is
\be
\label{2.291}
a_1^\dgr a_1 + a_2^\dgr a_2 = N \; .
\ee
Also suppose that the state, corresponding to energy $E_1$, is symmetric with
respect to spatial inversion, while the state, associated with energy $E_2$,
is antisymmetric. The standard situation that is modelled by this picture is
an ensemble of trapped atoms with discrete spectrum. When temperature is low
and atomic interactions are weak, one assumes that all atoms pile down to the
lowest energy levels, say, to the lowest two levels. These assumptions are
typical when considering cold atoms in a double-well trap [203].

The symmetric and antisymmetric states can be represented by linear combinations
of "left" and "right" field operators, $c_L$ and $c_R$, respectively, so that
\be
\label{2.292}
a_1 = \frac{1}{\sqrt{2}} ( c_L + c_R ) \; , \qquad
a_2 = \frac{1}{\sqrt{2}} ( c_L - c_R ) \; .
\ee
The operator $a_1$ is symmetric with respect to the interchange of $c_L$ and
$c_R$, while the operator $a_2$ is antisymmetric. The "left" and "right" field
operators satisfy the same commutation relations as the operators $a_1$ and
$a_2$, in particular,
$$
\left [ c_L, c_L^\dgr \right ] =
\left [ c_R, c_R^\dgr \right ] = 1 \; ,
$$
and other commutators being zero. Hence equations (\ref{2.292}) and their
converse,
\be
\label{2.293}
c_L = \frac{1}{\sqrt{2}} ( a_1 + a_2) \; , \qquad
c_R = \frac{1}{\sqrt{2}} ( a_1 - a_2) \; ,
\ee
represent canonical transformations. The new operators also obey the
$N$-polarity condition
\be
\label{2.294}
c_L^\dgr c_L + c_R^\dgr c_R = N \; .
\ee

By assumption, the states, describing the two considered energy levels,
correspond to natural orbitals. Consequently, the quantum-number conservation
condition is to be valid,
\be
\label{2.295}
< a_1^\dgr a_2 > \; = \; 0 \; .
\ee
Then, from Eq. (\ref{2.293}), we have
\be
\label{2.296}
< c_L^\dgr c_L > \; = \; < c_R^\dgr c_R > \; = \;
\frac{N}{2} \; ,
\ee
which means that the "left" and "right" sides are equally occupied. Also,
we find
\be
\label{2.297}
< c_L^\dgr c_R > \; = \; \frac{1}{2} \left ( < a_1^\dgr a_1 > -
< a_2^\dgr a_2 > \right ) \; .
\ee
In view of normalization (\ref{2.291}), we get
$$
< c_L^\dgr c_R > \; = \; < c_R^\dgr c_L > \; = \;
< a_1^\dgr a_1 > \; - \; \frac{N}{2} \; .
$$

The value $<a_1^\dgr a_1>$ depends on the strength of atomic interactions
and temperature. These two factors deplete $<a_1^\dgr a_1>$ from $N$, which
can be represented as
\be
\label{2.298}
<a_1^\dgr a_1> \; = \; N \left ( 1 \; - \;
\frac{\dlt}{2}\right ) \qquad (0 \leq \dlt \leq 1) \; ,
\ee
where $\dlt$ is a depletion factor. The occupation number of the second level
becomes
\be
\label{2.299}
<a_2^\dgr a_2> \; = \; \frac{N}{2}\; \dlt \; .
\ee
And Eq. (\ref{2.297}) yields
\be
\label{2.300}
<c_L^\dgr c_R> \; = \; \frac{N}{2} (1 - \dlt) \; .
\ee

In an ideal gas at zero temperature, there is no depletion, $\dlt=0$, so that
all atoms condense onto the lowest level,
$$
<a_1^\dgr a_1> \; = \; N \; , \qquad <a_2^\dgr a_2> \; = \; 0
\qquad (\dlt=0) \; .
$$
For high temperature or strong repulsive interactions, the depletion is
maximal, $\dlt=1$, which gives
$$
<a_1^\dgr a_1> \; = \; <a_2^\dgr a_2> \; = \;
\frac{N}{2} \qquad (\dlt=1) \; .
$$
The latter would mean that there appears the fragmented condensate. Such
a conclusion, however, should not be treated seriously, if one remembers
that the studied model has meaning only for very low temperatures and weak
interactions. High temperatures and strong interactions would destroy any
condensate as such, so that the arising fragmented condensate at those
conditions is nothing but an artifact of an oversimplified model.

One could notice that the occupation numbers (\ref{2.298}) and (\ref{2.299})
are macroscopic for any finite depletion $\dlt>0$. Thus one could hope that
there exist such low temperatures and weak interactions, when the depletion
is already nonzero, however the two-level model is still appropriate, hence,
the fragmented condensate could occur. But this hope seems to be invalid.
The problem is that in any confining potential, including the double-well
potential, the spectrum is countable, so that there are many other levels,
except the two considered. The stronger the interactions, the closer the
double-well spectrum to that of the harmonic oscillator [205]. No matter
how small the temperature and interactions, they will spread atoms over
higher levels, making all of them populated, so that the macroscopic
population could remain solely on the single lowest level. In any confining
potential, the fragmented condensate could exist only during finite time,
as a nonequilibrium substance, as it happens for coherent topological modes
[174--196]. But an equilibrium fragmented condensate, corresponding to two
or more levels, seems to be impossible in a confining potential.

One sometimes calls the fragmented condensate the localized parts of the
same condensate in a periodic potential, as in an optical lattice. This
terminology, however, is not justified, since the lattice-site indices are
not quantum numbers, in the same way as the "left"nd "right" indices in Eq.
(\ref{2.300}) are not good quantum numbers. Till there exists any tunneling
between the lattice sites, there is no fragmentation, but there is the sole
condensate with a periodic wave function. And if there is no tunneling, then
again there is no fragmentation, but there can be merely several spatial
separated condensates or no condensate at all, but an insulating Mott phase.

Concluding, equilibrium fragmented condensates could arise in model
considerations. However one should be cautious interpreting them as actually
existing in real physical systems. One should not forget the limitations of
oversimplified models, sometimes yielding artificial results having no
counterparts in real physical systems.

\section{Regular Optical Lattices}

\subsection{Optical Lattices}

Optical lattices are formed by standing waves created by laser beams. Optical
potentials are due to the interaction of the laser electric field with atomic
transition dipoles corresponding to transitions between two internal atomic
energy levels [16]. The laser frequency is taken to be far detuned from the
atomic resonance, which allows for the definition of an effective optical
potential. For large detuning, the excited level can be eliminated in the
adiabatic approximation [16--18,206].

The {\it optical potential}, created by laser beams, in three dimensions,
has the form
\be
\label{3.1}
V_L(\br) = \sum_{\al=1}^3 V_\al \sin^2\left ( k_0^\al r_\al\right ) \; ,
\ee
in which the wave vector $\bk_0\equiv\{ k_0^\al\}$ has the components
\be
\label{3.2}
k_0^\al = \frac{2\pi}{\lbd_\al} = \frac{\pi}{a_\al} \qquad
\left ( a_\al = \frac{\lbd_\al}{2} \right ) \; ,
\ee
related to the laser wavelength $\lbd_\al$. Potential (\ref{3.1}) can be
rewritten as

\be
\label{3.3}
V_L(\br) = \frac{3}{2} \; V_0 \; - \; \frac{1}{2} \;
\sum_{\al=1}^3 V_\al \cos \left ( 2k_0^\al r_\al\right ) \; ,
\ee
where
\be
\label{3.4}
V_0 \equiv \frac{1}{3} \; \sum_{\al=1}^3 V_\al \; .
\ee
The wave vector $\bk_0$ defines the {\it recoil energy}
\be
\label{3.5}
E_R \equiv \frac{k_0^2}{2m} \qquad ( k_0 \equiv |\bk_0| ) \; .
\ee
The lattice spacing in the $\al$-spatial direction is $a_\al$.

It is possible to create optical lattices in one and two dimensions as well.
For instance, the one-dimensional optical potential is
\be
\label{3.6}
V_L(z) = V_0\sin^2(k_0 z) \; .
\ee
Then the recoil energy (\ref{3.5}) becomes
\be
\label{3.7}
E_R = \frac{\pi^2}{2ma^2} \qquad
\left ( k_0 = \frac{\pi}{a} \right ) \; ,
\ee
where $a$ is the lattice spacing.

Keeping in mind, the local interaction potential (\ref{2.174}), the energy
Hamiltonian (\ref{2.156}) is
$$
\hat H = \int \hat\psi^\dgr(\br) \left ( -\;
\frac{\nabla^2}{2m} + U + V_L \right ) \hat\psi(\br) \; d\br  \; +
$$
\be
\label{3.8}
+ \frac{1}{2} \; \Phi_0 \int \hat\psi^\dgr(\br) \hat\psi^\dgr(\br)
\hat\psi(\br) \hat\psi(\br) \; d\br \; ,
\ee
in which $U=U(\br)$ is an external, say trapping, potential and
$V_L=V_L(\br)$ is an optical potential. The field operator $\hat\psi(\br)$
is the Bogolubov-shifted operator (\ref{2.86}).

For the system to be stable, the particle dispersion
\be
\label{3.9}
\Dlt^2(\hat N) = T \; \frac{\prt N}{\prt\mu} = N \;
\frac{T}{\rho} \; \frac{\prt\rho}{\prt\mu}
\ee
must satisfy the stability condition (\ref{2.249}), which guarantees that the
isothermal compressibility
\be
\label{3.10}
\kappa_T = \frac{\Dlt^2(\hat N)}{\rho TN} = \frac{1}{\rho^2}\;
\frac{\prt\rho}{\prt\mu}
\ee
is positive and finite.

In an optical lattice, the total number of particles $N$ does not necessarily
coincide with the number of the lattice sites $N_L$. Hence, the {\it filling
factor}
\be
\label{3.11}
\nu \equiv \frac{N}{N_L} \qquad ( 0 < \nu <\infty)
\ee
can be any positive finite number. Defining the {\it mean lattice spacing}
\be
\label{3.12}
a \equiv \left ( \frac{V}{N_L} \right )^{1/d}
\ee
for a $d$-dimensional lattice, the filling factor (\ref{3.11}) can be
represented as
\be
\label{3.13}
\rho a^d  =\nu \; .
\ee
Then, compressibility (\ref{3.10}) takes the form
\be
\label{3.14}
\kappa_T = \frac{a^d}{\nu^2} \; \frac{\prt\nu}{\prt\mu} \; .
\ee

The general consideration of Bose systems with broken gauge symmetry, given
in Sec. 2, is applicable to systems with arbitrary external potentials.
Therefore, similarly to Eq. (\ref{2.102}), the grand Hamiltonian for Bose
atoms in an optical lattice is
\be
\label{3.15}
H = \hat H - \mu_0 \hat N_0 - \mu_1 \hat N_1 - \hat \Lbd \; ,
\ee
where $\hat H$ is the energy Hamiltonian (\ref{3.8}) and $\hat\Lbd$ is the
linear killer (\ref{2.96}).

The lattice parameters can be varied in a rather wide range [16--18]. The
typical experimental values for the laser wavelengths, used in creating
optical lattices, are of the order $\lbd\sim 10^{-5}-10^{-4}$ cm; then the
mean lattice spacing is a $a\sim 10^{-5}-10^{-4}$ cm and the recoil energy
is $E_R\sim 10^4-10^5$ Hz. The ratio of the potential depth $V_0$ to the
recoil energy is $V_0/E_R\sim0.1-100$, so that quite deep and very shallow
lattices can be formed. The typical linear lattice size is $L\sim 10^{-3}$
cm.

Generally, it is possible to create not only periodic lattices, but also
quasiperiodic lattices, in which at least one of the spatial directions is
subject to the action of two or more periodic potentials with incommensurate
periods. Such quasiperiodic lattices are similar to quasicrystals [207,208].

\subsection{Periodic Structures}

The periodic structure of a lattice is characterized by the {\it lattice
vector} $\ba=\{ a_\al\}$.  The lattice is formed by the set $\{\ba_i\}$ of
the vectors
\be
\label{3.16}
\ba_i =\{ n_i a_\al | \; n_i =0,\pm 1, \pm 2, \ldots \} \; .
\ee
The optical potential is periodic with respect to vectors (\ref{3.16}),
\be
\label{3.17}
V_L(\br+\ba_i) = V_L(\br) \; .
\ee

For the description of periodic structures, one employs the {\it Bloch
functions}
\be
\label{3.18}
\vp_{nk}(\br) =  e^{i\bk\cdot\br} f_{nk}(\br) \; ,
\ee
labelled by the band index $n$ and quasimomentum $\bk$, with a periodic
factor function
\be
\label{3.19}
f_{nk}(\br+\ba_i) = f_{nk}(\br) \; .
\ee
The quasimomentum $\bk$ pertains to the {\it Brillouin zone}
\be
\label{3.20}
{\cal B} = \left\{ \bk :\; -\; \frac{\pi}{a_\al} \leq k_\al \leq
\frac{\pi}{a_\al} \right \} \; .
\ee
The number of $k$-points in the Brillouin zone (\ref{3.20}) equals the
number $N_L$ of the real-space cells in the total lattice,
\be
\label{3.21}
\sum_{k\in{\cal B}} 1 = N_L \; .
\ee

A uniform system can be treated as a degenerate case of the periodic one,
when the Bloch function $\vp_{nk}(\br)$ reduces to the plane wave
$e^{i\bk\cdot\br}/\sqrt{V}$ and the periodic factor function $f_{nk}(\br)$,
to the constant $1/\sqrt{V}$.

Without the loss of generality, the factor function (\ref{3.19}) can be
chosen so that the property
\be
\label{3.22}
f_{nk}^*(\br) = f_{n,-k}(\br)
\ee
be valid. As a result, the Bloch function (\ref{3.18}) satisfies the equation
\be
\label{3.23}
\vp^*_{nk}(\br) = \vp_{n,-k}(\br) \; .
\ee

The real-space lattice $\{\ba_i\}$ can be related to the reciprocal lattice
$\{{\bf g}_i\}$ formed by the vectors ${\bf g}_i$ defined by the condition
\be
\label{3.24}
{\bf g}_i \cdot \ba_i = {\bf g} \cdot \ba = 2\pi \; .
\ee
The vectors of the reciprocal lattice and of quasimomentum pertain to
different sets, since $\bk\in{\cal B}$ and ${\bf g}\in\{\bg_i\}$,
because of which they, generally, do not coincide. Consequently, the property
\be
\label{3.25}
\frac{1}{V} \; \int e^{i(\bk-\bp+\bg)\cdot\br} \; d\br =
\dlt_{kp} \dlt_{g0}
\ee
holds. Here and in what follows, the spatial integration is over the whole
system volume $V$.

The periodic function (\ref{3.19}) can be expanded over the reciprocal lattice
as
\be
\label{3.26}
f_{nk}(\br) = \frac{1}{\sqrt{V}} \;
\sum_g b_{gnk} e^{i\bg\cdot\br} \; ,
\ee
where the summation is over all reciprocal vectors $\bg\in\{\bg_i\}$. The
coefficient in Eq. (\ref{3.26}) is
$$
b_{gnk}  = \frac{1}{\sqrt{V}} \; \int f_{nk}(\br)
e^{-i\bg\cdot\br} \; d\br \; .
$$
Using expansion (\ref{3.26}), the Bloch function (\ref{3.18}) can be written
as
\be
\label{3.27}
\vp_{nk}(\br) = \frac{1}{\sqrt{V}} \; \sum_g b_{gnk}
e^{i(\bg+\bk)\cdot\br} \; ,
\ee
with
$$
b_{gnk}  = \frac{1}{\sqrt{V}} \; \int \vp_{nk}(\br)
e^{-i(\bg+\bk)\cdot\br} \; d\br \; .
$$

The Bloch functions are orthonormal,
\be
\label{3.28}
\int \vp^*_{mk}(\br) \vp_{np}(\br) \; d\br =
\dlt_{mn} \dlt_{kp} \; ,
\ee
and generate a complete basis, for which
\be
\label{3.29}
\sum_n \; \sum_k \; \vp_{nk}(\br) \vp_{nk}^*(\br') =
\dlt(\br -\br') \; .
\ee
Here and in what follows, summation over $\bk$ implies the summation over
the Brillouin zone (\ref{3.20}). From Eq. (\ref{3.27}), one has
$$
\sum_g b_{gmk}^* \; b_{gnp} = \dlt_{mn} \dlt_{kp} \; .
$$

Another basis, often employed for describing periodic structures, is
formed by Wannier functions, which are related to the Bloch functions
through the Fourier transformation
\be
\label{3.30}
w_{nj}(\br) = \frac{1}{\sqrt{N_L}} \; \sum_k \vp_{nk}(\br) \;
e^{-i\bk\cdot\ba_j} \; ,
\ee
with the summation over $\bk\in{\cal B}$, the inverse transform being
\be
\label{3.31}
\vp_{nk}(\br) = \frac{1}{\sqrt{N_L}} \;
\sum_j w_{nj}(\br) e^{i\bk\cdot\ba_j} \; ,
\ee
where the summation with respect to $j$ is over the whole lattice $\{ a_j\}$,
with $j=1,2,\ldots,N_L$.

The Wannier functions are defined up to a phase factor that can always
be chosen such that to make the Wannier functions real and well localized
[209,210]. The scalar product between the Bloch and Wannier functions
$$
\int \vp_{mk}^*(\br) w_{nj}(\br) \; d\br =
\frac{\dlt_{mn}}{\sqrt{N_L}}\; e^{-i\bk\cdot\ba_j}
$$
shows that these functions are asymptotically orthogonal for $N_L\ra\infty$,
\be
\label{3.32}
\lim_{N_L\ra\infty} \; \int \vp_{mk}^*(\br)
w_{nj}(\br) \; d\br = 0 \; .
\ee

Using the equations
$$
\frac{1}{N_L} \; \sum_k e^{i\bk\cdot(\ba_i-\ba_j)} = \dlt_{ij} \; ,
\qquad \frac{1}{N_L} \; \sum_j e^{i(\bk-\bp)\cdot\ba_j} =
\dlt_{kp} \; ,
$$
one can make it sure that Wannier functions are orthonormal,
\be
\label{3.33}
\int w_{mi}^*(\br) w_{nj}(\br) \; d\br = \dlt_{mn} \dlt_{ij} \; ,
\ee
and form a complete basis, since
\be
\label{3.34}
\sum_{nj} w_{nj}(\br) w_{nj}^*(\br') = \dlt(\br-\br') \; .
\ee

As is mentioned above, Wannier functions can be made real, so that
\be
\label{3.35}
w_{nj}^*(\br) = w_{nj}(\br) \; ,
\ee
which is connected with property (\ref{3.23}) of Bloch functions.

In view of Eqs. (\ref{3.30}) and (\ref{3.27}), one has
\be
\label{3.36}
w_{nj}(\br) = \frac{1}{\sqrt{VN_L}} \; \sum_{g,k} b_{gnk}
e^{i(\bg+\bk)\cdot(\br-\ba_j)} \; .
\ee
This shows that the Wannier function can be represented as
\be
\label{3.37}
w_{nj}(\br) \equiv w_n(\br - \ba_j) \; .
\ee
From here it follows that the coefficient
$$
b_{gnk} = \sqrt{\frac{N_L}{V} } \;
\int w_n(\br) \; e^{-i(\bg+\bk)\cdot\br} \; d\br
$$
enjoys the properties
$$
b_{gnk} \equiv b_n (\bg + \bk) \; , \qquad
b_n^* (\bg + \bk) = b_n ( -\bg -\bk) \; .
$$
The Bloch function (\ref{3.31}) can be represented as
\be
\label{3.38}
\vp_{nk}(\br) = \frac{1}{\sqrt{N_L}} \; \sum_j
w_n(\br-\ba_j) e^{i\bk\cdot\ba_j} = \frac{1}{\sqrt{V}} \;
\sum_g b_n (\bg + \bk) e^{i(\bg+\bk)\cdot\br} \; .
\ee
From here, we see that
\be
\label{3.39}
\vp_{n\; k+g} (\br) = \vp_{nk}(\br) \; .
\ee

The bases of Bloch functions, $\{\vp_{nk}(\br)\}$, and that of
Wannier functions, $\{ w_n(\br-\ba_j)\}$, seem to be equivalent for
characterizing periodic structures. This, however, is not completely
correct. As is discussed in Sec. 2.17, the indices of spatial
localization are not good quantum numbers. This concerns as well the
lattice indices $j=1,2,\ldots,N_L$. Because of the latter, Wannier
functions cannot serve as natural orbitals. But Bloch functions can.
Thus, the Bloch function, corresponding to the lowest band $n=0$ and
to the zero quasimomentum $\bk=0$, is the {\it condensate natural
orbital}
\be
\label{3.40}
\vp_0(\br) \equiv \lim_{k\ra 0} \vp_{0k}(\br) \; .
\ee
According to relation (\ref{3.18}),
$$
\vp_0(\br) = f_0(\br) \; ,
$$
hence, the condensate natural orbital is purely periodic. It is, by
definition, normalized,
\be
\label{3.41}
\int |\vp_0(\br)|^2 \; d\br =  1 \; ,
\ee
and, in compliance with Eq. (\ref{3.38}), is of the form
\be
\label{3.42}
\vp_0(\br) = \frac{1}{\sqrt{V}} \; \sum_g
b_0(\bg) e^{i\bg\cdot\br} \; .
\ee
Its relation to the Wannier functions, given by the equation
\be
\label{3.43}
\vp_0(\br) = \frac{1}{\sqrt{N_L}} \; \sum_j w_0 (\br-\ba_j) \; ,
\ee
demonstrates that, if $\vp_0(\br)$ is the condensate natural orbitals, then
the associated Wannier function
\be
\label{3.44}
w_0(\br-\ba_j) = \frac{1}{\sqrt{N_L}} \; \sum_k
\vp_{0k}(\br) e^{-i\bk\cdot\ba_j}
\ee
is not a condensate natural orbital, since the latter involves the Bloch
functions $\vp_{0k}(\br)$ of uncondensed atoms, with $\bk\neq 0$.

In thermodynamic limit, when $N\ra\infty$, $V\ra\infty$, and $N_L\ra\infty$,
the summation over quasimomenta is replaced by the integration according to
the rule
\be
\label{3.45}
\sum_k \; \longrightarrow \; V \int_{{\cal B}} \;
\frac{d\bk}{(2\pi)^d} \; ,
\ee
in which the integration is over the $d$-dimensional Brillouin zone
${\cal B}$.

\subsection{Condensate in Lattices}

Bloch functions can serve as natural orbitals. The lowest-band
zero-quasimomentum Bloch function corresponds to BEC. The condensate wave
function, entering the Bogolubov shift (\ref{2.86}), is
\be
\label{3.46}
\eta(\br) \; = \; \sqrt{N_0} \; \vp_0(\br) \; ,
\ee
with $\vp_0(\br)$ being the condensate natural orbital (\ref{3.40}). As usual,
the condensate function (\ref{3.46}) is normalized to the number of condensed
particles
\be
\label{3.47}
N_0 = \int |\eta(\br)|^2 \; d\br \; .
\ee
In line with expansion (\ref{3.43}), the condensate function (\ref{3.46}) reads
as
\be
\label{3.48}
\eta(\br) \; = \; \sqrt{ \frac{N_0}{N_L} } \;
\sum_j w_0(\br-\ba_j) \; .
\ee
This tells us again that $w_0(\br-\ba_j)$ cannot be treated as the condensate
wave function, but only combination (\ref{3.48}) forms the latter.

The field operators of uncondensed atoms can be expanded over the Bloch
functions,
\be
\label{3.49}
\psi_1(\br) = \sum_{nk} a_{nk} \vp_{nk} (\br) \; .
\ee
An expansion over Wannier functions is also admissible,
\be
\label{3.50}
\psi_1(\br) = \sum_{nj} c_{nj} w_n(\br - \ba_j) \; .
\ee

As is known from Chapter 2, the condensate function, by definition, is
orthogonal to the field operator of uncondensed atoms,
\be
\label{3.51}
\int \eta^*(\br) \psi_1(\br) \; d\br  = 0 \; .
\ee
Substituting here expansion (\ref{3.49}) requires that
\be
\label{3.52}
\lim_{k\ra 0} a_{0k} = 0 \; ,
\ee
which means that the term with $n=0$ and $\bk=0$ is excluded from sum
(\ref{3.49}).

From relations (\ref{3.30}) and (\ref{3.31}) between Bloch and Wannier
functions, it follows that
$$
\lim_{k\ra 0} a_{0k} = \frac{1}{\sqrt{N_L}} \; \sum_j c_{0j} \; .
$$
Therefore the property
\be
\label{3.53}
\sum_j c_{0j} = 0
\ee
must hold. Conditions (\ref{3.52}) and (\ref{3.53}) assure that the
operator of the number of uncondensed atoms
$$
\hat N_1 \equiv \int \psi_1^\dgr(\br) \psi_1(\br) \; d\br =
\sum_{nk} a_{nk}^\dgr a_{nk} = \sum_{nj} c_{nj}^\dgr c_{nj}
$$
enters additively into the operator
\be
\label{3.54}
\hat N = N_0 + \hat N_1
\ee
of the total number of particles.

Let us write explicitly the grand Hamiltonian (\ref{3.15}), assuming that
there are no external potentials disturbing the lattice, inserting there
the Bogolubov shift (\ref{2.86}), and involving expansion (\ref{3.49}) over
Bloch functions. This gives
\be
\label{3.55}
H = H^{(0)} + H^{(2)} + H^{(3)} + H^{(4)} \; ,
\ee
where the terms linear in $\psi_1$ are eliminated by the linear killer
(\ref{2.96}). The first term in sum (\ref{3.55}) is
$$
H^{(0)} = \int \eta^*(\br) \left ( - \; \frac{\nabla^2}{2m} + V_L -
\mu_0 \right ) \eta(\br) \; d\br \; +
$$
\be
\label{3.56}
 + \; \frac{1}{2} \; \Phi_0 \; \int |\eta(\br)|^4 \; d\br \; .
\ee
In order to avoid too cumbersome notation, let us combine the two indices
$\{ n,\bk\}$ into one index $k$. That is, the Bloch function $\vp_{nk}(\br)$
will be labelled simply as $\vp_k(\br)$, keeping in mind that here $k$ means
$n,\; \bk$. Then the second term in sum (\ref{3.55}) writes as
$$
H^{(2)} = \sum_{kp} \left [ \int \vp_k^*(\br) \left ( -\;
\frac{\nabla^2}{2m} + V_L - \mu_1 + 2\Phi_0 |\eta|^2 \right )
\vp_p(\br)\; d\br \right ] a_k^\dgr a_p \; +
$$
\be
\label{3.57}
+ \; \frac{1}{2} \; \sum_{kp} \left ( \Phi_{kp} a_k^\dgr a_p^\dgr +
\Phi_{kp}^* a_p a_k \right ) \; ,
\ee
where
$$
\Phi_{kp} \equiv \Phi_0 \; \int \vp_k^*(\br) \vp_p^*(\br)
\eta^2(\br) \; d\br \; .
$$
The third term is
\be
\label{3.58}
H^{(3)} = \sum_{kpq} \left ( \Phi_{kpq} a_k^\dgr a_p^\dgr a_q +
\Phi_{kpq}^* a_q^\dgr a_p a_k \right ) \; ,
\ee
in which
$$
\Phi_{kpq} \equiv \Phi_0 \; \int \vp_k^*(\br) \vp_p^*(\br)
\vp_q(\br) \eta(\br) \; d\br \; .
$$
And the last term in Eq. (\ref{3.55}) reads as
\be
\label{3.59}
H^{(4)} = \frac{1}{2} \; \sum_{kpql} \Phi_{kpql}
a_k^\dgr a_p^\dgr a_q a_l \; ,
\ee
with the notation
$$
\Phi_{kpql} \equiv \Phi_0 \; \int \vp_k^*(\br)
\vp_p^*(\br) \vp_q(\br) \vp_l(\br) \; d\br \; .
$$

Terms (\ref{3.58}) and (\ref{3.59}) can be simplified by invoking the
Hartree-Fock-Bogolubov (HFB) approximation. The linear in $a_k$ terms,
appearing in $H^{(3)}$, are assumed to be cancelled by the linear killer
(\ref{2.96}). The fourth term (\ref{3.59}) becomes
$$
H^{(4)} = \frac{1}{2} \; \sum_{kpq} \left (
4 \Phi_{kqqp} n_q a_k^\dgr a_p + \Phi_{kpqq} \sgm_q a_k^\dgr a^\dgr_p
+ \Phi^*_{kpqq} \sgm^*_q a_p a_k \right ) \; -
$$
\be
\label{3.60}
 - \; \frac{1}{2} \; \sum_{kp} \left ( 2 \Phi_{kppk} n_k n_p +
\Phi_{kkpp} \sgm_k^* \sgm_p \right ) \; ,
\ee
in which
$$
n_k \; \equiv \; < a_k^\dgr a_k > \; , \qquad
\sgm_k\; \equiv \; < a_k a_{-k} >
$$
and the quantum-number conservation conditions, valid for natural orbitals,
$$
< a_k^\dgr a_p > \; = \; \dlt_{kp} n_k \; , \qquad
< a_k a_p > \; = \; \dlt_{-kp} \sgm_k
$$
are taken into account, where $-k$ means $n$, $-\bk$.

By introducing the notation
$$
\om_{kp} \equiv \int \vp_k^*(\br) \left ( -\; \frac{\nabla^2}{2m} +
V_L + 2 \Phi_0 |\eta|^2 \right ) \vp_p(\br) \; d\br \; +
$$
\be
\label{3.61}
+ \; 2\sum_q \Phi_{kqqp} n_q \; - \; \mu_1 \dlt_{kp}
\ee
and
\be
\label{3.62}
\Dlt_{kp} \equiv \Phi_{kp} + \sum_q \Phi_{kpqq} \sgm_q \; ,
\ee
the grand Hamiltonian (\ref{3.55}) acquires the form
$$
H = E_{HFB} + \sum_{kp} \om_{kp} a_k^\dgr a_p \; +
$$
\be
\label{3.63}
+ \; \frac{1}{2} \; \sum_{kp} \left (
\Dlt_{kp} a_k^\dgr a_p^\dgr + \Dlt_{kp}^* a_p a_k \right ) \; ,
\ee
in which
\be
\label{3.64}
E_{HFB} \equiv H^{(0)} - \; \frac{1}{2} \; \sum_{kp} \left (
2 \Phi_{kppk} n_k n_p + \Phi_{kkpp} \sgm_k^* \sgm_p \right ) \; .
\ee

Hamiltonian (\ref{3.63}) can be diagonalized by a canonical transformation.
In order to simplify the consideration, we may assume that the diagonal
elements in the summation over $k$ and $p$ give the main contribution in
Eq. (\ref{3.63}). This amounts to using the diagonal approximation for
$\om_{kp}$ and $\Dlt_{kp}$, so that
\be
\label{3.65}
\om_{kp} = \dlt_{kp} \om_k \; , \qquad
\Dlt_{kp} = \dlt_{-kp} \Dlt_k \; ,
\ee
where $\dlt_{-kp}$ implies $\dlt_{mn}\dlt_{-kp}$, since $-k$ means $n,-\bk$.
Then Eq. (\ref{3.61}) reduces to
$$
\om_k = \int \vp_k^*(\br) \left ( -\; \frac{\nabla^2}{2m} + V_L +
2\Phi_0 |\eta|^2 \right ) \vp_k(\br) \; d\br \; +
$$
\be
\label{3.66}
+ \; 2 \sum_q \Phi_{kqqk} n_q \; - \; \mu_1
\ee
and Eq. (\ref{3.62}) gives
\be
\label{3.67}
\Dlt_k = \Phi_{-kk} + \sum_q \Phi_{-kkqq} \sgm_q \; .
\ee
Hamiltonian (\ref{3.63}) becomes
\be
\label{3.68}
H = E_{HFB} + \sum_k \om_k a_k^\dgr a_k \; + \;
\frac{1}{2} \; \sum_k \left ( \Dlt_k a_k^\dgr a_{-k}^\dgr +
\Dlt^*_k a_{-k} a_k \right ) \; .
\ee
This is in direct analogy with the grand Hamiltonian in the HFB
approximation for uniform systems [94--98], so that all calculations
can be done in the same way. The difference from the uniform case is
in different $\om_k$ and $\Dlt_k$ defined in Eqs. (\ref{3.66}) and
(\ref{3.67}) and in the fact that the quasimomentum pertains to the
Brillouin zone.

Following the same procedure as for the uniform system [94--98],
and restoring the double indexation $n,\bk$ for $k$, we obtain the
Bogolubov-type spectrum
\be
\label{3.69}
\ep_{nk} \; = \; \sqrt{\om_{nk}^2 -\Dlt^2_{nk} } \; ,
\ee
consisting of several branches labelled by the band index $n=0,1,2,\ldots$.

In agreement with the condensation condition (\ref{2.126}), we require that
\be
\label{3.70}
\lim_{k\ra 0} \ep_{0k} = 0 \; , \qquad \ep_{0k} \geq 0 \; .
\ee
This is equivalent to the condition
$$
\lim_{k\ra 0} ( \om_{0k} - \Dlt_{0k} ) = 0 \; .
$$
From here we find the Lagrange multiplier
$$
\mu_1 = \lim_{k\ra 0} \left \{ \int \vp_k^*(\br) \left ( - \;
\frac{\nabla^2}{2m} + V_L + 2\Phi_0 |\eta|^2 \right )
\vp_k(\br) \; d\br \; - \right.
$$
\be
\label{3.71}
\left. -\; \Phi_{-kk} + \sum_q \left ( 2 \Phi_{kqqk} n_q -
\Phi_{-kkqq}\sgm_q \right ) \right\} \; ,
\ee
where again the short-hand notation is used, with $k$ instead of $n,\bk$.

The condensate-function equation is derived similarly to Eq. (\ref{2.175}),
which in the HFB approximation for a lattice gives
\be
\label{3.72}
\left \{ -\; \frac{\nabla^2}{2m} + V_L(\br) + \Phi_0 [
\rho_0(\br) + 2\rho_1(\br)] \right \} \eta(\br) + \Phi_0 \sgm_1(\br)
\eta^*(\br) = \mu_0 \eta(\br) \; .
\ee
Here the densities of condensed and uncondensed atoms and the anomalous
average are
$$
\rho_0(\br) = |\eta(\br)|^2 \; , \qquad \rho_1(\br) =
\sum_{nk} n_{nk} |\vp_{nk}(\br)|^2 \; ,
$$
\be
\label{3.73}
\sgm_1(\br) = \sum_{nk} \sgm_{nk} \vp_{nk}(\br) \vp_{n,-k}(\br) \; .
\ee

The eigenproblem (\ref{3.72}) defines the condensate function $\eta(\br)$
and the Lagrange multiplier $\mu_0$ that guarantees the validity of the
normalization condition (\ref{3.47}). Using the latter yields
$$
\mu_0 = \frac{1}{N_0} \; \int \eta^*(\br) \left\{ -\;
\frac{\nabla^2}{2m} + V_L(\br) + \Phi_0 [ \rho_0(\br) +
2\rho_1(\br)] \right \} \eta(\br) \; d\br \; +
$$
\be
\label{3.74}
+ \; \frac{\Phi_0}{N_0} \; \int \sgm_1(\br) \left (
\eta^*(\br)\right )^2 \; d\br \; .
\ee
This is to be compared with multiplier (\ref{3.71}). Taking in the latter
the limit $k\ra 0$, we have
$$
\lim_{k\ra 0} \Phi_{-kk} = \Phi_0 \int |\eta(\br)|^4 \; d\br \; ,
$$
$$
\lim_{k\ra 0} \Phi_{kppk} = \frac{\Phi_0}{N_0} \; \int
\rho_0(\br) |\vp_p(\br)|^2 \; d\br \; ,
$$
$$
\lim_{k\ra 0} \Phi_{-kkpp} = \frac{\Phi_0}{N_0} \; \int
\left [ \eta^*(\br) \vp_p(\br) \right ]^2 \; d\br \; ,
$$
since
$$
\lim_{k\ra 0} \vp_k(\br) = \frac{\eta(\br)}{\sqrt{N_0}} \; .
$$
Taking into consideration Eqs. (\ref{3.73}), we find
$$
\mu_1 = \frac{1}{N_0} \; \int \eta^*(\br)\left \{ - \;
\frac{\nabla^2}{2m} + V_L(\br) + \Phi_0 [\; \rho_0(\br) +
2\rho_1(\br)\; ] \right \} \eta(\br) \; d\br \; -
$$
\be
\label{3.75}
- \; \frac{\Phi_0}{N_0} \; \int \sgm_1(\br) \left (
\eta^*(\br)\right )^2 \; d\br \; .
\ee
As is seen, $\mu_0\neq\mu_1$. They coincide only in the limit of
asymptotically weak interactions, when the Bogolubov approximation
becomes applicable. In this approximation
$$
\rho_1(\br) \; \ra \; 0 \; , \qquad \sgm_1(\br) \; \ra \; 0
\qquad (\Phi_0 \ra 0) \; ,
$$
hence
$$
\rho_0(\br) \; \ra \; \rho(\br) \equiv
\rho_0(\br) +\rho_1(\br) \; .
$$
As a result,
$$
\mu_0 \simeq \mu_1 \simeq \frac{1}{N_0} \; \int \eta^*(\br)
\left [ -\; \frac{\nabla^2}{2m} + V_L(\br) + \Phi_0 \rho(\br)
\right ] \eta(\br) \; d\br \qquad (\Phi_0\ra 0) \; .
$$
The Bogolubov approximation for weakly nonideal gas in tight-binding
bands was considered by Ramakumar and Das [211].

\subsection{Operator of Momentum}

The operator of momentum defines the dissipated heat (\ref{2.153}) and,
respectively, the superfluid fraction (\ref{2.155}). According to Eqs.
(\ref{2.138}) and (\ref{2.139}), it can be introduced through the relation
\be
\label{3.76}
\hat\bP \equiv \lim_{v\ra 0} \; \frac{\prt\hat H_v}{\prt\bv} \; ,
\ee
in which $\hat H_v$ is the energy Hamiltonian
\be
\label{3.77}
\hat H_v = \hat H + \int \hat\psi(\br) \left ( - i\bv \cdot \nabla +
\frac{mv^2}{2} \right ) \hat\psi(\br) \; d\br \; ,
\ee
obtained by substituting into $\hat H[\hat\psi]$ the Galilean-transformed
field operator (\ref{2.151}). This gives the standard form
\be
\label{3.78}
\hat\bP = \int \hat\psi^\dgr(\br) \left ( -i\nabla \right )
\hat\psi(\br) \; d\br \; ,
\ee
but with the Bogolubov-shifted field operator (\ref{2.86}).

We may notice that
$$
\int \eta^*(\br) \left ( -i\nabla \right )  \eta(\br) \; d\br =
N_0 \sum_g |b_0(\bg)|^2 \bg = 0 \; ,
$$
because of the property
$$
| b_0(-\bg)| = | b_0(\bg)| \; ,
$$
derived in Sec. 3.2. Also,
$$
\int \eta^*(\br) \left ( -i\nabla \right )
\psi_1(\br) \; d\br = 0 \; ,
$$
due to condition (\ref{3.52}). Therefore the operator of momentum
(\ref{3.78}) is defined only in terms of the field operators of uncondensed
atoms,
\be
\label{3.79}
\hat\bP = \int \psi_1^\dgr(\br) \left ( -i\nabla \right )
\psi_1(\br) \; d\br \; .
\ee

The field operator $\psi_1(\br)$ can be expanded either over Wannier
functions or over Bloch functions, as in Eqs. (\ref{3.49}) and (\ref{3.50}),
with the relations
\be
\label{3.80}
c_{nj} = \frac{1}{\sqrt{N_L}} \; \sum_k
a_{nk} e^{i\bk\cdot\ba_j} \; , \qquad a_{nk} =
\frac{1}{\sqrt{N_L}} \; \sum_j c_{nj} e^{-i\bk\cdot\ba_j} \; .
\ee
Defining the matrix elements over Wannier functions,
\be
\label{3.81}
\bp_{ij}^{mn} \equiv \int w_m(\br-\ba_i) ( -i\nabla)
w_n(\br-\ba_j) \; d\br \; ,
\ee
and over Bloch functions
\be
\label{3.82}
\bq_{kp}^{mn} \equiv \int \vp_{mk}^*(\br) ( -i\nabla)
\vp_{np}(\br) \; d\br \; ,
\ee
we get the representations of the momentum operator (\ref{3.79}) in terms
of the Wannier, $c_{nj}$, or Bloch, $a_{nk}$, operators as
\be
\label{3.83}
\hat\bP = \sum_{mn} \; \sum_{ij} \bp_{ij}^{mn} c_{mi}^\dgr c_{nj}
\ee
and, respectively,
\be
\label{3.84}
\hat\bP = \sum_{mn} \; \sum_{kp} \bq_{kp}^{mn} a_{mk}^\dgr a_{np} \; .
\ee
The matrix elements (\ref{3.81}) and (\ref{3.82}) are connected through the
transformations
$$
\bp_{ij}^{mn} = \frac{1}{N_L} \; \sum_{kp} \bq_{kp}^{mn} \;
e^{i\bk\cdot\ba_i -i\bp\cdot\ba_j} \; ,
$$
\be
\label{3.85}
\bq_{kp}^{mn} = \frac{1}{N_L} \; \sum_{ij} \bp_{ij}^{mn} \;
e^{-i\bk\cdot\ba_i +i\bp\cdot\ba_j} \; .
\ee

In Eq. (\ref{3.81}), we use the advantage of choosing Wannier functions as
being real. This equation can also be written as
\be
\label{3.86}
\bp_{ij}^{mn} = \int w_m(\br-\ba_{ij} ) (-i\nabla) w_n(\br) \; d\br \; ,
\ee
where $\ba_{ij}\equiv\ba_i-\ba_j$, which allows for the use of the
representation
\be
\label{3.87}
\bp_{ij}^{mn} \equiv \bp_{mn}(\ba_{ij}) \; .
\ee
Matrix elements (\ref{3.81}) have the property
\be
\label{3.88}
\left ( \bp_{ij}^{mn}\right )^* = \bp_{ji}^{nm} = - \bp_{ij}^{mn} \; ,
\ee
from which it follows that
\be
\label{3.89}
\bp_{jj}^{nn} = \bp_{nn}(0) = 0 \; .
\ee
Using Eq. (\ref{3.87}), the second of the matrix elements (\ref{3.85}) can
be rewritten as
$$
\bq_{kp}^{mn} = \frac{1}{N_L} \; \sum_{ij} \bp_{mn}(\ba_{ij}) \;
e^{- i \bk\cdot\ba_{ij}-i(\bk-\bp)\cdot\ba_j} \; .
$$
From here, we get
\be
\label{3.90}
\bq_{kp}^{mn} = \dlt_{kp} \bq_k^{mn} \; ,
\ee
where the diagonal element is
\be
\label{3.91}
\bq_k^{mn} = \sum_j \bp_{mn}(\ba_j)\; e^{- i \bk\cdot\ba_j} \; .
\ee
The latter enjoys the property
\be
\label{3.92}
\left ( \bq_k^{mn}\right )^* = \bq_k^{nm} = -\bq_{-k}^{mn} \; .
\ee
Thence, momentum (\ref{3.84}) takes the form
\be
\label{3.93}
\hat\bP = \sum_{mn} \; \sum_k \bq_k^{mn} a_{mk}^\dgr a_{nk} \; .
\ee
Using the quantum-number conservation condition
\be
\label{3.94}
< a_{mk}^\dgr a_{np} > \; = \; \dlt_{mn} \dlt_{kp}
< a_{nk}^\dgr a_{nk} > \; ,
\ee
we find that the total average momentum
\be
\label{3.95}
<\hat\bP> \; = \; \sum_{nk} \bq_k^{nm}
< a_{nk}^\dgr a_{nk}>\; = \; 0 \; ,
\ee
owing to property (\ref{3.92}) and assuming that $<a_{nk}^\dgr a_{nk}>$ is
symmetric with respect to the inversion $\bk$ to $-\bk$. This means that 
in the coordinate frame, coupled with the lattice, the total average 
momentum is zero, as it should be. If the lattice would be moving, then in 
the laboratory frame the distribution $<a_{nk}^\dgr a_{nk}>$ would not be 
symmetric with respect to the inversion of $\bk$.

\subsection{Tight-Binding Approximation}

Since Wannier functions can be made well localized [209,210], one can
assume that when atoms are close to the lattice site $\ba_j$, then they feel
the potential
\be
\label{3.96}
V_L(\br-\ba) \simeq \sum_\al \; \frac{m}{2}\; \om_\al^2
\left ( r_\al - a_j^\al \right )^2 \qquad (\br \approx \ba_j) \; ,
\ee
which is an expansion of the lattice potential (\ref{3.1}), so that
\be
\label{3.97}
\om_\al \equiv 2\; \sqrt{E_R V_\al} \; .
\ee
It is convenient to define the effective frequency
\be
\label{3.98}
\om_0 \equiv \left ( \prod_{\al=1}^d \om_\al \right )^{1/d} \; ,
\ee
in which $d$ is space dimensionality, and the effective localization length
\be
\label{3.99}
l_0 \equiv \frac{1}{\sqrt{m\om_0} } \; .
\ee

Good localization of atomic Wannier functions means that the localization
length (\ref{3.99}) is much smaller than the distance between the nearest
neighbors $a$, which can be expressed in several equivalent inequalities,
\be
\label{3.100}
\frac{l_0}{a} \ll 1 \; , \qquad k_0 l_0 \ll 1 \; ,
\ee
where $k_0$ is the modulus of the laser wave vector entering the recoil
energy (\ref{3.5}), which also allows us to write down the inequality
\be
\label{3.101}
\frac{E_R}{\om_0} \ll 1 \; .
\ee

For the harmonic potential (\ref{3.96}), the lowest-band localized Wannier
function can be approximated by the Gaussian form
\be
\label{3.102}
w(\br-\ba_j) = \left ( \frac{m\om_0}{\pi} \right )^{d/4} \;
\prod_\al \exp\left \{ -\; \frac{m}{2} \; \om_\al \left (
r_\al - a_j^\al \right )^2 \right \} \; .
\ee
For a cubic lattice, for which $\om_\al=\om_0$, this becomes
\be
\label{3.103}
w(\br-\ba_j) = \frac{1}{(\sqrt{\pi}\; l_0)^{d/2}} \;
\exp\left \{ - \; \frac{(\br-\ba_j)^2}{2l_0^2} \right \} \; .
\ee
The Wannier functions of higher bands could be approximated by the excited
wave functions of the harmonic oscillator. If one assumes that in the
lattice there are no such strong excitations that would transfer atoms to
higher excited bands, then one can limit oneself by considering only the
lowest band characterized by the approximate Wannier functions (\ref{3.102})
or (\ref{3.103}).

As a first example of using the tight-binding approximation, let us calculate
the momentum operator (\ref{3.93}), with the matrix element (\ref{3.91}), in
the single-band picture. Calculating the matrix element (\ref{3.81}), we meet
the integrals
$$
\int_0^\infty \; e^{-bx^2} \; {\rm sinh}\; x\; dx =
\sqrt{ \frac{\pi}{4b} }\; \exp\left ( \frac{1}{4b} \right )
\Phi\left ( \frac{1}{\sqrt{4b}} \right ) \; ,
$$
$$
\int_0^\infty \; x e^{-bx^2} \; {\rm cosh}\; x\; dx = \frac{1}{2b} +
\frac{1}{4b} \; \sqrt{ \frac{\pi}{b} }\; \exp\left ( \frac{1}{4b} \right )
\Phi\left ( \frac{1}{\sqrt{4b}} \right ) \; ,
$$
with the probability integral
$$
\Phi(x) \equiv \frac{2}{\sqrt{\pi}} \; \int_0^x \; e^{-t^2} \; dt \; .
$$
The latter enjoys the property
$$
\Phi(x) \simeq 1 \qquad (x \ra \infty) \; .
$$
Therefore, for small $b$, we find
$$
\int_0^\infty \; e^{-bx^2} \left ( x\; {\rm cosh}\; x - {\rm sinh}\; x
\right ) \; dx \simeq \frac{1}{4b} \; \sqrt{\frac{\pi}{b} } \;
\exp\left ( \frac{1}{4b} \right ) \qquad (b \ll 1) \; .
$$
Then, for element (\ref{3.81}), we obtain
\be
\label{3.104}
\bp_{ij} \equiv \bp(\ba_{ij} ) = i \; \frac{\ba_{ij}}{a_{ij}^2} \;
\exp \left ( - \; \frac{a_{ij}^2}{4l_0^2} \right ) \; ,
\ee
where $a_{ij}\equiv|\ba_{ij}|$ and $\ba_{ij}\equiv\ba_i-\ba_j$. The diagonal
element $\bp_{ii}=0$, in agreement with property (\ref{3.89}).

As is seen, the value of element (\ref{3.104}) exponentially decays for
increasing $a_{ij}$, which makes it possible to take into account only
the nearest neighbors. Then, for Eq. (\ref{3.91}), we have
$$
\bq_k = \sum_{<j>} \bp(\ba_j) \; e^{- i \bk\cdot\ba_j} \; ,
$$
where the summation is over the {\it nearest neighbors}.

For a $d$-dimensional {\it cubic lattice}, one has
$$
\sum_{<j>} \ba_j\; e^{-i\bk\cdot\ba_j} = - 2ia \sum_{\al=1}^d
\sin(k_\al a) {\bf e}_\al \; ,
$$
where ${\bf e}_\al$ is a unit vector, such that
$$
{\bf e}_\al^2 =  1 \; , \qquad
\sum_{\al=1}^d {\bf e}_\al^2 = d \; .
$$
Keeping in mind a cubic lattice and substituting Eq. (\ref{3.104}) into
Eq. (\ref{3.91}), we find
\be
\label{3.105}
\bq_k = \frac{2}{a} \; \exp\left ( -\; \frac{a^2}{4l_0^2}\right ) \;
\sum_\al \sin( k_\al a) {\bf e}_\al \; .
\ee
This is to be inserted into the operator of momentum (\ref{3.93}). Or element
(\ref{3.104}) can be used in Eq. (\ref{3.83}).

\subsection{Superfluidity in Lattices}

The general definition of the superfluid fraction is given by Eq.
(\ref{2.140}), which is valid for arbitrary systems. For an equilibrium
system, according to Eq. (\ref{2.155}), we come to the expression
\be
\label{3.106}
n_s = 1 \; - \; \frac{Q}{Q_0} \qquad
\left ( Q_0 \equiv \frac{3}{2}\; T\right ) \; ,
\ee
in which $Q$ is the dissipated heat
$$
Q  =  \frac{ <\bP^2>}{2mN} \; .
$$

To find the dissipated heat $Q$, let us treat the single-band case and use
the HFB approximation. As usual, the conservation conditions
$$
< a_k^\dgr a_p> \; = \; \dlt_{kp} n_k \; , \qquad
n_k \; \equiv \; < a_k^\dgr a_k > \; ,
$$
$$
< a_k a_p> \; = \; \dlt_{-kp} \sgm_k \; , \qquad
\sgm_k \; \equiv \; < a_k a_{-k} >
$$
are taken into account. Then, using the momentum operator (\ref{3.93}), we
obtain
\be
\label{3.107}
< \hat\bP^2 > \; = \; \sum_k \left ( n_k + n_k^2 - \sgm_k^2 \right )
\bq_k^2 \; ,
\ee
with the matrix element (\ref{3.91}). Employing here the explicit expressions
 for the normal $n_k$, and anomalous, $\sgm_k$, averages [57,94], we have
$$
n_k + n_k^2 - \sgm_k^2 = \frac{1}{4{\rm sinh}^2(\bt\ep_k/2)} \; ,
$$
where $\ep_k$ is the Bogolubov spectrum (\ref{3.69}). Using Eqs. (\ref{3.104})
and (\ref{3.105}), we have
\be
\label{3.108}
\bq_k^2 =  4| \bp(\ba)|^2 \; \sum_\al \sin^2(k_\al a) \; ,
\ee
where again a cubic lattice is considered and
\be
\label{3.109}
|\bp(\ba)|^2 = \frac{1}{a^2}\; \exp\left ( - \; \frac{a^2}{2l_0^2}
\right ) \; .
\ee
Thus, for the dissipated heat, entering the superfluid fraction (\ref{3.106}),
we derive
\be
\label{3.110}
Q = \frac{|\bp(\ba)|^2}{2m\rho} \; \int_{\cal B} \;
\frac{\sum_\al \sin^2(k_\al a)}{{\rm sinh}^2(\bt\ep_k/2)} \;
\frac{d\bk}{(2\pi)^d} \; .
\ee

The exponential factor in Eq. (\ref{3.109}) is small, since
$$
\frac{a^2}{l_0^2} = 3 \pi^2\; \sqrt{\frac{V_0}{E_R}} \; \gg \; 1
$$
by the condition of good localization (\ref{3.100}). But superfluidity can
exist in optical lattices even under well localized atomic Wannier functions.

\subsection{Transverse Confinement}

In experiments, one can create an optical lattice in one direction, while
atoms are tightly confined in two other directions. Then one obtains an
effective one-dimensional system in a periodic optical potential. To study
such a system, let us consider the case of weakly interacting atoms at zero
temperature, when practically all atoms are condensed, so that $N_0\approx N$.
Then the grand Hamiltonian (\ref{3.55}) reduces to form (\ref{3.56}). Let us
have a tight transverse trapping potential $U(\br_\perp)$, where $\br_\perp$
is the transverse vector. And let the optical lattice be imposed in the
$z$-direction. Then the Hamiltonian is
$$
H = \int \eta^*(\br) \left [ -\; \frac{\nabla^2}{2m} + U(\br_\perp) +
V_L(z) - \mu_0 \right ] \eta(\br) \; d\br \; +
$$
\be
\label{3.111}
+ \; \frac{1}{2} \; \Phi_0 \int |\eta(\br)|^4 \; d\br \; .
\ee

Let us assume that the transverse confinement is realized by a harmonic
potential $U(\br_\perp)$, with the transverse frequency $\om_\perp$. Then
the transverse localization length is
\be
\label{3.112}
l_\perp \equiv \frac{1}{\sqrt{m\om_\perp} } \; .
\ee
Let the lattice have length $L$ in the $z$-direction. Tight transverse
confinement implies that
\be
\label{3.113}
\frac{l_\perp}{L} \; \ll \; 1 \; .
\ee

In order to avoid considering atomic scattering on the transverse trapping
potential, we assume that the atomic scattering length is much shorter than
the transverse localization length (\ref{3.112}), such that
\be
\label{3.114}
\frac{|a_s|}{l_\perp} \; \ll \; 1 \; .
\ee

The equation
$$
i \; \frac{\prt}{\prt t} \; \eta(\br,t) = \frac{\dlt H}{\dlt\eta^*(\br,t)}
$$
gives the equation for the condensate function,
\be
\label{3.115}
i \; \frac{\prt}{\prt t} \; \eta(\br,t) = \left [ -\;
\frac{\nabla^2}{2m} + U(\br_\perp) + V_L(z) +
\Phi_0 |\eta(\br,t)|^2 - \mu_0 \right ] \; \eta(\br,t) \; .
\ee
We may look for the solution of this equation in the form
\be
\label{3.116}
\eta(\br,t) = \sqrt{N_0} \; \chi(\br_\perp) \; \vp(z,t) \;
e^{-i\om_\perp t} \; ,
\ee
with the factor functions normalized as
\be
\label{3.117}
\int |\chi(\br_\perp) |^2 \; d\br_\perp =  1 \; , \qquad
\int_{-L/2}^{L/2} | \vp(z,t)|^2 \; dz = 1 \; .
\ee
The transverse wave function $\chi(\br_\perp)$, under tight transverse
confinement, corresponds to the ground state, given by the equation
\be
\label{3.118}
\left [ - \; \frac{\nabla^2_\perp}{2m} + U(\br_\perp) \right ] \;
\chi(\br_\perp) = \om_\perp \; \chi(\br_\perp) \; ,
\ee
in which $\nabla_\perp$ is the transverse part of the Laplacian, entering
in $\nabla^2=\nabla_\perp^2+\prt^2/\prt z^2$.

Expression (\ref{3.116}) is substituted into Eq. (\ref{3.115}), which
is multiplied by $\chi^*(\br_\perp)$ and integrated over the transverse
variable $\br_\perp$. To this end, we define the effective one-dimensional
interaction parameter
\be
\label{3.119}
\Phi_1 \equiv \frac{N_0}{N_L} \;
\Phi_0 \int |\chi(\br_\perp)|^4 \; d\br_\perp \; .
\ee
In the case, when the transverse potential $U(\br_\perp)$ is harmonic, then
\be
\label{3.120}
\int | \chi(\br_\perp)|^4 \; d\br_\perp =
\frac{m\om_\perp}{2\pi} \; .
\ee

When almost all atoms are condensed, $N_0\approx N$, then the filling factor
is
\be
\label{3.121}
\nu \equiv \frac{N}{N_L} \approx \frac{N_0}{N_L} \; .
\ee
For a one-dimensional lattice, the number of lattice sites becomes
\be
\label{3.122}
N_L = \frac{L}{a} \; .
\ee
And the linear density of atoms is
\be
\label{3.123}
\rho \equiv \frac{N}{L} = \frac{\nu}{a} \; .
\ee
Using the above notation, the effective interaction parameter (\ref{3.119})
writes as
\be
\label{3.124}
\Phi_1 = \frac{\nu}{2\pi} \; \Phi_0 m\om_\perp = 2\nu a_s \om_\perp \; ,
\ee
where Eq. (\ref{2.174}) for $\Phi_0$ is taken into account.

Finally, we come to the nonlinear Schr\"odinger equation
\be
\label{3.125}
i \; \frac{\prt}{\prt t} \; \vp(z,t) = \left ( H_{NLS}[\vp] - \mu_0
\right ) \vp(z,t) \; ,
\ee
with the nonlinear Schr\"odinger Hamiltonian
\be
\label{3.126}
H_{NLS}[\vp] \equiv - \; \frac{1}{2m} \; \frac{\prt^2}{\prt z^2} +
V_L(z) + N_L \Phi_1 |\vp|^2 \; .
\ee
One also calls Eq. (\ref{3.125}) the Gross-Pitaevskii equation.

The one-dimensional nonlinear equation (\ref{3.125}) has been widely studied.
By discretizing the variable $z\in[-L/2,L/2]$, Eq. (\ref{3.125}) allows for
a convenient way of its numerical investigation [212].

Stationary solutions to Eq. (\ref{3.125}) have the form
\be
\label{3.127}
\vp(z,t) = \vp(z) e^{-i(E-\mu_0)t} \; ,
\ee
yielding the stationary nonlinear Schr\"odinger equation
\be
\label{3.128}
H_{NLS}[\vp] \vp(z) = E\vp(z) \; .
\ee

For a system in absolute equilibrium, one has
$$
\frac{\prt}{\prt t}\; \vp(z,t) = 0 \qquad
(absolute\; equilibrium) \; ,
$$
hence
\be
\label{3.129}
\mu_0 = \inf_\vp
\int \vp^*(z) H_{NLS} [\vp] \vp(z) \; dz \; .
\ee

The nonlinear equation (\ref{3.128}) may possess different types of
solutions. An important class is given by Bloch functions
\be
\label{3.130}
\vp_{nk}(z) = e^{ikz} f_{nk}(z) \; .
\ee
Here $n$ is the band index, $k$ is quasimomentum pertaining to the
one-dimensional Brillouin zone $[-k_B,k_B]$, with the Brillouin wave
number
\be
\label{3.131}
k_B \equiv \frac{\pi}{a} = k_0 \; .
\ee
The function $f_{nk}(z)$, being periodic,
$$
f_{nk}(z+a) = f_{nk}(z) \; ,
$$
can be expanded over the reciprocal wave numbers
$$
g_j \equiv \frac{2\pi}{a}\; j \qquad
(j=0,\pm 1, \pm 2,\ldots) \; ,
$$
which gives
$$
f_{nk}(z) = \frac{1}{\sqrt{L}} \; \sum_j \; b_{jnk} e^{ig_jz} \; .
$$
The coefficients $b_{jnk}$, because of normalization (\ref{3.117}), are
normalized as
$$
\sum_j \; | b_{jnk}|^2 =  1 \; .
$$

The Bloch function (\ref{3.130}) is a solution to the equation
\be
\label{3.132}
H_{NLS} [ \vp_{nk} ]\; \vp_{nk} (z)  =
E_{nk}\; \vp_{nk}(z) \; ,
\ee
while the periodic function $f_{nk}(z)$ satisfies the equation
\be
\label{3.133}
\left [ \frac{(\hat p + k)^2}{2m} \; + \; V_L(z) \; + \;
N_L \; \Phi_1 |f_{nk}(z)|^2 \right ] \; f_{nk}(z) =
E_{nk}\; f_{nk}(z) \; ,
\ee
in which $\hat p\equiv -i\prt/\prt z$. The eigenvalues $E_{nk}$ form
the {\it Bloch spectrum}. The lowest value of $E_{nk}$ defines the
chemical potential
\be
\label{3.134}
\mu_0 = \inf_{nk} E_{nk} \; ,
\ee
in agreement with Eq. (\ref{3.129}).

To find the compressibility (\ref{3.14}), we remember that we consider
the case, when almost all atoms are condensed, so that $N_0\approx N$,
hence, $N_1\ra 0$ and $n_0\ra 1$, $n_1\ra 0$. Therefore, according to
relation (\ref{2.116}),
$$
\mu_0 \approx \mu \qquad (N_0 \approx N) \; .
$$
Then, we may write
\be
\label{3.135}
\kappa_T = \frac{a}{\nu^2} \; \frac{\prt\nu}{\prt\mu} =
\frac{a}{\nu^2} \left ( \frac{\prt\mu}{\prt\nu} \right )^{-1} \; .
\ee

It is worth noting the following important point. In this
section, we have been considering the case of a practically completely
condensed system, when $N_0\approx N$. In the grand canonical ensemble
with broken gauge symmetry [37,63] the condensate does not fluctuate,
$\Dlt^2(\hat N_0)=0$. Since $N_0\approx N$ and $N$ is fixed, then
$\Dlt^2(\hat N)\approx 0$. However, as soon as $N$ is fixed, the
consideration is reduced to the canonical ensemble. In the latter,
the compressibility is not related to $\Dlt^2(\hat N)$, as in Eq.
(\ref{2.250}), but has to be calculated differently. In the canonical
ensemble, $\kappa_T$ can be expressed through the derivative of free
energy, according to Eq. (\ref{2.254}). Another ensemble, with a fixed
number of particles $N$, is the Gibbs ensemble in which the role of the
thermodynamic potential is played by the chemical potential. Then the
compressibility is expressed through the derivative of the chemical
potential, as in Eq. (\ref{3.155}). It would not be correct to say that,
when the number of particles is fixed, so that $\Dlt^2(\hat N_0)=0$,
then the compressibility would be zero because of relation (\ref{2.250}).
This relation has meaning only if the number of particles is not fixed.
But if $N$ is fixed, one has to invoke other definitions of $\kappa_T$,
such as Eqs. (\ref{2.254}) or (\ref{3.135}).

\subsection{Bloch Spectrum}

The study of the Bloch spectrum is the standard problem of quantum
solid-state physics [213,214]. The basic difficulty in the case of cold
atoms in optical lattices is the existence of interactions between atoms,
which makes the equations nonlinear. To illustrate the properties of the
Bloch spectrum, we shall analyze the quasi-one-dimensional optical
lattice of the previous Sec. 3.7.

The Bloch spectrum, defined in Eq. (\ref{3.132}), can be represented as
\be
\label{3.136}
E_{nk} = \int_{-L/2}^{L/2} \; \vp_{nk}^*(z) \;
H_{NLS}[\vp_{nk} ] \; \vp_{nk}(z)\; dz \; .
\ee
Its lowest value (\ref{3.134}) gives the condensate chemical potential
\be
\label{3.137}
\mu_0 = \lim_{k\ra 0} \; \min_n \; E_{nk} \; ,
\ee
which, in the considered case of a fully condensed system, equals the
system chemical potential, $\mu_0=\mu$.

The Bloch spectrum is a single-particle spectrum, contrary to the
Bogolubov spectrum of collective elementary excitations. For atoms
in a lattice, these spectra are different [215].

One defines the particle {\it group velocity}
\be
\label{3.138}
v_{nk} \equiv \frac{\prt E_{nk}}{\prt k}
\ee
and the {\it effective mass} $m_{nk}^*$,
\be
\label{3.139}
\frac{1}{m_{nk}^*} \equiv \frac{\prt^2 E_{nk}}{\prt k^2} \; .
\ee
The long-wave expression of the Bloch spectrum is
\be
\label{3.140}
E_{nk} \simeq \mu_0 + v_{n0} k + \frac{k^2}{2m_{n0}^*}
\qquad (k\ra 0) \; ,
\ee
where
$$
v_{n0} \equiv \lim_{k\ra 0}  v_{nk} \; , \qquad
m^*_{n0} \equiv \lim_{k\ra 0}  m^*_{nk} \; .
$$

Let us consider just one band. Using Eq. (\ref{3.136}) and expanding
the Bloch functions over Wannier functions, as in Eq. (\ref{3.31}), we
have
\be
\label{3.141}
E_k = \frac{1}{N_L} \; \sum_{ij} \; \int_{-L/2}^{L/2} \;
w_i(z) \; H_{NLS} [\vp_k] \; w_j(z) e^{-ika_{ij}} \; dz \; ,
\ee
where $a_{ij}\equiv a_i-a_j$. For an ideal lattice, invoking
representation (\ref{3.37}), that is, $w_j(z)=w(z-a_j)$, we get
\be
\label{3.142}
E_k = \sum_j \; \int_{-L/2}^{L/2} \; w(z) \;
H_{NLS} [\vp_k] \; w(z-a_j) e^{ika_j} \; dz \; .
\ee
The nonlinear Schr\"odinger Hamiltonian (\ref{3.126}) can be written as
\be
\label{3.143}
H_{NLS}[\vp] =  H_L(z) + N_L \Phi_1 |\vp|^2 \; ,
\ee
where the {\it linear lattice term}
\be
\label{3.144}
H_L(z) \equiv - \; \frac{1}{2m} \;
\frac{\prt^2}{\prt z^2} \; + \; V_L(z)
\ee
is separated out.

Keeping in mind the tight-binding approximation of Sec. 3.5, we shall
consider only the nearest-neighbor sites. Then, for the linear term
(\ref{3.144}), there are two types of matrix elements over Wannier
functions, the single-site integral
\be
\label{3.145}
h_0 \equiv \int_{-L/2}^{L/2} \; w(z) H_L(z) w(z) \; dz
\ee
and the nearest-neighbor overlap integral
\be
\label{3.146}
h_1 \equiv \int_{-L/2}^{L/2} w(z) H_L(z) w(z-a) \; dz \; .
\ee

The general form of the matrix element of the nonlinear term in Eq.
(\ref{3.143}) is proportional to the integral
\be
\label{3.147}
I_{j_1j_2j_3j_4} \equiv \int_{-L/2}^{L/2} w_{j_1}(z)
w_{j_2}(z) w_{j_3}(z) w_{j_4}(z) \; dz \; .
\ee
Treating again only the nearest neighbors, we have three kinds of
integrals. The single-site integral is
\be
\label{3.148}
I_0 \equiv \int_{-L/2}^{L/2} w^4(z) \; dz \; .
\ee
The first-order overlap integral is
\be
\label{3.149}
I_1 \equiv \int_{-L/2}^{L/2} w^3(z) w(z-a)\; dz \; .
\ee
And also, we have the second-order overlap integral
\be
\label{3.150}
I_2 \equiv \int_{-L/2}^{L/2} w^2(z) w^2(z-a)\; dz \; .
\ee

Taking only account of the nearest neighbors, Eq. (\ref{3.142}) reads
as
\be
\label{3.151}
E_k = h_0 + 2h_1\cos ka \; + \; \Phi_1 \sum_{j_1j_2j_3} \;
I_{0j_1j_2j_3} \; \exp\{ -ik(a_{j_1} - a_{j_2} - a_{j_3} ) \} \; .
\ee
The second-order overlap integral (\ref{3.150}) is smaller than the
zero-and first-order integrals (\ref{3.148}) and (\ref{3.149}).
Therefore, retaining only the overlap integrals up to first order, one
has the following terms. The single-site term is
\be
\label{3.152}
\ep_0 \equiv h_0 + \Phi_1 I_0 \; .
\ee
The first-order overlap integrals define the {\it tunneling parameter}
\be
\label{3.153}
J \equiv -h_1 -4 I_1\Phi_1 \; .
\ee
So that the {\it Bloch spectrum} (\ref{3.151}) becomes
\be
\label{3.154}
E_k = \ep_0 - 2J\cos(ka) \; .
\ee
Taking account of the second-order overlap integral (\ref{3.150}) would
result in an additional term containing $\cos(2ka)$.

The chemical potential (\ref{3.137}) is
\be
\label{3.155}
\mu_0 \equiv \lim_{k\ra 0} E_k = \ep_0 - 2J \; ,
\ee
provided that the lowest band has been considered. This, with Eqs.
(\ref{3.152}) and (\ref{3.153}), gives
\be
\label{3.156}
\mu_0 =  h_0 + 2h_1 + \Phi_1 (I_0 + 8 I_1) \; .
\ee

The isothermal compressibility, given in Eq. (\ref{3.135}), can be
approximately defined by taking into account only the dependence of
the coupling parameter (\ref{3.124}) on the filling factor $\nu$, but
neglecting the possible dependence on $\nu$ of the Wannier functions.
Then
\be
\label{3.157}
\kappa_T = \frac{ma\;l_\perp^2}{2a_s\nu^2(I_0+8I_1)} =
\frac{1}{\rho(I_0+8I_1)\Phi_1} \; .
\ee
We may notice that, similarly to the case of a uniform system, bosons
in a lattice can be stable only in the presence of nonzero repulsive
interactions. If atomic interactions would be attractive, such that
$a_s<0$, then compressibility (\ref{3.157}) would be negative. And
if the Bose gas would be ideal, such that $\Phi_1\ra 0$, then
$\kappa_T$ would be infinite. In both these cases, the system would
be unstable [93,145,146]. In the case of attractive atomic interactions,
when $a_s<0$, the system stability can be restored for a {\it finite}
number of atoms by imposing a trapping potential in all directions
[9,174,216--218].

With the Bloch spectrum (\ref{3.154}), the group velocity (\ref{3.138})
is
\be
\label{3.158}
v_k = 2J a \sin(ka)
\ee
and the effective mass, given by Eq. (\ref{3.139}), reads as
\be
\label{3.159}
m_k^* = \frac{1}{2Ja^2\cos(ka)} \; .
\ee
In the limit of $k\ra 0$, the effective mass is
\be
\label{3.160}
m^* \equiv m_0^* = \frac{1}{2Ja^2} \qquad (k=0) \; ,
\ee
while the group velocity (\ref{3.158}) behaves as
$$
v_k \simeq 2J a^2 k \qquad (k\ra 0) \; .
$$

\subsection{Spectrum Parameters}

To estimate the parameters entering the Bloch spectrum, let us consider
the lowest band in the tight-binding approximation. The corresponding
Wannier function for a one-dimensional lattice is
\be
\label{3.161}
w(z) = \frac{1}{(\sqrt{\pi}\;l_0)^{1/2}} \;
\exp\left ( - \; \frac{z^2}{2l_0^2} \right ) \; .
\ee
Using this, together with notation (\ref{3.99}), we find the single-site
integral (\ref{3.148}),
\be
\label{3.162}
I_0 = \frac{1}{\sqrt{2\pi}\;l_0} = \sqrt{\frac{m\om_0}{2\pi} } \; ,
\ee
the first-order overlap integral (\ref{3.149}),
\be
\label{3.163}
I_1 =  I_0 \exp \left ( - \; \frac{3a^2}{8l_0^2} \right ) \; ,
\ee
and the second-order overlap integral (\ref{3.150}),
\be
\label{3.164}
I_2 = I_0 \exp \left ( -\; \frac{a^2}{2l_0^2} \right ) =
I_1 \exp\left ( -\; \frac{a^2}{8l_0^2} \right ) \; .
\ee
From these expressions, we see that
\be
\label{3.165}
I_2 \ll I_1 \ll I_0 \qquad \left ( \frac{l_0}{a} \ll 1
\right ) \; ,
\ee
which justifies the neglect of the second-order overlap integral
(\ref{3.164}).

Calculating Eqs. (\ref{3.145}) and (\ref{3.146}), we meet the integrals
$$
\int_0^\infty \; \sin^2(bx) e^{-x^2} \; dx = \frac{\sqrt{\pi}}{4}
\left ( 1 - e^{-b^2} \right ) \; ,
$$
$$
\int_{-\infty}^{+\infty} \; \cos[p(x+a)] \; e^{-x^2} \; dx =
\sqrt{\pi} \; cos(pa) \; \exp\left ( -\; \frac{p^2}{4} \right ) \; .
$$
Then, for the single-site integral (\ref{3.145}), we get

\be
\label{3.166}
h_0 = \frac{\om_0}{4} + \frac{V_0}{2} \left [ 1 -
\exp \left ( -k_0^2 l_0^2 \right ) \right ] \; ,
\ee
and for the overlap integral (\ref{3.146}), we have
\be
\label{3.167}
h_1 = -\; \frac{\om_0a^2}{8l_0^2} \left \{ 1 \; - \;
\frac{2l_0^2}{a^2} \; - \; \frac{4V_0l_0^2}{\om_0a^2} \left [
1 + \exp\left ( -k_0^2 l_0^2 \right ) \right ] \right \} \;
\exp\left ( - \; \frac{a^2}{4l_0^2}\right ) \; .
\ee
These can be simplified remembering the conditions (\ref{3.100})
of good localization, when $l_0\ll a$ and $k_0l_0\ll 1$. Then Eq.
(\ref{3.166}) reduces to
\be
\label{3.168}
h_0 \cong \frac{\om_0}{4} + \frac{V_0 E_R}{\om_0} \; ,
\ee
where
$$
E_R \equiv \frac{k_0^2}{2m} = \frac{\pi^2}{2ma^2} \; .
$$
And Eq. (\ref{3.167}) simplifies to
\be
\label{3.169}
h_1 \cong - \left ( \frac{\pi^2\om_0^2}{16E_R}\; - \; V_0
\right ) \; \exp \left ( - \; \frac{\pi^2\om_0}{8E_R}
\right ) \; .
\ee

The local energy term (\ref{3.152}) becomes
\be
\label{3.170}
\ep_0 = \frac{\om_0}{4} + \frac{V_0E_R}{\om_0}  +
\frac{\Phi_1}{2a}\; \sqrt{ \frac{\pi\om_0}{E_R} } \; .
\ee
The effective frequency $\om_0$ can be defined, in first approximation,
as in Eq. (\ref{3.97}),
\be
\label{3.171}
\om_0 \approx 2\; \sqrt{V_0E_R} \qquad
\left ( \frac{a_s}{a} \ll 1 \right ) \; ,
\ee
which does not take into account atomic interactions. To find the
dependence of $\om_0$ on the interaction of atoms, we may treat the
effective frequency $\om_0$ as a trial parameter defined in line with
the optimized perturbation theory [219--221]. Accepting the optimization
condition as
\be
\label{3.172}
\frac{\prt\ep_0}{\prt\om_0} = 0 \; ,
\ee
from Eq. (\ref{3.170}), we get the equation
\be
\label{3.173}
\om_0^2 \left ( 1 + \frac{\Phi_1}{a}\; \sqrt{\frac{\pi}{\om_0 E_R}}
\right ) = 4 V_0 E_R \; .
\ee
Invoking Eq. (\ref{3.171}), we see that
\be
\label{3.174}
\frac{E_R}{V_0} \approx 4 \left ( \frac{E_R}{\om_0}
\right )^2 \ll 1 \; ,
\ee
according to condition (\ref{3.101}). Using this in Eq. (\ref{3.173}),
we find
\be
\label{3.175}
\om_0 \cong 2 \; \sqrt{V_0E_R} \left [ 1 \; - \;
\frac{\Phi_1}{2aE_R} \left ( \frac{\pi^2E_R}{4V_0}
\right )^{1/4} \right ] \; .
\ee
Taking account of atomic interactions diminishes the effective frequency
$\om_0$.

Recall that we consider weak interactions, since in the other case,
the system could not be almost completely condensed. Hence, to a good
approximation, one can use the effective frequency estimated in Eq.
(\ref{3.171}). Then expression (\ref{3.168}) becomes
\be
\label{3.176}
h_0 \; \cong \; \sqrt{V_0E_R} \; ,
\ee
while Eq. (\ref{3.169}) reduces to
\be
\label{3.177}
h_1 \;\cong \; - \left ( \frac{\pi^2}{4} \; - \; 1 \right )
V_0 \exp\left ( - \; \frac{\pi^2}{4} \; \sqrt{\frac{V_0}{E_R} }
\right ) \; .
\ee
For the tunneling parameter (\ref{3.153}), we obtain
$$
J \cong \left ( \frac{\pi^2}{4} \; - \; 1 \right ) V_0 \exp
\left ( -\; \frac{\pi^2}{4}\; \sqrt{\frac{V_0}{E_R}} \right )\; -
$$
\be
\label{3.178}
- 2\sqrt{2\pi} \; \frac{\Phi_1}{a} \left ( \frac{V_0}{E_R}
\right )^{1/4} \exp\left ( - \;
\frac{3\pi^2}{8}\; \sqrt{\frac{V_0}{E_R}} \right ) \; .
\ee
The chemical potential (\ref{3.156}) reads as
$$
\mu_0 \cong \sqrt{V_0E_R} \; - \; \left ( \frac{\pi^2}{2} \; -
\; 2 \right ) V_0 \exp\left ( - \; \frac{\pi^2}{4} \;
\sqrt{\frac{V_0}{E_R}} \right ) \; +
$$
\be
\label{3.179}
+\; \frac{\Phi_1}{\sqrt{2\pi}\; l_0}
\left [ 1 + 8 \exp\left ( -\; \frac{9\pi^2}{8}\;
\sqrt{\frac{V_0}{E_R} } \right ) \right ] \; .
\ee
And compressibility (\ref{3.157}) can be simplified to
\be
\label{3.180}
\kappa_T \cong \frac{\sqrt{2\pi}\;l_0}{\rho\Phi_1} \; ,
\ee
which means that the stability condition $0<\kappa_T<\infty$ implies
that interactions are repulsive and finite, $\Phi_1>0$.

\subsection{Elementary Excitations}

Elementary excitations characterize small deviations from the stationary
solutions of Eq. (\ref{3.128}) or Eq. (\ref{3.132}). Suppose that $\vp(z)$,
is an arbitrary stationary solution of Eq. (\ref{3.128}), with an energy
$E$. Small deviations from the stationary solution are described by the
wave function
\be
\label{3.181}
\vp(z,t) = \left [ \vp(z) + u(z) e^{-i\ep t} +
v^*(z) e^{i\ep t} \right ] e^{-i(E-\mu_0)t} \; .
\ee
Substituting this form into the nonlinear Schr\"odinger equation
(\ref{3.125}) gives the {\it Bogolubov equations}
$$
\left ( H_{NLS}[\vp] - E + N_L \Phi_1 |\vp(z)|^2 - \ep \right )
u(z) + N_L \Phi_1 \vp^2(z) v(z) = 0 \; ,
$$
\be
\label{3.182}
\left ( H_{NLS}[\vp] - E + N_L \Phi_1 |\vp(z)|^2 + \ep \right )
v(z) + N_L \Phi_1 (\vp^*(z))^2 u(z) = 0 \; .
\ee
As a stationary solution $\vp(z)$ with an energy $E$ one can take any
Bloch function $\vp_{nq}$ with an energy $E_{nq}$.

For an equilibrium system, BEC corresponds to the lowest-energy Bloch
function $\vp_0(z)$ with the energy $E=\mu_0$. Considering the elementary
excitations above the condensate requires to set as $\vp(z)$ in Eq.
(\ref{3.182}) the  condensate Bloch function
\be
\label{3.183}
\vp_0(z) = \frac{1}{\sqrt{N_L}} \; \sum_j w(z-a_j) \; .
\ee
The Bogolubov functions $u(z)$ and $v(z)$ should be proportional to Bloch
functions $\vp_k(z)$ with nonzero $k$. Hence, we take
\be
\label{3.184}
u(z) \equiv u_k \vp_k(z) \; , \qquad v(z) \equiv v_k \vp_k(z) \; .
\ee

Let us introduce the notation
\be
\label{3.185}
\om_k \equiv \int_{-L/2}^{L/2} \;
\vp_k^*(z) H_L(z) \vp_k(z) \; dz \; + \; 2\Dlt_k - \mu_0 \; ,
\ee
in which $H_L(z)$ is the linear lattice Hamiltonian (\ref{3.144}) and
\be
\label{3.186}
\Dlt_k \equiv \Phi_1 N_L \int_{-L/2}^{L/2} \;
|\vp_k(z)|^2 \vp_0^2(z) \; dz \; .
\ee
The Bogolubov equations (\ref{3.182}) reduce to
$$
(\om_k - \ep) u_k + \Dlt_k v_k = 0 \; ,
$$
\be
\label{3.187}
\Dlt_k u_k + (\om_k + \ep) v_k = 0 \; ,
\ee
which defines the Bogolubov spectrum of collective excitations
\be
\label{3.188}
\ep_k = \sqrt{\om_k^2 - \Dlt_k^2} \; .
\ee

Calculating quantities (\ref{3.185}) and (\ref{3.186}), we employ the
tight-binding approximation. The Bloch functions $\vp_k(z)$ are expanded
over the Wannier functions $w(z-a_j)$, as in Eq. (\ref{3.31}), and $w(z)$
is taken in form (\ref{3.161}). This yields for Eq. (\ref{3.185})
\be
\label{3.189}
\om_k = \Dlt + 4J \sin^2 \left ( \frac{ka}{2} \right ) \; ,
\ee
where
\be
\label{3.190}
\Dlt \equiv ( I_0 + 8I_1) \Phi_1 \; ,
\ee
while for Eq. (\ref{3.186}),
\be
\label{3.191}
\Dlt_k = \Dlt - 8 I_1 \Phi_1 \sin^2 \left (
\frac{ka}{2} \right ) \; .
\ee
Then the Bogolubov spectrum (\ref{3.188}) is
\be
\label{3.192}
\ep_k = \left [ 4 \; \frac{c^2}{a^2}\; \sin^2\left (
\frac{ka}{2}\right ) + 16 \left ( J^2 - 4I_1^2 \Phi_1^2\right )
\sin^4 \left ( \frac{ka}{2}\right ) \right ]^{1/2} \; ,
\ee
where $c$ is the {\it sound velocity}
\be
\label{3.193}
c \equiv \sqrt{2\Dlt(J + 2I_0\Phi_1)a^2} \; .
\ee
In the long-wave limit, when $k\ra 0$, one has
$$
\om_k \simeq \Dlt + J (ka)^2 \; , \qquad
\Dlt_k \simeq \Dlt - 2I_1 \Phi_1 (ka)^2 \; .
$$
And Eq. (\ref{3.192}) gives the gapless phonon spectrum
\be
\label{3.194}
\ep_k \simeq ck \qquad (k\ra 0) \; .
\ee

Comparing the Bloch spectrum (\ref{3.154}) and the Bogolubov spectrum
(\ref{3.192}), we see that they are quite different [215,222,223]. To
stress their difference, we may rewrite the Bloch spectrum (\ref{3.154})
in the form
\be
\label{3.195}
E_k = \mu_0 + 4J \sin^2 \left ( \frac{ka}{2} \right ) \; ,
\ee
where $\mu_0$ is the chemical potential (\ref{3.155}). In the long-wave
limit, the Bloch spectrum (\ref{3.195}) is
$$
E_k \simeq \mu_0 + J (ka)^2 \qquad (k\ra 0) \; ,
$$
which is, clearly, a single-particle spectrum with the gap $\mu_0$,
contrary to the gapless phonon spectrum (\ref{3.194}).

It is important to emphasize that for equilibrium bosons in lattices, in
the presence of BEC, when the gauge symmetry is broken, the single-particle
spectrum of {\it uncondensed} atoms (\ref{3.69}) coincides with the spectrum
of elementary excitations (\ref{3.188}), both of them being the Bogolubov
spectra.

Of the Bloch spectrum (\ref{3.154}), or (\ref{3.195}), solely one point,
where $n=0$ and $k=0$, and $E_k=\mu_0$, corresponds to an equilibrium BEC.
But the Bloch spectrum, in general, describes {\it nonequilibrium condensates}
that are the analog of the {\it coherent modes} [174--176]. This is why the
Bloch spectrum does not need to coincide with the Bogolubov spectrum

\subsection{Wave Stability}

The condensate wave function $\vp_0(z)$ for an equilibrium system
is assumed to correspond to a stable system. The related condition
of thermodynamic stability is that compressibility (\ref{3.135}), or
(\ref{3.157}), be positive and finite, $0\leq\kappa_T<\infty$.

Other Bloch functions $\vp_{nk}(z)$, which are solutions to Eq. (\ref{3.132}),
as is stressed above, do not correspond to a thermodynamically equilibrium
BEC. Though these functions $\vp_{nk}(z)$ are stationary solutions, but a
statistical system with a condensate, characterized by such a function is
not in absolute equilibrium. It is, therefore, useful to study the stability
of the Bloch functions $\vp_{nk}(z)$.

There are, in general, several kinds of stability for solutions to
differential equations [224--227]. The most often used is the notion of
Lyapunov stability [228]. Let us recall this notion in general terms.
Suppose we consider functions of the type $\vp(x,t)$, where $x$ is a
variable of arbitrary nature, which can pertain to the continuous manifold
$\mathbb{R}^d$, or to a discrete manifold, or to their combination, and
where $t\in[0,\infty)$. Treating $x$ as an enumeration index, one can
define the column function $\vp(t)\equiv[\vp(x,t)]$. Let $\vp(t)$ pertain
to a Banach space (normed, complete space), where a norm $||\vp(t)||$ is
defined. When $\vp(t)$ pertains to a Hilbert space, the norm is naturally
generated by the scalar product. Let us consider the evolution equation
\be
\label{3.196}
\frac{\prt\vp(t)}{\prt t} = F[\vp] \; ,
\ee
in which $F[\vp]$ is an operator functional in the same Banach space. It
is assumed that, for a given initial condition $\vp(0)$, the Cauchy problem
(\ref{3.196}) enjoys a unique solution.

A solution $\overline\vp(t)$ is {\it Lyapunov stable}, if for any
other solution $\vp(t)$, such that
\be
\label{3.197}
|| \vp(0) - \overline\vp(0) || < \dlt_0 \; ,
\ee
with any $\dlt_0>0$, there exists a positive number $\dlt$, for which
\be
\label{3.198}
|| \vp(t) - \overline\vp(t) || < \dlt \qquad ( t > 0 ) \; .
\ee
The solution $\overline\vp(t)$ is {\it asymptotically stable}, when
there can be found such $\dlt_0$ in Eq. (\ref{3.197}) that
\be
\label{3.199}
\lim_{t\ra\infty} || \vp(t) - \overline\vp(t) || = 0 \; .
\ee
The solution $\overline\vp(t)$ is {\it exponentially stable}, if there
exists such $\dlt_0$ in Eq. (\ref{3.197}) that
\be
\label{3.200}
\lim_{t\ra\infty} \; \frac{1}{t} \; \ln
 || \vp(t) - \overline\vp(t) || < 0 \; .
\ee
In particular, $\overline\vp(t)$ can be a stationary solution, for which
$F[\overline\vp]=0$. Exponential stability is a special case of asymptotic
stability.

Conversely, if, under condition (\ref{3.197}): inequality (\ref{3.198}) is
not valid, the solution $\overline\vp(t)$ is Lyapunov unstable; when limit
(\ref{3.199}) does not follow for any $\dlt_0$, $\overline\vp(t)$ is
asymptotically unstable; and if the limit (\ref{3.200}) becomes positive,
then $\overline\vp(t)$ is exponentially unstable.

From these definitions, it is clear that the Lyapunov stability does
not lead to the asymptotic stability and, vice versa, the asymptotic
stability does not imply the Lyapunov stability. Also, the Lyapunov
instability does not forbid the asymptotic stability for some $\dlt_0$.
And the asymptotic instability does not contradict to the Lyapunov
stability.

Lyapunov developed [228] two methods of controlling stability, the
direct method and the method of linearization.

{\it Lyapunov direct method} is based on the existence of the Lyapunov
functional, such that
\be
\label{3.201}
L[\vp] \geq 0
\ee
for all $\vp(t)$ from the considered Banach space and which does not
increase,
\be
\label{3.202}
\frac{\prt}{\prt t} \; L[\vp] \leq 0 \; ,
\ee
on the trajectories of Eq. (\ref{3.196}). If such a Lyapunov functional
exists, then the solution $\vp(t)$ is {\it Lyapunov stable}.

The Lyapunov direct method is global, requiring the validity of condition
(\ref{3.201}) on the whole Banach space. In many cases, one is interested
not in the global stability, but in the local stability in the vicinity of
a known solution $\overline\vp(t)$. Then the direct Lyapunov method is
reformulated as follows.

{\it Lyapunov local method} assumes the existence of a functional $L[\vp]$,
which does not increase on the trajectories, as in Eq. (\ref{3.202}), and
which is minimal in the small vicinity of a given function $\overline\vp(t)$,
that is, when for $\vp=\overline\vp+\dlt\vp$, one has
\be
\label{3.203}
\dlt L[\vp] = 0 \; , \qquad \dlt^2 L[\vp] > 0 \; .
\ee
If such a functional exists, then the solution $\overline\vp(t)$ is {\it
locally stable}.

In many cases, the role of the Lyapunov function is played by energy or by
an effective energy, if the energy is complimented by additional constraints.
When the energy functional $E[\vp]$ is an integral of motion, then $\prt
E[\vp]/\prt t=0$, so that condition (\ref{3.202}) holds. Hence, if $E[\vp]
\geq 0$, the motion is Lyapunov stable. This does not mean that the motion
is locally stable in the vicinity of a given $\overline\vp$. To study
the local stability near $\overline\vp$, one has to satisfy conditions
(\ref{3.203}) for the Lyapunov functional $E[\vp]$. One says that a solution
is {\it energetically stable}, if
\be
\label{3.204}
\frac{\prt}{\prt t}\; E[\vp] \leq 0 \; , \qquad
\dlt E[\vp] = 0 \; , \qquad \dlt^2 E[\vp] > 0
\ee
for $\vp=\overline\vp+\dlt\vp$. That is, the {\it energetic stability}
is just an example of the {\it local stability}, when the Lyapunov
functional is represented by an energy functional.

Above, we have considered the variants of the Lyapunov direct method
for analysing the stability of solutions to the evolution equation
(\ref{3.196}). The second Lyapunov method is based on the linearization
of Eq. (\ref{3.196}).

{\it Lyapunov linearization method} requires the linearization of the
evolution equation (\ref{3.196}) with respect to small deviations from the
given solution $\overline\vp$. Taking $\vp=\overline\vp+\dlt\vp$, one obtains
linear equations for $\dlt\vp$, which are subject to the standard stability
analysis [228,229]. When all Lyapunov exponents are negative, the solution
$\overline\vp$ is {\it asymptotically stable}. If at least one of them is
positive, then $\overline\vp$ is asymptotically unstable. And when some of
the Lyapunov exponents are zero, while others being negative, the considered
solution is {\it neutrally stable}, provided it remains finite for all $t>0$.
If the linearized equations show that the solution $\overline\vp$ is either
asymptotically stable or neutrally stable, then such a solution is termed {\it
dynamically stable}.

Suppose that one is interested in the stability of a stationary solution
$\overline\vp$, for which $\prt\overline\vp/\prt t=0$. Considering small
deviations from $\overline\vp$, when  $\vp(t)=\overline\vp+\dlt\vp(t)$,
one can check he local, or energetic stability by means of Eqs.
(\ref{3.202}) and (\ref{3.203}), or (\ref{3.204}). Alternatively, one can
use the linearization method for the evolution equation (\ref{3.196}). Then
there exists the following relation between different types of stability.

\vskip 2mm

{\bf Theorem}. {\it The local stability of a stationary solution yields
its dynamic stability}.

\vskip 2mm

{\it Proof}. Let $\overline\vp$ be a stationary solution of Eq.
(\ref{3.196}), such that $\prt\overline\vp/\prt t=0$. And let this solution
be locally stable, which implies that there exists a Lyapunov functional
$L[\vp]$ satisfying conditions (\ref{3.202}) and (\ref{3.203}). Expanding
the Lyapunov functional for the perturbed solution $\vp(t)=\overline\vp+
\dlt\vp(t)$, we have
$$
L[\vp] =  L[\overline\vp] + \dlt L[\vp] + \dlt^2 L[\vp] \; .
$$
From here, in view of Eqs. (\ref{3.202}) and (\ref{3.203}), it follows
$$
\frac{\prt}{\prt t}\; \dlt^2 L[\vp] =
\frac{\prt}{\prt t} \; L[\vp] \leq 0 \; .
$$
Taking
$$
|| \dlt\vp(0) || \leq \dlt^2 L[\vp(0)] \; ,
$$
we find
$$
||\dlt\vp(t) || \leq \dlt^2 L[\vp(t)] \leq \dlt^2 L [\vp(0) ] \; ,
$$
which shows that the linear deviation $\dlt\vp(t)$ does not increase
with time. Hence, $\overline\vp$ is dynamically stable.

\vskip 2mm

When the Lyapunov functional is the energy functional, the theorem
tells us that the energetic stability of a stationary solution yields
its dynamic stability.

The inverse, however, is not true. The dynamic stability does not
necessarily yield the local, or energetic, stability. To illustrate
this, let us consider  a two-component field $\vp=\{\vp_1,\vp_2\}$,
where $\vp_j=\vp_j(x,t)$, with the energy functional
$$
E[\vp] = \frac{1}{2} \; \int \left ( \vp_1|^2 - |\vp_2|^2 -
g|\vp_1|^2 |\vp_2|^2 \right ) \; dx \; .
$$
The evolution equations are defined in the usual way as
$$
i\; \frac{\prt\vp_j}{\prt t} =
\frac{\dlt E[\vp]}{\dlt\vp_j^*} \; ,
$$
which gives
$$
i\; \frac{\prt\vp_1}{\prt t} = \frac{1}{2}
\vp_1 \left ( 1 - g|\vp_2|^2 \right ) \; ,
\qquad  i\; \frac{\prt\vp_2}{\prt t} = -\; \frac{1}{2} \vp_2
\left ( 1 + g|\vp_1|^2 \right ) \; .
$$
The stationary solutions of these equations are
$\overline\vp_1=\overline\vp_2=0$. The linearized equations give
$$
\dlt\vp_1 = c_1 e^{-i\om t} \; , \qquad
\dlt\vp_2 = c_2 e^{i\om t} \; ,
$$
where $c_j=c_j(x)$ and $\om=1/2$. Hence, the deviations $\dlt\vp_j$
do not increase with time, which means that the stationary solutions
$\overline\vp_1$ and $\overline\vp_2$ are {\it dynamically stable}.

For the energy functional, we have
$$
\frac{\prt}{\prt t} \; E[\vp] = 0 \; , \qquad \dlt E[\vp] = 0 \; .
$$
However, the second variation
$$
\dlt^2 E[\vp] = \frac{1}{2} \; \int \left ( |c_1|^2 - |c_2|^2
\right ) \; dx
$$
is not positive defined, that is, the stationary solutions
$\overline\vp_j=0$ do not provide a minimum of $E[\vp]$. Hence these
stationary solutions are {\it energetically unstable}.

For an optical lattice, discussed in the previous sections, we may
define the energy functional
\be
\label{3.205}
E[\vp] \equiv N \int \vp^*(z) [ H_L(z) - E] \vp(z) \; dz \; + \;
\frac{N}{2} \; N_L \Phi_1 \int |\vp(z)|^4 dz \; .
\ee
The stationarity condition
\be
\label{3.206}
\frac{\dlt E[\vp]}{\dlt\vp^*(z)} = 0
\ee
results in the stationary nonlinear Schr\"odinger equation (\ref{3.128}).
The latter, due to its nonlinearity, can possess different types of
solutions, including Bloch waves, localized solitons, as well as density
waves with a period differing from that of the optical potential [212].
Limiting ourselves by the class of Bloch functions, we come to Eq.
(\ref{3.132}).

The Bloch spectrum $E_{nk}$, defined by Eq. (\ref{3.132}), displays
rather nontrivial behavior, caused by the nonlinearity of the
eigenproblem. For sufficiently strong nonlinearity, there appears
the swallow-tail structure of $E_{nk}$, when it is not uniquely
defined as a function of $k$ [230--234]. The swallow tails can appear
at the edge of the lowest band, with $n=0$ and $k=\pi/a$, and also in
the middle of upper bands, with $n\geq 1$ and $k=0$. This happens when
the interaction is sufficiently strong, so that
\be
\label{3.207}
\frac{aV_0}{\Phi_1} < 1 \; .
\ee

In the presence of nonlinearity, the optical lattice may provoke
instability of Bloch waves $\vp_{nk}(z)$ for some $k$. To find the
region of stability, one considers small deviations in the vicinity
of $\vp_{nn}(z)$ by setting
\be
\label{3.208}
\vp_{nk}(z,t) = \vp_{nk}(z) + \dlt \vp(z,t) \; ,
\ee
where
\be
\label{3.209}
\dlt\vp(z,t) = \left [ u_{nq}(z) e^{i(qz-\ep t)} +
v^*_{nq}(z) e^{-i(qz-\ep t)} \right ] \; .
\ee
The energetic, or static, stability is defined by substituting Eq.
(\ref{3.208}) into the energy functional (\ref{3.205}) and taking
$\ep=0$ in Eq. (\ref{3.209}). The dynamic stability is analyzed by
linearizing the evolution equation
\be
\label{3.210}
i \; \frac{\prt}{\prt t} \; \vp_{nk}(z,t) = \left (
H_{NLS}[ \vp_{nk} ] - E_{nk} \right ) \vp_{nk}(z,t)
\ee
with respect to the small deviation (\ref{3.209}).
The functions $\vp_{nk}(z)$, $u_{nq}(z)$, and $v_{nq}(z)$ are Bloch
waves.

Linearizing Eq. (\ref{3.210}) gives for the energy $\ep$ the spectrum
of collective excitations $\ep_{nkq}$ around the Bloch spectrum $E_{nk}$.
Dynamic instability occurs when $\ep_{nkq}$ becomes complex, since then
there appears an exponentially increasing term in Eq. (\ref{3.209}).

From the general theory, expounded at the beginning of this section,
it follows that the energetic stability yields the dynamic stability.
This means that, if the solution is dynamically unstable, it is also
energetically unstable. However, the solution can be dynamically stable,
while being energetically unstable. All this is, of course, valid for
optical lattices [230--234], as well as for vortex states [235].

It is worth recalling that the Bloch functions $\vp_{nk}$, with $n>0$
and $k>0$, correspond to excited nonequilibrium condensates. Therefore
the stability, in any sense, of such nonequilibrium condensates should
be neither required nor expected. What is required is the stability of
the equilibrium ground-state BEC, corresponding to the natural Bloch
orbital $\vp_0$, with $n=0$ and $k=0$. The stability of the latter is
guaranteed by compressibility (\ref{3.135}) being positive and finite,
which, in view of Eqs. (\ref{3.157}) and (\ref{3.180}), requires that
atomic interactions be repulsive and finite, that is, $\Phi_1>0$.

\subsection{Moving Lattices}

Different nonequilibrium states of BEC can be crated by moving the
optical lattice. To obtain an effective equation for BEC in a moving
lattice, let us assume that the latter moves with velocity $v=v(t)$
along the $z$-axis. This means that, in the frame of the lattice, the
condensate moves with velocity $-v(t)$. The wave function of a moving
condensate $\vp_v(z,t)$ satisfies the nonlinear Schr\"odinger equation
\be
\label{3.211}
i\; \frac{\prt}{\prt t} \; \vp_v(z,t) = \left (
H_{NLS} [ \vp_v] - E \right ) \vp_v(z,t)
\ee
and is related to the wave function of an immovable condensate through
the Galilean transformation
\be
\label{3.212}
\vp_v(z,t) = \vp(z+vt,t) \exp \left \{ -i \left (
mvz + \frac{mv^2}{2}\; t\right ) \right \} \; .
\ee
Substituting function (\ref{3.212}) into Eq. (\ref{3.211}) and
neglecting the term
$$
\left (\hat p -mv\right ) \vp(z+vt,t) \approx 0 \qquad
\left (\hat p \equiv -i\; \frac{\prt}{\prt z}\right )
$$
yields
\be
\label{3.213}
i\; \frac{\prt}{\prt t} \; \vp(z+vt,t) = \left ( H_{NLS} [\vp]
- E - m\dot{v} z \right ) \vp(z+vt,t) \; ,
\ee
where $\dot{v}\equiv\prt v/\prt t$.

There exist two different length scales in the system. One is the
intersite distance $a$, being the lattice period. The typical variation
of Bloch functions is on the lattice scale $a$. And another scale is
the effective size $L$ of the studied atomic cloud, being much larger
than $a$,
\be
\label{3.214}
\frac{a}{L} \; \ll \; 1 \; .
\ee
The existence of such very different scales allows for the use of
averaging techniques [236--239] and of the scale separation approach
[240--243]. A similar procedure in electrodynamics is called the slowly
varying amplitude approximation [244--249]. In line with such techniques,
we may look for the solution of Eq. (\ref{3.213}) in the form
\be
\label{3.215}
\vp(z+vt,t) = A(z,t) \vp_q(z) \; ,
\ee
where, for simplicity, we consider a single band and assume that the
quasimomentum $q = q(t)$ is, generally speaking, a function of time. The
factor $A(z,t)$ in Eq. (\ref{3.215}) is a slowly varying amplitude, such
that
\be
\label{3.216}
\left | \frac{\prt A}{\prt z} \right | \; \ll \;
\left | \frac{\prt\vp_q}{\prt z} \right | \; .
\ee
While $\vp_q(z)$ is a Bloch function given by the equation
\be
\label{3.217}
H_L(z) \vp_q(z) = E_q \vp_q(z) \; ,
\ee
with the linear lattice Hamiltonian (\ref{3.144}). According to
condition (\ref{3.216}), the function $\vp_q(z)$ is fastly varying
in space, as compared to the slow amplitude $A(z,t)$. The latter is
also called the envelope. It is normalized as
\be
\label{3.218}
\frac{1}{L} \; \int_{-L/2}^{L/2} \; | A(z,t)|^2 \; dz = 1 \; .
\ee
The Bloch function $\vp_q(z)$ satisfies the usual normalization
condition
\be
\label{3.219}
\int_{-L/2}^{L/2} \; |\vp_q(z)|^2 \; dz = 1 \; .
\ee

An important point is the ansatz
\be
\label{3.220}
H_L(z) A(z,t) \vp_q(z) = \left [ E_{q+\hat p} A(z,t) \right ]
\vp_q(z) \; ,
\ee
whose justification [250] is based on the averaging techniques.

Substituting form (\ref{3.215}) into Eq. (\ref{3.213}) yields the
equation
\be
\label{3.221}
i\; \frac{\prt A}{\prt t} = ( E_{q+\hat p} - E_q) A +
\al_q |A|^2 A + (\dot{q} - m\dot{v} ) zA \; ,
\ee
where $A=A(z,t)$, the overdot means time derivative, and
\be
\label{3.222}
\al_q \equiv N_L \Phi_1 \; \int_{-L/2}^{L/2} \;
|\vp_q(z)|^4 \; dz \; .
\ee
Deriving (\ref{3.221}), we have also used the approximate equality
$$
\frac{\prt}{\prt q}\; \vp_q(z) \approx iz \vp_q(z)
$$
following from the fact that $\vp_q\propto e^{iqz}$.

Expanding
\be
\label{3.223}
E_{q+\hat p} \simeq E_q + v_q \hat p + \frac{1}{2m_q^*} \;
\hat p^2
\ee
in powers of $\hat p=-i\prt/\prt z$, with the group velocity $v_q$
and effective mass $m_q^*$ defined by the relations
\be
\label{3.224}
v_q \equiv \frac{\prt E_q}{\prt q} \; \qquad
\frac{1}{m_q^*} \equiv \frac{\prt^2 E_q}{\prt q^2} \; ,
\ee
we come to the evolution equation for the amplitude
\be
\label{3.225}
i \left ( \frac{\prt A}{\prt t} + v_q \;
\frac{\prt A}{\prt z} \right ) + \frac{1}{2m_q^*} \;
\frac{\prt^2 A}{\prt z^2} = \al_q |A|^2 A +
( \dot{q} - m\dot{v} ) zA \; .
\ee

\subsection{Soliton Formation}

Suppose that, after moving the lattice and reaching a Bloch state
with a quasimomentum $q$, the motion has been stopped, so that
$\dot{q}=\dot{v}=0$. Then Eq. (\ref{3.225}) reduces to
\be
\label{3.226}
i \left ( \frac{\prt A}{\prt t} + v_q \; \frac{\prt A}{\prt z}
\right ) + \frac{1}{2m_q^*} \; \frac{\prt^2 A}{\prt z^2} =
\al_q |A|^2 A \; .
\ee
This is a nonlinear Schr\"odinger equation supporting soliton
solutions [251]. Similar equations are met in laser physics [252],
in the theory of turbulent plasma [253,254], in the description
of magnetic matter [255], and in the theory of many other nonlinear
materials [256,257]. For Bose condensates, Eq. (\ref{3.226}) was
derived by Lenz et al. [258].

Equation (\ref{3.226}) can be simplified by changing the variable to
\be
\label{3.227}
x \equiv \frac{z-v_qt}{\xi} \; ,
\ee
where
\be
\label{3.228}
\xi = \frac{1}{\sqrt{2|m_q^*\ep|}}
\ee
is the healing length and $\ep$ is the soliton energy to be defined
by the normalization condition (\ref{3.218}). Let us introduce a
function $f(x)$ through the relation
\be
\label{3.229}
A(z,t) \equiv \sqrt{ \frac{\ep}{\al_q} } \; f(x) e^{-i\ep t} \; .
\ee
The function $f(x)$ can be chosen real, since Eq. (\ref{3.226}) is
invariant under the global gauge transformation $A\ra Ae^{i\al}$.
And let us define
\be
\label{3.230}
\zeta \equiv {\rm sgn} ( m_q^* \ep) \; .
\ee
Then Eq. (\ref{3.226}) can be reduced to
\be
\label{3.231}
\frac{d^2f}{dx^2} + \zeta \left ( 1 - f^2 \right ) f = 0 \; .
\ee
Depending on the sign of $\zeta$ in Eq. (\ref{3.230}), there are the
following possibilities.

\vskip 2mm

{\it Dark solitons} correspond to $\zeta=1$ and the boundary conditions
\be
\label{3.232}
\lim_{x\ra\pm\infty} \; f(x) = \pm 1 \; .
\ee
The name comes from the fact that the density distribution $|f(x)|^2$
has the lowest value at $x=0$. Dark solitons are called cavitons in the
theory of plasma [253,254] and in laser physics [252]. There can be two
types of dark solitons.

\vskip 2mm

{\it Normal dark soliton} is formed by atoms with a positive effective
mass, repulsive interactions, and with a positive soliton energy,
\be
\label{3.233}
m_q^* > 0 \; , \qquad \al_q > 0 \; , \qquad \ep > 0 \; .
\ee

\vskip 2mm

{\it Dark gap soliton} is characterized by a negative effective mass,
attractive interactions, and a negative soliton energy,
\be
\label{3.234}
m_q^* < 0 \; , \qquad \al_q < 0 \; , \qquad \ep < 0 \; .
\ee
It is worth stressing that the signs of the effective interaction
$\al_q$ and the soliton energy $\ep$ are chosen to be the same in
order that the expression $\sqrt{\ep/\al_q}$ in Eq. (\ref{3.229})
be real. This does not limit the generality, but simply takes into
account that the amplitude (\ref{3.229}) is defined up to a phase
factor.

The name of the gap soliton is due to the fact that, to achieve a
negative effective mass, the atomic cloud has to be shifted to the
edge of the Brillouin zone. This shift can be realized by the
appropriate motion of the lattice.

The form of the dark soliton, being the solution of Eq. (\ref{3.231}),
with $\zeta=1$, under the boundary conditions (\ref{3.232}), is
\be
\label{3.235}
f(x) = {\rm tanh}\; \frac{x}{\sqrt{2}} \; .
\ee
The normalization condition (\ref{3.218}) for amplitude (\ref{3.229}),
with $f(x)$ from Eq. (\ref{3.235}), gives
\be
\label{3.236}
\ep = \al_q = \frac{1}{2m_q^*\xi^2} \; ,
\ee
the healing length (\ref{3.228}) being
\be
\label{3.237}
\xi = \frac{1}{\sqrt{2m_q^*\al_q} } \; .
\ee
Hence, the total dark envelope (\ref{3.229}) becomes
\be
\label{3.238}
A(z,t) = {\rm tanh} \left (
\frac{z-v_q t}{\sqrt{2}\; \xi} \right ) e^{-i\ep t} \; ,
\ee
with the healing length (\ref{3.237}) and soliton energy (\ref{3.236}).

{\it Bright solitons} arise for $\zeta=-1$, under the boundary conditions
\be
\label{3.239}
\lim_{x\ra\pm\infty} f(x) = 0 \; .
\ee
Such solitons correspond to the maximum of the density distribution
$|f(x)|^2$ at $x=0$. They are called as well the bell solitons. There
are again two possibilities.

\vskip 2mm

{\it Normal bright soliton} is described by a positive effective mass,
though attractive interactions and negative soliton energy,
\be
\label{3.240}
m_q^* > 0 \; , \qquad \al_q < 0 \; , \qquad \ep < 0 \; .
\ee

\vskip 2mm

{\it Bright gap soliton} possesses a negative effective mass, but
repulsive interactions and positive soliton energy,
\be
\label{3.241}
m_q^* < 0 \; , \qquad \al_q > 0 \; , \qquad \ep > 0 \; .
\ee

Again, to make the effective mass negative, it is necessary to move
the lattice so that the atomic cloud would acquire a quasimomentum
at the edge of the first Brillouin zone.

The bright soliton solution, resulting from Eq. (\ref{3.231}), with
$\zeta=-1$, under the boundary conditions (\ref{3.239}), is
\be
\label{3.242}
f(x) = \frac{\sqrt{2}}{{\rm cosh}x} \; .
\ee
The normalization condition (\ref{3.218}) yields the soliton energy
\be
\label{3.243}
\ep = - \; \frac{m_q^*}{8} \left (\al_q L\right )^2 = -\;
\frac{1}{2m_q^*\xi^2} \; ,
\ee
with the healing length
\be
\label{3.244}
\xi = \frac{2}{|m_q^*\al_q L|} \; .
\ee
The total bright soliton envelope (\ref{3.229}) takes the form
\be
\label{3.245}
A(z,t) = \sqrt{ \frac{L}{2\xi} } \; {\rm sech} \left (
\frac{z-v_qt}{\xi} \right ) e^{-i\ep t} \; ,
\ee
with the healing length (\ref{3.244}) and soliton energy (\ref{3.243}).
This form is analogous to Langmur solitons in plasma [253,254].

One sometimes distinguishes solitons by their topological charge,
defined as
$$
\lim_{x\ra\infty} [ f(x) - f(-x) ] \; .
$$
When the latter is nonzero, the solitons are termed topological. Thus,
dark solitons are topological. If the topological charge is zero, the
solitons are called nontopological. Hence, bright solitons are
nontopological.

Dark solitons of BEC were generated for repulsive $^{78}$Rb atoms
[259,260] and bright solitons in BEC were formed with attractive $^7$Li
atoms [261,262]. The specific feature of gap solitons is that they can
be created only in the presence of a periodic lattice [263]. Gap
solitons were observed for $^{87}$Rb atoms [264].

\subsection{Transverse Resonance}

When creating gap solitons, it is necessary to shift an atomic cloud
to the boundary of the Brillouin zone, where the effective mass becomes
negative. But then the Bloch energy increases, and it may happen that
the transverse modes of atomic motion could be excited. In such a case,
the quasi-one-dimensional picture for treating a single wave packet is
not anymore appropriate, and one has to take into account transverse
excitations [265]. This can be done in the following way [250].

When, despite of a strong transverse harmonic confinement, nevertheless,
some transverse modes can be excited, then it is necessary, instead of
Eq. (\ref{3.118}), to consider the transverse motion described by the
equation
\be
\label{3.246}
\left ( - \; \frac{\nabla_\perp^2}{2m} + \frac{m}{2} \;
\om_\perp^2 r_\perp^2 \right ) \chi_n(\br_\perp) =
E_n^\perp \chi_n(\br_\perp) \; .
\ee
Under the harmonic transverse confinement, the eigenenergy spectrum
of Eq. (\ref{3.246}) is
\be
\label{3.247}
E_n^\perp = ( 2 n_r + |m_a| + 1) \om_\perp \; ,
\ee
where $n_r=0,1,2,\ldots$ is a radial quantum number,
$m_a=0,\pm 1,\pm 2,\ldots$ is an azimuthal quantum number, and the
multi-index $n=\{ n_r,m_a\}$ includes both of these numbers. The
lowest energy (\ref{3.247}) is $E_0^\perp=\om_\perp$.

When trapped atoms are in the lowest-energy transverse state, then
their total spectrum is the sum of the lowest energy of transverse
motion, $\om_\perp$, and of the Bloch energy $E_q$. For an excited
transverse state, with an energy $E_n^\perp$, the total spectrum
is the sum $E_n^\perp+E_p$. Although $\om_\perp<E_n^\perp$, for
$n\neq 0$, but if $q\approx\pi/a$, there exists such a quasimomentum
$p\in[-\pi/a,\pi/a]$ that $E_q>E_p$. And the resonance condition
\be
\label{3.248}
\om_\perp + E_q = E_n^\perp + E_p
\ee
can become valid. Then the transverse modes, with the energy
$E_n^\perp$, become excited.

In the presence of the excited transverse modes, the condensate
wave function $\eta(\br,t)$, satisfying Eq. (\ref{3.115}), has to
be written as
\be
\label{3.249}
\eta(\br,t) = \sum_i \chi_{n_i}(\br_\perp) B_i(z,t)
\vp_{q_i}(z) \exp\left \{ -i \left ( E_{n_i}^\perp -
\mu_0 \right ) t \right \} \; ,
\ee
which generalizes Eqs. (\ref{3.116}), (\ref{3.127}), and
(\ref{3.215}). The transverse wave functions $\chi_n(\br_\perp)$
are normalized to one, as in Eq. (\ref{3.117}). The Bloch functions
$\vp_q(z)$ are normalized as in Eq. (\ref{3.219}). But the envelopes
$B_i(z,t)$ satisfy the normalization conditions
\be
\label{3.250}
\frac{1}{L} \; \int_{-L/2}^{L/2} \; |B_i(z,t)|^2 \; dz =
N_i \; ,
\ee
where $N_i=N_i(t)$, generally, are functions of time. This
normalization differs from that in Eq. (\ref{3.218}).

Now, instead of one interaction parameter (\ref{3.222}), there are
several interaction parameters
\be
\label{3.251}
\al_{ijkl} \equiv N_L \Phi_0
\int \chi_{n_i}^* \chi_{n_j}^* \chi_{n_k} \chi_{n_l} \; d\br_\perp
\; \int \vp_{q_i}^* \vp_{q_j}^* \vp_{q_k} \vp_{q_l} \; dz \; .
\ee
Following the same way as in the previous sections, instead of Eq.
(\ref{3.226}), we obtain the set of equations
\be
\label{3.252}
i\left ( \frac{\prt B_i}{\prt t} + v_i \;
\frac{\prt B_i}{\prt z} \right ) + \frac{1}{2m_i^*} \;
\frac{\prt^2 B_i}{\prt z^2} = \frac{1}{N_L} \;
\sum_{jkl} \al_{ijkl} B_j^* B_k B_l
\ee
for the mode envelopes $B_i(z,t)$, with $i=0,1,2$. The mode
envelope $B_0(z,t)$ corresponds to the central mode of a gap soliton,
with $q=\pi/a$, while the envelopes $B_1(z,t)$ and $B_2(z,t)$, to the
two side modes, for which the resonance condition (\ref{3.248}) is
valid. There are two transverse modes, since the resonance condition
(\ref{3.248}) holds for two Bloch energies $E_{q_1}$ and $E_{q_2}$,
for which
\be
\label{3.253}
E_{q_1} = E_{q_2}  \qquad \left ( q_1 = \frac{\pi}{a}-q \; , \;\;
q_2 = \frac{\pi}{a} + q\right ) \; .
\ee

The form of parameters (\ref{3.251}) shows that
\be
\label{3.254}
\al_{ijkl} = \al_{jikl} = \al_{ijlk} \; .
\ee
Also, for the quasimomenta $q_1$ and $q_2$, related as in Eq.
(\ref{3.253}), one has
\be
\label{3.255}
\vp_{q_1}^*(z) = \vp_{q_2}(z) \; .
\ee
For the transverse wave functions, according to condition
(\ref{3.248}), one has
$$
\chi_{n_1}(\br_\perp) = \chi_{n_2}(\br_\perp) \; .
$$
Since the functions $\chi_0(\br_\perp)$ and $\chi_1(\br_\perp)$
possess different symmetries with respect to the inversion of
$\br_\perp$, not all integrals in Eq. (\ref{3.251}) are nonzero.
These are $\al_{0000}$ and
$$
\al_{1111} = \al_{2222} = \al_{1212} \; , \qquad
\al_{0101} = \al_{0202} = \al_{0012} \; ,
$$
as well as all those that are obtained from the above ones using
symmetry (\ref{3.254}) and property (\ref{3.255}).

An interesting solution of Eqs. (\ref{3.252}) is represented by
a {\it triple solution} [250], which is a triplet of solitons, one of
which corresponds to $q=\pi/a$, that is, to a gap soliton, while two
others are the side transverse modes. All three modes are bound with
each other, so that they stay localized in space. Bound triplets of
solitons are also called tritons [256].

It is worth mentioning that quasi-one-dimensional gap solitons,
as those that are considered in Sec. 3.13, are usually unstable
with respect to the formation of transverse modes [256]. Hence,
gap solitons are, strictly speaking, quasisolitons, that is, the
soliton-like solutions that in the long run are unstable, but can
live sufficiently long to be observable. However the triple gap
soliton can be stable [250].

\subsection{Lagrange Variation}

Instead of solving the system of partial differential equations,
it is possible to reduce the problem to the solution of a set of
ordinary differential equations by means of the Lagrange variational
method. To illustrate the latter, let us consider the system of
three partial differential equations (\ref{3.252}).

Let us define the energy functional
\be
\label{3.256}
E [B] \; \equiv \; \sum_j \; \int B_j^* \; i \;
\frac{\prt}{\prt t} \; B_j \; dz \; ,
\ee
the Hamiltonian functional
\be
\label{3.257}
H [B]\; \equiv \; \sum_j \; \int B_j^* \left ( v_j \hat p +
\frac{{\hat p}^2}{2m} \right ) B_j \; dz \; + \;
\frac{1}{2L} \; \sum_{ijkl} \al_{ijkl}
\int B_i^* B_j^* B_k B_l \; dz \; ,
\ee
and the Lagrangian
\be
\label{3.258}
L [B] \equiv E [B] - H[B] \; .
\ee
Equations (\ref{3.252}) follow from the Lagrange variational
equations
\be
\label{3.259}
\frac{d}{dt} \; \frac{\dlt L[B]}{\dlt\dot{B}_j} \; - \;
\frac{\dlt L[B]}{\dlt B_j} \; .
\ee

Approximate solutions to Eqs. (\ref{3.252}) can be constructed by
invoking trial forms for the wave packets $B_i$, for instance, as
the Gaussian envelopes

\be
\label{3.260}
B_j = \frac{C_j}{(\sqrt{\pi}\; b_j)^{1/2}} \;
\exp\left\{ - \; \frac{(z-z_j)^2}{2b_j} \right \} \;
\exp\left\{ - i \left ( \al_j t - \bt_j z - \gm_j z^2
\right ) \right \} \; ,
\ee
where all variables $C_j$, $b_j$, $z_j$, $\al_j$, $\bt_j$, and
$\gm_j$ are treated as functions of time. The evolution equations
for all these variables are obtained by applying the Lagrange
equations to each of the variables $C_j$, $b_j$, $z_j$, $\al_j$,
$\bt_j$, and $\gm_j$. Then, instead of three equations (\ref{3.252})
in partial derivatives, one gets a set of 18 equations in ordinary
derivatives [250]. The latter are much easier to solve numerically,
as well as to analyze the stability of their solutions.

\section{Boson Hubbard Model}

\subsection{Wannier Representation}

In Sec. 3.3 the representation of the grand Hamiltonian (\ref{3.15})
is given by expanding the field operators over Bloch functions. This
results in the Bloch representation of the Hamiltonian (\ref{3.55})
specified in Eq. (\ref{3.56}) to (\ref{3.59}). The field operators
could also be expanded over the basis of Wannier functions. Such a
Wannier representation, leading to the Hubbard model [266], is widely
employed for treating electrons in solid-state lattices and the related
metal-insulator phase transition [267--269]. In a particular case of
half filling and neglecting double occupancies the Hubbard model can
be reduced to the so-called $t-J$ model [270].

For a periodic Bose system with BEC, the field operator can be expanded
over Wannier functions,
\be
\label{4.1}
\hat\psi(\br) = \sum_{nj} \hat c_{ij} w_n(\br-\ba_j) \; .
\ee
Keeping in mind the Bogolubov-shifted field operator (\ref{2.86}), that
is
\be
\label{4.2}
\hat\psi(\br) = \eta(\br) + \psi_1(\br) \; ,
\ee
we have the expansion
\be
\label{4.3}
\eta(\br) = \sqrt{\frac{N_0}{N_L}} \; \sum_j w_0(\br-\ba_j)
\ee
for the condensate wave function and the expansion
\be
\label{4.4}
\psi_1(\br) = \sum_{nj} c_{nj} w_n (\br -\ba_j)
\ee
for the operator of uncondensed atoms. This means that
\be
\label{4.5}
\hat c_{nj} =\sqrt{\frac{N_0}{N_L} } \; \dlt_{n0} + c_{nj} \; .
\ee
Summing this over the lattice yields
\be
\label{4.6}
\sum_{nj} \hat c_{nj}  =\sqrt{N_0 N_L} \;  + \;
\sum_{nj} c_{nj} \; .
\ee
Remembering the orthogonality property (\ref{3.51}) and Eq.
(\ref{3.53}), we have
\be
\label{4.7}
\sum_j \hat c_{0j}  =\sqrt{N_0N_L} \; , \qquad
\sum_j c_{0j} = 0 \; .
\ee

Substituting expansion (\ref{4.1}) into the energy Hamiltonian
(\ref{2.156}), we meet the following matrix elements: the single-site
term
\be
\label{4.8}
h_i^{mn} \equiv \int w_m^*(\br-\ba_i) \; H_L(\br) \;
w_n(\br-\ba_i) \; d\br \; ,
\ee
the hopping, or tunneling, term
\be
\label{4.9}
J_{ij}^{mn} \equiv - \int w_m^*(\br-\ba_i) \;
H_L(\br) \; w_n(\br-\ba_j) \; d\br \; ,
\ee
where  $i\neq j$, and the interaction term
\be
\label{4.10}
U_{j_1 j_2 j_3 j_4}^{n_1 n_2 n_3 n_4} \equiv
\Phi_0 \int w_{n_1}^*(\br-\ba_{j_1}) w_{n_2}^*(\br-\ba_{j_2})
w_{n_3}(\br-\ba_{j_3})  w_{n_4}(\br-\ba_{j_4}) \; d\br \; ,
\ee
where the local interaction potential (\ref{2.174}) is assumed and
$$
H_L (\br) \equiv -\; \frac{\nabla^2}{2m} + V_L(\br)
$$
is the linear lattice Hamiltonian. Then Hamiltonian (\ref{2.156})
becomes
$$
\hat H = - \sum_{i\neq j} \; \sum_{mn}
J_{ij}^{mn} \hat c_{mi}^\dgr \hat c_{nj} \; + \;
\sum_j \; \sum_{mn}
h_{j}^{mn} \hat c_{mj}^\dgr \hat c_{nj} \; +
$$
\be
\label{4.11}
 + \; \frac{1}{2} \; \sum_{ \{j\} } \;
\sum_{ \{n\} } U_{j_1 j_2 j_3 j_4}^{n_1 n_2 n_3 n_4}
\hat c_{n_1j_1}^\dgr \hat c_{n_2j_2}^\dgr \hat c_{n_3j_3}
 \hat c_{n_4j_4} \; .
\ee

To simplify Eq. (\ref{4.11}), one supposes that the main contribution
here comes from the lowest band, so that the single-band approximation
can be employed. In so doing, one omits the band indices. Implying
that Wannier functions are well localized, one retains in the hopping
term only the nearest neighbors, with the tunneling parameter $J$, and
in the interaction term, one keeps only the on-site interaction, with
an interaction parameter $U$. Thus, one arrives at the {\it Hubbard
model}
\be
\label{4.12}
\hat H = - J \sum_{<ij>} \hat c_i^\dgr \hat c_j \; + \;
h_0 \sum_j \hat c_j^\dgr \hat c_j \; + \;
\frac{U}{2} \; \sum_j \hat c_j^\dgr \hat c_j^\dgr
\hat c_j \hat c_j \; ,
\ee
in which $<ij>$ means the summation over nearest neighbors and the
parameters $J$, $h_0$, and $U$ do not depend on the lattice-site
indices because of the lattice regularity, when other external fields,
except the lattice one, are absent.

The Hubbard Hamiltonian (\ref{4.12}) is widely used for treating
electrons in condensed matter [266--270]. The principal difference of
the case of bosons, in the presence of BEC, from the case of fermions,
is that the Wannier field operators here are Bogolubov-shifted, such
that
\be
\label{4.13}
\hat c_j \equiv \sqrt{\frac{N_0}{N_L} } \; + \; c_j \; ,
\ee
in agreement with Eq. (\ref{4.5}).

The Hamiltonian parameters can be calculated in the same way as it
was done for the quasi-one-dimensional case of Sec. 3.9. Considering
now a three-dimensional lattice in the tight-binding approximation
yields
$$
h_0 \cong 3E_R \; \sqrt{\frac{V_0}{E_R} } \; , \qquad
J \cong \frac{3}{4} \left ( \pi^2 - 4\right ) V_0 \exp\left (
-\; \frac{\pi^2}{4}\; \sqrt{\frac{V_0}{E_R} }\right ) \; ,
$$
\be
\label{4.14}
U \cong \sqrt{\frac{8}{\pi} } \; k_0 a_s E_R \left (
\frac{V_0}{E_R} \right )^{3/4} \; .
\ee
The ratio of the on-site interaction to the hopping parameter is
\be
\label{4.15}
\frac{U}{J} \cong 0.362 k_0 a_s \left (
\frac{E_R}{V_0} \right )^{1/4} \exp\left ( \frac{\pi^2}{4} \;
\sqrt{\frac{V_0}{E_R}} \right ) \; ,
\ee
where $k_0=|\bk_0|$ is the laser wave vector modulus, being for
a cubic lattice $k_0=\pi/a$. As is seen from ratio (\ref{4.15}),
making the lattice deeper by rising the lattice depth $V_0$ increases
the influence of the on-site interaction.

\subsection{Grand Hamiltonian}
For a lattice system with BEC, the Hubbard Hamiltonian (\ref{4.12})
is a part of the grand Hamiltonian (\ref{3.15}). It is necessary to
deal with a grand Hamiltonian, since the Wannier field operators,
as any other field operators, do not conserve the number of particles
[271] as well as other normalization conditions. Such conditions and
other additional constraints are to be taken into account for
constructing a representative statistical ensemble guaranteeing theory
self-consistency [56,57,93--99].

Recalling the definition of the condensate fraction, $n_0$, the
fraction of uncondensed atoms $n_1$, and that of the lattice filling
factor $\nu$,
\be
\label{4.16}
n_0 \equiv \frac{N_0}{N} \; , \qquad n_1 \equiv \frac{N_1}{N} \; ,
\qquad \nu \equiv \frac{N}{N_L} \; ,
\ee
we can denote the filling factors of condensed and uncondensed atoms
as
\be
\label{4.17}
\frac{N_0}{N} = \nu n_0 \; , \qquad \frac{N_1}{N_L} = \nu n_1 \; .
\ee
Then the Bogolubov shift (\ref{4.13}) writes as
\be
\label{4.18}
\hat c_j = \sqrt{\nu n_0} \; + \; c_j \; .
\ee
The orthogonality condition (\ref{4.7}) yields
\be
\label{4.19}
\sum_j c_j = 0 \qquad (n_0 > 0 ) \; .
\ee
The quantum-number conservation condition $<\psi_1>=0$ requires that
\be
\label{4.20}
<c_j >\; = \; 0 \; .
\ee
The latter means that there should be no linear in $c_j$ terms in
the grand Hamiltonian [57]. Averaging operator (\ref{4.18}) gives
the condensate order parameter
\be
\label{4.21}
 < \hat c_j > \; = \; \sqrt{\nu n_0} \; .
\ee

The grand Hamiltonian (\ref{3.15}) includes the Hubbard part
(\ref{4.12}), the number of condensed atoms $N_0$, the number
operator for uncondensed atoms
\be
\label{4.22}
\hat N_1 = \sum_j c_j^\dgr c_j \; ,
\ee
and the linear killer
\be
\label{4.23}
\hat\Lbd = \sum_j \left ( \lbd_j c_j^\dgr +
\lbd_j^* c_j \right ) \; ,
\ee
which guarantees the absence in $H$ of the terms linear in $c_j$,
if such occur.

The grand Hamiltonian (\ref{3.15}), in the Wannier representation,
acquires the form
\be
\label{4.24}
H = H^{(0)} + H^{(2)} + H^{(3)} + H^{(4)} \; ,
\ee
analogously to its Bloch representation (\ref{3.55}). In the Wannier
representation, we have
\be
\label{4.25}
H^{(0)} = -J z_0 n_0 N + \frac{V}{2}\; \nu n_0^2 N -
(\mu_0-h_0) n_0 N \; ,
\ee
where $z_0$ is the number of nearest neighbors,
\be
\label{4.26}
z_0 \equiv \frac{1}{N} \; \sum_{<ij>} 1 \; .
\ee
The linear in $c_j$ term is absent from Eq. (\ref{4.24}) due to the
orthogonality condition (\ref{4.19}). The second-order term is
\be
\label{4.27}
H^{(2)} = - J \sum_{<ij>} c_i^\dgr c_j \; + \;
\left ( 2U \nu n_0 - \mu_1 + h_0 \right ) \sum_j c_j^\dgr c_j \; +
\; \frac{U}{2}\; \nu n_0 \sum_j \left ( c_j^\dgr c_j^\dgr + c_j c_j
\right ) \; .
\ee
The third- an fourth-order terms are
\be
\label{4.28}
H^{(3)} = U \sqrt{\nu n_0} \; \sum_j \left ( c_j^\dgr c_j^\dgr c_j +
c_j^\dgr c_j c_j \right )
\ee
and, respectively,
\be
\label{4.29}
H^{(4)} = \frac{U}{2} \; \sum_j c_j^\dgr c_j^\dgr c_j c_j \; .
\ee

The condition of equilibrium,
\be
\label{4.30}
< \frac{\prt H}{\prt N_0} > \; = \; 0 \; ,
\ee
defines the Lagrange multiplier $\mu_0$. In the Wannier representation,
the normal fraction reduces to
\be
\label{4.31}
n_1 = \frac{1}{N} \; \sum_j < c_j^\dgr c_j > \; ,
\ee
while the anomalous average becomes
\be
\label{4.32}
\sgm \equiv \frac{1}{N} \; \sum_j < c_j c_j > \; .
\ee
For an ideal lattice, the averages in the right-hand side of Eqs.
(\ref{4.31}) and (\ref{4.32}) do not depend on the site index $j$.
Therefore one has
\be
\label{4.33}
 < c_j^\dgr c_j > \; = \; \nu n_1 \; , \qquad
< c_j c_j > \; = \; \nu \sgm \; .
\ee
The constant $h_0$ can be incorporated in the notation of the
Lagrange multipliers $\mu_0$ and $\mu_1$. In that way, the condition
of equilibrium (\ref{4.30}) results in
\be
\label{4.34}
\mu_0 =  -J z_0 + \nu n_0 U + \nu U \left [ 2n_1 +
\frac{1}{2} \left ( \sgm^* + \sgm \right ) \right ] \; + \;
\frac{U}{2\sqrt{\nu n_0}} \; \sum_j < c_j^\dgr c_j^\dgr c_j +
c_j^\dgr c_j c_j > \; .
\ee

The equations of motion for the operators $c_j$ are given by the
equalities
$$
i \; \frac{\prt c_j}{\prt t} = \frac{\prt H}{\prt c_j^\dgr} =
[ c_j, H ] \; ,
$$
which yield
\be
\label{4.35}
i\; \frac{\prt c_j}{\prt t} = \left ( - J + 2\nu n_0 U - \mu_1
\right ) c_j + \nu n_0 U c_j^\dgr + \sqrt{\nu n_0} \; U \left (
2c_j^\dgr c_j + c_j c_j \right ) + U c_j^\dgr c_j c_j \; .
\ee

Let us note that condition (\ref{4.30}) defines an equilibrium state.
When one is interested in a nonequilibrium situation, one should go
back to the condensate-function equation (\ref{2.172}). The condensate
wave function $\eta(\br,t)$ can be expanded over Wannier functions. In
a particular case of a purely coherent system, one obtains the Wannier
representation for the nonlinear Schr\"odinger equation [272], which
does not take account of either uncondensed atoms or anomalous averages.
For more realistic cases, one should deal with the total Eq.
(\ref{2.172}).

\subsection{Bose-Condensed System}

The description of BEC in an optical lattice, based on the Wannier
representation of the grand Hamiltonian (\ref{4.24}) and on the
Hubbard model (\ref{4.12}), can be done similarly to the way used
when dealing with the Bloch representation in Sec. 3.3. The
consideration now becomes simpler, since in deriving the Hubbard
model (\ref{4.12}) some simplifications were invoked, which were
justified by good localization of Wannier functions.

The field operator of uncondensed atoms, in the single-band
approximation, can be expanded either over Wannier or over Bloch
functions,
\be
\label{4.36}
\psi_1(\br) = \sum_j c_j w(\br-\ba_j) = \sum_k a_k \vp_k(\br) \; .
\ee
Therefore the field operators in the Wannier and Bloch representations
are related as
\be
\label{4.37}
c_j = \frac{1}{\sqrt{N_L}} \; \sum_k a_k e^{i\bk\cdot\ba_j} \; ,
\ee
following from the relation
$$
\vp_k(\br) = \frac{1}{\sqrt{N_L}} \;
\sum_j w(\br-\ba_j) e^{i\bk\cdot\ba_j}
$$
between the Bloch and Wannier functions. Let us recall that from the
orthogonality condition (\ref{3.51}) one has conditions (\ref{3.52})
and (\ref{3.53}), which, for the single-band case, reduce to condition
(\ref{4.19}) and to
\be
\label{4.38}
\lim_{k\ra 0} a_k = 0 \qquad (n_0 > 0) \; .
\ee

Substituting relation (\ref{4.37}) into Eq. (\ref{4.27}), assuming
a cubic $d$-dimensional lattice, and using the equality
$$
\sum_{<ij>} e^{i\bk\cdot\ba_{ij}} = 2 N_L
\sum_{\al=1}^d \cos(k_\al a) \; ,
$$
we have
$$
H^{(2)} = \sum_k \left [ - 2J \sum_\al \cos (k_\al a) \; + \;
2U\nu n_0 - \mu_1 \right ] a_k^\dgr a_k \; +
$$
\be
\label{4.39}
+ \; \frac{U}{2}\; \nu n_0 \; \sum_k \left (
a_k^\dgr a_{-k}^\dgr + a_{-k} a_k \right ) \; .
\ee
For the third-order term (\ref{4.28}), we get
\be
\label{4.40}
H^{(3)} = U \; \sqrt{\frac{\nu n_0}{N_L} } \;
\sum_{kp} \left ( a_k^\dgr a_p a_{k+p} +
a_{k+p}^\dgr a_p a_k \right ) \; .
\ee
And the fourth-order term (\ref{4.29}) becomes
\be
\label{4.41}
H^{(4)} = \frac{U}{2N_L} \;
\sum_{kpq} a_k^\dgr a_p^\dgr a_{k+q} a_{p-q} \; .
\ee

In the Bogolubov approximation, one omits the third-and fourth-order
terms (\ref{4.40}) and (\ref{4.41}), as has been done in Refs. [273,274].
A more general consideration is based on the Hartree-Fock-Bogolubov (HFB)
approximation, in line with the self-consistent approach developed in
Refs. [57,93--99].

Applying the HFB approximation to Eqs. (\ref{4.40}) and (\ref{4.41})
requires to recall the definition of the normal and anomalous averages
\be
\label{4.42}
n_k \; \equiv \; < a_k^\dgr a_k > \; , \qquad
\sgm_k \; \equiv \; < a_{-k} a_k >
\ee
and of their integral forms
\be
\label{4.43}
n_1 = \frac{1}{N} \; \sum_k n_k \; , \qquad \sgm= \frac{1}{N} \;
\sum_k \sgm_k \; .
\ee

The third-order term (\ref{4.40}) in the HFB approximation is zero
because of the orthogonality condition (\ref{4.38}). And the fourth-order
term (\ref{4.41}) becomes
\be
\label{4.44}
H^{(4)} = \frac{\nu}{2}\; U \; \sum_k \left (
4 n_1 a_k^\dgr a_k + \sgm a_k^\dgr a_{-k}^\dgr +
\sgm^* a_{-k} a_k \right ) \; - \; \frac{\nu}{2} \;
U N \left ( 2n_1^2 +|\sgm|^2\right ) \; .
\ee
Combining the terms of the grand Hamiltonian (\ref{4.24}), we employ
the notation
\be
\label{4.45}
\om_k \equiv - 2J \sum_\al \cos( k_\al a) + 2\nu U -\mu_1
\ee
and
\be
\label{4.46}
\Dlt \equiv \nu U ( n_0 +\sgm) \; .
\ee
Then Eq. (\ref{4.24}) reduces to the form
\be
\label{4.47}
H = E_{HFB} + \sum_k \om_k a_k^\dgr a_k \; + \;
\frac{1}{2} \; \sum_k \left ( \Dlt a_k^\dgr a_{-k}^\dgr +
\Dlt^* a_{-k} a_k \right ) \; ,
\ee
in which
\be
\label{4.48}
E_{HFB} \equiv H^{(0)} - \nu N \; \frac{U}{2}\left ( 2n_1^2 +
|\sgm|^2 \right ) \; .
\ee
Similarly to the uniform case [57,94], the anomalous average $\sgm$
can be made real. The Lagrange multiplier (\ref{4.34}), in the HFB
approximation, is
\be
\label{4.49}
\mu_0 = - z_0 J + \nu U ( 1 + n_1 + \sgm) \; ,
\ee
where the normalization $n_0+n_1=1$ is used.

The HFB Hamiltonian (\ref{4.47}) can be diagonalized by means of
the Bogolubov canonical transformation in the same way as it is done
for a uniform system [94--98]. This reduces Hamiltonian (\ref{4.47})
to the Bogolubov form
\be
\label{4.50}
H_B = E_B + \sum_k \ep_k b_k^\dgr b_k \; ,
\ee
in which
\be
\label{4.51}
E_B = E_{HFB} \; + \; \frac{1}{2} \; \sum_k ( \ep_k - \om_k )
\ee
and the Bogolubov spectrum is
\be
\label{4.52}
\ep_k = \sqrt{\om_k^2 - \Dlt^2} \; .
\ee
As is explained in Sec. 2.9, the condensate existence requires the
condensation condition
\be
\label{4.53}
\lim_{k\ra 0} \ep_k = 0 \; , \qquad {\rm Re}\; \ep_k \geq 0 \; .
\ee
The latter defines the Lagrange multiplier
\be
\label{4.54}
\mu_1 = - z_0 J + \nu U ( 1 + n_1 - \sgm ) \; ,
\ee
which is different from Eq. (\ref{4.49}). Recall that $z_0$ is
the coordination number that for a $d$-dimensional cubic lattice is
$z_0=2d$. Substituting Eq. (\ref{4.54}) into notation (\ref{4.45})
yields
\be
\label{4.55}
\om_k = \Dlt + 2J \sum_\al [ 1 - \cos(k_\al a) ] =
\Dlt + 4J \sum_\al \sin^2\left ( \frac{k_\al a}{2}\right ) \; .
\ee

In the long-wave limit, when $k\equiv|\bk|\ra 0$,
\be
\label{4.56}
\om_k \simeq \Dlt + J (ka)^2 \qquad (k\ra 0) \; .
\ee
Therefore, spectrum (\ref{4.52}) is of the phonon type,
\be
\label{4.57}
\ep_k \simeq ck \qquad (k\ra 0) \; ,
\ee
with the sound velocity
\be
\label{4.58}
c = \sqrt{ 2J a^2 \Dlt} \; .
\ee

When the lattice is not cubic, it is characterized by $d$ lattice
spacings $a_\al$, with $\al=1,2,\ldots,d$. Hence, generally, for a
$d$-dimensional lattice, there can be defined $d$ sound velocities
$$
c_\al \equiv \lim_{k\ra 0} \; \frac{\prt\ep_k}{\prt k_\al} =
\sqrt{2J\Dlt}\; a_\al \; .
$$

Similarly to the uniform case [94--98], for Eqs. (\ref{4.42}), we get
the momentum distribution
\be
\label{4.59}
n_k = \frac{\om_k}{2\ep_k}\;
{\rm coth}\left ( \frac{\ep_k}{2T} \right ) \; - \;
\frac{1}{2}
\ee
and the anomalous average
\be
\label{4.60}
\sgm_k = -\; \frac{\Dlt_k}{2\ep_k}\;
{\rm coth}\left ( \frac{\ep_k}{2T} \right ) \; .
\ee
Transforming sums (\ref{4.43}) to integrals according to Eq.
(\ref{3.45}), we find the fraction of uncondensed atoms
\be
\label{4.61}
n_1 = \frac{1}{\rho}\; \int_{\cal B} n_k \;
\frac{d\bk}{(2\pi)^3}
\ee
and the anomalous average
\be
\label{4.62}
\sgm = \frac{1}{\rho}\; \int_{\cal B} \sgm_k \;
\frac{d\bk}{(2\pi)^3} \; ,
\ee
where the integration is over the Brillouin zone.

In the center of the Brillouin zone, functions (\ref{4.59}) and
(\ref{4.60}) behave as
$$
n_k \simeq \frac{T\Dlt}{(ck)^2} \; , \qquad \sgm_k \simeq -\;
\frac{T\Dlt}{(ck)^2} \qquad (k\ra 0) \; .
$$
At the boundary of the Brillouin zone, one has
$$
\om_k \simeq \Dlt + 2z_0 J \; , \qquad
\ep_k \simeq 2\sqrt{z_0 J(\Dlt+z_0J)} \qquad \left ( k_\al\ra
\frac{\pi}{a_\al} \right ) \; .
$$
This shows that both $n_k$ and $\sgm_k$ are integrable, so that $n_1$
as well as $\sgm$ in Eqs. (\ref{4.61}) and (\ref{4.62}) are finite.
That is, in the case of a lattice, there are no problems with a
divergent anomalous average, as for a uniform system with the local
interaction potential, which is discussed in Sec. 2.13. The lattice
regularizes $\sgm$ making it always finite.

The condensate fraction $n_0=1-n_1$ exists below the BEC phase
transition $T_c$. This is a second-order phase transition between
the normal phase and Bose-condensed phase. At $T\ra T_c$, one has
$n_0\ra 0$, $\sgm\ra 0$, and $\Dlt\ra 0$. Also, $\ep_k\ra\om_k$,
when $T\ra T_c$. At the critical temperature, Eq. (\ref{4.55})
becomes
\be
\label{4.63}
\om_k =  4J \sum_\al \sin^2\left ( \frac{k_\al a}{2} \right )
\qquad ( T = T_c) \; .
\ee
The critical temperature $T_c$, where $n_0=0$ and $n_1=1$, is given
by the equation
\be
\label{4.64}
\rho = \frac{1}{2} \; \int_{\cal B} \left [ {\rm coth}
\left ( \frac{\om_k}{2T_c} \right ) -1 \right ] \;
\frac{d\bk}{(2\pi)^d} \; ,
\ee
in which $\om_k$ is defined in Eq. (\ref{4.63}) and a $d$-dimensional
lattice us considered.

To estimate the critical temperature, let us keep in mind a cubic
lattice, for which the filling factor can be written as
\be
\label{4.65}
\nu \equiv \frac{N}{N_L} = \rho a^d \; .
\ee
By introducing the dimensionless quasimomentum vector
$$
{\bf x} \equiv \left \{ x_\al = \frac{k_\al a}{\pi} : \;
\al=1,2,\ldots,d \right \} \; ,
$$
Eq. (\ref{4.64}) can be represented in the form
\be
\label{4.66}
2^{d+1}\nu = \int_{\overline{\cal B}} \left \{ {\rm coth} \left [
\frac{2J}{T_c} \; \sum_{\al=1}^d \sin^2\left (
\frac{\pi}{2}\; x_\al \right )  \right ] -1 \right \} \;
d{\bf x} \; ,
\ee
in which the integration over each $x_\al$ is between $-1$ and $1$,
the dimensionless Brillouin zone being
$$
\overline{\cal B} \equiv \{ x_\al \in [-1,1] \; : \;
\al=1,2,\ldots, d \} \; .
$$
Noticing that the main contribution to integral (\ref{4.66}) comes
from the central region of the Brillouin zone, we can approximate
this integral by considering the asymptotic behaviour of the integrand
at small $x_\al$. This gives
\be
\label{4.67}
T_c \cong 2^d \pi^2 \left (
\int_{\overline{\cal B}} \frac{d{\bf x}}{\sum_\al x_\al^2}
\right )^{-1} J \nu \; .
\ee
In that approximation, BEC does not happen for one- and two-dimensional
lattices,
\be
\label{4.68}
T_c \; \leq \; 0 \qquad (d\leq 2) \; .
\ee
For larger dimensionality $d>2$, we can evaluate the integral in Eq.
(\ref{4.67}) by invoking the Debye-like approximation. To this end,
the integral over the Brillouin zone is replaced by the Debye sphere,
whose radius is chosen so that to retain the normalization condition
\be
\label{4.69}
\int_{\cal B} \frac{d\bk}{(2\pi)^d} =
\frac{N_L}{V} = \frac{\rho}{\nu} \; .
\ee
The latter, in dimensionless units, reads as
\be
\label{4.70}
\int_{\overline{\cal B}} d{\bf x} = 2^d \; .
\ee
Then, the Debye approximation implies
\be
\label{4.71}
\int_{\overline{\cal B}} \;d{\bf x} =
\frac{2\pi^{d/2}}{\Gm(d/2)} \;
\int_0^{x_D} \; x^{d-1} \; dx = 2^d \; .
\ee
From here, the Debye radius is
\be
\label{4.72}
x_D = \frac{2}{\sqrt{\pi}} \left [
\frac{d}{2} \; \Gm\left (\frac{d}{2}\right )
\right ]^{1/d} \; .
\ee
In this approximation,
$$
\int_{\overline{\cal B} } \;
\frac{d{\bf x}}{\sum_\al x_\al^2} =
\frac{2\pi^{d/2}x_D^{d-2}}{(d-2)\Gm(d/2)} \; .
$$
Equation (\ref{4.67}) yields
\be
\label{4.73}
T_c = 2^d \pi^2 \; \frac{(d-2)\Gm(d/2)}{2\pi^{d/2}x_D^{d-2}} \;
J\nu \; .
\ee
For a three-dimensional lattice, one has
\be
\label{4.74}
T_c = \frac{2\pi}{x_D}\; J\nu \; , \qquad x_D = \left (
\frac{6}{\pi}\right )^{1/3} \qquad (d=3) \; .
\ee
This results in the BEC temperature
\be
\label{4.75}
T_c \cong 5 J\nu \qquad (d=3) \; .
\ee
A close estimate for the critical temperature of condensation in a
lattice follows from the Bogolubov approximation [275].

Thus, in the HFB approximation, the BEC does not occur in one- and
two-dimensional lattices. In a three-dimensional lattice, the BEC
exists below $T_c$ given by Eq. (\ref{4.75}). The transition
temperature does not depend on the on-site interaction $U$. However,
one should expect that, for a sufficiently strong repulsion $U$,
the system could go to an insulating state. This means that the
HFB approximation for the boson Hubbard model is applicable only
for the Bose-condensed phase, but is not suitable for the insulating
state. Other approximations will be discussed in the following
sections.

\subsection{Thermodynamic Characteristics}

All thermodynamic characteristics can be derived from the system grand
potential. For example, when the system is in the Bose-condensed phase,
for which the HFB approximation is applicable, the grand potential is
\be
\label{4.76}
\Om = E_B + TV \int_{\cal B} \; \ln \left ( 1 - e^{-\bt \ep_k}
\right ) \frac{d\bk}{(2\pi)^3} \; ,
\ee
where $\ep_k$ is the Bogolubov spectrum (\ref{4.52}) and
\be
\label{4.77}
E_B = H^{(0)} - \nu N \; \frac{U}{2} \left ( 2n_1^2 +\sgm^2
\right ) \; + \; \frac{1}{2} \; \sum_k ( \ep_k - \om_k)
\ee
follows from Eq. (\ref{4.51}).

The system free energy is
\be
\label{4.78}
F = \Om + \mu N \; ,
\ee
with the chemical potential
\be
\label{4.79}
\mu= \mu_0 n_0 + \mu_1 n_1 \; .
\ee
The latter, using the Lagrange multipliers (\ref{4.49}) and
(\ref{4.54}), becomes
\be
\label{4.80}
\mu =  -z_0 J + \nu U ( 1 + n_1 + \sgm - 2n_1 \sgm ) \; .
\ee
The internal energy is $E\equiv<H_B>+\mu N$. Here $H_B$ is given by
Eq. (\ref{4.50}). Also, from Eqs. (\ref{4.25}) and (\ref{4.49}), we
have
\be
\label{4.81}
H^{(0)} = \nu n_0 N \; \frac{U}{2} [ n_0 -2( 1+ n_1 +\sgm) ]\; .
\ee
And Eq. (\ref{4.48}) gives
\be
\label{4.82}
E_{HFB} = \nu N \; \frac{U}{2} \left [ n_0^2 - \sgm^2 -
2 (1 + n_0\sgm) \right ] \; .
\ee
Then for the ground-state energy
$$
E_0 \equiv E_B + \mu N = E_{HFB} + \frac{1}{2} \;
\sum_k (\ep_k - \om_k) + \mu N \; ,
$$
we find
\be
\label{4.83}
\frac{E_0}{N} = - z_0 J + \nu \; \frac{U}{2} \left ( 1 + n_1^2 -
\sgm^2 - 2n_1 \sgm \right ) \; + \; \frac{1}{2\rho} \;
\int_{\cal B} (\ep_k - \om_k ) \; \frac{d\bk}{(2\pi)^3} \; .
\ee

Particle fluctuations are characterized by the dispersion
$\Dlt^2(\hat N)$ of the number-of-particle operator, as defined
in Eq. (\ref{2.248}). For a Bose system, with the broken gauge
symmetry by means of the Bogolubov shift, the condensate fluctuations
are negligible [57,93], that is, $\Dlt^2(\hat N_0)\ra 0$. Hence, all
fluctuations are due to uncondensed particles,
\be
\label{4.84}
\Dlt^2(\hat N) = \Dlt^2(\hat N_1) \qquad
\left ( \hat N_1 = \sum_k a_k^\dgr a_k \right ) \; .
\ee
Since the HFB approximation results in the Hamiltonian
(\ref{4.47}), quadratic with respect to the filed operators $a_k$,
then calculating $\Dlt^2(\hat N_1)$, one has to be in the frame of
the quadratic approximation, that is, neglecting the terms with
$n_k^2$ and $\sgm_k^2$. Similarly to the uniform case [10,93,145,146],
for BEC in a lattice, we have
\be
\label{4.85}
\Dlt^2(\hat N_1) =  N [ 1 + 2\lim_{k\ra 0}
(n_k + \sgm_k ) ] \; .
\ee
From Eqs. (\ref{4.59}) and (\ref{4.60}), we find
$$
n_k \simeq \frac{T\Dlt}{\ep_k^2} \; + \; \frac{\Dlt}{12T} \;
+ \; \frac{T}{2\Dlt} \; - \; \frac{1}{2} \; ,
$$
\be
\label{4.86}
\sgm_k \simeq -\; \frac{T\Dlt}{\ep_k^2} \; - \;
\frac{\Dlt}{12T} \qquad (\ep_k \ra 0 ) \; .
\ee
Therefore,
\be
\label{4.87}
\lim_{k\ra 0} ( n_k + \sgm_k) = \frac{1}{2} \left (
\frac{T}{\Dlt} \; - \; 1 \right ) \; .
\ee
Then Eq. (\ref{4.85}) gives
\be
\label{4.88}
\Dlt^2(\hat N_1) = \frac{NT}{\Dlt} \; .
\ee
According to Eq. (\ref{4.46}), one gets
\be
\label{4.89}
\Dlt^2(\hat N) = \frac{NT}{\nu U(n_0+\sgm)} \; .
\ee
The isothermal compressibility reads as
\be
\label{4.90}
\kappa_T = \frac{\Dlt^2(\hat N)}{\rho T N} =
\frac{1}{\rho \nu U(n_0+\sgm)} \; .
\ee

Particle fluctuations are, of course, thermodynamically normal and
the compressibility is finite. The latter diverges only at the critical
point $T_c$, where $n_0\ra 0$ and $\sgm\ra 0$. But below $T_c$, the
compressibility is finite everywhere for $T<T_c$, provided that there
is a finite interaction $U$.

\subsection{Superfluid Fraction}

The superfluid fraction can be calculated by employing Eq. (\ref{3.106})
which has been derived in Sec. 2.10. Equation (\ref{3.106}) is general and
exact. For a lattice, the operator of momentum can be represented in forms
(\ref{3.83}) or (\ref{3.84}). The latter reduces to Eq. (\ref{3.93}). In
tight-binding approximation, one can invoke Eq. (\ref{3.105}). The
dissipated heat in the HFB approximation is given in Eq. (\ref{3.110}).

Another possibility is to rederive the superfluid fraction using
explicitly the Hubbard Hamiltonian (\ref{4.12}), in this derivation
being based on the general definitions (\ref{2.140}) or (\ref{2.144}).
For an equilibrium system, both these definitions yield Eq. (\ref{2.143}).
Generalizing the latter for a $d$-dimensional system, we have
\be
\label{4.91}
n_s = \frac{1}{mNd} \left [ \lim_{v\ra 0} < \frac{\prt}{\prt \bv}
\cdot \hat\bP_v > - \; \bt \Dlt^2 ( \hat\bP ) \right ] \; .
\ee

In order to use Eq. (\ref{4.91}), with the Hubbard Hamiltonian
(\ref{4.12}), it is necessary to find how the Wannier field operators
$\hat c_j$ change under the velocity boost. Generally, if the system
is boosted with the velocity $\bv$, the field operator of the moving
system can be expanded over the Wannier functions as
\be
\label{4.92}
\hat\psi_v(\br,t) = \sum_j \hat c_j (\bv,t) w(\br - \ba_j) \; .
\ee
Being in the frame of the Hubbard model, the single-band case is
considered here. Inverting expansion (\ref{4.92}) gives
\be
\label{4.93}
\hat c_j(\bv,t) = \int w^*(\br-\ba_j) \hat\psi_v(\br,t)\;
d\br \; .
\ee
Substituting in Eq. (\ref{4.93}) the Galilean transformation
(\ref{2.151}), with the expansion
\be
\label{4.94}
\hat\psi(\br-\bv t,t) = \sum_j \hat c_j(0,t)
w(\br-\bv t -\ba_j) \; ,
\ee
we get the relation
\be
\label{4.95}
\hat c_i(\bv,t) = \sum_j \hat c_j(0,t)
\exp\left ( -i \; \frac{mv^2}{2} \; t \right ) \;
\int w^*(\br-\ba_i) w(\br-\bv t-\ba_j) e^{im\bv\cdot\br}
\; d\br \; .
\ee
This is a general relation connecting the Wannier field operators
$\hat c_i(\bv,t)$ for a moving system with these operators $\hat
c_j(0,t)$ for an immovable lattice.

Keeping in mind an equilibrium system, we can set time to zero,
introducing the simplified notation
\be
\label{4.96}
\hat c_j(\bv) \equiv \hat c_j(\bv,0) \; .
\ee
Diminishing the velocity to zero, we return to the old notation of
the Wannier field operators of an immovable lattice,
\be
\label{4.97}
\hat c_j \equiv \lim_{v\ra 0} \hat c_j(\bv) =
\lim_{v\ra 0} \hat c_j(\bv,0) \; .
\ee
Then relation (\ref{4.95}) becomes
\be
\label{4.98}
\hat c_i(\bv) = \sum_j  \hat c_j \;
\int w^*(\br-\ba_i) w(\br-\ba_j) e^{im\bv\cdot\br} \; d\br \; .
\ee
Since in the Hubbard model, one assumes well localized Wannier
functions, one can use the approximation
\be
\label{4.99}
\int w^*(\br-\ba_j) w(\br-\ba_j) e^{im\bv\cdot\br} \; d\br
\cong \dlt_{ij} e^{im\bv\cdot\ba_j} \; .
\ee
Therefore, relation (\ref{4.98}) simplifies to
\be
\label{4.100}
\hat c_j(\bv) = \hat c_j e^{im\bv\cdot\ba_j} \; .
\ee

The Hamiltonian (\ref{4.12}) of the Hubbard model for an immovable
system is the functional $\hat H=\hat H[\hat c_j]$. For a moving
lattice, the latter becomes $\hat H_v=\hat H[\hat c_j(\bv)]$, which
gives
\be
\label{4.101}
\hat H_v = - J \sum_{<ij>} \hat c_i^\dgr
\hat c_j e^{-im\bv\cdot\ba_{ij}} \; + \;
h_0 \sum_j \hat c_j^\dgr \hat c_j \; + \;
\frac{U}{2} \; \sum_j \hat c_j^\dgr \hat c_j^\dgr
\hat c_j \hat c_j \; ,
\ee
where $\ba_{ij}\equiv\ba_i-\ba_j$.

The operator of momentum, according to definition (\ref{2.138}), is
\be
\label{4.102}
\hat\bP_v \equiv \frac{\prt\hat H_v}{\prt\bv} = im J
\sum_{<ij>} \ba_{ij} \hat c_i^\dgr \hat c_j
e^{-im\bv\cdot\ba_{ij}} \; .
\ee
From here it follows that
$$
\frac{\prt}{\prt\bv} \cdot \hat\bP_v = m^2 J \sum_{<ij>}
\ba_{ij}^2 \hat c_i^\dgr \hat c_j e^{-im\bv\cdot\ba_{ij}} \; .
$$
The operator of momentum for an immovable lattice is
\be
\label{4.103}
\hat\bP \equiv \lim_{v\ra 0} \hat\bP_v = im J \sum_{<ij>}
\ba_{ij} \hat c_i^\dgr \hat c_j \; .
\ee
Substituting here the Bogolubov shift (\ref{4.18}) and using the
orthogonality condition (\ref{4.19}), we have
\be
\label{4.104}
\hat\bP = im J \sum_{<ij>} \ba_{ij} c_i^\dgr c_j \; .
\ee
That is, only the operators of uncondensed atoms contribute to the
momentum operator, in agreement with Eq. (\ref{3.83}). Limiting in
the latter the summation by nearest neighbors gives
$$
\hat\bP = \sum_{<ij>} \bp_{ij} c_i^\dgr c_j \qquad
(\bp= imJ\ba_{ij} ) \; .
$$
Taking into account that in equilibrium $<\hat\bP>=0$, for the
superfluid fraction (\ref{4.91}) we find
\be
\label{4.105}
n_s = \frac{mJ}{Nd} \; \sum_{<ij>} a_{ij}^2
< \hat c_i^\dgr \hat c_j> \; - \; \frac{<\bP^2>}{mTNd} \; ,
\ee
where $a_{ij}\equiv|\ba_{ij}|$. From Eq. (\ref{4.104}), we have
\be
\label{4.106}
<\hat\bP^2> \; = \; m^2 J^2 \sum_{<ij>} a_{ij}^2 \left (
< c_i^\dgr c_j^\dgr c_j c_i > - < c_i^\dgr c_i^\dgr c_j c_j > +
< c_i^\dgr c_i > \right ) \; .
\ee
Taking into account the Bogolubov shift (\ref{4.18}) gives
$$
< \hat c_i^\dgr \hat c_j > \; = \;
\nu n_0 + < c_i^\dgr c_j > \; .
$$
Then for the superfluid fraction (\ref{4.105}) we get
$$
n_s = \frac{mJ}{Nd} \sum_{<ij>} a_{ij}^2 ( \nu n_0 +
< c_i^\dgr c_j > ) \; -
$$
\be
\label{4.107}
 -\; \frac{mJ^2}{TNd} \sum_{<ij>} a_{ij}^2 \left (
< c_i^\dgr c_i > + < c_i^\dgr c_j^\dgr c_j c_i > -
< c_i^\dgr c_i^\dgr c_j c_j > \right ) \; .
\ee

On the other hand, the superfluid fraction can be defined by means of
Eq. (\ref{2.155}) that is as general as Eq. (\ref{2.143}). Generalizing
Eq. (\ref{2.155}) to a $d$-dimensional system yields
\be
\label{4.108}
n_s =  1 \; - \; \frac{Q}{Q_0} \; , \qquad
Q \equiv \frac{<\bP^2>}{2mN} \; , \qquad
Q_0 \equiv \frac{d}{2}\; T \; .
\ee
The dissipated heat, according to Eq. (\ref{4.106}), is
\be
\label{4.109}
Q = \frac{mJ^2}{2N} \; \sum_{<ij>} a_{ij}^2 \left (
< c_i^\dgr c_i > + < c_i^\dgr c_j^\dgr c_j c_i > -
< c_i^\dgr c_i^\dgr c_j c_j > \right ) \; .
\ee
Comparing Eqs. (\ref{4.105}) and (\ref{4.107}) with Eqs.
(\ref{4.108}) and (\ref{4.109}) shows that the expressions for the
superfluid fractions differ by the first term. This difference comes
from the use of the approximations involved in the derivation of the
Hubbard model. However, in the frame of the same approximation scheme,
these two terms should be approximately equal. To prove that this is
really so, let is notice that
\be
\label{4.110}
< c_i^\dgr c_j> \; = \; \frac{1}{N_L} \;
\sum_k n_k e^{-i\bk\cdot\ba_{ij}} \; , \qquad
< c_i c_j> \; = \; \frac{1}{N_L} \;
\sum_k \sgm_k e^{i\bk\cdot\ba_{ij}} \; ,
\ee
where $n_k$ and $\sgm_k$ are the same as in Eq. (\ref{4.42}) and enjoy
the properties
$$
n_k = n_{-k} \; , \qquad \sgm_k = \sgm_{-k} \; .
$$
We introduce the mean distance between the nearest neighbors, $a$,
by the formula
\be
\label{4.111}
a^2 \equiv \frac{1}{z_0 N_L} \; \sum_{<ij>} a_{ij}^2 \; .
\ee
And let us define the effective mass
\be
\label{4.112}
m^* \equiv \frac{1}{2Ja^2}
\ee
by analogy with Eq. (\ref{3.160}). Also, we may notice that
$$
\sum_{<ij>} \; \sum_k a_{ij}^2 n_k e^{i\bk\cdot\ba_{ij}}  \cong
\sum_{<ij>} \; \sum_k a_{ij}^2 n_k \; .
$$
Keeping in mind that $n_0+n_1=1$, for the first term of Eq. (\ref{4.107}),
we finally obtain
\be
\label{4.113}
\frac{mJ}{Nd} \; \sum_{<ij>} a_{ij}^2 \left ( \nu n_0 +
<c_i^\dgr c_j> \right ) \cong \frac{mz_0}{m^* 2d} \; .
\ee
In order that expressions (\ref{4.107}) and (\ref{4.108}) be approximately
equal, it is necessary that
\be
\label{4.114}
\frac{mz_0}{m^* 2d} \cong 1 \; .
\ee
The latter holds, for instance, when $m^*\approx m$ and $z_0\approx
2d$. For a cubic lattice, the coordination number is exactly $z_0=2d$.

In this way, the superfluid fraction, defined by Eq. (\ref{2.144}), for
the Hubbard model is
\be
\label{4.115}
n_s = \frac{mJ}{Nd} \; \sum_{<ij>} a_{ij}^2 \left ( \nu n_0 +
< c_i^\dgr c_j > \right ) \; - \; \frac{2Q}{Td} \; ,
\ee
with the dissipated heat (\ref{4.109}). At the same time, it can also be
defined by Eq. (\ref{4.108}), since the first term in Eq. (\ref{4.115}),
according to Eqs. (\ref{4.113}) and (\ref{4.114}), is close to one.

For a $d$-dimensional cubic lattice, one has
$$
\sum_{<ij>} a_{ij}^2 e^{i\bk\cdot\ba_{ij}} =
2a^2 N_L \sum_{\al=1}^d \cos(k_\al a) \; .
$$
From here
$$
\sum_{<ij>} a_{ij}^2 < c_i^\dgr c_j > \; = \; 2a^2 \;
\sum_k \; \sum_{\al=1}^d n_k cos(k_\al a) \; .
$$

If one resorts to the HFB approximation, then
$$
< c_i^\dgr c_j^\dgr c_j c_i > \; = \; < c_i^\dgr c_j >^2 +
(\nu n_1)^2 + | < c_i c_j >|^2\; ,
$$
$$
< c_i^\dgr c_i^\dgr c_j c_j > \; = \; 2 < c_i^\dgr c_j >^2 +
|\nu\sgm|^2 \; .
$$
And the dissipated heat (\ref{4.109}) becomes
\be
\label{4.116}
Q = \frac{mJ^2}{2N} \; \sum_{<ij>} a_{ij}^2 \left (
\nu n_1 + \nu^2 n_1^2 - < c_i^\dgr c_j>^2 - \nu^2|\sgm|^2 +
|< c_i c_j>|^2 \right ) \; ,
\ee
where the properties
$$
< c_i^\dgr c_j > \; = \; < c_j^\dgr c_i > \; , \qquad
< c_i c_j > \; = \; < c_j c_i > \; ,
$$
$$
< c_j^\dgr c_j > \; = \; \nu n_1 \; , \qquad
< c_j c_j> \; = \; \nu\sgm
$$
are used. Taking into account that $\sgm_k$ in the Fourier transform
(\ref{4.110}) can be made real, for a cubic lattice, one gets
$$
\sum_{<ij>} a_{ij}^2 < c_i c_j > \; = \; 2a^2 \sum_k \;
\sum_{\al=1}^d \sgm_k \cos(k_\al a) \; ,
$$
$$
\sum_{<ij>} a_{ij}^2 < c_i^\dgr c_j >^2 \; = \; \frac{2a^2}{N_L} \;
\sum_{kp} \; \sum_{\al=1}^d n_k n_p \cos(k_\al +p_\al) a \; ,
$$
$$
\sum_{<ij>} a_{ij}^2 |< c_i c_j >|^2 \; = \; \frac{2a^2}{N_L} \;
\sum_{kp} \; \sum_{\al=1}^d \sgm_k
\sgm_p \cos(k_\al +p_\al) a \; .
$$
Then Eq. (\ref{4.116}) reads as
\be
\label{4.117}
Q = \frac{mJ\nu}{2m^* N^2} \; \sum_{kp} \; \sum_{\al=1}^d
\left [ n_k + 2 ( n_k n_p - \sgm_k \sgm_p ) \sin^2 \left (
\frac{k_\al+p_\al}{2}\; a\right ) \right ] \; .
\ee
This can be rewritten in a more symmetric form taking into account that
$$
2\sum_p \sin^2(p_\al a) = N_L \; , \qquad
2\sum_p \sin^2 \left ( \frac{k_\al+p_\al}{2}\; a\right ) = N_L \; .
$$
Using the latter equalities allow us to rewrite Eq. (\ref{4.117}) as
\be
\label{4.118}
Q = \frac{1}{N_L} \; \sum_{kp} \; \frac{q_{kp}^2}{2mN} \left (
n_k + n_k n_p - \sgm_k \sgm_p \right ) \; ,
\ee
where
$$
q_{kp}^2 \equiv \frac{2m^2J}{m^*} \; \sum_{\al=1}^d \sin^2
\left ( \frac{k_\al+p_\al}{2}\; a\right ) \; .
$$
Assuming that the main contribution to sum (\ref{4.118}) comes from
the diagonal terms results in
\be
\label{4.119}
Q = \sum_k \; \frac{q_k^2}{2mN} \left ( n_k + n_k^2 - \sgm_k^2
\right ) \; ,
\ee
with
$$
q_k^2 \equiv \frac{2m^2J}{m^*} \;
\sum_{\al=1}^d \sin^2(k_\al a) \; .
$$
Expression (\ref{4.119}) is in agreement with Eq. (\ref{3.107}). For the
superfluid fraction (\ref{4.115}), we obtain
\be
\label{4.120}
n_s = \frac{m}{m^*} \left [ n_0 + \frac{1}{N} \;
\sum_k \frac{1}{d} \; \sum_{\al=1}^d n_k \cos(k_\al a)
\right ] \; - \; \frac{2Q}{Td} \; ,
\ee
where the dissipated heat is given by Eq. (\ref{4.118}) or
(\ref{4.119}). Recall that the first term in Eq. (\ref{4.120}) is
approximately equal to one.

Often, employing definition (\ref{2.144}) for the superfluid fraction,
one transforms the Wannier field operators not as in Eq. (\ref{4.100}),
but by means of a phase factor, that is, replacing $\hat c_j$ by $\hat
c_je^{i\vartheta_j}$. And, considering the hopping term, one takes
$\vartheta\equiv\vartheta_i-\vartheta_j$. As a result [276,277], the hopping
term in Hamiltonian (\ref{4.101}) acquires the factor $e^{-i\vartheta}$,
instead of $e^{-im\bv\cdot\ba_{ij}}$. Hence, one assumes that $\vartheta
=m\bv\cdot\ba_{ij}$, which is equivalent to considering a cubic lattice.

\subsection{Single-Site Approximation}

The tunneling term in the Hubbard Hamiltonian (\ref{4.12}) can be
simplified by means of the single-site approximation [278], assuming
the decoupling
\be
\label{4.121}
\hat c_i^\dgr \hat c_j \; \cong \; < \hat c_i^\dgr > \;
\hat c_j + \hat c_i^\dgr < \hat c_j > -
< \hat c_i^\dgr >< \hat c_j^\dgr >  \qquad
(i \neq j) \; .
\ee
This decoupling is equivalent to the Gutzwiller approximation [279],
in which the system state is characterized by a product of single-site
wave functions [279--282].

According to representation (\ref{4.18}), the average
\be
\label{4.122}
< \hat c_j > \; = \; \sqrt{\nu n_0}
\ee
is an order parameter for the Bose system in a lattice, and it does
not depend on the lattice-site index, if the lattice is ideal.

To better understand the meaning of the single-site approximation, let
us compare the exact expression
\be
\label{4.123}
\hat c_i^\dgr \hat c_j = \nu n_0 +\sqrt{\nu n_0} \left (
c_i^\dgr + c_j \right ) + c_i^\dgr c_j
\ee
with that one following from decoupling (\ref{4.121}), which gives
\be
\label{4.124}
\hat c_i^\dgr \hat c_j \cong \nu n_0 + \sqrt{\nu n_0}
\left ( c_i^\dgr + c_j \right ) \qquad (i \neq j) \; .
\ee
This shows that the single-site approximation neglects the tunneling
of the uncondensed atoms,
\be
\label{4.125}
c_i^\dgr c_j \cong 0 \qquad (i\neq j,\; n_0> 0 ) \; .
\ee
Hence, instead of the exact expression
$$
\sum_{<ij>} \hat c_i^\dgr \hat c_j =  z_0 n_0 N  +
\sum_{<ij>} c_i^\dgr c_j
$$
in the tunneling term, one has
$$
\sum_{<ij>} \hat c_i^\dgr \hat c_j
 \cong z_0 n_0 N
$$
in the single-site approximation.

Using decoupling (\ref{4.121}) allows us to present the Hubbard
Hamiltonian (\ref{4.12}) as a sum
\be
\label{4.126}
\hat H = \sum_j \hat H_j^{eff}
\ee
of the single-site terms
\be
\label{4.127}
\hat H_j^{eff} = z_0 J \left [ \nu n_0 - \sqrt{\nu n_0}
\left ( c_j^\dgr + \hat c_j \right ) \right ] +
\frac{U}{2} \; \hat c_j^\dgr \hat c_j^\dgr \hat c_j \hat c_j  \; .
\ee
This is where the name of the single-site approximation comes from.

In view of the Bogolubov shift (\ref{4.18}) and the orthogonality
condition (\ref{4.19}), we have
\be
\label{4.128}
\sum_j \hat c_j = \sqrt{\nu n_0} \; N_L \; .
\ee
Then, instead of sum (\ref{4.126}), we get
\be
\label{4.129}
\hat H = \sum_j \hat H_j
\ee
with
\be
\label{4.130}
\hat H_j = - z_0 J \nu n_0 + \frac{U}{2} \;
\hat c_j^\dgr \hat c_j^\dgr \hat c_j \hat c_j  \; .
\ee
Substituting here the Bogolubov shift (\ref{4.18}) yields
\be
\label{4.131}
\hat H_j = \sum_{n=0}^4 \hat H_j^{(n)} \; .
\ee
The zero-order term of Eq. (\ref{4.131}) is
\be
\label{4.132}
\hat H_j^{(0)} = - z_0 J \nu n_0 + \frac{U}{2} \; (\nu n_0)^2 \; .
\ee
The first-order term, owing to condition (\ref{4.19}), is zero,
$H^{(1)}=0$. The second-order term becomes
\be
\label{4.133}
H_j^{(2)} = \frac{U}{2} \; \nu n_0 \left (
4 c_j^\dgr c_j + c_j^\dgr c_j^\dgr + c_j c_j \right ) \; .
\ee
The third-order term is
\be
\label{4.134}
\hat H_j^{(3)} = U \sqrt{\nu n_0} \left (
c_j^\dgr c_j^\dgr c_j + c_j^\dgr c_j c_j \right ) \; ,
\ee
and the fourth-order one is
\be
\label{4.135}
H_j^{(4)} = \frac{U}{2} \; c_j^\dgr c_j^\dgr c_j c_j =
\frac{U}{2} \; c_j^\dgr c_j \left ( c_j^\dgr c_j -1
\right ) \; .
\ee

The grand Hamiltonian
\be
\label{4.136}
H = \hat H - \mu_0 N_0 - \mu_1 \hat N_1
\ee
is the sum of four terms, as in Eq. (\ref{4.24}). The zero-order
term is the same as in Eq. (\ref{4.25}), with the constant $h_0$ being
incorporated into the Lagrange multiplier $\mu_0$. The second-order
term is
\be
\label{4.137}
H^{(2)} = \sum_j \hat H_j^{(2)} - \mu_1 \hat N_1 =
( 2U \nu n_0 - \mu_1 ) \sum_j c_j^\dgr c_j + \frac{U}{2} \;
\nu n_0 \sum_j \left ( c_j^\dgr c_j^\dgr + c_j c_j \right ) \; ,
\ee
where again $h_0$ is incorporated into $\mu_1$. And the third- and
fourth-order terms are
\be
\label{4.138}
H^{(3)} = \sum_j \hat H_j^{(3)} \; , \qquad
H^{(4)} = \sum_j \hat H^{(4)} \; ,
\ee
with $H^{(3)}$ and $H^{(4)}$ from Eqs. (\ref{4.134}) and (\ref{4.135}).
Equation (\ref{4.137}), as compared to Eq. (\ref{4.27}), does not
contain the tunneling of uncondensed atoms.

Because of the no-tunneling condition (\ref{4.125}), the momentum
operator (\ref{4.104}) vanishes,
\be
\label{4.139}
\hat\bP = 0 \qquad (c_i^\dgr c_j = 0, \; i\neq j) \; .
\ee
Thence, there is no the dissipated heat, $Q=0$. The condensate
fraction (\ref{4.107}), with notations (\ref{4.111}) and (\ref{4.112}),
becomes
\be
\label{4.140}
n_s = \frac{mz_0}{m^* 2d} \; n_0 \; .
\ee
For a cubic lattice, $z_0=2d$. Since $m\approx m^*$, then $n_s\approx
n_0$. The fact that the superfluid and condensate fractions practically
coincide is the defect of the single-site approximation, which neglects
the tunneling of uncondensed atoms. Comparing Eqs. (\ref{4.108}) and
(\ref{4.140}) tells us that this approximation is valid, when almost
all atoms are condensed, so that $n_0\approx 1$. The latter, in turn,
happens when atomic interactions are sufficiently weak. Strong
interactions, as is known [94--98], deplete the condensed fraction,
while enhance the superfluid fraction, resulting in the inequality $n_s
\gg n_0$. In the presence of a lattice, strong interactions should lead
to the destruction of both BEC and superfluidity and to the appearance
of a localized insulating state.

\subsection{Localized State}

When atomic interactions are much larger than the tunneling rate, so
that
\be
\label{4.141}
\frac{U}{J} \; \gg \; 1 \; ,
\ee
than the intersite hopping of atoms is completely suppressed. This can
be symbolized by the {\it localization condition}
\be
\label{4.142}
c_i^\dgr c_j = \dlt_{ij} c_i^\dgr c_i \; .
\ee
When all atoms are localized, global coherence cannot develop. There is
no BEC and no gauge symmetry breaking,
\be
\label{4.143}
n_0 = 0 \; , \qquad n_1 = 1 \; .
\ee
The Hubbard Hamiltonian (\ref{4.12}) reduces to
\be
\label{4.144}
\hat H = \frac{U}{2} \; \sum_j c_j^\dgr c_j^\dgr c_j c_j \; .
\ee
Using the site-number operator, or {\it filling operator}
\be
\label{4.145}
\hat n_j \equiv c_j^\dgr c_j \; ,
\ee
Hamiltonian (\ref{4.144}) can be rewritten as
\be
\label{4.146}
\hat H = \frac{U}{2} \; \sum_j \hat n_j
\left ( \hat n_j - 1\right ) \; .
\ee

Since there is just one kind of atoms, the sole chemical potential is
sufficient. So, the grand Hamiltonian is
\be
\label{4.147}
H \equiv \hat H - \mu \hat N =  \sum_j H_j \; ,
\ee
where the site Hamiltonians are
\be
\label{4.148}
H_j = \frac{U}{2} \; \hat n_j^2 - \left (
\frac{U}{2} + \mu \right ) \hat n_j \; ,
\ee
the number-of-particle operator being
\be
\label{4.149}
\hat N = \sum_j \hat n_j \; .
\ee

For an ideal lattice, the average of the site-number operator
(\ref{4.145}) gives the filling factor
\be
\label{4.150}
\nu \; = \; < \hat n_j > \; .
\ee
Explicitly, the latter reads as
\be
\label{4.151}
\nu =
\frac{{\rm Tr}\hat n_j e^{-\bt H} }{{\rm Tr}e^{-\bt H} } \; .
\ee
This equation defines the chemical potential $\mu=\mu(\nu,T)$. Owing
to the additive form of the grand Hamiltonian (\ref{4.147}), expression
(\ref{4.151}) becomes
\be
\label{4.152}
\nu =
\frac{{\rm Tr}\hat n_j e^{-\bt H_j} }{{\rm Tr}e^{-\bt H_j} } \; .
\ee

The eigenproblem for the site-number operators (\ref{4.145}),
\be
\label{4.153}
\hat n_j| n > \; = \; n | n > \; ,
\ee
enjoys, as eigenvalues, the integers $n=0,1,2,\ldots$, while $|n>$
being the occupation-number state [55]. The eigenproblem for the site
Hamiltonians (\ref{4.148}),
\be
\label{4.154}
H_j | n > \; = \; e_n | n > \; ,
\ee
gives the energy levels
\be
\label{4.155}
e_n = \frac{U}{2}\; n^2 - \left ( \frac{U}{2} + \mu \right ) n \; .
\ee
The levels are discrete. One can define the energy gap
\be
\label{4.156}
\Dlt e_n \equiv e_{n+1} + e_{n-1} - 2e_n \; ,
\ee
which gives
\be
\label{4.157}
\Dlt e_n =  U \; .
\ee
This is why one tells that the insulator energy spectrum possesses
a gap.

With the basis $\{|n>\}$ of the occupation-number states, the filling
factor (\ref{4.152}) takes the form
\be
\label{4.158}
\nu =
\frac{\sum_{n=0}^\infty ne^{-\bt e_n}}
{\sum_{n=0}^\infty e^{-\bt e_n}} \; .
\ee
For high temperature $(T\gg U)$, the sums in Eq. (\ref{4.158}) can
be replaced by integrals. But for low temperature $(T\ll U)$, the
main contribution to the sums comes from the term with the lowest
energy $e_n$.

Let us consider low temperatures, such that
\be
\label{4.159}
\frac{T}{U} \; \ll \; 1 \; .
\ee
The minimum of $e_n$ is defined by the conditions
\be
\label{4.160}
\frac{\prt e_n}{\prt n} = 0 \; , \qquad
\frac{\prt^2 e_n}{\prt n^2} > 0 \; .
\ee
This yields the effective number
\be
\label{4.161}
n_{eff} = \frac{2\mu+U}{2U} \qquad ( U > 0) \; .
\ee
For low temperatures, conditioned by inequality (\ref{4.159}), the
filling factor (\ref{4.158}) is
\be
\label{4.162}
\nu \simeq n_{eff} = \frac{2\mu+U}{2U} \; .
\ee
This defines the chemical potential
\be
\label{4.163}
\mu \simeq (2\nu -1 ) \;\frac{U}{2} \; .
\ee
The grand thermodynamic potential is
\be
\label{4.164}
\Om \equiv - T \ln \sum_{n=0}^\infty e^{-\bt e_n}
\simeq e_\nu \; ,
\ee
with
\be
\label{4.165}
e_\nu \equiv \frac{U}{2} \; \nu^2 - \left (
\frac{U}{2} + \mu \right ) \nu \; .
\ee
Substituting here the chemical potential (\ref{4.163}) yields
\be
\label{4.166}
e_\nu = -\; \frac{U}{2}\; \nu^2 \; .
\ee
The internal energy reads as
\be
\label{4.167}
E \; \equiv \; < H > + \mu N \simeq e_\nu N_L + \mu N \; .
\ee
Hence, the ground-state energy per atom is
\be
\label{4.168}
\frac{E}{N} =  ( \nu -1 )\; \frac{U}{2} \qquad ( T = 0 ) \; .
\ee
The fluctuations of particles are characterized by the dispersion
$\Dlt^2(\hat N)$, as is described in Sec. 2.14. With the
number-of-particle operator (\ref{4.149}), we have [93,94]
$$
\Dlt^2(\hat N) = \sum_j \Dlt^2 (\hat n_j) \; + \;
\sum_{i\neq j} {\rm cov} (\hat n_i, \hat n_j ) \; .
$$
For the considered localized state,
$$
< \hat n_i \hat n_j > \; = \; < \hat n_i> < \hat n_j >
\qquad (i\neq j) \; ,
$$
because of which ${\rm cov}(\hat n_i,\hat n_j )=0$. The dispersion of
the filling operator (\ref{4.145}) is
\be
\label{4.169}
\Dlt^2(\hat n_j) \; \equiv \; < \hat n_j^2> -
< \hat n_j >^2 =  T \; \frac{\prt\nu}{\prt\mu} \; .
\ee
This, according to Eq. (\ref{3.14}), defines the compressibility
\be
\label{4.170}
\kappa_T = \frac{a^d}{\nu^2} \; \frac{\prt\nu}{\prt\mu} =
\frac{a^d\Dlt^2(\hat n_j)}{\nu^2T} \; .
\ee
At low temperatures, satisfying inequality (\ref{4.159}), we have
$$
< \hat n_j > \; \simeq \; \nu \; , \qquad
< \hat n_j^2 > \; \simeq \; \nu^2 \; .
$$
Therefore $\Dlt^2(\hat n_j)\simeq 0$ and $\kappa_T\simeq 0$. In that
sense, one tells that the localized state is incompressible.

\section{Phase States and Transitions}

\subsection{Existence of Pure Phases}

To locate the phase transition between the purely delocalized and
localized states, one can compare the corresponding thermodynamic potentials.
For instance, one can consider the grand potential of the Bose-condensed
superfluid phase, $\Om_{sup}=\Om_{sup}(\nu,J,U,T)$, given, e.g., by Eq.
(\ref{4.76}). From another side, one has the grand potential of the localized
phase, $\Om_{loc}=\Om_{loc}(\nu,U,T)$, given, e.g., by Eq. (\ref{4.164}).
The phase boundary could be described by the equality $\Om_{sup}=\Om_{loc}$.
However, for finite temperature, the system cannot be completely localized,
but contains a portion of wandering atoms [283--285]. The same concerns the
noninteger filling factors $\nu$, for which the system is not absolutely
localized [286]. Complete localization can occur for integer filling factors
$\nu=1,2,\ldots$ at zero temperature. Such a completely localized state is
called the Mott insulator [267,268]. The phase transition between the
superfluid state and Mott insulator, occurring for an integer filling at
zero temperature, has been studied in various approximations. One usually
starts with the Hubbard Hamiltonian
\be
\label{5.1}
\hat H = - J \sum_{<ij>} c_i^\dgr c_j \; + \;
\frac{U}{2} \; \sum_j c_j^\dgr c_j^\dgr c_j c_j \; +
\; \sum_j h_j c_j^\dgr c_j \; ,
\ee
where, for generality, the site-dependent term with $h_j$ is included to.
The quantity $h_j$ represents some external fields, like a confining field.
It may also represent additionally imposed fields, regular or random. Pure
superfluid or insulating phases can exist only under special values of the
system parameters.

\subsection{Hard-Core Lattice Gas}

In the limit of an infinite on-site interaction, $U\ra\infty$, one comes
to the hard-core lattice-gas model. In that limit, there can be at most
one particle at each site, which is symbolized by the hard-core condition
\be
\label{5.2}
\left ( c_j^\dgr \right )^2 = 0 \; , \qquad
c_j^2 = 0 \; .
\ee
The Hamiltonian (\ref{5.1}) reduces to
\be
\label{5.3}
\hat H =  -J \sum_{<ij>} c_i^\dgr c_j \; + \;
\sum_j h_j c_j^\dgr c_j \; .
\ee
The hard-core boson operators, with condition (\ref{5.2}), can be
transformed as
$$
c_j = S_j^x - i S_j^y \equiv S_j^- \; , \qquad
c_j^\dgr = S_j^x + i S_j^y \equiv S_j^+ \; ,
$$
\be
\label{5.4}
c_j^\dgr c_j = \frac{1}{2} + S_j^z \; ,
\ee
being expressed through the quasispin operators
$$
S_j^x = \frac{1}{2} \left ( c_j + c_j^\dgr \right ) \; , \qquad
S_j^y = \frac{i}{2} \left ( c_j - c_j^\dgr \right ) \; , \qquad
S_j^z = c_j^\dgr c_j \; - \; \frac{1}{2} \; .
$$
This transforms Hamiltonian (\ref{5.3}) to the quasispin representation
\be
\label{5.5}
\hat H = - J \sum_{<ij>} S_i^+ S_j^- \; + \;
\sum_j h_j S_j^z \; + \; \frac{h}{2} \; N_L \; ,
\ee
in which
\be
\label{5.6}
h \equiv \frac{1}{N_L} \; \sum_j h_j \; .
\ee
The hopping term in Eq. (\ref{5.5}) can be rewritten using the equality
$$
\sum_{<ij>} S_i^+ S_j^- = \sum_{<ij>} \left ( S_i^x S_j^x +
S_i^y S_j^y \right ) \; .
$$
For three-dimensional lattices, at sufficiently large tunneling
parameter $J$, there can arise the BEC state, while for low $J$, the
Mott insulator state can develop [287--289]. For a one-dimensional
lattice, there can arise only quasi-long-range order, that is, not the
true BEC but a quasicondensate [290].

The ground state of hard-core lattice bosons has also been analyzed by
involving the Girardeau Bose-Fermi mapping [13,291,292] by Lin and Wu
[293]. They found that in a one-dimensional lattice there can exist the
Mott insulator and a normal nonlocalized Bose system if the filling
factor is noninteger, but no real BEC, though quasicondensate can exist.

\subsection{Effective Interaction Parameter}

In the more realistic case, when the on-site interaction is finite,
there should exist a critical value of this interaction at which the
superfluid-insulator phase transition occurs. It is convenient to define
the dimensionless parameter
\be
\label{5.7}
u \equiv \frac{U}{z_0 J} \; .
\ee
The critical value of this parameter can be simply estimated as follows.
The tunneling energy of each particle approximately is $z_0J$. The energy
of a pair interaction of two particles in each lattice site is $U/2$,
hence, the potential energy per particle at one site is $U/4$. The total
potential energy for the lattice is $N_LU/4$. So, the potential energy
per particle is $N_LU/4N=U/4\nu$. The phase transition happens when the
tunneling energy $z_0J$ equals the potential energy $U/4\nu$. This
defines the critical parameter (\ref{5.7}) as
\be
\label{5.8}
u_c = 4\nu \; .
\ee
Here it has been assumed that there are no external fields.

\subsection{Gutzwiller Single-Site Approximation}

This approximation involves the use of variational wave functions of the
Gutzwiller type, which are represented as products over the lattice sites
[278--282,294]. At zero temperature, the critical parameter (\ref{5.8})
is found to be
\be
\label{5.9}
u_c =\left ( \sqrt{\nu} + \sqrt{1+\nu} \right )^2 \; .
\ee
For the unity filling $\nu=1$, this gives
\be
\label{5.10}
u_c = 5.8 \qquad (\nu =1) \; .
\ee
The critical value (\ref{5.10}) does not depend on the lattice
dimensionality, which is usual for mean-field-type approximations.
Generally, the microscopic dynamics of the localization-delocalization
transition is influenced by the space dimensionality [295].

\subsection{Dynamical Mean-Field Approximation}

Another type of a mean-field approximation was employed by Amico and
Penna [296]. They found that, at zero temperature, the superfluid-insulator
transition is located at
\be
\label{5.11}
u_c = 4\nu \; ,
\ee
in exact agreement with the simple estimate (\ref{5.8}). Again, the
result does not depend on the lattice dimensionality. One may notice
that the Gutzwiller-approximation value (\ref{5.9}) reduces to either
Eq. (\ref{5.8}) or Eq. (\ref{5.11}) for large coordinate numbers $\nu$.

\subsection{Small-System Numerical Diagonalization}

Direct numerical diagonalization can be done for small one-dimensional
lattices, of about 10 sites, at zero temperature [276]. This gives, for
the unity filling factor, the critical value
\be
\label{5.12}
u_c = 2.3 \qquad (\nu=1,\; d=1) \; .
\ee
For one-dimensional lattices, value (\ref{5.12}) is lower than the values
predicted by mean-filed estimates because of the stronger influence of
fluctuations that are underestimated in mean-field approximations.

\subsection{Density-Matrix Renormalization Group}

This is another numerical method that can be applied to small
one-dimensional lattices [297] giving
\be
\label{5.13}
u_c = 1.7 \qquad (\nu=1, \; d=1) \; .
\ee
This is close to value (\ref{5.12}) found by the numerical diagonalization.

\subsection{Strong-Coupling Perturbation Theory}

Perturbation theory in powers of $J/U$ can be used, with the following
summation of series, e.g., by means of Pad\'e approximants [298-300]. One-,
two-, and three-dimensional cubic (square) lattices have been considered.
The results are
$$
u_c = 1.9 \qquad (\nu=1,\; d=1 \; [299]) \; ,
$$
$$
u_c = 4.2 \qquad (\nu=1,\; d=2 \; [299]) \; ,
$$
\be
\label{5.14}
u_c = 4.9 \qquad (\nu=1,\; d=3 \; [298]) \; .
\ee
The value for the one-dimensional lattice is close to the numbers obtained
by the small-system numerical diagonalization (\ref{5.12}) and by the
density-matrix renormalization group (\ref{5.13}).

\subsection{Monte Carlo Simulations}

The simulations are accomplished for a finite number of bosons, which can
reach $N\simeq 10^3$ atoms. One-, two-, and three-dimensional rectangular
lattices have been investigated [301--305]. The most recent results are
presented below:
$$
u_c = 1.8 \qquad (\nu=1,\; d=1 \; [302]) \; ,
$$
$$
u_c = 4.2 \qquad (\nu=1,\; d=2 \; [305]) \; ,
$$
\be
\label{5.15}
u_c = 4.9 \qquad (\nu=1,\; d=3 \; [304]) \; .
\ee
All these quantities agree well with the strong-coupling perturbation theory.
The finite-temperature phase diagram at filling factor $\nu=1$ is also found
[304,305]. The critical temperature $T_c=T_c(U)$ as a function of the on-site
interaction displays a nonmonotonic behavior, as has been suggested by
Kleinert et al. [275]. When $U$ increases from zero, $T_c$, first rises,
reaches the maximum at around $U/J\approx 5$, and then diminishes to zero at
the critical value $U_c/J$.

\subsection{Order of Phase Transition}

The transition between the superfluid and Mott insulator phases, occurring
at zero temperature and integer filling factors, is an example of the quantum
phase transitions. This is a continuous transition, that is, a second-order
phase transition. The role of an order parameter is played by the condensate
fraction $n_0$. Considering the latter as a function $n_0=n_0(u)$ of the
dimensionless parameter (\ref{5.7}), one has the following behavior. In the
absence of interaction, $n_0(0)=1$. Then, the condensate fraction diminishes
with increasing $u$, and drops continuously to zero at the critical value
$u_c$. The continuous nature of the superfluid-Mott insulator quantum phase
transition in any dimension follows from the fact that the $d$-dimensional
Hubbard model with BEC pertains to the universality class of a
$d+1$-dimensional $XY$-model [306].

\subsection{Experiments on Superfluid-Insulator Transition}

Several experiments observing the superfluid-insulator phase transition of
cold bosons in optical lattices have been accomplished. Actually, because
of the finiteness of the lattices, it is not a sharp phase transition that
has been observed, but a gradual crossover between the superfluid and Mott
insulator states, occurring around the critical values $u_c$ predicted by
theory.

The first experiment was by Greiner et al. [307] with $^{87}$Rb BEC at
zero temperature. A three-dimensional lattice was formed by three optical
standing waves aligned orthogonal to each other. The laser beams operated
at a wavelength $\lbd=852$ nm, forming the cubic three-dimensional lattice
with the lattice spacing $a=4.26\times 10^{-5}$ cm. The existence of
coherence in the superfluid Bose-condensed phase and its absence in the
insulating localized phase was analysed by studying the level of interference
after suddenly turning off the trapping potential and allowing atoms to expand
freely. For the unity filling factor, the superfluid-insulator crossover was
localized around $u_c\approx 6$.

In the experiment by St\"oferle et al. [308], a one-dimensional optical
lattice was realized with $^{87}$Rb atoms. The optical lattice was formed
by laser beams at a wavelength $\lbd=826$ nm, which translated into the
lattice spacing $a=4.13\times 10^{-5}$ cm. Bragg spectroscopy was employed
for investigating the excitation spectrum. The superfluid insulator crossover,
at the unity filling factor, was observed close to $u_c\approx 5.8$, though
for one-dimensional lattices it should happen at $u_c\approx 1.8$. This
disagreement could be due to the finite size of the trap. A very important
finding was that for strongly interacting Bose systems in optical lattices
the superfluid fraction could be significantly different from the coherent
BEC fraction.

Different types of lattices, one-, two-, and three-dimensional optical
lattices were created by K\"ohl et al. [309] for $^{87}$Rb atoms. The
optical lattices were formed by retro-reflected laser beams at a wavelength
$\lbd=826$ nm, which corresponded to the lattice spacing $a=4.13\times
10^{-5}$ cm. The superfluid-insulator crossover was investigated by using
Bragg spectroscopy for studying the excitation spectra. The appearance of
the discrete spectrum structure, associated with the Mott insulating phase,
was observed between $u=4$ and $u=8$.

A three-dimensional lattice, filled by sodium $^{23}$Na atoms, was formed by
Xu et al. [310]. A dye laser operated at $\lbd=594.7$ nm. Hence, the lattice
spacing was $a=2.97\times 10^{-5}$ cm. The system properties were studied
by the time-off-flight images. The filling factors could be varied between
$\nu=1$ to $\nu=5$. The superfluid-insulator crossovers were observed around
the critical values $u_c$ given by the single-site approximation (\ref{5.9}).

In the experiment by Spielman et al. [311] the BEC of $^{87}$Rb atoms was
loaded into a two-dimensional optical lattice formed by laser light of
$\lbd=820$ nm. This gave the lattice spacing $a=4.1\times 10^{-5}$ cm. The
system properties were analysed by studying the time-of-flight images. For
the unity filling factor, the superfluid-insulator crossover was located
around the critical value $u_c\approx 4$, which was in agreement with the
Monte Carlo simulations as could be seen from Eq. (\ref{5.15}).

\subsection{Layered Superfluid-Insulator Structure}

When, in addition to the optical lattice potential, there is an external
trapping potential, the system becomes nonuniform, so that the lattice
is no longer ideal. Then in some spatial parts the conditions could be
created for the occurrence of the insulating phase, while in other spatial
locations the superfluid phase would be preferable. This results in the
formation of a shell structure, where the layers of Mott insulating phases
alternate with the layers of superfluid phases. This layered structure
was studied by Monte Carlo methods [312], by employing a pseudospin
approximation [313], and was observed in experiment [314].

\subsection{Models with Neighbor Interactions}

The Hubabrd model (\ref{5.1}) contains only the on-site atomic interaction
$U$. This is, of course, a simplification. When deriving the Hubabrd
Hamiltonian in Sec. 4.1, we could see that the Wannier representation
(\ref{4.11}) of the general system Hamiltonian (\ref{2.156}) includes
interactions between all lattice sites. Strictly speaking, all these
interactions are nonzero even for the local interaction potential, as
follows from the matrix element (\ref{4.10}). Considering solely the on-site
interaction assumes that the interaction potential is of short-range type
and Wannier functions are well localized at their lattice sites. But when
the lattice optical potential is shallow, the corresponding Wannier functions
may be not so well localized. Or, if the pair atomic interaction potential
is not short-range, then the interactions of atoms at neighboring sites can
be sufficiently important and not negligible. For example, the dipolar
quantum gas of $^{52}$Cr possesses long-range dipolar interactions [315,316].
Therefore, there are realistic situations, when the different-site
interactions could be important. This requires to consider the {\it
extended Hubbard model}
\be
\label{5.16}
\hat H = -J \sum_{<ij>} c_i^\dgr c_j \; + \; \frac{U}{2} \;
\sum_j c_j^\dgr c_j^\dgr c_j c_j \; + \; \frac{1}{2} \;
\sum_{i\neq j} U_{ij} c_i^\dgr c_j^\dgr c_i c_j  \; + \;
\sum_j h_j c_j^\dgr c_j \; ,
\ee
in which, in addition to the on-site term, intersite atomic interactions
are taken into account. One usually considers the nearest-neighbor $(U_1)$
and next-nearest-neighbor $(U_2)$ interactions. Because of the complexity
of Hamiltonian (\ref{5.16}), such extended models are mainly studied by means
of numerical techniques, e.g., by using the density-matrix renormalization
group method [317,318] or quantum Monte Carlo simulations [319--322]. With
additional parameters, the phase diagram of the extended model becomes
essentially more rich. There can exist pure superfluid and insulator phases,
there can arise a spatially separated mixture of these phases, striped solid
phases can develop, and staggered patterns of alternating empty and multiply
occupied sites can occur. For {\it noninteger fillings}, the states can
appear with simultaneous diagonal order, corresponding to a solid, and
off-diagonal long-range order, typical of the BEC state, accompanied by
superfluidity. The latter state, combining solid, superfluid, and BEC
properties, is an example of the {\it coherent superfluid solid}, according
to the classification of Sec. 1.4.

\subsection{Quasiperiodic Optical Lattices}

In experiment, it is possible to create not only periodic optical potentials,
as in Eq. (\ref{3.1}), but also quasiperiodic potentials [323,324]. The
general form of a quasiperiodic potential is a superposition of several
periodic functions with incommensurate periods, as in the expression
$$
V_L(\br) = \sum_{\al=1}^d \; \sum_\mu V_{\al\mu}
\sin^2\left ( k_\mu^\al r_\al \right ) \; .
$$
One usually considers one-dimensional quasiperiodic lattices with the
potential
\be
\label{5.17}
V_L(z) = V_0 \sin^2(k_0 z) + V_1\sin^2(q_0 z) \; ,
\ee
in which neither $k_0/q_0$ nor $q_0/k_0$ are integers. The physics of
the one-dimensional systems with the quasiperiodic lattice potential
(\ref{5.17}) has been studied in different approximations [325--328]
as well as numerically [329--331]. Quasiperiodic systems were found to
have an intermediate behavior between periodic ones and random systems.
Depending on the system parameters, there can exist the superfluid phase,
Mott-insulator phase, as well as the coexisting superfluid and normal
phases. In addition, an incommensurate density wave can arise, representing
the incommensurate insulator, whose period is incommensurate with the
lattice periods. Also, the Bose glass phase can develop, in which there
is BEC, $n_0>0$, but there is no superfluidity, $n_s=0$.

In order to demonstrate how the Hubbard model for a quasiperiodic lattice
could be constructed, let us consider the one-dimensional case with the
bichromatic optical potential (\ref{5.17}). The derivation of the Hubbard
model can be done in the same way as in Sec. 4.1, if one treats one of the
sublattices as primary and accomplishes the expansions over Wannier functions
associated with this primary sublattice. For instance, the first term in
potential (\ref{5.17}) can be treated as primary, hence, the primary
sublattice having the period $a$, related to the laser wavevector $k_0=\pi/a$.

The lattice local Hamiltonian can be separated into two parts,
\be
\label{5.18}
H_L(z) = -\; \frac{\nabla^2}{2m} + V_L(z) = H_L^0(z) +
\Dlt H_L(z) \; ,
\ee
the first part including the primary potential,
\be
\label{5.19}
H_L^0(z) \equiv - \; \frac{\nabla^2}{2m} + V_0 \sin^2(k_0 z) \; ,
\ee
and the addition being
\be
\label{5.20}
\Dlt H_L(z) \equiv V_1 \sin^2(q_0 z) \; .
\ee
The primary term (\ref{5.19}) is periodic over the sublattice $\{ a_j\}$,
so that
$$
H_L^0(z+ a_j) = H_L^0(z) \; .
$$
The Hubbard parameters are defined in Eqs. (\ref{4.8}), (\ref{4.9}), and
(\ref{4.10}). As in Eq. (\ref{4.12}), we shall consider the single-band
Hubbard model.

For the tunneling term, we have
$$
J_{ij} = J_{ij}^0 + \Dlt J_{ij} \; , \qquad
J_{ij}^0 \equiv - \int w(z-a_i) H_L^0(z) w(z-a_j) \; dz \; ,
$$
\be
\label{5.21}
\Dlt J_{ij} \equiv - \int w(z-a_i)
\Dlt H_L(z) w(z-a_j) \; dz \; .
\ee
Respectively, the single-site term $h_j$ becomes
$$
h_j = h_0 + \Dlt h_j \; , \qquad
h_0 \equiv \int w(z) H_L^0(z) w(z) \; dz \; ,
$$
\be
\label{5.22}
\Dlt h_j \equiv \int w(z-a_j)
\Dlt H_L(z) w(z-a_j) \; dz \; .
\ee

Using for Wannier functions the Gaussian approximation (\ref{3.161}), we
follow calculations similar to those in Sec. 3.9. Then we have the primary
parameters
\be
\label{5.23}
J = \left ( \frac{\pi^2}{4} \; - \; 1 \right )
V_0 \exp\left ( - \; \frac{a^2}{4l_0^2} \right ) \; ,
\qquad h_0 =\sqrt{V_0 E_R} \; ,
\ee
where $J$ is $J_{ij}^0$ for nearest neighbors. The additional term for the
tunneling parameter is
\be
\label{5.24}
\Dlt J_{ij} = -\; \frac{V_1}{2} \; \exp
\left ( - \; \frac{a^2}{4l_0^2} \right )
\left\{ 1 - \exp\left (- q_0^2 l_0^2\right )
\cos [ q_0 (a_i + a_j)] \right \} \; ,
\ee
and the addition to the single-site term is
\be
\label{5.25}
\Dlt h_j = \frac{V_1}{2} \left [ 1  -\exp(-q_0^2 l_0^2)
\cos (2q_0 a_j) \right ] \; .
\ee

The latter equations can be simplified, when there is good localization,
not only with respect to the primary sublattice, such that $k_0l_0\ll 1$,
but also with respect to the secondary sublattice, so that
\be
\label{5.26}
q_0 l_0 \ll 1 \; .
\ee
Then Eq. (\ref{5.24}) becomes
\be
\label{5.27}
\Dlt J_{ij} \cong - V_1 \exp
\left ( - \; \frac{a^2}{4l_0^2} \right )
\sin^2 \left ( q_0 \; \frac{a_i+a_j}{2} \right ) \; ,
\ee
while Eq. (\ref{5.25}) reduces to
\be
\label{5.28}
\Dlt h_j \cong V_1 \sin^2( q_0 a_j) \; .
\ee
The on-site interaction parameter $U$ remains the same as in Eq.
(\ref{4.14}).

Thus, the Hubbard parameters $J$ and $h_0$ acquire the site-dependent terms
(\ref{5.27}) and (\ref{5.28}). Comparing these additional terms with the
primary values $J$ and $h_0$, we have
\be
\label{5.29}
\left | \frac{\Dlt J_{ij}}{J} \right | \leq
0.7 \; \frac{V_1}{V_0} \; , \qquad
\frac{\Dlt h_j}{h_0} \leq
\frac{V_1}{2V_0} \; \sqrt{ \frac{V_0}{E_R} } \; .
\ee
And the Hubbard Hamiltonian for the quasiperiodic bichromatic potential
takes the form
\be
\label{5.30}
\hat H = - \sum_{<ij>} J_{ij} c_i^\dgr c_j \; + \;
\sum_j h_j c_j^\dgr c_j \; + \; \frac{U}{2} \; \sum_j
c_j^\dgr c_j^\dgr c_j c_j \; .
\ee

If the amplitude of the second term in potential (\ref{5.17}) is comparable
with that of the first term, then, as follows from ratio (\ref{5.29}),
Hamiltonian (\ref{5.30}) describes a very nonideal lattice with strong
dependence of $J_{ij}$ and $h_j$ on site numbers. Therefore, one usually
considers the case, when the second term in potential (\ref{5.17}) plays the
role of perturbation, so that $V_1\ll V_0$. In such a case, $|\Dlt J_{ij}|\ll
J$, and one can neglect the variation of the tunneling parameter $J$. However,
even when $V_1\ll V_0$, but the primary sublattice is sufficiently deep, so
that $E_R\ll V_0$, then $\Dlt h_j$ can be of order of $h_0$, as it is seen
from Eq. (\ref{5.22}). Therefore, the variation of the single-site parameter
$h_j$ cannot be neglected. In that situation, one can replace in the Hubbard
Hamiltonian (\ref{5.30}) the tunneling $J_{ij}$ by the site-independent value
$J$, while keeping $h_j$ as the sum $h_0+\Dlt h_j$, with the site-varying part
(\ref{5.28}). The properties of quasiperiodic optical lattices have much in
common with those of quasicrystals [207,208].

\subsection{Rotating Optical Lattices}

When a Bose system with BEC is rotated, vortices appear after the rotation
frequency ${\bf\Om}$ reaches the critical value. An optical lattice, in
principle, also can be rotated sufficiently fast in order that vortices
could grow in it. Two interesting situations can occur, when rotating
one-dimensional lattices, with the lattice direction either along the
vortex, that is, along the rotation frequency ${\bf\Om}$, or perpendicular
to the latter, hence, perpendicular to the vortex. Both these cases were
studied theoretically in Refs. [332,333] and [334,335], respectively.
Experimentally these cases have not yet been realized.

When an axisymmetric BEC with a single vortex on the axis of symmetry
is subject to a lattice optical potential along the same axis, then this
one-dimensional lattice would slice the rotating BEC into many circular
disks [332,333,336]. The resulting periodic array forms a set of
quasi-two-dimensional condensate layers. Each effectively two-dimensional
condensate becomes a single pancake vortex.

A rotating Bose system is conveniently described in the rotating
reference frame. Then the system Hamiltonian acquires an additional
centrifugal term
$$
\hat U_{rot} = - \int \psi^\dgr(\br)\; {\bf\Om} \cdot \hat{\bf L} \;
\psi(\br) \; d\br \; ,
$$
where ${\bf\Om}$ is the angular rotation frequency and
$\hat{\bf L}=\br\times\bp$ is angular momentum. Passing to the Wannier
representation transforms the centrifugal term into
$$
\hat U_{rot} = - \sum_{i\neq j} J_{ij}^{rot} c_i^\dgr c_j \; +
\; \sum_j h_j^{rot} c_j^\dgr c_j \; ,
$$
with
$$
J_{ij}^{rot} \equiv \int w^*(\br-\ba_i)
( {\bf\Om} \cdot \hat{\bf L}) w(\br-\ba_j) \; d\br \; ,
$$
$$
h_j^{rot} \equiv - \int w^*(\br-\ba_j)
( {\bf\Om} \cdot \hat{\bf L}) w(\br-\ba_j) \; d\br \; .
$$
This means that rotating a lattice induces additional atomic tunneling
and a single-site energy term.

Thus, optical lattices provide the possibility of creating Bose systems
with a wide variety of properties, which can be employed in different
applications.

\section{Optical Lattices with Disorder}

\subsection{Random Potentials}

The presence of an externally incorporated disorder can essentially
change the system properties. And the possibility of varying the level
of disorder presents a powerful tool for achieving different states of
matter. This problem has been considered in solid state physics for many
years, with applications to conducting properties of materials [337--340],
 magnetic properties of spin glasses [341,342], to real-space glasses
[343,344], amorphous alloys [345,346], and to physics of caking [347].
Disorder can also be introduced by the boundaries of finite systems
[101,348] and by domain walls [349,350].

Recently, several experiments [351--354] have studied $^{87}$Rb in random
optical potentials created by optical speckles. It has been observed that
the speckle randomness induces damping of collective excitations [351] and
inhibition of expansion [352--354]. The effects of a disordered optical
potential on the transport and phase coherence of a BEC of $^7$Li atoms
has also been studied, and inhibition of transport and damping of dipole
excitations have been observed [355]. An ultracold bosonic gas of
$^{87}$Rb in a 3-dimensional optical lattice was investigated, with
disorder induced by a small contribution of fermionic $^{40}$K impurity
atoms [356]. The random admixture was found to favor the localization of
bosonic atoms. It was suggested [357] that a one-dimensional BEC in a weak
random potential can exhibit Anderson localization [358].

The Hamiltonian of atoms in an optical lattice potential $V_L(\br)$, and
also subject to the action of an additional random potential $\xi(\br)$,
has the form
\be
\label{6.1}
\hat H = \int \psi^\dgr(\br) \left [  -\;
\frac{\nabla^2}{2m} + V_L(\br) + \xi(\br) \right ]
\psi(\br) \; d\br \; + \; \frac{1}{2} \; \Phi_0 \;
\int \psi^\dgr(\br) \psi^\dgr(\br) \psi(\br)\psi(\br) \;
d\br \; ,
\ee
where $\Phi_0\equiv 4\pi a_s/m$. The random potential is real
\be
\label{6.2}
\xi^*(\br) = \xi(\br) \; .
\ee
Without the loss of generality, it can be taken as zero-centered, so that
its stochastic averaging be
\be
\label{6.3}
\ll \xi(\br) \gg \; = \; 0 \; .
\ee
The stochastic averaging of the pair product
\be
\label{6.4}
\ll \xi(\br) \xi(\br') \gg \; = \; R(\br-\br')
\ee
is characterized by a random-potential correlation function that is real
and symmetric, such that
\be
\label{6.5}
R^*(\br) = R(\br) = R(-\br) \; .
\ee
Details of defining the stochastic averaging can be found in the book
[359].

When using the Bloch representation for the field operator
$$
\psi(\br) = \sum_{nk} a_{nk} \vp_{nk}(\br) \; ,
$$
it is convenient to introduce the matrix element of the random potential
\be
\label{6.6}
\bt_{kp}^{mn} \equiv \int
\vp_{mk}^*(\br) \xi(\br) \vp_{np}(\br) \; d\br \; .
\ee
Then the term of Hamiltonian (\ref{6.1}), containing the random potential,
is
$$
\int \psi^\dgr(\br) \xi(\br) \psi(\br) \; d\br =
\sum_{mn} \; \sum_{kp} \bt_{kp}^{mn} a_{mk}^\dgr a_{np} \; .
$$

If, when considering a lattice, one prefers the Wannier representation
$$
\psi(\br) = \sum_{nj} c_{nj} w_n(\br-\ba_j) \; ,
$$
then one meets the matrix element
\be
\label{6.7}
\gm_{ij}^{mn} \equiv \int
w_m^*(\br-\ba_i) \xi(\br) w_n(\br-\ba_j) \; d\br \; .
\ee
And the random term in Eq. (\ref{6.1}) becomes
$$
\int \psi^\dgr(\br) \xi(\br) \psi(\br) \; d\br =
\sum_{mn} \; \sum_{ij} \gm_{ij}^{mn} c_{mi}^\dgr c_{nj} \; .
$$
The matrix elements (\ref{6.6}) and (\ref{6.7}) are connected through
the Fourier transformations
\be
\label{6.8}
\bt_{kp}^{mn} = \frac{1}{N_L} \; \sum_{ij}
\gm_{ij}^{mn} e^{-i\bk\cdot\ba_i + i\bp\cdot\ba_j} \; ,
\qquad
\gm_{ij}^{mn} = \frac{1}{N_L} \; \sum_{kp}
\bt_{kp}^{mn} e^{i\bk\cdot\ba_i - i\bp\cdot\ba_j} \; .
\ee

In the uniform limit, when the Bloch functions $\vp_{nk}(\br)$ tend to
the plane waves $e^{i\bk\cdot\br}/\sqrt{V}$, it is convenient to define
the Fourier transform of the random potential
\be
\label{6.9}
\xi_k \equiv \frac{1}{\sqrt{V}} \;
\int \xi(\br) e^{-i\bk\cdot\br}\; d\br \; , \qquad
\xi(\br) \equiv \frac{1}{\sqrt{V}} \;
\sum_k \xi_k e^{i\bk\cdot\br} \; .
\ee
Then one has
\be
\label{6.10}
\bt_{kp}^{mn} = \frac{1}{\sqrt{V}} \; \xi_{k-p} \; .
\ee

In the presence of BEC, one has to break the global gauge symmetry [63],
which can be done by the Bogolubov shift of the field operator
\be
\label{6.11}
\psi(\br)\; \ra \; \hat\psi(\br) \equiv \eta(\br) + \psi_1(\br) \; ,
\ee
as is described in Sec. 2.7. This implies, in agreement with Sec. 3.3,
the corresponding shift of the Bloch field operator
\be
\label{6.12}
a_{nk} \; \ra \; \hat a_{nk} \equiv \dlt_{n0}\; \dlt_{k0}
\sqrt{N_0} \; + \; a_{nk}
\ee
and of the Wannier field operator
\be
\label{6.13}
c_{nj}  \; \ra \; \hat c_{nj} \equiv \dlt_{n0}\; \sqrt{\nu n_0} \; +
\; c_{nj} \; .
\ee
The Bogolubov shift realizes unitary nonequivalent operator representations
[99,360--362].

The grand Hamiltonian of a system with BEC is
\be
\label{6.14}
H = \hat H - \mu_0 N_0 - \mu_1 \hat N_1 \; .
\ee
The type of disorder induced by an external random potential is termed
the {\it frozen} disorder or {\it quenched} disorder. The corresponding
grand thermodynamic potential is defined as
\be
\label{6.15}
\Om = - T \ll \ln\; {\rm Tr}\; e^{-\bt H} \gg \; .
\ee
And the free energy is
$$
F = \Om + \mu N \; , \qquad \mu = \mu_0 n_0 + \mu_1 n_1 \; .
$$

It is important to emphasize that there are two kinds of averaging for
quantum random systems. One type of averaging is the stochastic averaging,
denoted by the double angle brackets $\ll\ldots\gg$, characterizing the
averaging over the distribution of random potentials. And there is the
quantum statistical averaging, which for an operator $\hat A$ is defined
as
\be
\label{6.16}
< \hat A>_H \; \equiv \; {\rm Tr}\; \hat\rho \hat A \; ,
\ee
with the statistical operator
\be
\label{6.17}
\hat\rho = \frac{\exp(-\bt H)}{{\rm Tr}\exp(-\bt H)} \; .
\ee
The total averaging
\be
\label{6.18}
< \hat A > \; \equiv \; \ll {\rm Tr}\; \hat\rho \hat A \gg
\ee
includes both, the quantum as well as stochastic averaging procedures.

The condensate fraction $n_0=1-n_1$ is expressed through the fraction of
uncondenced atoms
\be
\label{6.19}
n_1 \equiv \frac{1}{N} \;
\int < \psi_1^\dgr(\br)\psi_1(\br)> d\br \; = \;
\frac{1}{N} \; \sum_{nk} < a_{nk}^\dgr a_{nk} > \; = \;
\frac{1}{N} \; \sum_{nj} < c_{nj}^\dgr c_{nj} > \; .
\ee
The anomalous average is
\be
\label{6.20}
\sgm \equiv \frac{1}{N} \;
\int < \psi_1(\br)\psi_1(\br)> d\br \; = \;
\frac{1}{N} \; \sum_{nk} < a_{nk} a_{n,-k} > \; = \;
\frac{1}{N} \; \sum_{nj} < c_{nj} c_{nj} > \; .
\ee

For random systems, it is possible to define an additional order
parameter, the {\it glassy fraction}
\be
\label{6.21}
n_G \; \equiv \; \frac{1}{N} \; \int
\ll | < \psi_1(\br) >_H |^2 \gg d\br \; ,
\ee
which is analogous to the Edwards-Anderson order parameter for spin
glasses [341,342]. Though the total average $<\psi_1(\br)>=0$, the partial,
quantum, average $<\psi_1(\br)>_H$ is not necessarily zero. Respectively,
the quantum averages
\be
\label{6.22}
\al_{nk} \; \equiv \; < a_{nk} >_H \; , \qquad
\overline\al_{nj} \; \equiv \; < c_{nj} >_H
\ee
are not zero, contrary to the total averages
\be
\label{6.23}
< \psi_1(\br) > \; = \; < a_{nk} > \; = \;
< c_{nj} > \; = \; 0 \; .
\ee
From here it follows that
\be
\label{6.24}
\ll \al_{nk} \gg \; = \;
\ll \overline\al_{nj} \gg \; = \; 0 \; .
\ee
The glassy fraction (\ref{6.21}) can be rewritten as
\be
\label{6.25}
n_G = \frac{1}{N} \; \sum_{nk} \ll |\al_{nk}|^2 \gg \; = \;
\frac{1}{N} \; \sum_{nj} \ll | \overline\al_{nj} |^2 \gg \; .
\ee

The presence of random fields induces in the system additional
fluctuations, which is typical of random and chaotic systems [363].
Without disordering fields, the condensate is depleted by atomic
interactions and temperature. The inclusion of random fields depletes
the condensate even more and creates the glassy fraction (\ref{6.21}).

\subsection{Uniform Limit}

A uniform system is a limiting case of a very shallow lattice. Then,
in the Bloch representation, the Bloch functions are plane waves.
Uniform Bose-condensed systems in random potentials have been studied
for asymptotically weak atomic interactions and asymptotically weak
strength of disorder [364-366]. A theory for arbitrary strong random
potentials and interactions was developed in Refs. [367,368]. When a
disordered potential is created inside a trap, then the BEC properties
can be investigated by means of time-of -flight experiments [369].

For a system, uniform on average, the condensate fraction is
\be
\label{6.26}
n_0 = \frac{1}{\rho} \; | \eta(\br) |^2 \qquad \left (
\eta(\br) =\sqrt{\rho_0} \right ) \; .
\ee
The fraction of uncondensed atoms (\ref{6.19}) can be written as
\be
\label{6.27}
n_1 = \frac{1}{N} \; \sum_k n_k \qquad \left ( n_k \; \equiv \;
< a_k^\dgr a_k > \right ) \; .
\ee
The anomalous average (\ref{6.20}) takes the form
\be
\label{6.28}
\sgm = \frac{1}{N} \; \sum_k \sgm_k \qquad \left ( \sgm_k \;
\equiv \; < a_k a_{-k} > \right ) \; .
\ee
And the glassy fraction (\ref{6.21}) becomes
\be
\label{6.29}
n_G = \frac{1}{N} \; \sum_k \ll | \al_k|^2 \gg \; ,
\ee
where the notation
\be
\label{6.30}
\al_k \; \equiv \; < a_k >_H
\ee
is introduced.

Accomplishing in Eq. (\ref{6.1}) the Bogolubov shift (\ref{6.11})
and passing to the Fourier-transformed field operators $a_k$, for
the grand Hamiltonian (\ref{6.14}), we obtain
\be
\label{6.31}
H = \sum_{n=0}^4 H^{(n)} \; + \; H_{ext} \; ,
\ee
in which the terms in the sum are
$$
H^{(0)} = \left ( \frac{1}{2} \; \rho_0 \Phi_0 -
\mu_0\right ) N_0 \; , \qquad H^{(1)} = 0 \; ,
$$
$$
H^{(2)} = \sum_{k\neq 0} \left ( \frac{k^2}{2m} +
2\rho_0 \Phi_0 - \mu_1 \right ) a_k^\dgr a_k \; + \;
\frac{1}{2} \; \sum_{k\neq 0} \rho_0 \Phi_0 \left (
a_k^\dgr a_{-k}^\dgr + a_{-k} a_k \right ) \; ,
$$
$$
H^{(3)} = \sqrt{\frac{\rho_0}{V} } \;
\sum_{kp(\neq 0)} \Phi_0 \left ( a_k^\dgr a_{k+p} a_{-p} +
a_{-p}^\dgr a_{k+p}^\dgr a_k \right ) \; ,
$$
\be
\label{6.32}
H^{(4)} = \frac{1}{2V} \; \sum_q \;
\sum_{kp(\neq 0)} \Phi_0 a_k^\dgr a_p^\dgr a_{k-q} a_{p+q}
\ee
and the last term, caused by the random potential, is
\be
\label{6.33}
H_{ext} = \rho_0 \xi_0 \; \sqrt{V} \; + \;
\sqrt{\rho_0} \; \sum_{k\neq 0} \left ( a_k^\dgr \xi_k +
\xi_k^* a_k \right ) \; + \; \frac{1}{\sqrt{V}} \;
\sum_{kp(\neq 0)} a_k^\dgr a_p \xi_{k-p} \; .
\ee

The correlation function of a disordered potential, defined in Eq.
(\ref{6.4}), is assumed to possess a Fourier transform
\be
\label{6.34}
R_k = \int R(\br) e^{-i\bk\cdot\br} \; d\br \; , \qquad
R(\br) = \frac{1}{V} \; \sum_k R_k e^{i\bk\cdot\br} \; .
\ee
From properties (\ref{6.2}) and (\ref{6.5}), it follows that
\be
\label{6.35}
\xi_k^* = \xi_{-k} \; , \qquad R_k^* = R_{-k} = R_k \; .
\ee
And the stochastic averaging (\ref{6.4}) gives
\be
\label{6.36}
\ll \xi_k^* \xi_p \gg \; = \; \dlt_{kp} R_k \; .
\ee

In the case of white noise, one has
\be
\label{6.37}
R(\br) = R_0\dlt(\br) \; , \qquad R_k = R_0 \; .
\ee
Thence, Eq. (\ref{6.36}) reduces to
\be
\label{6.38}
\ll \xi_k^* \xi_p \gg \; = \; \dlt_{kp} R_0 \; .
\ee

The Hamiltonian terms in Eq. (\ref{6.32}) can be simplified by using the
Hartree-Fock-Bogolubov approximation, as in Refs. [57,94--99]. But the
random Hamiltonian term (\ref{6.33}) should be treated with care. If one
would use the standard mean-field decoupling for the last term in Eq.
(\ref{6.33}), one would kill in this term all quantum effects because of
Eqs. (\ref{6.3}) and (\ref{6.23}). Therefore it is necessary to invoke a
more delicate decoupling procedure. For this purpose, it is convenient to
employ the {\it stochastic mean-field approximation} suggested and used
earlier for other physical systems [241,242,370--373]. This approximation
in the present case yields
\be
\label{6.39}
a_k^\dgr a_p \xi_{k-p} = \left ( a_k^\dgr \al_p + \al_k^* a_p -
\al^*_k \al_p \right ) \xi_{k-p} \; .
\ee

Then Hamiltonian (\ref{6.31}) can be diagonalized by the use of the {\it
nonuniform nonlinear} canonical transformation
\be
\label{6.40}
a_k = u_k b_k + v_{-k}^* b_{-k}^\dgr + w_k \vp_k \; ,
\ee
which generalizes the standard uniform Bogolubov transformation. The
diagonalization implies that the resulting Hamiltonian should be diagonal
in terms of the operators $b_k^\dgr$ and $b_k$, so that
$$
< b_k >_H \; = \; < b_k b_p >_H \; = \; 0 \; .
$$
Hence, relation (\ref{6.40}) gives
\be
\label{6.41}
< a_k >_H \; = \; w_k \vp_k \; .
\ee
This diagonalization results in
$$
u_k^2 = \frac{\om_k+\ep_k}{2\ep_k} \; , \qquad
v_k^2  = \frac{\om_k-\ep_k}{2\ep_k} \; , \qquad
w_k = -\; \frac{1}{\om_k+mc^2} \; ,
$$
\be
\label{6.42}
\om_k = \frac{k^2}{2m} + mc^2 \; , \qquad
\om_k^2 = \ep_k^2 + (mc^2)^2 \; ,
\ee
where
\be
\label{6.43}
\ep_k =\sqrt{(ck)^2 + \left ( \frac{k^2}{2m}\right )^2 }
\ee
is the Bogolubov spectrum, with the sound velocity $c$ defined by the
equation
\be
\label{6.44}
mc^2 =  (n_0 + \sgm) \rho \Phi_0 \; .
\ee
The random variable $\vp_k$ in transformation (\ref{6.40}) satisfies the
Fredholm equation
\be
\label{6.45}
\vp_k = \sqrt{\rho_0} \; \xi_k \; - \;
\frac{1}{\sqrt{V}} \; \sum_p \;
\frac{\xi_{k-p}\vp_p}{\om_p+mc^2} \; .
\ee
The nonuniform nonlinear transformation (\ref{6.40}) reduces the grand
Hamiltonian (\ref{6.31}) to the form
\be
\label{6.46}
H = E_B + \sum_k \ep_k b_k^\dgr b_k  + \ \vp_0\; \sqrt{N_0} \; ,
\ee
with
\be
\label{6.47}
E_B = \frac{1}{2}\; \sum_k (\ep_k-\om_k) \; - \;
\left [ 1 - n_0^2 + \frac{1}{2}\; (n_0 +\sgm )^2 \right ] \;
\rho^2 \Phi_0 N \; .
\ee

The diagonal Hamiltonian (\ref{6.46}) allows us to find the momentum
distribution
\be
\label{6.48}
n_k = \frac{\om_k}{2\ep_k} \; {\rm coth}
\left ( \frac{\ep_k}{2T} \right ) \; - \;
\frac{1}{2} \; + \; \ll |\al_k|^2 \gg
\ee
and the anomalous average
\be
\label{6.49}
\sgm_k = -\; \frac{mc^2}{2\ep_k} \; {\rm coth} \left (
\frac{\ep_k}{2T} \right ) + \ll |\al_k|^2 \gg \; .
\ee
The random variable (\ref{6.30}), in view of Eqs. (\ref{6.41}) and
(\ref{6.42}), is
\be
\label{6.50}
\al_k = w_k \vp_k = -\; \frac{\vp_k}{\om_k+mc^2} \; .
\ee
Therefore
\be
\label{6.51}
\ll |\al_k|^2 \gg \; = \;
\frac{\ll |\vp_k|^2 \gg}{(\om_k+mc^2)^2} \; .
\ee

The fraction of uncondensed atoms (\ref{6.27}) becomes a sum of two terms,
\be
\label{6.52}
n_1 = n_N + n_G \; ,
\ee
in which the first term
\be
\label{6.53}
n_N = \frac{1}{2\rho} \; \int \left [ \frac{\om_k}{\ep_k} \;
{\rm coth} \left ( \frac{\ep_k}{2T} \right ) - 1 \right ] \;
\frac{d\bk}{(2\pi)^3}
\ee
is the fraction of uncondensed atoms, due to finite temperature and
interactions, while the second term
\be
\label{6.54}
n_G = \frac{1}{\rho} \; \int \;
\frac{\ll |\vp_k|^2 \gg}{(\om_k+mc^2)^2} \; \frac{d\bk}{(2\pi)^3}
\ee
is the glassy fraction (\ref{6.29}), caused by the random potential. The
anomalous average (\ref{6.28}) is also a sum of two terms,
\be
\label{6.55}
\sgm =  \sgm_N + n_G \; ,
\ee
with the first term
\be
\label{6.56}
\sgm_N = -\; \frac{1}{2\rho} \; \int \; \frac{mc^2}{\ep_k} \;
{\rm coth} \left ( \frac{\ep_k}{2T} \right ) \;
\frac{d\bk}{(2\pi)^3}
\ee
and the second term (\ref{6.54}).

To find the superfluid fraction, we resort to the general definition of
Sec. 2.10,
\be
\label{6.57}
n_s = 1 - \; \frac{2Q}{3T} \qquad
\left ( Q \equiv \frac{<\bP^2>}{2mN} \right ) \; .
\ee
The dissipated heat consists of two parts,
\be
\label{6.58}
Q = Q_N + Q_G \; .
\ee
The first part
\be
\label{6.59}
Q_N = \frac{1}{8m\rho} \; \int \;
\frac{k^2}{{\rm sinh}^2(\ep_k/2T)} \; \frac{d\bk}{(2\pi)^3}
\ee
is the heat dissipated by uncondensed atoms related to finite temperature
and interactions. And the second part
\be
\label{6.60}
Q_G = \frac{1}{2m\rho} \; \int \;
\frac{k^2\ll|\vp_k|^2\gg}{\ep_k(\om_k+mc^2)} \;
{\rm coth} \left ( \frac{\ep_k}{2T} \right ) \;
\frac{d\bk}{(2\pi)^3}
\ee
is the heat dissipated by the glassy fraction of atoms.

The analysis [367,368] of the derived equations shows that for weak
interactions and sufficiently strong disorder, the superfluid fraction can
become smaller than the condensate fraction, in agreement with Monte Carlo
simulations [374]. But at relatively strong interactions, the superfluid
fraction is larger than the condensate fraction for any strength of
disorder. The condensate and superfluid fractions, and the glassy fraction
always coexist, being together either nonzero or zero. In the presence of
disorder, the condensate fraction becomes a nonmonotonic function of the
interaction  strength, displaying an antidepletion effect caused by the
competition between the stabilizing role of the atomic interaction and
the destabilizing role of the disorder. An ideal Bose gas with BEC is
stochastically unstable, the BEC being destroyed by infinitesimally weak
disorder. Finite atomic interactions stabilize the system. When the strength
of disorder is increased, reaching a critical value $\zeta_c$ of the disorder
parameter
$$
\zeta \equiv \frac{7m^2R_0}{4\pi\rho^{1/3}} = \frac{a}{l_L} \; ,
$$
in which
$$
l_L \equiv \frac{4\pi}{7m^2R_0} \; \qquad a \equiv
\frac{1}{\rho^{1/3}} \; ,
$$
the condensate and superfluid fractions drop to zero by a first-order
phase transition. The characteristic length $l_L$ practically coincides
with the Larkin length [375]. The critical value of the disorder parameter
$\zeta_c$, where the first-order phase transition occurs, depends on the
interaction strength and temperature. More details on the properties of
the disordered superfluid can be found in Refs. [367,368].

After the strength of disorder reaches its critical value, characterized
by the critical value of the disorder parameter $\zeta_c$, the global
coherence becomes destroyed and there is no global BEC occupying the whole
system. Also, there is no superfluidity. But for $\zeta>\zeta_c$, the
condensate can decay into fragments each of the size of the Larkin length
$l_L$, so that the phase transition at $\zeta_c$ could be interpreted as
a spatial condensate fragmentation [376]. For $\zeta>\zeta_c$, the local
remnants of the condensate could remain, while superfluidity being absent.
Such a state, according to the classification of Sec. 1.4, corresponds to
the Bose glass. The destruction of superfluidity, with increasing disorder,
through a first-order transition, was also found in Ref. [377].

\subsection{Disordered Lattice}

Passing from the initial Hamiltonian (\ref{6.1}) to the Wannier
representation, we may, as usual, consider the single-band case. Then the
matrix element (\ref{6.7}) of the random potential $\xi(\br)$ reads as
\be
\label{6.61}
\gm_{ij} = \int w^*(\br-\ba_i) \xi(\br) w(\br-\ba_j) \; d\br \; .
\ee
For the single-site effective potential, we have
\be
\label{6.62}
h_j = h_0 + \gm_j \; ,
\ee
in which the constant term
$$
h_0 \equiv \int w^*(\br) H_L(\br) w(\br)\; d\br
$$
in what follows can be neglected without the loss of generality. The
random term in Eq. (\ref{6.62}) is
\be
\label{6.63}
\gm_j \equiv \gm_{jj} = \int | w(\br-\ba_j)|^2
\xi(\br) \; d\br \; .
\ee
This diagonal term is real because of Eq. (\ref{6.2}),
\be
\label{6.64}
\gm_j^* = \gm_j \; .
\ee

From Hamiltonian (\ref{6.1}), the Hubbard Hamiltonian follows
\be
\label{6.65}
\hat H = - \sum_{<ij>} \tilde J_{ij} c_i^\dgr c_j \; + \;
\sum_j \gm_j c_j^\dgr c_j \; + \;
\frac{U}{2} \; \sum_j c_j^\dgr c_j^\dgr c_j c_j \; ,
\ee
with the tunneling integral
\be
\label{6.66}
\tilde J_{ij} = J - \gm_{ij}
\ee
renormalized by the random variable (\ref{6.61}).

To understand how substantial this change of the tunneling parameter
could be, let us consider the stochastic average
\be
\label{6.67}
\ll \gm_i \gm_j \gg \; = \; \int | w(\br-\ba_i)|^2\; R(\br-\br')\;
| w(\br-\ba_j) |^2 \; d\br d\br' \; .
\ee
As is common, we may assume that the correlation length of the correlation
function $R(\br)$ is short, such that
$$
\left | \frac{\int r^2 R(\br)\; d\br}{\int R(\br)\; d\br}
\right |\; \ll \; a^2 \; .
$$
Then the white-noise approximation for the random potential is applicable.
In that case, Eq. (\ref{6.67}) reduces to
\be
\label{6.68}
\ll \gm_i \gm_j \gg \; = \; R_0 \int
| w(\br-\ba_{ij} ) w(\br) |^2 \; d\br \; .
\ee
Using the tight-binding approximation of Sec. 3.5, we have
\be
\label{6.69}
\ll \gm_i \gm_j \gg \; = \; R_0 \left ( \frac{m\om_0}{2\pi}
\right )^{3/2} \; \exp \left ( - \; \frac{3a_{ij}^2}{2l_0^2}
\right ) \; ,
\ee
where
$$
\om_0 = 2\sqrt{E_R V_0} \; , \qquad
l_0 \equiv \frac{1}{\sqrt{m\om_0}} \; .
$$
For the white-noise case,
\be
\label{6.70}
\ll \gm_{ij} \gm_{ij} \gg \; = \; \ll \gm_i \gm_j \gg \; ,
\ee
therefore it is sufficient to consider the stochastic averages
(\ref{6.68}) and (\ref{6.69}).

Taking the nearest neighbors for $i\neq j$ in Eq. (\ref{6.69}) and
considering a cubic lattice gives
\be
\label{6.71}
\ll \gm_i \gm_j \gg \; = \; \frac{R_0 k_0^3}{(2\pi)^{3/2}} \;
\left ( \frac{V_0}{E_R} \right )^{3/4} \;
\exp \left ( - \; \frac{3\pi^2}{2} \; \sqrt{\frac{V_0}{E_R} }
\right ) \; .
\ee
This is to be compared with the tunneling parameter $J$ defined in Eq.
(\ref{4.14}). Using the latter yields
\be
\label{6.72}
\frac{\ll \gm_i \gm_j \gg}{J^2} = \frac{\zeta}{136}
\left ( \frac{E_R}{V_0} \right )^{5/4}  \;
\exp \left ( - \pi^2 \; \sqrt{\frac{V_0}{E_R} }
\right ) \; ,
\ee
where the relation
$$
\frac{R_0}{a^3E_R^2} = \frac{16}{7\pi^3}\; \zeta
$$
is employed and
\be
\label{6.73}
\zeta \equiv \frac{a}{l_L} \qquad \left ( l_L \equiv
\frac{4\pi}{7m^2R_0} \right )
\ee
being the disorder parameter.

From Eq. (\ref{6.72}) it is seen that if the disorder is not extremely
strong, then
\be
\label{6.74}
\frac{\ll \gm_i \gm_j \gg}{J^2} \ll 1 \; , \qquad
(\zeta \ll \zeta_{max} ) \; ,
\ee
where
$$
\zeta_{max} \equiv 136
\left ( \frac{V_0}{E_R} \right )^{5/4} \;
\exp \left ( \pi^2 \; \sqrt{\frac{V_0}{E_R} }
\right ) \; .
$$
For deep lattices, when $E_R\ll V_0$, $\zeta_{max}$ is very large. Even
for quite shallow lattices, one has
$$
\zeta_{max} \gg 1 \qquad \left (
\frac{E_R}{V_0} \ll 100 \right ) \; .
$$
If the disorder is so strong that $\zeta>\zeta_{max}$, then the lattice,
as such, can be destroyed, and one would return to the uniform limit.
Hence considering a lattice model is meaningful only for $\zeta<\zeta_{max}$.
Under condition (\ref{6.74}), the random part $\gm_{ij}$ in the renormalized
tunneling integral (\ref{6.66}) can be neglected, so that
$$
\tilde J_{ij} \cong J \qquad (\zeta\ll \zeta_{max} )
$$
for $i\neq j$ pertaining to the nearest neighbors.

In this way, for the realistic cases of optical lattices, Hamiltonian
(\ref{6.65}) simplifies to
\be
\label{6.75}
\hat H = - J \sum_{<ij>} c_i^\dgr c_j \; + \;
\sum_j \gm_j c_j^\dgr c_j \; + \; \frac{U}{2} \;
\sum_j c_j^\dgr c_j^\dgr c_j c_j \; .
\ee
For treating the lattice with BEC, one has to break the gauge symmetry
by means of the Bogolubov shift (\ref{6.13}), replacing the operators
$c_j$ by $\hat c_j$.

\subsection{Disordered Superfluid}

To study the property of the superfluid state in a disordered lattice,
we have to follow the standard prescriptions. In the Hubbard Hamiltonian
(\ref{6.75}), we make the Bogolubov shift (\ref{6.13}) and then define the
grand Hamiltonian (\ref{6.14}). This yields
\be
\label{6.76}
H = - J  \sum_{<ij>} \hat c_i^\dgr \hat c_j \; + \; \frac{U}{2} \;
\sum_j \hat c_j^\dgr \hat c_j^\dgr \hat c_j \hat c_j \; + \;
H_{ext} - \mu_0 N_0 - \mu_1 \hat N_1 \; ,
\ee
where $\hat N_1$ is given in Eq. (\ref{4.22}) and the term
\be
\label{6.77}
H_{ext} \equiv \sum_j \gm_j \hat c_j^\dgr \hat c_j
\ee
is due to the external random potential. The last term (\ref{6.77}), with
the Bogolubov shift (\ref{6.13}), reads as
\be
\label{6.78}
H_{ext} = \sum_j \gm_j \left [ \nu n_0 + \sqrt{\nu n_0} \;
\left ( c_j^\dgr + c_j \right ) + c_j^\dgr c_j \right ] \; .
\ee

Expanding the random variable over the lattice, we have
\be
\label{6.79}
\gm_j = \frac{1}{\sqrt{N_L}}\;
\sum_k \bt_k e^{i\bk\cdot\ba_j} \; , \qquad
\bt_k = \frac{1}{\sqrt{N_L}}\; \sum_j
\gm_j e^{-i\bk\cdot\ba_j} \; .
\ee
Because of property (\ref{6.64}),
\be
\label{6.80}
\bt_k^* = \bt_{-k} \qquad (\gm_j^* =\gm_j ) \; .
\ee
The relation of the variable $\bt_k$ with $\bt_{kp}$, defined in Eq.
(\ref{6.8}), follows from the fact that $\gm_j\equiv\gm_{jj}$, hence
$$
\gm_j = \frac{1}{N_L} \; \sum_{kp} \bt_{kp}
e^{i(\bk-\bp)\cdot\ba_j} \; .
$$
Then Eqs. (\ref{6.79}) give
$$
\bt_k = \frac{1}{\sqrt{N_L}} \; \sum_p \bt_{k+p,p} \; .
$$

The random variables $\gm_j$ enjoy the properties
\be
\label{6.81}
\ll \gm_j \gg \; = \; 0 \; , \qquad
\ll \gm_i \gm_j \gg \; = \; R_{ij} \; ,
\ee
where the correlation function $R_{ij}$ is given by Eq. (\ref{6.67}). In
the particular case of white noise, according to Eq. (\ref{6.68}), one has
\be
\label{6.82}
R_{ij} = R_0 \int | w(\br-\ba_{ij}) w(\br) |^2 \; d\br \; .
\ee
For the tight-binding approximation, this reduces to Eq. (\ref{6.69}).

The variables $\bt_k$ have the properties
\be
\label{6.83}
\ll \bt_k \gg \; = \; 0 \; , \qquad
\ll \bt^*_k \bt_p \gg \; = \; \tilde R_{kp} \; ,
\ee
with
\be
\label{6.84}
\tilde R_{kp} \equiv \frac{1}{N_L} \;
\sum_{ij} R_{ij} e^{i\bk\cdot\ba_i-i\bp\cdot\ba_j} \; .
\ee

Considering the random Hamiltonian (\ref{6.78}), we take into account that
$$
\frac{1}{\sqrt{N_L}} \; \sum_j \gm_j = \bt_0 \; , \qquad
\sum_j \gm_j c_j = \sum_k \bt_k^* a_k \; , \qquad
\sum_j \gm_j c_j^\dgr c_j  =
\frac{1}{\sqrt{N_L}} \; \sum_{kp} \bt_{k-p} a_k^\dgr a_p \; .
$$
Thence Eq. (\ref{6.78}) takes the form
\be
\label{6.85}
H_{ext} = \nu n_0 \bt_0\; \sqrt{N_L} \; + \;
\sqrt{\nu n_0}\; \sum_k \left (
a_k^\dgr \bt_k + \bt_k^* a_k \right ) \; + \;
\frac{1}{\sqrt{N_L}} \; \sum_{kp} \bt_{k-p} a_k^\dgr a_p \; .
\ee
Following the same procedure as in Sec. 6.2, we can employ the stochastic
mean-field approximation, simplifying the last term in Eq. (\ref{6.85}) as
\be
\label{6.86}
a_k^\dgr a_p \bt_{k-p} = \left ( a_k^\dgr \al_p + \al_k^* a_p -
\al_k^* \al_p \right ) \bt_{k-p} \; ,
\ee
which is analogous to Eq. (\ref{6.39}), and where
\be
\label{6.87}
\al_k \; \equiv \; < a_k>_H \; , \qquad
\ll \al_k \gg \; = \; < a_k>\; = \; 0 \; .
\ee
For the third- and fourth-order terms, with respect to the operators $a_k$,
the HFB approximation can be used.

Minimizing the grand potential (\ref{6.15}) over $N_0$ defines
\be
\label{6.88}
\mu_0 = - z_0 J + \nu U ( 1+ n_1 +\sgm) + \mu_G \; ,
\ee
where the additional term
\be
\label{6.89}
\mu_G \equiv \frac{1}{2\sqrt{\nu n_0}\; N_L} \; \sum_k
\ll \al_k^* \bt_k + \bt_k^* \al_k \gg
\ee
is due to the presence of the random potential.

Similarly to Eq. (\ref{6.45}), we introduce a random variable $\vp_k$
satisfying the Fredholm equation
\be
\label{6.90}
\vp_k = \sqrt{\nu n_0}\; \bt_k + \frac{1}{\sqrt{N_L}} \;
\sum_p \al_p \bt_{k-p} \; .
\ee
With this variable, Hamiltonian (\ref{6.85}) transforms into
\be
\label{6.91}
H_{ext} = \sqrt{N_0} \; \vp_0 \; + \; \sum_k \left ( a_k^\dgr \vp_k
+ \vp_k^* a_k - \al_k^* \vp_k\right ) \; .
\ee
Owing to properties (\ref{6.80}), the random variable $\vp_k$ satisfies
the conditions
\be
\label{6.92}
\vp_k^* = \vp_{-k} \; , \qquad \vp_0^* = \vp_0 \; .
\ee
The variable $\vp_0$, according to Eq. (\ref{6.90}), obeys the equation
\be
\label{6.93}
\vp_0 =\sqrt{\nu n_0}\; \bt_0 \; + \;
\frac{1}{\sqrt{N_L}} \; \sum_p \al_p^* \bt_p \; .
\ee
Using the latter reduces Eq. (\ref{6.89}) to the form
\be
\label{6.94}
\mu_G = \frac{\ll\vp_0\gg}{\sqrt{N_0}} \; .
\ee
Since
\be
\label{6.95}
\ll \vp_k \gg \; = \; 0 \qquad ( \ll \al_k \gg\; = \; 0 ) \; ,
\ee
one gets
\be
\label{6.96}
\mu_G = 0 \; .
\ee

Keeping in mind a cubic lattice, for which
$$
\sum_{<ij>} c_i^\dgr c_j =  2 \sum_k \sum_\al \cos(k_\al a) a_k^\dgr a_k \; ,
$$
let us define the quantities
\be
\label{6.97}
\om_k \equiv - 2J \sum_\al \cos(k_\al a) \; + \; 2\nu U - \mu_1
\ee
and
\be
\label{6.98}
\Dlt \equiv \nu ( n_0 + \sgm ) U \; .
\ee
Then the grand Hamiltonian (\ref{6.76}) transforms into
\be
\label{6.99}
H = E_{HFB} + \sum_k \om_k a_k^\dgr a_k \; + \;
\frac{1}{2}\; \sum_k \Dlt \left ( a_k^\dgr a_{-k}^\dgr +
a_{-k} a_k \right ) \; + \; H_{ext} \; ,
\ee
where the random part is given by Eq. (\ref{6.91}) and the first term is
$$
E_{HFB} = - z_0 J n_0 N  + \frac{U}{2} \; \nu n_0^2 N -
\mu_0 n_0 N - \; \frac{U}{2} \; \left ( 2n_1^2 + \sgm^2
\right ) N \; ,
$$
in which
$$
n_1  = \frac{1}{N} \; \sum_k n_k \; , \qquad
\sgm = \frac{1}{N} \; \sum_k \sgm_k \; .
$$

To diagonalize Hamiltonian (\ref{6.99}), we employ the nonuniform nonlinear
transformation (\ref{6.40}), for which we find
$$
u_k^2  =\frac{\om_k+\ep_k}{2\ep_k} \; , \qquad
v_k^2  = \frac{\om_k-\ep_k}{2\ep_k} \; , \qquad
w_k = - \; \frac{1}{\om_k+\Dlt} \; .
$$
Here $\om_k$ and $\Dlt$ are defined in Eqs. (\ref{6.97}) and (\ref{6.98})
and the Bogolubov spectrum is
\be
\label{6.100}
\ep_k =\sqrt{\om_k^2-\Dlt^2} \; .
\ee
As in Sec. 4.3, the Lagrange multiplier $\mu_1$ is defined by the condensation
condition (\ref{4.53}), which yields
\be
\label{6.101}
\mu_1 = - z_0 J + \nu U ( 1 + n_1 - \sgm ) \; .
\ee
Substituting the latter into Eq. (\ref{6.97}) gives
\be
\label{6.102}
\om_k = \Dlt + 4J \sum_\al \sin^2
\left ( \frac{k_\al a}{2} \right ) \; ,
\ee
where a cubic lattice is assumed.

The diagonalized form of Hamiltonian (\ref{6.99}) is
\be
\label{6.103}
H = E_B \; + \; \sum_k \ep_k b_k^\dgr b_k \; + \; H_\vp \; ,
\ee
in which $E_B$ is the same as in Eq. (\ref{4.51}), but Hamiltonian
(\ref{6.103}) differs from Eq. (\ref{4.50}) by the random term
\be
\label{6.104}
H_\vp = \vp_0\; \sqrt{N_0} \; - \; \sum_k \left ( \al_k^* +
\frac{\vp_k^*}{\om_k+\Dlt} \right ) \vp_k \; .
\ee
Since the quantum averaging with Hamiltonian (\ref{6.103}) gives
$$
< b_k>_H \; = \; < b_k b_p >_H \; = \; 0 \; ,
$$
from transformation (\ref{6.40}), we have
\be
\label{6.105}
\al_k = -\; \frac{\vp_k}{\om_k+\Dlt} \; .
\ee
Using this reduces the random part (\ref{6.104}) to the simple form
\be
\label{6.106}
H_\vp = \vp_0 \; \sqrt{N_0} \; .
\ee
We may notice that because of condition (\ref{6.95}),
\be
\label{6.107}
< H_\vp> \; = \; \ll H_\vp \gg \; = \; 0 \; .
\ee

For the normal and anomalous averages
$$
n_k \; \equiv \; < a_k^\dgr a_k > \; , \qquad
\sgm_k \; \equiv \; < a_k a_{-k} > \; ,
$$
we find
\be
\label{6.108}
n_k = n_k^N + \ll |\al_k|^2 \gg \; , \qquad
\sgm_k = \sgm_k^N + \ll |\al_k|^2 \gg \; ,
\ee
where
\be
\label{6.109}
n_k^N = \frac{\om_k}{2\ep_k} \; {\rm coth}
\left ( \frac{\ep_k}{2T} \right ) -\; \frac{1}{2} \; ,
\qquad \sgm_k^N = -\; \frac{\Dlt}{2\ep_k} \;
{\rm coth}\left ( \frac{\ep_k}{2T} \right ) \; .
\ee
According to Eq. (\ref{6.105}),
\be
\label{6.110}
\ll |\al_k|^2 \gg \; = \;
\frac{\ll|\vp_k|^2\gg}{(\om_k+\Dlt)^2} \; .
\ee
The fraction of uncondensed atoms is the sum
\be
\label{6.111}
n_1 = n_N + n_G \; ,
\ee
in which the first term
\be
\label{6.112}
n_N = \frac{1}{\rho} \; \int_{\cal B} \; n_k^N \;
\frac{d\bk}{(2\pi)^d}
\ee
is due to finite interactions and temperature, while the second term
\be
\label{6.113}
n_G = \frac{1}{\rho} \; \int_{\cal B} \;
\frac{\ll|\vp_k|^2\gg}{(\om_k+\Dlt)^2} \; \frac{d\bk}{(2\pi)^d}
\ee
is caused by the random potential.

The anomalous average (\ref{6.28}) becomes
\be
\label{6.114}
\sgm = \sgm_N + n_G \; ,
\ee
where
\be
\label{6.115}
\sgm_N = \frac{1}{\rho} \; \int_{\cal B} \; \sgm_k^N \;
\frac{d\bk}{(2\pi)^d} \; ,
\ee
and $n_G$ is given by Eq. (\ref{6.113}).

To find the superfluid fraction, we follow the discussion of Sec. 4.5. For
the dissipated heat (\ref{4.109}), we get
\be
\label{6.116}
Q = Q_N + Q_G \; ,
\ee
similarly to Eq. (\ref{6.58}). Keeping in mind a $d$-dimensional cubic
lattice and envoking the same approximations as is Sec. 4.5, we obtain
\be
\label{6.117}
Q_N = \frac{1}{8m\rho} \; \int_{\cal B} \;
\frac{q_k^2}{{\rm sinh}^2(\ep_k/2T)} \; \frac{d\bk}{(2\pi)^d}
\ee
and
\be
\label{6.118}
Q_G = \frac{1}{2m\rho} \; \int_{\cal B} \;
\frac{q_k^2\ll|\vp_k|^2\gg}{\ep_k(\om_k+\Dlt)} \; {\rm coth}
\left ( \frac{\ep_k}{2T} \right ) \; \frac{d\bk}{(2\pi)^d} \; .
\ee
Here
$$
q_k^2 \equiv \left ( \frac{m}{m^*} \right )^2 \;
\frac{1}{a^2} \; \sum_{\al=1}^d \sin^2( k_\al a) \; ,
$$
with $m^*\equiv1/2Ja^2$. In the long-wave limit, $q_k^2$ behaves as
$$
q_k^2 \simeq \left ( \frac{m}{m^*} \right )^2 \; k^2 \qquad
(k\ra 0 ) \; .
$$

To realize actual calculations, one has to solve Eq. (\ref{6.90}), which,
taking account of relation (\ref{6.105}), acquires the form
\be
\label{6.119}
\vp_k = \sqrt{\nu n_0}\; \bt_k \; - \; \frac{1}{\sqrt{N_L}} \; \sum_p
\frac{\bt_{k-p}\vp_p}{\om_p+\Dlt} \; .
\ee
The solution of Eq. (\ref{6.119}) can be accomplished following the way
of Ref. [367].

The superfluid fraction can be calculated using either definition
(\ref{4.108}) or Eq. (\ref{4.120}). The condensate fraction is defined by
the normalization condition $n_0+n_1=1$, which in view of Eq. (\ref{6.111}),
reads as
\be
\label{6.120}
n_0 + n_N + n_G = 1 \; .
\ee

The approximations, involved in the present section, describe well the
superfluid state of disordered optical lattices. However, with increasing
interactions and the strength of disorder, the system can undergo phase
transitions into an insulating or Bose-glass phases. The description of
the latter phases requires the use of other approximations or numerical
calculations.

\subsection{Phase Diagram}

The influence of disorder on the boson Hubbard model has been studied
for lattices of different dimensionalities and in several approaches.
Scaling arguments [378--381] and renormalization group techniques
[382--387] were used [306,388,389]. There have been employed numerical
calculations for small $(4\leq N\leq 25)$ systems [390], strong-coupling
expansions [298], density matrix renormalization group [391], mean-field
single-site approximation [392--394], and Monte Carlo simulations
[395--404].

The general physical properties of disordered boson lattices have been
described by Fisher et al. [306]. Depending on the system parameters,
several states can be realized. There exists the usual superfluid
coherent phase, where $n_s>0$ and $n_0>0$. This phase possesses a finite
compressibility and has no gap in the spectrum of collective excitations.
The Mott insulating phase can occur at zero temperature and integer
filling. This phase has zero compressibility and a gap for particle-hole
excitations. The superfluid-Mott insulator phase transition is a
second-order transition being in the universality class of the
$(d+1)$-dimensional $XY$ model. When an additional trapping potential
is imposed on the lattice, the phase transition point could be slightly
shifted [405].

Except these two phases that also exist in regular optical lattices,
for disordered lattices, there can arise a novel phase called the Bose
glass [306]. This third phase, due to the presence of disorder, is
insulating because of the localization effects of the randomness, it
is characterized by a finite compressibility and no gap in the spectrum.
The Bose glass exhibits no superfluidity $(n_s=0)$, but possesses local
remnants of BEC $(n_0>0)$. Such a Bose-glass phase can also develop in
a quasiperiodic bichromatic lattice [406].

At nonzero temperature and noninteger filling, there can be no Mott
insulator, but localized and delocalized atoms coexist. The itinerant
component can be either superfluid or normal. To realize pure phases,
one usually considers the zero-temperature and integer-filling case.
The lattice disorder is commonly introduced as a random term in the
Hubbard model (\ref{6.75}), with a site dependent, uniformly
distributed, on-site potential, $-D<\gm_j<D$, so that the related
stochastic distribution is
\begin{eqnarray}
p(\gm_j) =\left \{ \begin{array}{ll}
\frac{1}{2D} \; , & |\gm_j| < D \\
\\ \nonumber
0, & |\gm_j| > D \; .
\end{array} \right.
\end{eqnarray}

The most often discussed phase diagram concerns the case of zero
temperature and integer filling, when the pure Mott insulating phase
can exist. The disorder is introduced through a uniformly distributed
on-site potential, as is explained above. It is generally accepted that
when increasing disorder, the Mott insulating phase transforms, through
a first-order phase transition, into the Bose glass phase. While
increasing the tunneling parameter yields the superfluid phase with
BEC. Under sufficiently strong disorder, the Bose glass transforms
into superfluid through a second-order transition, provided the uniformly
distributed disorder is bounded by a finite $D$.

There exist, however, a controversy with regard to the transformation
of the Mott insulator into superfluid. Fisher et al. [306] argued that,
in the presence of any finite disorder, the transition from the Mott
insulator to superfluid occurs only through the intermediate Bose-glass
phase. This picture has been supported by several numerical calculations
[298,391,399,401]. The corresponding quantitative phase diagram is shown
in Fig. 1. Precise numerical values depend on the lattice dimensionality
and can be found in the cited references.

\begin{figure}[ht]
\center
\includegraphics[width=9cm]{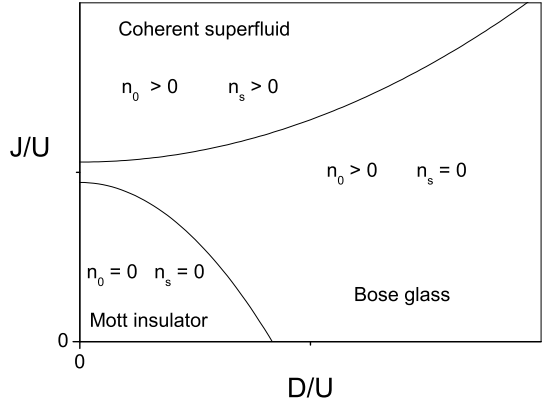}
\caption{Possible qualitative phase portrait for a disordered
lattice with an integer filling factor at zero temperature.}
\label{fig:Fig.1}
\end{figure}

Other researches claim that there are two different regimes in a
disordered boson Hubbard model. For weak disorder, the Mott insulating
phase is sustained up to the {\it direct} transition into a superfluid.
Strong disorder changes the nature of the transition to that of the
Bose glass to superfluid transition. Thus, contrary to the above cited
works, it is stated that at weak disorder the direct Mott insulator to
superfluid phase transition does occur, without an intervening Bose glass
phase. This picture is also based on several numerical investigations
[389,393,396,398,400,402,403]. The related qualitative phase diagram
is presented in Fig. 2.

\begin{figure}[ht]
\center
\includegraphics[width=9cm]{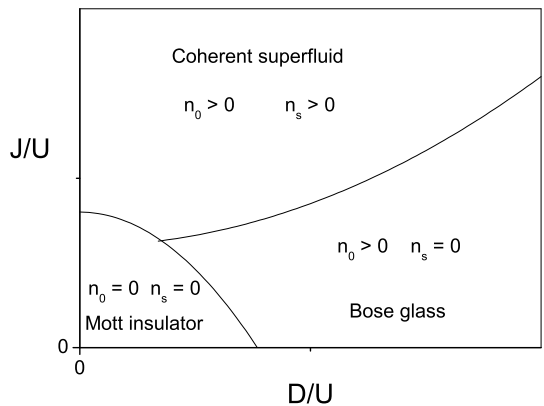}
\caption{Possible qualitative phase diagram for a disordered lattice
with an integer filling factor at zero temperature.}
\label{fig:Fig.2}
\end{figure}

\section{Nonstandard Lattice Models}

\subsection{Coexisting States}

For noncommenssurate filling, a mixture of localized and delocalized
atoms exists in the lattice [407] at all temperatures, including zero.
Such a mixture of coexisting localized and delocalized states occurs at
finite temperatures for any filling, including integer. Finally, there
are indications [408] that, even at zero temperature and integer filling,
close to the boundary between the superfluid and Mott insulating phases,
there can arise an itinerant-localized dual structure, where the localized
and itinerant states coexist.

The coexistence of two different states could be described
phenomenologically, involving the Ginzburg-Landau functional [409]
generalized to the mixture of several states [54]. But, of course, it
is more important to have a microscopic model characterizing such a
coexistence of delocalized and localized atoms.

Let us assume that a superfluid state coexists with a portion of
localized atoms. Then, in addition to the Bogolubov shifted field
operator $\eta(\br)+\psi_1(\br)$, there exists the field operator
$\psi_2(\br)$, so that $\eta(\br)$ is the condensate wave function,
$\psi_1(\br)$ is the field operator of uncondensed delocalized atoms,
and $\psi_2(\br)$ is the field operator of uncondensed localized atoms.
Each of these operators can be expanded over Wannier functions,
$$
\eta(\br) = \sqrt{\nu n_0} \; \sum_j w_0(\br-\ba_j) \; ,
\qquad \psi_1(\br) =  \sum_j c_j w(\br-\ba_j) \; ,
$$
\be
\label{7.1}
\psi_2(\br) =  \sum_j b_j w(\br-\ba_j) \; .
\ee
The field operators of delocalized and localized atoms commute with
each other,
\be
\label{7.2}
[\psi_1(\br),\psi_2(\br')] =
[\psi_1(\br),\psi_2^\dgr(\br')] = 0 \; ,\qquad
[c_i,b_j] = [c_i,b_j^\dgr] = 0 \; .
\ee
The Bose commutation relations are valid for each type of the field
operators.

Thus, the starting point is the assumption of the existence of two
types of atoms differing from each other by their localization property.
The delocalized atoms are characterized by an operator $c_j$, while the
localized atoms, by $b_j$. This is somewhat close to the consideration
of a two-component system [288]. Another analogy is the picture of a
crystal where a portion of atoms are localized, while a part of atoms
can jump between lattice sites [220,410]. Probably, the most direct
interpretation of the existence of two types of atoms is the
consideration of the multiband Hamiltonian (\ref{4.11}), in which two
bands are taken into account. One band is the {\it conducting} band,
whose atoms are itinerant, being characterized by the field operator
$\hat c_j$. Another band is that of {\it bound} states, with localized
atoms described by an operator $b_j$.

The localized atoms of bound states are considered as normal, such that
\be
\label{7.3}
< b_j> \; = \; 0 \; .
\ee
This requires [57] that the related Hamiltonian be invariant under the
global gauge transformation $b_j\ra b_je^{i\al}$, with a real $\al$. So,
the two-band Hubabrd Hamiltonian can be written down as
\be
\label{7.4}
\hat H = -\sum_{<ij>} \left ( J \hat c_i^\dgr \hat c_j +
J_2 b_i^\dgr b_j \right ) \; + \; \sum_j \left (
\frac{U}{2} \; \hat c_j^\dgr \hat c_j^\dgr \hat c_j \hat c_j +
U_1 \hat c_j^\dgr \hat c_j  b_j^\dgr b_j +
\frac{U_2}{2} \; b_j^\dgr b_j^\dgr b_j b_j \right ) \; .
\ee

A correct introduction of different degrees of freedom has to be such
that to exclude the double counting [411]. This requires that the field
operators of different types be orthogonal to each other,
\be
\label{7.5}
\int \eta^*(\br) \psi_1(\br) \; d\br = 0 \; , \qquad
\int \eta^*(\br) \psi_2(\br) \; d\br = 0 \; , \qquad
\int \psi_1^\dgr(\br) \psi_2(\br) \; d\br = 0 \; .
\ee
With expansions (\ref{7.1}), this yields
$$
\sum_j c_j = 0 \; , \qquad \sum_j b_j = 0 \; , \qquad
\sum_j c_j^\dgr b_j = 0 \; ,
$$
\be
\label{7.6}
\sum_j \hat c_j^\dgr b_j = 0 \; , \qquad
\left ( \hat c_j \equiv \sqrt{\nu n_0} + c_j \right ) \; .
\ee

To distinguish the bound states of localized atoms from the
conducting-band states of nonlocalized atoms, it is necessary to
impose the {\it localization condition}
\be
\label{7.7}
b_i^\dgr b_j = \dlt_{ij} b_j^\dgr b_j \; .
\ee
This removes the second tunneling term in Eq. (\ref{7.4}). In other
respects, the atoms can be treated as similar, possessing the same
interaction parameter $U=U_1=U_2$.

Then the two-band Hamiltonian (\ref{7.4}), describing the conducting
and bound-state bands, takes the form
\be
\label{7.8}
\hat H = - J \sum_{<ij>}
\hat c_i^\dgr \hat c_j \; + \; \frac{U}{2} \;
\sum_j \left ( \hat c_j^\dgr \hat c_j^\dgr \hat c_j \hat c_j +
2 \hat c_j^\dgr \hat c_j  b_j^\dgr b_j +
 b_j^\dgr b_j^\dgr b_j b_j \right ) \; .
\ee
For the conducting-band operators, the Bogolubov shift (\ref{6.13})
is assumed in order to take into account the possible appearance of BEC.

The delocalized atoms can include a Bose-condensed fraction and a part
of uncondensed atoms, while all localized atoms, by their definition,
are not condensed. The operators of uncondensed atoms are
\be
\label{7.9}
\hat N_1 = \sum_j c_j^\dgr c_j \; , \qquad
\hat N_2 =\sum_j b_j^\dgr b_j \; .
\ee
For the corresponding numbers of particles, we have
$$
N_0 = \nu n_0 N_L \; , \qquad N_1 = \sum_j < c_j^\dgr c_j> \; ,
$$
\be
\label{7.10}
N_2 = \sum_j < b_j^\dgr b_j > \; , \qquad
N_0 + N_1 =  \sum_j < \hat c_j^\dgr \hat c_j > \; .
\ee
The related atomic fractions are defined as
\be
\label{7.11}
n_0 \equiv \frac{N_0}{N} \; , \qquad
n_1 \equiv \frac{N_1}{N} \; , \qquad
n_2 \equiv \frac{N_2}{N} \; ,
\ee
with the normalization condition
\be
\label{7.12}
n_0 + n_1 + n_2 = 1 \; .
\ee
For an ideal lattice, Eqs. (\ref{7.10}) give
\be
\label{7.13}
< c_j^\dgr c_j > \; = \; \nu n_1 \; , \qquad
< b_j^\dgr b_j > \; = \; \nu n_2 \; , \qquad
< \hat c_j^\dgr \hat c_j > \; = \; \nu ( n_0 + n_1 ) \; .
\ee
Anomalous averages can exist for delocalized atoms, while the former
are absent for localized bound-state atoms,
\be
\label{7.14}
< c_j c_j> \; \equiv \; \nu\sgm \; , \qquad
< b_j b_j > \; \equiv \; 0 \; .
\ee

The grand Hamiltonian for the system is
\be
\label{7.15}
H = \hat H - \mu_0 N_0 - \mu_1 \hat N_1  - \mu_2 \hat N_2 \; .
\ee
The grand potential and free energy are
\be
\label{7.16}
\Om = - T \ln\; {\rm Tr} \; e^{-\bt H} \; , \qquad
F = \Om + \mu N \; ,
\ee
where $\mu$ is the system chemical potential. The latter is defined as
$$
\mu = \frac{\prt F}{\prt N} =
\frac{\prt F}{\prt N_0} \; \frac{\prt N_0}{\prt N} +
\frac{\prt F}{\prt N_1} \; \frac{\prt N_1}{\prt N} +
\frac{\prt F}{\prt N_2} \; \frac{\prt N_2}{\prt N} \; .
$$
Under relations
$$
\frac{\prt F}{\prt N_0}  = \mu_0 \; , \qquad
\frac{\prt F}{\prt N_1}  = \mu_1 \; , \qquad
\frac{\prt F}{\prt N_2}  = \mu_2 \; ,
$$
$$
\frac{\prt N_0}{\prt N}  = n_0 \; , \qquad
\frac{\prt N_1}{\prt N}  = n_1 \; , \qquad
\frac{\prt N_2}{\prt N}  = n_2 \; ,
$$
we get
\be
\label{7.17}
\mu = \mu_0 n_0 + \mu_1 n_1 + \mu_2 n_2 \; .
\ee

The Lagrange multiplier $\mu_0$ is defined by the minimization of the
grand thermodynamical potential $\Om$ with respect to the number of
condensed atoms $N_0$, under the fixed numbers of atoms $N_1$ and $N_2$.
The multiplier $\mu_1$ is defined from the condition of the BEC existence,
as is explained in Sec. 2.9, and which is equivalent to the condition of
the gapless spectrum. But what defines $\mu_2$? The latter could be found
if the number $N_2$ of localized atoms would be fixed. This, however, is
not the case, since only the total number $N$ can be fixed.

The Lagrange multiplier $\mu_2$ can be found from the condition that the
system is in stable equilibrium, when $\dlt F=0$, with the variation over
the numbers of atoms, so that
\be
\label{7.18}
\frac{\prt F}{\prt N_0} \; \dlt N_0 +
\frac{\prt F}{\prt N_1} \; \dlt N_1 +
\frac{\prt F}{\prt N_2} \; \dlt N_2 = 0\; .
\ee
Using the relation $N_2=N-N_0-N_1$, one has
$$
\dlt N_2 = -\dlt N_0 - \dlt N_1 \; .
$$
And Eq. (\ref{7.18}) gives
$$
(\mu_0 - \mu_2) \dlt N_0 + (\mu_1 - \mu_2) \dlt N_1 = 0 \; ,
$$
or, equivalently,
$$
[ (\mu_0 - \mu_2) n_0 + (\mu_1 - \mu_2) n_1 ] \dlt N = 0 \; .
$$
This is valid for an arbitrary variation of $N$, hence
$$
(\mu_0 - \mu_2) n_0 + (\mu_1 - \mu_2) n_1 = 0 \; .
$$
From here
\be
\label{7.19}
\mu_2 = \frac{\mu_0 n_0 + \mu_1 n_1}{n_0 + n_1} \; .
\ee
Substituting the latter into Eq. (\ref{7.17}) yields
\be
\label{7.20}
\mu = \mu_2  =\frac{\mu_0 n_0 + \mu_1 n_1}{n_0 + n_1} \; .
\ee
When there exist only delocalized atoms, hence $N_2=0$ and $n_0+n_1=1$,
Eq. (\ref{7.20}) acquires the standard form (\ref{2.116}).

Hamiltonian (\ref{7.8}) consists of the parts corresponding to the atoms
of the conducting band, the atoms of the bound-state band, and contains
the term
\be
\label{7.21}
\hat H_{int} \equiv U \sum_j \hat c_j^\dgr \hat c_j b_j^\dgr b_j
\ee
associated with the interband atomic interaction. With the operator
$\hat c_j$, given in Eq. (\ref{7.6}), we have
$$
\hat c_j^\dgr \hat c_j b_j^\dgr b_j  = \nu n_0 b_j^\dgr b_j +
\sqrt{\nu n_0} \; \left ( c_j^\dgr + c_j\right ) b_j^\dgr b_j +
c_j^\dgr c_j b_j^\dgr b_j \; .
$$
The operators of the different types can be decoupled as follows:
$$
c_j^\dgr b_j^\dgr b_j \; = \; < c_j^\dgr> b_j^\dgr b_j +
c_j^\dgr < b_j^\dgr b_j> - < c_j^\dgr><b_j^\dgr b_j> \; ,
$$
\be
\label{7.22}
c_j^\dgr c_j b_j^\dgr b_j = c_j^\dgr c_j < b_j^\dgr b_j> +
< c_j^\dgr c_j> b_j^\dgr b_j - < c_j^\dgr c_j><b_j^\dgr b_j> \; .
\ee
Since $<c_j>=0$ and because of conditions (\ref{7.6}) one gets
$$
\sum_j c_j^\dgr b_j^\dgr b_j = \nu n_2 \sum_j c_j^\dgr = 0 \; .
$$
The interaction term (\ref{7.21}) transforms into
\be
\label{7.23}
\hat H_{int} = \nu U \sum_j \left [ n_2 c_j^\dgr c_j +
(n_0 +n_1) b_j^\dgr b_j \right ] \; -  \nu n_1 n_2 U N \; .
\ee

The grand Hamiltonian (\ref{7.15}) reduces to the sum
\be
\label{7.24}
H = H_{del} + H_{loc} - \nu U n_1 n_2 N \; ,
\ee
in which the first term describes delocalized atoms, while the second,
localized atoms. The Hamiltonian of delocalized atoms reads as
\be
\label{7.25}
H_{del} = \sum_{n=0}^4 H^{(n)} \; .
\ee
The zero- and the second-order terms are
$$
H^{(0)} = \left ( - z_0 J + \frac{U}{2} \; \nu n_0 -
\mu_0 \right ) n_0 N \; ,
$$
\be
\label{7.26}
H^{(2)} = - J \sum_{<ij>} c_i^\dgr c_j + ( 2\nu U n_0 +
\nu U n_2 - \mu_1) \sum_j c_j^\dgr c_j \; + \;
\frac{U}{2} \; \nu n_0 \sum_j \left (
c_j^\dgr c_j^\dgr + c_j c_j \right ) \; .
\ee
The first-order term is zero, because of conditions (\ref{7.6}), that is,
$H^{(1)}=0$. The third- and fourth-order terms are
\be
\label{7.27}
H^{(3)} = \sqrt{\nu n_0} \; U \sum_j \left ( c_j^\dgr c_j^\dgr c_j +
c_j^\dgr c_j c_j \right )
\ee
and, respectively,
\be
\label{7.28}
H^{(4)} = \frac{U}{2} \; \sum_j c_j^\dgr c_j^\dgr c_j c_j \; .
\ee
The term in Eq. (\ref{7.24}), describing the localized atoms, can be
written as
\be
\label{7.29}
H_{loc} = \sum_j H_j \; ,
\ee
where
\be
\label{7.30}
H_j = \frac{U}{2} \; b_j^\dgr b_j \left ( b_j^\dgr b_j -1\right ) +
[ \nu U ( n_0 + n_1 ) - \mu b_j^\dgr b_j \; .
\ee

The Hamiltonian (\ref{7.25}) of delocalized atoms can be treated in the
HFB approximation, as is done in Secs. 4.2 and 4.3. Following the same
procedure, and minimizing the grand potential over $N_0$, we get
\be
\label{7.31}
\mu_0 = - z_0 J + \nu U ( n_0 + 2n_1 + \sgm ) \; .
\ee

Introducing the notation
\be
\label{7.32}
\om_k \equiv - 2J \sum_\al \cos(k_\al a) \; +
\nu U ( 1 + n_0 + n_1 ) - \mu_1 \; , \qquad
\Dlt \equiv \nu U ( n_0 + \sgm) \; ,
\ee
we keep in mind a cubic lattice and use the relation
$$
2 ( n_0 + n_1 ) + n_2 =  1 + n_0 + n_1 \; .
$$
The condition (\ref{4.53}) of condensate existence can be represented as
\be
\label{7.33}
\lim_{k\ra 0} ( \om_k - \Dlt) = 0 \; .
\ee
The latter yields
\be
\label{7.34}
\mu_1 = - z_0 J + \nu U ( 1 + n_1  -\sgm ) \; .
\ee

In this way, we obtain the same forms of the expressions as in Eq.
(\ref{4.55}), of the spectrum (\ref{4.52}), and of Eq. (\ref{4.59}) and
(\ref{4.60}) defining the momentum distribution $n_k$ and the anomalous
average $\sgm_k$, respectively.

The subsystem of localized atoms, with Hamiltonian (\ref{7.29}) can be
considered as in Sec. 4.7. The eigenvalues of the single-site Hamiltonian
(\ref{7.30}) are
\be
\label{7.35}
e_n = \frac{U}{2}\; n^2 \left [ \nu U ( n_0 + n_1 )\; - \;
\frac{U}{2} \; - \; \mu \right ] n \; ,
\ee
where $n=0,1,2,\ldots$ Minimizing $e_n$ with respect to $n$ gives the
effective number
\be
\label{7.36}
n_{eff} = \frac{2\mu+U[1-2\nu(n_0+n_1)]}{2U} \; .
\ee
Since $<b_j^\dgr b_j>=\nu n_2$, for an ideal lattice, we have
\be
\label{7.37}
\nu n_2 =
\frac{{\rm Tr}\;\hat n_j e^{-\bt H_j}}{{\rm Tr}e^{-\bt H_j}} \; .
\ee
At low temperature, this gives
\be
\label{7.38}
\nu n_2 \simeq n_{eff} \qquad (T \ll U ) \; .
\ee
Hence,
\be
\label{7.39}
n_2 \simeq \frac{2\mu+U[1-2\nu(n_0+n_1)]}{2\nu U} \; .
\ee
In view of the normalization condition (\ref{7.12}), the chemical
potential, following from Eq. (\ref{7.39}), becomes
\be
\label{7.40}
\mu \simeq (2\nu -1)\; \frac{U}{2} \qquad (T \ll U) \; .
\ee
At the same time, from Eqs. (\ref{7.20}), (\ref{7.31}), and (\ref{7.34}),
we have
\be
\label{7.41}
\mu =  -z_0 J + \nu U \left [ n_0 + n_1 + \sgm +
\frac{(1-2\sgm)n_1}{n_0+n_1} \right ] \; .
\ee
Expressions (\ref{7.40}) and (\ref{7.41}) give the equality
\be
\label{7.42}
\left [ \left ( \nu \; - \; \frac{1}{2}\right ) U + z_0 J
\right ] (1 - n_2) = \nu U \left [ ( 1 - n_2 +\sgm) (1 - n_2) +
n_1 (1 -2\sgm) \right ]
\ee
connecting $n_2$ with $n_1$ and $\sgm$. The latter are defined in Eqs.
(\ref{4.61}) and (\ref{4.62}), respectively.

The single-particle spectrum (\ref{7.35}), with the chemical potential
(\ref{7.40}), reads as
\be
\label{7.43}
e_n = nU \left ( \frac{n}{2}\; - \; \nu n_2 \right ) \; .
\ee
The effective energy level is
\be
\label{7.44}
e_{eff} \equiv \lim_{n\ra n_{eff}} e_n = -\;
\frac{U}{2} \; ( \nu n_2)^2 \; .
\ee

When delocalized atoms coexist with localized ones, there are two
types of spectra. The spectrum of collective excitations, caused by
the delocalized atoms, is given by the Bogolubov form (\ref{4.52}). The
spectrum is gapless, displaying in the long-wave limit the asymptotic
behavior $\ep_k\simeq ck$, as in Eq. (\ref{4.57}), with the sound velocity
\be
\label{7.45}
c =\sqrt{2 J a^2 \Dlt} = \sqrt{\frac{\Dlt}{m^*} } \; ,
\ee
according to Eq. (\ref{4.58}) and the notation $m^*=1/2Ja^2$ for the
effective mass. At the same time, there exists the single-particle
spectrum (\ref{7.43}) possessing the gap $\Dlt e_n=U$, which is defined
as in Eq. (\ref{4.156}). Both these spectra can be experimentally observed.
The spectrum of collective excitations describes density fluctuations due
to delocalized atoms in the conducting band. And the single-particle
spectrum describes discrete energy levels of localized atoms in the
bound-state band. The energy gap in the single-particle spectrum
characterizes the quantity of energy that is necessary for transferring
an atom from the bound-state band to the conducting band.

\subsection{Vibrational Excitations}

In the extended Hubbard model (\ref{5.16}), one considers
atomic interactions between different lattice sites. Such intersite
interactions can become important when atoms interact through long-range
forces, for instance, through dipolar interactions [315,316,412]. Then
one should consider the extended Hubbard model, which has been treated
in a number of papers [317--322].

It is worth noting that for short-range interactions, the intersite
forces can also be rather strong, when the effective scattering length
becomes very large due to Feshnach resonance [10,24] or to geometric
resonances in waveguides [413,414]. But, anyway, the intersite
interactions are much smaller than the on-site interactions. Thus,
for a three-dimensional cubic lattice in the tight-binding approximation,
as follows from Secs. 3.5. and 3.9, we have the on-site interaction
$$
U = \frac{\Phi_0}{(2\pi)^{3/2}l_0^3} \; ,
$$
with the localization length
$$
l_0 = \frac{1}{\sqrt{m\om_0} } = \frac{a}{\pi} \left (
\frac{E_R}{V_0} \right )^{1/4} \; .
$$
While for the nearest-neighbor interactions, we get
$$
U_{ij} =  U \exp\left ( - \; \frac{3a^2}{2l_0^2} \right ) \; .
$$
For well-localized atoms, for which $l_0\ll a$, one has
$|U_{ij}/U|\ll 1$. So that the intersite interactions are negligible
as compared to the on-site interactions. These short-range interactions
could be comparable only for very shallow lattices, for which, however,
the Hubbard model as such would be not a good approximation to reality.
But for long-range interactions, the extended Hubbard model is well
justified.

As soon as there are atomic interactions between different sites,
there appear collective vibrational excitations, that is, phonons.
Their introduction into the extended Hubbard model can be done similarly
to the quantization of collective coordinates for extended quantum systems
[415,416].

In the Hubbard model (\ref{5.16}), there is the single-site term
containing the quantity $h_j$, defined in Eq. (\ref{4.8}), which for
a single-band model becomes
\be
\label{7.46}
h_j \equiv \int w^*(\br-\ba_j) \left [ -\; \frac{\nabla^2}{2m} +
V_L(\br) \right ] w(\br-\ba_j) \; d\br \; .
\ee
Let us introduce the notation
\be
\label{7.47}
\bp_j^2 \equiv \int w^*(\br-\ba_j) (-\nabla^2)
w(\br-\ba_j) \; d\br \; .
\ee
Then Eq. (\ref{7.46}) writes as
\be
\label{7.48}
h_j = \frac{\bp_j^2}{2m} \; + \; \int w^*(\br) V_L(\br)
w(\br) \; d\br \; .
\ee
The second term in Eq. (\ref{7.48}) is a constant and can be omitted.
The extended Hubbard model takes the form
\be
\label{7.49}
\hat H = - J \sum_{<ij>} c_i^\dgr c_j \; + \;
 \sum_j \; \frac{\bp_j^2}{2m} \; c_j^\dgr c_j \; + \;
\frac{U}{2} \; \sum_j c_j^\dgr c_j^\dgr c_j c_j \; + \;
\frac{1}{2} \; \sum_{i\neq j} U_{ij} c_i^\dgr c_j^\dgr c_j c_i \; .
\ee

Accomplishing the quantization of collective variables assumes that the
quantities $\bp_j$ and $U_{ij}$ become operators, so that
\be
\label{7.50}
\bp_j \; \ra \; \hat\bp_j \; , \qquad
U_{ij} \; \ra \; \hat U_{ij} \; ,
\ee
where
\be
\label{7.51}
\hat U_{ij} \equiv U(\hat\br_{ij}) \qquad
(\hat\br_{ij} \equiv \hat\br_i - \hat\br_j ) \; .
\ee
The coordinate operator
\be
\label{7.52}
\hat\br_j = \ba_j + \hat{\bf u}_j
\ee
is such that
\be
\label{7.53}
< \hat\br_j > \; = \; \ba_j \; ,
\ee
hence
\be
\label{7.54}
< \hat{\bf u}_j > \; = \; 0\; .
\ee
The operator $\hat{\bf u}_j$ describes the deviation from the
average equilibrium position $\ba_j$. It is assumed that $\hat{\bf u}_j=
\{\hat u^\al_j\}$ characterizes small fluctuations around the lattice site
$\ba_j$.

The operator function (\ref{7.51}) can be expanded in powers of $u_j^\al$.
Such a second-order expansion reads as
\be
\label{7.55}
U(\hat\br_{ij}) \simeq U_{ij} + \sum_\al \Phi_{ij}^\al
\left ( \hat u_i^\al - \hat u_j^\al\right ) \; + \;
\sum_{\al\bt} \Phi_{ij}^{\al\bt} \left ( \hat u_i^\al \hat u_j^\bt -\;
\frac{1}{2}\; \hat u_i^\al \hat u_i^\bt - \; \frac{1}{2}\;
\hat u_j^\al \hat u_j^\bt \right ) \; ,
\ee
where
$$
U_{ij} \equiv U(\ba_{ij}) \qquad
(\ba_{ij} \equiv \ba_i - \ba_j) \; ,
$$
\be
\label{7.56}
\Phi_{ij}^\al \equiv \frac{\prt U_{ij}}{\prt a_i^\al} \; ,
\qquad \Phi_{ij}^{\al\bt} \equiv
\frac{\prt^2 U_{ij}}{\prt a_i^\al\prt a_j^\bt} \; .
\ee
Atomic interactions, as usual, are symmetric with respect to spatial
inversion,
\be
\label{7.57}
U(\ba_{ji}) = U(\ba_{ij}) \; , \qquad
U_{ji} = U_{ij} \; .
\ee
From here, it follows that
\be
\label{7.58}
\Phi_{ij}^\al = - \Phi_{ji}^\al \; , \qquad
\Phi_{ij}^{\al\bt} = \Phi_{ij}^{\bt\al} =
\Phi_{ji}^{\bt\al} = \Phi_{ji}^{\al\bt} \; .
\ee

For an ideal lattice, the sum
$$
\sum_{j(\neq i)} U_{ij} = \sum_{j(\neq i)} U(\ba_{ij})
$$
does not depend on $\ba_i$. Therefore
\be
\label{7.59}
\sum_{j(\neq i)} \Phi_{ij}^\al = \frac{\prt}{\prt a_i^\al} \;
\sum_{j(\neq i)} U_{ij} = 0 \; , \qquad
\sum_{j(\neq i)} \Phi_{ij}^{\al\bt} = -\;
\frac{\prt^2}{\prt a_i^\al\prt a_i^\bt} \;
\sum_{j(\neq i)} U_{ij} = 0 \; ,
\ee
where the property
$$
\frac{\prt U(\ba_{ij})}{\prt a_i^\al} = - \;
\frac{\prt U(\ba_{ij})}{\prt a_j^\al}
$$
is employed. Consequently, Eq. (\ref{7.55}) satisfies the relation
\be
\label{7.60}
\sum_{i\neq j} U(\hat\br_{ij})  =
\sum_{i\neq j} U_{ij} \; + \; \sum_{i\neq j} \;
\sum_{\al\bt} \Phi_{ij}^{\al\bt} \hat u_i^\al \hat u_j^\bt \; .
\ee
Also, because of condition (\ref{7.54}), one has
\be
\label{7.61}
< \hat U_{ij} > \; = \; U_{ij} \; + \;
\sum_{\al\bt} \Phi_{ij}^{\al\bt}
\left ( < \hat u_i^\al \hat u_j^\bt> -
< \hat u_j^\al \hat u_j^\bt> \right ) \; .
\ee

To simplify the problem, let us resort to the mean-field decoupling
\be
\label{7.62}
< c_i^\dgr c_i c_j^\dgr c_j> \; = \;
< c_i^\dgr c_i> < c_j^\dgr c_j> \qquad (i\neq j)
\ee
for the atomic operators. And let us decouple the atomic and phonon
variables as follows:
$$
\hat\bp_j^2 c_j^\dgr c_j = \hat\bp_j^2 < c_j^\dgr c_j > +
<\hat\bp_j^2> c_j^\dgr c_j -
< \hat\bp_j^2> < c_j^\dgr c_j > \; ,
$$
\be
\label{7.63}
\hat U_{ij} c_i^\dgr c_j^\dgr c_j c_i =
\hat U_{ij} < c_i^\dgr c_j^\dgr c_j c_i > +
< \hat U_{ij}> c_i^\dgr c_j^\dgr c_j c_i -
< \hat U_{ij}> < c_i^\dgr c_j^\dgr c_j c_i > \; .
\ee
Also, let us recall the notation for the filling factor
\be
\label{7.64}
\nu \equiv \frac{N}{N_L} = \frac{1}{N_L} \;
\sum_j < c_j^\dgr c_j> \; = \; < c_j^\dgr c_j> \; ,
\ee
where again an ideal lattice is assumed.

Accomplishing quantization (\ref{7.50}) in Hamiltonian (\ref{7.49}) and
decoupling the atomic and phonon variables according to Eq. (\ref{7.63}),
we come to the Hamiltonian
\be
\label{7.65}
\hat H = E_{sh} + \hat H_{at} + \hat H_{ph}
\ee
with the separated variables. Here the first term shifts Hamiltonian
(\ref{7.65}) by the nonoperator quantity
\be
\label{7.66}
E_{sh} = -\nu \sum_j \; \frac{<\hat\bp_j^2>}{2m} \; - \;
\frac{\nu^2}{2} \; \sum_{i\neq j} <\hat U_{ij} > \; .
\ee
The second term in Eq. (\ref{7.65}) is an effective atomic Hamiltonian
\be
\label{7.67}
\hat H_{at} = - J\sum_{<ij>} c_i^\dgr c_j \; + \;
\sum_j \; \frac{<\hat\bp_j^2>}{2m} \; c_j^\dgr c_j \; + \;
\frac{U}{2} \; \sum_j c_j^\dgr c_j^\dgr c_j c_j \; + \;
\frac{1}{2} \; \sum_{i\neq j} <\hat U_{ij}>
c_i^\dgr c_j^\dgr c_j c_i \; .
\ee
And the last term in Eq. (\ref{7.65}) is an effective phonon Hamiltonian
\be
\label{7.68}
\hat H_{ph} =\nu \sum_j \; \frac{\hat\bp_j^2}{2m} \; + \;
\frac{\nu^2}{2} \; \sum_{i\neq j} \; \sum_{\al\bt}
\Phi_{ij}^{\al\bt} u_i^\al u_j^\bt \; .
\ee

The quantity $\Phi_{ij}^{\al\bt}$ is called the dynamical matrix. The
eigenproblem
\be
\label{7.69}
\frac{\nu^3}{m} \; \sum_{j(\neq i)} \;
\sum_\bt \Phi_{ij}^{\al\bt}\; e^{i\bk\cdot\ba_{ij}}\; e_{ks}^\bt =
\om_{ks}^2 e_{ks}^\al
\ee
defines the phonon spectrum $\om_{ks}$ and the polarization vectors
${\bf e}_{ks}=\{ e_{ks}^\al\}$, in which $s=1,2,\ldots,d$ is a
polarization index. The phonon spectrum and polarization vectors can be
chosen to be symmetric with respect to the inversion of the wave vector
$\bk$, so that
$$
\om_{-ks} = \om_{ks}\; , \qquad {\bf e}_{-ks} = {\bf e}_{ks} \; .
$$
The polarization vectors enjoy the properties
$$
{\bf e}_{ks} \cdot {\bf e}_{ks'} = \dlt_{ss'} \; , \qquad
\sum_s e_{ks}^\al e_{ks}^\bt = \dlt_{\al\bt} \; ,
$$
meaning that they form a complete orthonormal basis.

The variables $\hat{\bf u}_j$ and $\hat\bp_j$ can be expanded over the
polarization-vector basis,
$$
\hat{\bf u}_j = \sum_{ks} \;
\frac{\nu{\bf e}_{ks}}{\sqrt{2mN\om_{ks}}} \;
\left ( b_{ks} + b^\dgr_{-ks}\right )
e^{i\bk\cdot\ba_j} \; ,
$$
\be
\label{7.70}
\hat\bp_j = - i \sum_{ks} \;
\sqrt{ \frac{m\om_{ks}}{2N} } \; {\bf e}_{ks}
\left ( b_{ks} - b^\dgr_{-ks}\right ) e^{i\bk\cdot\ba_j} \; ,
\ee
with the quasimomentum $\bk$ in the Brillouin zone. Here $b_{ks}$
and $b_{ks}^\dgr$ are the phonon field operators satisfying the Bose
commutation relations
$$
[ b_{ks}, b_{ps'} ] = 0 \; , \qquad
[ b_{ks}, b^\dgr_{ps'} ] = \dlt_{kp}\dlt_{ss'} \; .
$$
This guarantees that $\hat{\bf u}_j$ and $\hat\bp_j$ are mutually
conjugate variables and obey the commutation relations
$$
[ \hat u_i^\al, \hat u_j^\bt] = 0 \; , \qquad
[ \hat p_i^\al, \hat p_j^\bt] = 0 \; , \qquad
[ \hat u_i^\al, \hat p_j^\bt] = i\dlt_{\al\bt} \dlt_{ij} \; .
$$

With expansion (\ref{7.70}), the phonon Hamiltonian (\ref{7.68}) becomes
\be
\label{7.71}
\hat H_{ph} =\sum_{ks} \om_{ks}
\left ( b_{ks}^\dgr b_{ks} + \frac{1}{2}\right ) \; .
\ee
The phonon-spectrum equation follows from eigenproblem (\ref{7.69}),
which yields
\be
\label{7.72}
\om_{ks}^2 = \frac{\nu^3}{m} \; \sum_{j(\neq i)} \;
\sum_{\al\bt} \Phi_{ij}^{\al\bt} e_{ks}^\al e_{ks}^\bt \;
e^{i\bk\cdot\ba_{ij}} \; .
\ee
In the long-wave limit, this gives
\be
\label{7.73}
\om_{ks}^2 \simeq - \; \frac{\nu^3}{m} \;
\sum_{j(\neq i)} \; \sum_{\al\bt}
\Phi_{ij}^{\al\bt} e_{ks}^\al e_{ks}^\bt (\bk\cdot\ba_{ij} )^2 \; ,
\ee
where $k\ra 0$. We may note that
$$
\Phi_{ij}^{\al\bt} \equiv
\frac{\prt^2 U(\ba_{ij})}{\prt a_i^\al\prt a_j^\bt} = - \;
\frac{\prt^2 U(\ba_{ij})}{\prt a_i^\al\prt a_i^\bt} \; .
$$
Therefore $\om_{ks}^2\geq 0$, tending to zero in the limit $k\ra 0$. For
a $d$-dimensional lattice, the polarization index $s=1,2,\ldots,d$. Hence
Eq. (\ref{7.73}) defines $d$ phonon branches.

For Hamiltonian (\ref{7.71}), we have
\be
\label{7.74}
< b_{ks}^\dgr b_{ks} > \; = \; \left [ \exp
\left ( \frac{\om_{ks}}{T} \right ) - 1 \right ]^{-1} \; .
\ee
This yields
$$
< \hat\bp_j^2 > \; = \; \frac{m}{2N} \; \sum_{ks}
\om_{ks} {\rm coth} \left ( \frac{\om_{ks}}{2T} \right ) \; .
$$
Then the mean phonon kinetic energy per particle is
\be
\label{7.75}
K \; \equiv \; \frac{<\bp_j^2>}{2m} = \frac{1}{4N} \;
\sum_{ks} \om_{ks} {\rm coth}
\left ( \frac{\om_{ks}}{2T} \right ) \; .
\ee
For the correlation function of atomic displacements, we have
\be
\label{7.76}
< \hat u_i^\al \hat u_j^\bt > \; = \;
\frac{\dlt_{ij}\nu^2}{2N} \; \sum_{ks} \;
\frac{e_{ks}^\al e_{ks}^\bt}{m\om_{ks}} \; {\rm coth}
\left ( \frac{\om_{ks}}{2T} \right ) \; .
\ee
The average total phonon energy is
\be
\label{7.77}
< \hat H_{ph} > \; = \; 2NK = \frac{1}{2}\; \sum_{ks} \om_{ks}
{\rm coth} \left ( \frac{\om_{ks}}{2T} \right ) \; .
\ee

Denoting the interaction average (\ref{7.61}) as
\be
\label{7.78}
\tilde U_{ij} \; \equiv \; < \hat U_{ij} > \; = \;
U_{ij} + \Dlt U_{ij} \; ,
\ee
we have
\be
\label{7.79}
\Dlt U_{ij} = - \; \frac{\nu^2}{2N} \;
\sum_{ks} \; \sum_{\al\bt} \Phi_{ij}^{\al\bt} \;
\frac{e_{ks}^\al e_{ks}^\bt}{m\om_{ks}} \;
{\rm coth} \left ( \frac{\om_{ks}}{2T} \right ) \; .
\ee
The latter quantity shows how atomic interactions change in the presence
of phonon excitations.

The energy shift (\ref{7.66}), in view of properties (\ref{7.59}), becomes
\be
\label{7.80}
E_{sh} = - \left ( K  + \frac{\nu}{2}\; \Phi \right ) N \; ,
\ee
where the notation
$$
\Phi \; \equiv \; \frac{1}{N_L} \; \sum_{i\neq j} U_{ij}
$$
is employed.

\subsection{Phonon-Induced Interactions}

The atomic Hamiltonian (\ref{7.67}), with Eqs. (\ref{7.75}), (\ref{7.78}),
and (\ref{7.79}), takes the form
\be
\label{7.81}
\hat H_{at} = - J \sum_{<ij>} c_i^\dgr c_j \; + \;
K \sum_j c_j^\dgr c_j \; + \; \frac{U}{2} \;
\sum_j c_j^\dgr c_j^\dgr c_j c_j \; + \; \frac{1}{2} \;
\sum_{i\neq j} \tilde U_{ij} c_i^\dgr c_j^\dgr c_j c_i \; .
\ee
Comparing Eqs. (\ref{7.49}) and (\ref{7.81}), we see that phonon
excitations increase the total energy of atoms by the second term in
Eq. (\ref{7.81}), containing the kinetic phonon energy (\ref{7.75}). The
second term here is the energy of the vibrational atomic motion. In the
grand canonical ensemble, the energy $K$ can be incorporated into the
chemical potential.

Atomic vibrations renormalize the interaction potential of atoms. Now
the effective atomic interaction is shifted according to Eq. (\ref{7.78}).
In order to evaluate how substantial this renormalization is, let us
resort to the isotropic approximation, keeping in mind a cubic lattice,
when the phonon spectrum is the same for all polarizations, so that
$\om_{ks}$ can be replaced by the average $\om_k$, defined as
\be
\label{7.82}
\om_k^2 \; \equiv \; \frac{1}{d} \; \sum_{s=1}^d \om_{ks}^2 \; .
\ee
And let us introduce the {\it effective dynamical matrix}
\be
\label{7.83}
D_{ij} \; \equiv \; \frac{1}{d} \; \sum_{\al=1}^d \;
\frac{\prt^2 U(\ba_{ij})}{\prt a_i^\al\prt a_i^\al} = -\;
\frac{1}{d} \; \sum_{\al=1}^d \Phi_{ij}^{\al\al} \; .
\ee
Since $D_{ij}$ enters everywhere in the sums with $i\neq j$, for the
simplicity of notation, we can set
\be
\label{7.84}
D_{jj} \equiv 0 \; .
\ee
Recall that, due to the properties described in Eq. (\ref{7.59}), one
has
$$
\sum_j D_{ij} = 0 \; .
$$

From Eq. (\ref{7.72}), the phonon spectrum in the isotropic approximation
is given by
\be
\label{7.85}
\om_k^2 = - \; \frac{\nu^3}{m} \;
\sum_j D_{ij} e^{i\bk\cdot\ba_{ij} } \; .
\ee
In the long-wave limit, this yields
\be
\label{7.86}
\om_k^2 \simeq \frac{\nu^3}{2m} \;
\sum_j D_{ij} (\bk \cdot\ba_{ij})^2 \qquad (k\ra 0) \; .
\ee
Taking into account only the nearest neighbors reduces Eq. (\ref{7.85})
to
\be
\label{7.87}
\om_k^2 = \frac{4\nu^3}{m} \; D_0
\sum_\al \sin^2\left ( \frac{k_\al a_\al}{2} \right ) \; ,
\ee
where $D_0$ is $D_{ij}$ for the nearest neighbors,
\be
\label{7.88}
D_0 \; \equiv \; \frac{1}{d} \; \sum_{\al=1}^d \;
\frac{\prt^2 U(\ba)}{\prt a_\al^2} \; ,
\ee
with $\ba=\{ a_\al\}$ being the elementary lattice vector. Then the
long-wave limit (\ref{7.86}) becomes
\be
\label{7.89}
\om_k \simeq c_0 k \qquad (k\ra 0) \; ,
\ee
with the sound velocity
\be
\label{7.90}
c_0 \; \equiv \; \sqrt{\frac{\nu^3}{m} \; D_0 a^2 } \; .
\ee
Passing from summation to integration, according to the rule
$$
\sum_k \; \ra \; V \int_{\cal B} \; \frac{d\bk}{(2\pi)^d} \; ,
$$
for the interaction shift (\ref{7.79}), we obtain
\be
\label{7.91}
\Dlt U_{ij} = \frac{\nu^2 d}{2m\rho} \;
D_{ij} \int_{\cal B} \; \frac{1}{\om_k} \;
{\rm coth} \left ( \frac{\om_k}{2T} \right ) \;
\frac{d\bk}{(2\pi)^d} \; .
\ee

The main contribution to this integral comes from the region of long
waves, when $k\ra 0$. It is, therefore, possible to invoke the Debye
approximation, in which the phonon spectrum is taken in the long-wave
form (\ref{7.89}), while the integration over the Brillouin zone is
replaced by the integration over the Debye sphere by means of the
substitution
\be
\label{7.92}
\int_{\cal B} \; d\bk \; \ra \;
\frac{2\pi^{d/2}}{\Gm(d/2)} \; \int_0^{k_D} k^{d-1}\; dk \; .
\ee
The {\it Debye radius} $k_D$ is defined so that to preserve the correct
normalization
$$
V \int_{\cal B} \; \frac{d\bk}{(2\pi)^d} = N_L \; ,
\qquad \rho a^d = \nu \; ,
$$
that is, from the equation
$$
\frac{2\pi^{d/2}V}{\Gm(d/2)(2\pi)^d} \;
\int_0^{k_D} \; k^{d-1} \; dk = N_L \; .
$$
The latter results in
\be
\label{7.93}
k_D^d = \frac{2^d\pi^{d/2}d}{2\nu} \; \Gm
\left ( \frac{d}{2}\right ) \rho \; .
\ee
In particular, in three dimensions,
\be
\label{7.94}
k_D^3 = \frac{6\pi^2}{\nu} \; \rho =
\frac{6\pi^2}{a^3} \qquad (d=3) \; .
\ee
A characteristic lattice temperature, related to the Debye radius $k_D$,
is the {\it Debye temperature}
\be
\label{7.95}
T_D \equiv c_0 k_D =\sqrt{4\pi}
\left [ \frac{d}{2} \; \Gm\left ( \frac{d}{2} \right )
\right ]^{1/d} \; \frac{c_0}{a} \; .
\ee
This temperature separates the region of low temperatures $(T\ll T_D)$
from that of high temperatures $(T\gg T_D)$ with respect to phonon
characteristics.

Another important quantity is the mean-square atomic displacement
\be
\label{7.96}
r_0^2 \; \equiv \frac{1}{d} \; \sum_{\al=1}^d < \hat u_j^\al
\hat u_j^\al > \; .
\ee
Using Eq. (\ref{7.76}) for a cubic lattice gives
\be
\label{7.97}
r_0^2 = \frac{\nu^2}{2m\rho} \; \int_{\cal B} \;
\frac{1}{\om_k} \; {\rm coth}
\left ( \frac{\om_k}{2T} \right ) \;
\frac{d\bk}{(2\pi)^d} \; .
\ee
Then the phonon-induced interaction (\ref{7.91}) can be represented as
\be
\label{7.98}
\Dlt U_{ij} = D_{ij} r_0^2 d \; .
\ee
Hence, the phonon-induced interaction strongly depends on the
amplitude of atomic vibrations described by the mean-square deviation
(\ref{7.96}) or (\ref{7.97}).

First of all, we may notice that integral (\ref{7.97}) diverges for
any finite temperature $T>0$, if the space dimensionality is $d\leq 2$.
This means that, for these low dimensions, atoms cannot be localized
in a lattice. Their mean-square deviations (\ref{7.97}) become infinite.
And the phonon-induced interaction (\ref{7.98}) is also infinite. It is
important to stress that interaction (\ref{7.98}) becomes infinite for
any finite $D_{ij}$. And the dynamical matrix $D_{ij}$ is always finite
for any nonvanishing atomic interactions. As is discussed in Sec. 7.2,
the matrix $D_{ij}$ is finite, though may be small, even for the local
interactions, proportional to $\dlt(\br)$. Thus, for such local
interactions and well localized atoms, for which $l_0\ll a$, one has
$$
U(\ba_{ij}) \approx U\exp \left ( -\;
\frac{\ba_{ij}^2 d}{2l_0^2} \right ) \; ,
$$
from where
$$
D_{ij} \approx \frac{a^2 d^2}{l_0^4} \;
U(\ba_{ij}) \; .
$$
Therefore the localized states of the low-dimensional lattices, with
$d\leq 2$, at finite temperature $T>0$, are unstable with respect to
vibrational excitations. That is, in such lattices, a purely insulating
stable phase cannot exist.

At zero temperature, the localized state in a one-dimensional
lattice also cannot exist, since integral (\ref{7.97}) diverges
for $d=1$ even when $T=0$. Thence, the Mott insulating phase cannot
be a stable phase in a one-dimensional lattice. However, it can
exist in quasi-one-dimensional lattices, which, actually, are just
three-dimensional lattices elongated in one direction and tightly
confined in two other directions. What one realizes in experiments are
always quasi-low-dimensional lattices, but never purely one-dimensional
or purely two-dimensional ones. So, what is measured in experiments with
low-dimensional lattices does not need to exactly coincide with numerical
calculations accomplished for purely low-dimensional lattices.

For $d$-dimensional lattices at zero temperature, the Debye
approximation gives
$$
\int_{\cal B} \; \frac{1}{\om_k} \;
\frac{d\bk}{(2\pi)^d} = \frac{\rho d}{(d-1)\nu T_D} \; ,
$$
where $T_D$ is the Debye temperature (\ref{7.95}). Then Eq. (\ref{7.97})
yields
\be
\label{7.99}
r_0^2 = \frac{\nu d}{2(d-1)mT_D} \qquad (T=0) \; .
\ee
Hence the phonon-induced interaction (\ref{7.98}) is
\be
\label{7.100}
\Dlt U_{ij} = \frac{\nu d^2}{2(d-1)mT_D} \; D_{ij} \qquad (T=0) \; .
\ee
These formulas again confirm that the localized states in one-dimensional
lattices cannot exist. But the localized states in two-dimensional lattices
can occur at zero temperature. Three-dimensional lattices with localized
atoms are also stable at zero temperature.

At finite temperatures, such that $T\gg T_D$, the Debye approximation
gives
$$
\int_{\cal B} \; \frac{1}{\om_k} \; {\rm coth}
\left ( \frac{\om_k}{2T} \right ) \;
\frac{d\bk}{(2\pi)^d} \simeq
\frac{2\rho Td}{(d-2)\nu T_D^2} \; .
$$
And the mean-square deviation (\ref{7.97}) becomes
\be
\label{7.101}
r_0^2 \simeq \frac{\nu T d}{(d-2)m T_D^2} \; .
\ee
Then the phonon-induced interaction (\ref{7.98}) is
\be
\label{7.102}
\Dlt U_{ij} \simeq \frac{\nu T d^2}{(d-2)m T_D^2} \; D_{ij} \; .
\ee
From Eqs. (\ref{7.101}) and (\ref{7.102}), we again see that the
two-dimensional lattices with localized atoms are unstable at nonzero
temperature. But such localized states can arise for three-dimensional
lattices.

To evaluate how strong the phonon-induced interaction $\Dlt U_{ij}$ is,
being compared to the bare interaction $U_{ij}$, let us consider a cubic
lattice with nearest-neighbor interactions, when
\be
\label{7.103}
D_{ij} = D_0 = \frac{\prt^2 U(\ba)}{\prt a^2} \; .
\ee
Suppose that atoms interact through dipole forces, for which
$U(\ba)\sim a^{-3}$. As a result,
\be
\label{7.104}
D_0 = \frac{12}{a^2} \; U(\ba) \; .
\ee
Then the phonon sound velocity (\ref{7.90}) is
\be
\label{7.105}
c_0 = \sqrt{\frac{12}{m} \; \nu^3 U(\ba) } \; .
\ee

At zero temperature, the phonon-induced interaction (\ref{7.100})
writes as
\be
\label{7.106}
\Dlt U_{ij} = \frac{d^2}{(d-1)k_D a^2} \;
\sqrt{\frac{3}{\nu m} \; U(\ba) } \; .
\ee
Using the relation
$$
E_R = \frac{k_0^2}{2m} = \frac{\pi^2}{2ma^2} \; ,
$$
we find
\be
\label{7.107}
\frac{\Dlt U_{ij}}{U_{ij}} =
\frac{\sqrt{6}\; d^2}{\pi(d-1) k_D a} \;
\sqrt{\frac{E_R}{\nu U_{ij} } } \; ,
\ee
where
$$
k_D a = \sqrt{4\pi} \left [ \frac{d}{2} \; \Gm
\left ( \frac{d}{2} \right ) \right ]^{1/d} \; .
$$
For two-and three-dimensional lattices we have
$$
k_D a =\sqrt{4\pi} = 3.545 \qquad (d=2) \; ,
$$
$$
k_D a  = 3.898 \qquad (d=3) \; .
$$
Therefore Eq. (\ref{7.107}) yields
\be
\label{7.108}
\frac{\Dlt U_{ij}}{U_{ij}} = 0.9 \;
\sqrt{\frac{E_R}{\nu U_{ij} } }
\qquad ( T=0;\; d=2,3 ) \; .
\ee

At finite temperatures, such that $T\gg T_D$, the phonon-induced
interaction is given by Eq. (\ref{7.102}), from where
\be
\label{7.109}
\frac{\Dlt U_{ij}}{U_{ij}} \simeq
\frac{d^2}{(d-2)(k_D a)^2}
\left ( \frac{T}{\nu U_{ij} }\right ) \; .
\ee
For a three-dimensional lattice, this gives
\be
\label{7.110}
\frac{\Dlt U_{ij}}{U_{ij}} \simeq  0.6 \; \frac{T}{\nu U_{ij} }
\qquad (T > T_D , \; d=3) \; .
\ee
Equations (\ref{7.108}) and (\ref{7.110}) show that phonons can
substantially renormalize atomic interactions.

The extended Hubbard model (\ref{5.16}) has been studied theoretically
for one-dimensional [317--322] and two-dimensional [417] lattices.
However, it is necessary to be cautious interpreting the results
of numerical calculations. As follows from the above analysis, the
localized states in one-dimensional lattices are unstable with respect
to phonon excitations at any temperature. The Mott insulating phase,
strictly speaking, cannot be realized in such lattices even at zero
temperature. In two- and three-dimensional lattices, the Mott insulating
phase can happen at zero temperature. But, investigating the phase diagram,
one should take into account the phonon-induced renormalization of atomic
interactions.

The boundary between the insulating and Bose-condensed phases can be
defined by studying the behavior of the condensate fraction [418]. One
should keep in mind that this boundary can be shifted because of the
influence of the phonon-induced interactions.

For optical lattices with disorder, the phase diagram essentially depends
on the presence of the order-parameter fluctuations [419]. The existence
of the vibrational atomic fluctuations can also strongly influence the
phase portrait of disordered optical lattices. For the latter, the phonon
excitations can occur to be even more dramatic than for ideal lattices.

\section{Double-Well Optical Lattices}

\subsection{Effective Hamiltonians}

Recently, a double-well optical lattice was realized experimentally
[420], being a lattice each site of which is represented by a
double-well potential. Dynamics of cold atoms in a separate double-well
has been considered in several publications [421--423]. But to study the
properties of the whole double-well lattice, it is necessary to have an
appropriate lattice Hamiltonian.

We should start with the general Hamiltonian in the Wannier representation,
given by Eq. (\ref{4.11}). Contrary to the case of the standard Hubbard model
(\ref{4.12}), for the double-well lattice, it is impossible to resort to the
single-band approximation. This is because the tunneling of atoms between the
wells of a double well results in the splitting of the ground-state level onto
two energy levels that can be very close to each other [205]. Without taking
account of this splitting there would be no atomic tunneling between the
wells. Since, in addition, atoms interact with each other, this tunneling
is essentially nonlinear [424,425].

Thus, for the double-well lattices, we have to retain not less than two
lowest energy levels, that is, we have to deal with at least a two-band case.
It is important to stress that in the expansion of the field operator
\be
\label{8.111}
\psi(\br) = \sum_{nj} c_{nj} w_n(\br-\ba_j)
\ee
the index $n$ enumerates the quantum bands, that is, the quantum energy
levels, but not "left" or "right" positions. The latter, as is explained
in Sec. 2.17, are not good quantum numbers. The necessity of taking into
account several energy levels is typical of atoms in complex multi-well
configurations [426--428] as well as can occur for some metastable
systems [429].

Introducing the notation
\be
\label{8.112}
E_{ij}^{mn} \equiv
\int w_m^*(\br-\ba_i) H_L(\br) w_n(\br-\ba_j) \; d\br \; ,
\ee
in which
$$
H_L(\br) \equiv -\; \frac{\nabla^2}{2m} + V_L(\br) \; ,
$$
Hamiltonian (\ref{4.11}) can be rewritten as
\be
\label{8.113}
\hat H = \sum_{ij} \; \sum_{mn}
E_{ij}^{mn} c_{mi}^\dgr c_{nj} \; + \;
\frac{1}{2} \; \sum_{ \{ j\} } \; \sum_{ \{ n\} }
U_{j_1j_2j_3j_4}^{n_1n_2n_3n_4}
c_{n_1j_1}^\dgr c_{n_2j_2}^\dgr c_{n_3j_3} c_{n_4j_4} \; .
\ee
The indices $m$ and $n$ here have to take at least two values, $m,n=1,2$.
When there is BEC in the lattice, so that the gauge symmetry becomes broken,
the field operators $c_{nj}$ should be replaced by the Bogolubov-shifted
operators (\ref{4.5}).

In order to find the relation between the matrix element (\ref{8.112}) and
the Bloch energy $E_{nk}$, which is the eigenvalue of the equation
$$
H_L(\br) \vp_{nk}(\br) =  E_{nk} \vp_{nk}(\br) \; ,
$$
we can employ the expansion of the Bloch functions $\vp_{nk}(\br)$ over the
Wannier functions. Then the above eigenproblem transforms into the equation
$$
H_L(\br) w_n(\br-\ba_i) = \frac{1}{N_L} \;
\sum_{jk} E_{nk} e^{-i\bk\cdot\ba_{ij} } w_n(\br-\ba_j) \; .
$$
This shows that the Wannier functions, strictly speaking, are not the
eigenfunctions of the lattice Hamiltonian $H_L(\br)$. Using this equation,
for the matrix element (\ref{8.112}), we find
$$
E_{ij}^{mn} = \dlt_{mn} E_{ij}^n \; ,
$$
where
$$
E_{ij}^n = \frac{1}{N_L} \;
\sum_k E_{nk} e^{i\bk\cdot \ba_{ij}} \; .
$$
The latter expression can be represented in the form
\be
\label{8.114}
E_{ij}^n = \dlt_{ij} E_n + ( 1 - \dlt_{ij} ) J_{ij}^n \; ,
\ee
in which
$$
E_n = \int w_n^*(\br-\ba_j) H_L(\br) w_n(\br-\ba_j) \; d\br =
\frac{1}{N_L} \; \sum_k E_{nk} \; ,
$$
$$
J_{ij}^n =
\int w_n^*(\br-\ba_i) H_L(\br) w_n(\br-\ba_j) \; d\br =
\frac{1}{N_L} \; \sum_k E_{nk} e^{i\bk\cdot\ba_{ij}} \; ,
\qquad (i\neq j) \; .
$$
There is a temptation to reduce the number of these parameters by assuming
that the above quantities do not depend on the band indices. This, however,
is not a good idea. The tunneling between the wells of a single double-well
depends on the nonzero value of the difference $E_{jj}^{22}-E_{jj}^{11}$.
But this difference would be zero, if the values $E_{ij}^{mn}$ would not
depend on the band indices. In oder not to kill the tunneling between the
wells of a double well, one has to retain the dependence on the band indices.
The consideration of model (\ref{8.113}), with many independent parameters,
is rather complicated and can be done by setting some of these parameters
to zero [430].

There is, however, a case, when Hamiltonian (\ref{8.113}) can be essentially
simplified. This is when the filling factor is strictly fixed to one, so that
the double occupancy of a lattice site is prohibited, which is manifested by
the {\it unipolarity conditions}
\be
\label{8.115}
\sum_n c_{nj}^\dgr c_{nj} = 1 \; , \qquad c_{nj} c_{nj} = 0 \; .
\ee

Let us also assume that the lattice is in the insulating state, such that
the atomic hopping between different lattice sites is negligible,

\be
\label{8.116}
\left | \frac{J_{ij}}{E_n} \right | \ll 1 \qquad (i\neq j) \; .
\ee
Under this condition, Wannier functions become approximate eigenfunctions
of $H_L(\br)$, in the sense that
$$
H_L(\br) w_n(\br-\ba_j) \simeq E_n w_n(\br-\ba_j) \; .
$$

Under conditions (\ref{8.115}) and (\ref{8.116}), Hamiltonian (\ref{8.113})
reduces to the form
\be
\label{8.117}
\hat H = \sum_{nj} E_n c_{nj}^\dgr c_{nj} \; + \;
\frac{1}{2}\; \sum_{i\neq j} \; \sum_{mnm'n'}
V_{ij}^{mnm'n'} c_{mi}^\dgr c_{nj}^\dgr c_{m'j} c_{n'i} \; ,
\ee
in which
$$
V_{ij}^{mnm'n'} \equiv V_{ijji}^{mnm'n'} + V_{ijij}^{mnn'm'} \; .
$$
Retaining only the two lowest bands implies that $n=1,2$. Then the
unipolarity conditions (\ref{8.115}) are
$$
c_{1j}^\dgr c_{1j} + c_{2j}^\dgr c_{2j} = 1 \; , \qquad
c_{1j} c_{1j} = c_{2j} c_{2j} = 0 \; .
$$

For this two-band case, it is possible to resort to the pseudospin
representation, similar to that used for some ferroelectrics [431,432]. The
pseudospin operators are defined as
$$
S_j^x = \frac{1}{2} \left ( c_{1j}^\dgr c_{1j} -
c_{2j}^\dgr c_{2j} \right ) \; , \qquad
S_j^y = \frac{i}{2} \left ( c_{1j}^\dgr c_{2j} -
c_{2j}^\dgr c_{1j} \right ) \; ,
$$
\be
\label{8.118}
S_j^z = \frac{1}{2} \left ( c_{1j}^\dgr c_{2j} +
c_{2j}^\dgr c_{1j} \right ) \; .
\ee
This gives
$$
c_{1j}^\dgr c_{1j} = \frac{1}{2} + S_j^x \; , \qquad
c_{2j}^\dgr c_{2j} = \frac{1}{2} - S_j^x \; ,
$$
\be
\label{8.119}
c_{1j}^\dgr c_{2j} = S_j^z - i S_j^y \; , \qquad
c_{2j}^\dgr c_{1j} = S_j^z + i S_j^y \; .
\ee
To clarify the physical meaning of the pseudospin operators (\ref{8.118}),
one can introduce the left, $c_{jL}$, and the right, $c_{jR}$, operators by
the relations
$$
c_{1j} = \frac{1}{\sqrt{2}}\; ( c_{jL} + c_{jR} ) \; , \qquad
c_{2j} = \frac{1}{\sqrt{2}}\; ( c_{jL} - c_{jR} ) \; ,
$$
\be
\label{8.120}
c_{jL} = \frac{1}{\sqrt{2}}\; ( c_{1j} + c_{2j} ) \; , \qquad
c_{jR} = \frac{1}{\sqrt{2}}\; ( c_{1j} - c_{2j} ) \; .
\ee
Then operators (\ref{8.118}) become
$$
S_j^x = \frac{1}{2} \left ( c_{jL}^\dgr c_{jR} +
c_{jR}^\dgr c_{jL} \right ) \; , \qquad
S_j^y = -\;\frac{i}{2} \left ( c_{jL}^\dgr c_{jR} -
c_{jR}^\dgr c_{jL} \right ) \; ,
$$
\be
\label{8.121}
S_j^z = \frac{1}{2} \left ( c_{jL}^\dgr c_{jL} -
c_{jR}^\dgr c_{jR} \right ) \; .
\ee
These equations demonstrate that $S_j^x$ describes the tunneling intensity
between the left and right wells of a double-well potential centered at the
$j$-site; $S_j^y$ characterizes the Josephson current between the left and
right wells; and $S_j^z$ is the displacement operator for the imbalance
between the wells.

The ground-state wave function and that of the first excited state in a
double-well possess different symmetry properties and differing topology
[205]. For instance, the ground-state wave function $w_1(\br)$ is symmetric
with respect to spatial inversion, while the excited-state wave function
$w_2(\br)$ is antisymmetric,
\be
\label{8.122}
w_1(-\br) = w_1(\br) \; , \qquad w_2(-\br) = - w_2(\br) \; .
\ee
Due to the symmetry properties (\ref{8.122}), some of the matrix
elements $V_{ij}^{mnm'n'}$ can become zero. Concretely, these are the matrix
elements that are nondiagonal with respect to the band indices.

Let us consider the integral
$$
\Phi_{ij}^{1112} \equiv \int
\Phi(\br-\br') w_1^*(\br-\ba_i)\; | w_1(\br'-\ba_j)|^2 \;
w_2(\br-\ba_i) \; d\br d\br' \; ,
$$
in which $\Phi(-\br)=\Phi(\br)$ is any symmetric pair-interaction potential
and $i\neq j$. Shifting here $\br$ by $\ba_i$ and $\br'$ by $\ba_j$ gives an
equivalent form
$$
\Phi_{ij}^{1112} \equiv \int
\Phi(\br-\br' + \ba_{ij}) w_1^*(\br)\; | w_1(\br')|^2 \;
w_2(\br) \; d\br d\br' \; .
$$
Accomplishing here the spatial inversion of $\br$ and $\br'$, using the
symmetry of the interaction potential, and invoking the symmetry properties
(\ref{8.122}), we have
$$
\Phi_{ij}^{1112} = - \int \Phi(\br-\br'-\ba_{ij} ) w_1^*(\br) \;
|w_1(\br')|^2 \; w_2(\br) \; d\br d\br' \; .
$$
From the last two equations, keeping in mind that the Wannier functions
form an orthonormal basis and can be made well localized, we find
$$
\Phi_{ij}^{1112} \cong -\Phi_{ij}^{1112} \cong 0 \; .
$$
In the same way, it is easy to show that other nondiagonal matrix elements
of the interaction potential are practically zero. As a result, we obtain
\be
\label{8.123}
V_{ij}^{1112} = V_{ij}^{1121} = V_{ij}^{1211} =
V_{ij}^{2111} = V_{ij}^{2221} = V_{ij}^{2212} =
V_{ij}^{2122} = V_{ij}^{1222} = 0\; .
\ee
Choosing real Wannier functions yields
\be
\label{8.124}
V_{ij}^{1212} = V_{ij}^{2121} = V_{ij}^{2211} =
V_{ij}^{1122} \; , \qquad  V_{ij}^{2112} = V_{ij}^{1221} \; .
\ee
These matrix elements are nonzero.

Let us introduce the notation
\be
\label{8.125}
E_0 \equiv \frac{1}{2} \left ( E_1 + E_2 \right )
\ee
and the following combinations of the interaction matrix elements:
$$
A_{ij} \equiv \frac{1}{4} \left ( V_{ij}^{1111} + V_{ij}^{2222}
+ 2 V_{ij}^{1221} \right ) \; , \qquad
B_{ij} \equiv \frac{1}{2} \left ( V_{ij}^{1111} + V_{ij}^{2222}
- 2 V_{ij}^{1221} \right ) \; ,
$$
\be
\label{8.126}
C_{ij} \equiv \frac{1}{2} \left (  V_{ij}^{2222}
- V_{ij}^{1111} \right ) \; , \qquad
I_{ij} \equiv -2 V_{ij}^{1122} \; .
\ee
Also, let us define the quantity
\be
\label{8.127}
\Om \equiv E_2 - E_1 + \sum_{j(\neq i)} C_{ij} \; ,
\ee
playing the role of a tunneling parameter characterizing the tunneling
between the wells of a double-well potential.

As a result, Hamiltonian (\ref{8.117}) reduces to the pseudospin form
\be
\label{8.128}
\hat H = E_0 N + \frac{1}{2} \; \sum_{i\neq j} A_{ij} \; - \;
\Om \sum_j S_j^x \; + \; \sum_{i\neq j} B_{ij}S_i^x S_j^x \; - \;
\sum_{i\neq j} I_{ij} S_i^z S_j^z \; .
\ee
By their definitions, the diagonal matrix elements $V_{ij}^{1111}$,
$V_{ij}^{2222}$, and $V_{ij}^{1221}$, by their absolute values, can be close
to each other, but much larger than the exchange matrix element $I_{ij}$, so
that
\be
\label{8.129}
| V_{ij}^{1122} | \ll | V_{ij}^{1111} | \; .
\ee
Then, from Eqs. (\ref{8.126}), it follows that
\be
\label{8.130}
| C_{ij} | \ll | A_{ij} | \; , \qquad
| I_{ij} | \ll | A_{ij} | \; .
\ee
However the term, containing $A_{ij}$, is not an operator, hence can be
omitted from the Hamiltonian, as well as the term $E_0N$. The remaining
terms, with $B_{ij}$, $I_{ij}$, and $\Om$, can be of the same order. By
varying the shape of a double well, it is possible to make the energy
difference $E_2-E_1$ quite large or exponentially small [205]. Thence
the tunneling parameter $\Om$ in Eq. (\ref{8.127}) can be varied in a wide
range. That is, in general, the term with $B_{ij}$ cannot be omitted. It can
be neglected only when the tunneling parameter $\Om$ is sufficiently large,
such that it is much larger than $B_{ij}$. Note that the tunneling between
different lattice sites can be modulated by shaking the lattice [433]. In
a similar way, one could also modulate the effective tunneling between the
adjacent wells of a double-well potential.

\subsection{Phase Transitions}

To study what kind of phase transitions occurs for Hamiltonian (\ref{8.128}),
let us resort to the mean-field approximation
\be
\label{8.131}
S_i^\al S_j^\bt \; = \; < S_i^\al> S_j^\bt +
S_i^\al < S_j^\bt> - < S_i^\al><S_j^\bt> \qquad (i\neq j) \; .
\ee
And let us introduce the notation
\be
\label{8.132}
A \equiv \frac{1}{N_L} \; \sum_{i\neq j} A_{ij} \; , \qquad
B \equiv \frac{1}{N_L} \; \sum_{i\neq j} B_{ij} \; , \qquad
I \equiv \frac{1}{N_L} \; \sum_{i\neq j} I_{ij} \; .
\ee Under the unipolarity conditions (\ref{8.115}), the filling factor
is strictly one, and $N_L=N$. Also, we define the effective tunneling
\be
\label{8.133}
\overline\Om \equiv \Om - 2B < S_j^x> \; .
\ee
Then Hamiltonian (\ref{8.128}) acquires the form
\be
\label{8.134}
\hat H = H_{non} - \overline\Om \sum_j S_j^x -
2I \sum_j < S_i^z> S_j^z \; ,
\ee
in which the first term is the nonoperator quantity
\be
\label{8.135}
H_{non} = NE_0 + \frac{N}{2} \left ( A - 2B < S_i^x>^2 +
2I < S_i^z>^2 \right ) \; .
\ee
By introducing an effective "magnetic" field
\be
\label{8.136}
{\bf H}_{eff} \equiv \{ H_{eff}^\al \} \equiv
\{ \overline\Om, \; 0, \; 2I<S_j^z> \} \; ,
\ee
Hamiltonian (\ref{8.134}) can be rewritten as
\be
\label{8.137}
\hat H = H_{non} - \sum_j {\bf H}_{eff} \cdot \bS_j \; .
\ee
The corresponding free energy is
\be
\label{8.138}
F = H_{non} - NT \ln \left ( 2 {\rm cosh}\;
\frac{H_{eff}}{2T} \right ) \; ,
\ee
where $H_{eff}\equiv|{\bf H}_{eff}|$, which gives
\be
\label{8.139}
H_{eff} = \sqrt{\overline{\Om}^2 + 4I^2 < S_j^z>^2 } \; .
\ee

The average values $<S_j^\al>$ can be found from the equation
$$
< S_j^\al> \; = \; -\; \frac{1}{N} \;
\frac{\prt F}{\prt H_{eff}^\al} \; .
$$
This yields the equations for the $x$-component (tunneling intensity)
\be
\label{8.140}
< S_j^x > \; = \; \frac{\overline\Om}{2H_{eff}} \;
{\rm tanh}\left ( \frac{H_{eff}}{2T} \right ) \; ,
\ee
the $y$-component (Josephson current)
\be
\label{8.141}
< S_j^y > \; = \; 0 \; ,
\ee
and the $z$-component (well imbalance)
\be
\label{8.142}
< S_j^z > \; = \; < S_j^z> \; \frac{I}{H_{eff}} \; {\rm tanh}
\left ( \frac{H_{eff}}{2T} \right ) \; .
\ee

Let us define the averages
\be
\label{8.143}
x \equiv 2 < S_j^x> \; , \qquad y \equiv 2 < S_j^y> \; ,
\qquad z \equiv 2 < S_j^z > \; .
\ee
It is convenient to introduce the dimensionless quantities
\be
\label{8.144}
\om \equiv \frac{\Om}{I+B} \; , \qquad
b \equiv \frac{B}{I+B} \; .
\ee
Using these, we have
\be
\label{8.145}
\frac{\overline\Om}{I+B} = \om - bx \; , \qquad
\frac{I}{I+B} = 1 - b \; .
\ee
Also, let us define the dimensionless quantity
\be
\label{8.146}
h \equiv \frac{H_{eff}}{I+B} \; ,
\ee
which is
\be
\label{8.147}
h = \sqrt{(\om-bx)^2 + (1-b)^2 z^2 } \; .
\ee
The nonoperator part (\ref{8.135}) of Hamiltonian (\ref{8.134}) reads as
\be
\label{8.148}
\frac{H_{non}}{N} = E_0 + \frac{A}{2} + \frac{I+B}{4}
\left [ (1 - b) z^2 - bx^2 \right ] \; .
\ee
Employing the dimensionless quantities, defined above, the temperature
$T$ will be measured in units of $I+B$.

In the dimensionless notation, the averages (\ref{8.140}), (\ref{8.141}),
and (\ref{8.142}) yield the tunneling intensity
\be
\label{8.149}
x = \frac{\om-bx}{h} \; {\rm tanh} \left ( \frac{h}{2T} \right ) \; ,
\ee
the Josephson current
\be
\label{8.150}
y = 0 \; ,
\ee
and the well imbalance
\be
\label{8.151}
z = z \; \frac{1-b}{h} \; {\rm tanh} \left ( \frac{h}{2T}
\right ) \; .
\ee
These variables satisfy the condition
\be
\label{8.152}
x^2 + y^2 + z^2 = {\rm tanh}^2 \left ( \frac{h}{2T}
\right ) \; .
\ee

Equations (\ref{8.149}) and (\ref{8.151}) are invariant under the
replacement
$$
x \ra - x \; , \qquad \om \ra - \om \; , \qquad z \ra - z \; .
$$
Therefore, without the loss of generality, we can consider only the case,
when $x\geq 0$, $\om\geq 0$, and $z\geq 0$. The inequality $\om\geq 0$ is
in agreement with the accepted enumeration of the energy levels in a
double-well potential, when $E_1<E_2$.

Equation (\ref{8.151}) shows that there can be two types of solutions, when
$z\neq 0$ and when $z=0$. The well-imbalance $z$ plays the role of an order
parameter. If $z\neq 0$, this means that atoms are mainly shifted to one of
the double wells. While if $z=0$, then atoms on the average equally populate
both wells. The thermodynamic phase with $z\neq 0$ is called ordered, while
that with $z=0$ is termed disordered.

For the ordered phase, when $z\neq 0$, Eq. (\ref{8.151}) gives
\be
\label{8.153}
\frac{1-b}{h} \; {\rm tanh} \left (
\frac{h}{2T} \right ) =  1 \; ,
\ee
which defines $z$. Using Eq. (\ref{8.153}) in Eq. (\ref{8.149}) yields
$$
x = \frac{\om-bx}{1-b} \; ,
$$
from where it follows
\be
\label{8.154}
x = \om \; .
\ee
Since, by definition, the variable $x$ is positive and less than one,
we see that the ordered phase can exist if
\be
\label{8.155}
0 \leq \om \leq 1 \qquad ( 0\leq x \leq 1) \; .
\ee
Equation (\ref{8.153}), with
\be
\label{8.156}
h = ( 1 - b) \; \sqrt{\om^2 + z^2 } \; ,
\ee
defines $z>0$ for $T<T_c$. The critical temperature is
\be
\label{8.157}
T_c = \frac{(1-b)\om}{2{\rm artanh}\;\om} \; ,
\ee
where
$$
{\rm artanh}\;\om = \frac{1}{2} \;
\ln \; \frac{1+\om}{1-\om} \; .
$$
When $\om\ra 0$, then
$$
T_c \simeq \frac{1-b}{2} \qquad (\om \ll 1) \; ,
$$
which in dimensional units becomes $T_c\simeq I/2$. And if $\om\ra 1$, then
$T_c\ra 0$. Thus, the ordered phase, with $z>0$, can exist when both $T<T_c$
and $\om\leq 1$.

In the disordered phase,
\be
\label{8.158}
z = 0 \; .
\ee
Equation (\ref{8.147}) gives
\be
\label{8.159}
h = \om - bx \; .
\ee
Then Eq. (\ref{8.149}) yields
\be
\label{8.160}
x = {\rm tanh} \left ( \frac{\om-bx}{2T} \right ) \; .
\ee
For $x$ to be non-negative, it should be that
\be
\label{8.161}
\om \geq bx \qquad (x \geq 0 ) \; .
\ee
The disordered phase arises, when the ordered phase cannot exist, that
is, when either $T>T_c$ or $\om>1$.

The transition between the ordered and disordered phases happens when
either the temperature or the tunneling parameter are varied. For example,
at zero temperature $T=0$, the ordered phase has
\be
\label{8.162}
x =\om \; , \qquad z =\sqrt{1-\om^2} \qquad (\om < 1) \; .
\ee
While the disordered phase is described by
\be
\label{8.163}
x = 1 \; , \qquad z = 0 \; , \qquad ( \om > 1 ) \; .
\ee
At the value $\om=1$, the quantum phase transition occurs.

The reduced internal energy
$$
E \equiv \frac{1}{N} < \hat H >
$$
is
\be
\label{8.164}
E = \frac{H_{non}}{N} \; - \; \frac{I+B}{2} \left [
( \om - bx ) x + ( 1 - b ) z^2 \right ] \; .
\ee
In view of Eq. (\ref{8.148}), this gives
\be
\label{8.165}
E = E_0 + \frac{A}{2} + \frac{I+B}{4} \left [ bx^2 - 2\om -
( 1 - b) z^2 \right ] \; .
\ee
The energy of the ordered phase,
$$
E = E_0 + \frac{A}{2} \; - \; \frac{I+B}{4} \left ( 1 - b +
\om^2 \right ) \qquad (\om < 1)
$$
is always lower than the energy of the disordered phase,
$$
E = E_0 + \frac{A}{2} + \frac{I+B}{4} \; ( b - 2\om )
\qquad ( \om > 1) \; .
$$
These energies coincide at the critical value $\om=1$.

\subsection{Collective Excitations}

The dynamics of the pseudospin operators follows from the Heisenberg
equations
$$
i \; \frac{d S_j^\al}{dt} =
\left [ S_j^\al, \hat H \right ] \; ,
$$
with the commutation relations
$$
[ S_i^x, S_j^y ] = i \dlt_{ij} S_j^z \; , \qquad
[ S_i^y, S_j^z ] = i \dlt_{ij} S_j^x \; , \qquad
[ S_i^z, S_j^x ] = i \dlt_{ij} S_j^y \; .
$$
Using the pseudospin Hamiltonian (\ref{8.128}) results in the equations
of motion
$$
\frac{dS_i^x}{dt} = 2 S_i^y \;
\sum_{j(\neq i)} I_{ij} S_j^z \; , \qquad
\frac{dS_i^y}{dt} = \Om S_i^z - 2 S_i^x \;
\sum_{j(\neq i)} I_{ij} S_j^z - 2 S_i^z \;
\sum_{j(\neq i)} B_{ij} S_j^x \; ,
$$
\be
\label{8.166}
\frac{dS_i^z}{dt} = - \Om S_i^y + 2 S_i^y \;
\sum_{j(\neq i)} B_{ij} S_j^x \; .
\ee

Collective excitations in the random-phase approximation can be found by
representing the pseudospin operators as
\be
\label{8.167}
S_j^\al \; = \; < S_j^\al > + \dlt S_j^\al
\ee
and considering $\dlt S_j^\al$ as a small deviation from an equilibrium
average. Equation (\ref{8.167}) is to be substituted into Eqs. (\ref{8.166}),
which are linearized with respect to $\dlt S_j^\al$. In the zero order, one
gets
\be
\label{8.168}
\frac{dx}{dt} = 0 \; , \qquad
\frac{dy}{dt} = (\overline\Om - Ix ) z \; , \qquad
\frac{dz}{dt} = 0 \; .
\ee
For the ordered phase, when $z\neq 0$, one has $x=\om$; and for the
disordered phase, $z=0$. Therefore the second equation in Eqs. (\ref{8.168})
becomes $dy/dt=0$. So, all averages, $x,y$, and $z$, do not depend on time,
as it should be for equilibrium quantities. In the first order, one has
$$
\frac{d}{dt} \; \dlt S_i^x =  Iz \dlt S_i^y \; ,
$$
$$
\frac{d}{dt} \; \dlt S_i^y = - I z \dlt S_i^x -
z \sum_{j(\neq i)} B_{ij} \dlt S_j^x \; + \; \
\overline\Om \dlt S_i^z \; - \; x
\sum_{j(\neq i)} I_{ij} \dlt S_j^z \; ,
$$
\be
\label{8.169}
\frac{d}{dt} \; \dlt S_i^z = - \overline\Om \dlt S_i^y \; .
\ee
One defines the Fourier transforms
\be
\label{8.170}
\dlt S_j^\al = \frac{1}{N_L} \;
\sum_k \sgm_k^\al e^{i(\bk\cdot\ba_j -\ep t)} \; ,
\qquad \sgm_k^\al  = \sum_j
\dlt S_j^\al e^{-i(\bk\cdot\ba_j -\ep t)} \; .
\ee
Similarly, Fourier transforms are introduced for the interaction functions,
$$
B_{ij} = \frac{1}{N_L} \; \sum_k B_k e^{i\bk\cdot\ba_{ij}} \; ,
\qquad B_k = \sum_i B_{ij} e^{-i\bk\cdot\ba_{ij}} \; ,
$$
\be
\label{8.171}
I_{ij} = \frac{1}{N_L} \; \sum_k I_k e^{i\bk\cdot\ba_{ij}} \; ,
\qquad I_k = \sum_i I_{ij} e^{-i\bk\cdot\ba_{ij}} \; .
\ee
Then Eqs. (\ref{8.169}) yield
$$
i \ep \sgm_k^x + z I \sgm_k^y = 0 \; , \qquad
( I + B_k ) z \sgm_k^x - i\ep \sgm_k^y +
( x I_k - \overline\Om) \sgm_k^z = 0 \; ,
$$
\be
\label{8.172}
\overline\Om \sgm_k^y  - i\ep \sgm_k^z = 0 \; .
\ee
The condition for the existence of nontrivial solutions to Eqs.
(\ref{8.172}) gives the equation
\be
\label{8.173}
\ep \left [ \ep^2  - ( \overline\Om - x I_k ) -
I ( I + B_k ) z^2 \right ] = 0 \; ,
\ee
in which
$$
\overline\Om = \Om - Bx \; .
$$
Equation (\ref{8.173}) defines three branches for the spectrum of collective
excitations. One branch is trivial,
\be
\label{8.174}
\ep_1(\bk ) = 0 \; .
\ee
And two other branches are given by the equation
\be
\label{8.175}
\ep_{2,3}^2(\bk) = \overline\Om ( \overline\Om - xI_k) +
I( I + B_k) z^2 \; .
\ee
The branches of spectrum (\ref{8.175}) describe the pseudospin oscillations.
These branches possess gaps for both ordered as well as disordered phases,
and, in the long-wave limit, they vary as $k^2$.

\subsection{Nonequilibrium States}

Nonequilibrium states in lattices appear, when atoms are subject to
temporal external fields. This is possible to accomplish by varying the
shape of the lattice, for instance, by changing the configuration of the
double-well potential in each lattice site. Atomic interactions can be
modulated by means of the Feshbach resonance techniques. A nonequilibrium
situation arises in the process of loading atoms into the lattice.

The nonequilibrium behavior of atoms in a double well lattice can be
characterized by considering the temporal evolution of the average
quantities (\ref{8.143}). The evolution equations for these quantities
are to be obtained from averaging the operator equations (\ref{8.166}).
When accomplishing such an averaging, it is customary to invoke the
mean-field approximation (\ref{8.131}). This standard way has a principal
defect of not taking into account atomic collisions resulting in the
appearance of damping. Not taking account of the latter can lead to
incorrect dynamics and wrong physical conclusions.

The existence of atomic collisions yielding the arising attenuation, can
be included in the dynamics by employing the local-field approximation.
This approximation is based on the fact that a kind of local equilibrium
exists even in strongly nonequilibrium systems [434-436]. Then one can
consider atomic collisions as occurring in an effective local field of
other particles [437]. The resulting attenuation effects are included into
the evolution equations through the damping parameters, characterizing the
longitudinal, $\gm_1$, and transverse, $\gm_2$, relaxation. The values of
these parameters can be calculated in the same way as is done for magnetic
systems [438].

The local fields for variables (\ref{8.143}) are defined through
expressions (\ref{8.149}), (\ref{8.150}), and (\ref{8.151}) as
\be
\label{8.176}
x_t \equiv \frac{\om-bx}{h} \;
{\rm tanh} \left ( \frac{h}{2T} \right ) \; ,
\qquad y_t= 0 \; , \qquad
z_t \equiv \frac{1-b}{h} \; z\;
{\rm tanh} \left ( \frac{h}{2T} \right ) \; ,
\ee
where $h$ is given by Eq. (\ref{8.147}). Averaging Eqs. (\ref{8.166}) in
the local-field approximation results in the evolution equations for the
tunneling intensity
\be
\label{8.177}
\frac{dx}{dt} =  ( 1 - b) yz - \gm_2 (x - x_t) \; ,
\ee
Josephson current
\be
\label{8.178}
\frac{dy}{dt} = ( \om - x) z - \gm_2 ( y -y_t ) \; ,
\ee
and the well imbalance
\be
\label{8.179}
\frac{dz}{dt} = ( bx - \om) y - \gm_1 ( z- z_t) \; ,
\ee
with the local fields (\ref{8.176}). These equations describe the time
evolution of $x=x(t)$, $y=y(t)$, and $z=z(t)$ under the given initial
conditions
\be
\label{8.180}
x_0 = x(0) \; , \qquad y_0 = y(0) \; , \qquad z_0 = z(0) \; .
\ee
Here the dimensionless parameters (\ref{8.144}) are employed and time is
measured in units of $1/(I+B)$.

The evolution equations (\ref{8.177}), (\ref{8.178}), and (\ref{8.179}),
depending on the parameters $b$ and $\om$, can show two types of behavior,
attenuating to two different fixed points, one corresponding to the ordered
stationary solution and another, to the disordered stationary solution.

\subsection{Heterophase Lattices}

It may happen that a double-well lattice is neither completely ordered
nor completely disordered, but consists of the ordered regions intermixed
with disordered parts. The spatial distribution of these differently
ordered regions is random, as well as their sizes and shapes. Such an
object composed of a random mixture of different phases is called the {\it
heterophase system}. In condensed matter, there are plenty of examples of
such systems, as has been reviewed in the articles [54,66]. A schematic
picture of a heterophase double-well lattice is shown in Fig. 3, where the
ordered regions are marked by arrows, while the disordered regions are left
empty. Because of their random distribution in space and because they often
arise randomly, that is, in a fluctuational way, in time, the heterophase
regions inside a statistical system are called heterophase fluctuations.
Such fluctuations can be provoked by the environmental randomness [439],
even when the latter corresponds to a very weak noise. But they can also be
due to intrinsic causes, such as local fluctuations of entropy or temperature
[440], fluctuations of density or composition fluctuations [441], and other
internal perturbations generated by the system itself [442]. The nuclei of
one phase inside another are also termed droplets or clusters [443]. The
system with heterophase fluctuations can be more stable than a pure-phase
sample. What kind of internal structure is more profitable for a system is
chosen by the system itself, which self-organizes for reaching an optimal
state [444,445]. A thorough description of possible origins of heterophase
fluctuations is given in the review articles [54,66]. A general microscopic
 theory of statistical systems with heterophase fluctuations has been
developed [67,446--453] and reviewed in Refs. [54,66]. Below, this theory
is applied to describing the heterophase double-well lattices.

\begin{figure}[ht]
\center
\frame{\includegraphics[width=4cm]{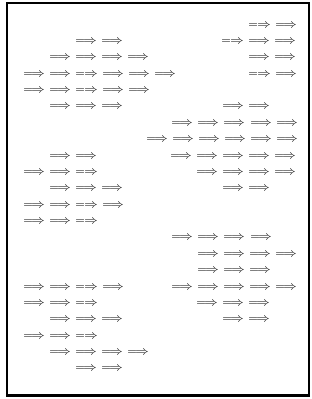}}
\vskip 5mm
\caption{Schematic picture of a heterophase double-well lattice. Arrows
show the regions of the ordered phase. Empty space corresponds to the
regions of the disordered phase}
\label{fig:Fig.3}
\end{figure}

Each thermodynamic phase is characterized by its typical states forming
a Hilbert space. The coexistence of several phases corresponds to the
direct sum of the Hilbert spaces, related to particular phases. The Fock
space over the direct sum of Hilbert spaces is identified with the tensor
product of Fock spaces over each of the Hilbert space [454].

Suppose that the considered system is a mixture of several thermodynamic
phases enumerated by the index $f=1,2,\ldots$ Each phase is characterized
by a Fock space $\cF_f$ of microscopic states typical of the given phase
[54]. The total Fock space of the whole system is the fiber space
\be
\label{8.181}
\tilde\cF = \bigotimes_f \cF_f \; .
\ee
After averaging over heterophase configurations [54], the system is
described by a renormalized Hamiltonian
\be
\label{8.182}
\tilde H = \bigoplus_f \hat H_f \; ,
\ee
which is a direct sum of the partial Hamiltonians associated with the
related phases. In a double-well lattice, there can be two different
thermodynamic phases, the ordered phase and the disordered one, as is
described in Sec. 8.2. Therefore the fiber space is
\be
\label{8.183}
\tilde\cF =\cF_1 \bigotimes \cF_2 \; ,
\ee
while the renormalized Hamiltonian is
\be
\label{8.184}
\tilde H = \hat H_1 \bigoplus \hat H_2 \; .
\ee

The ordered and disordered phases are characterized by different values of
the averages of the imbalance operator $S_j^z$. The average of an operator
$\hat A$, related to an $f$-phase, is defined as
\be
\label{8.185}
< \hat A>_f \; \equiv \; {\rm Tr}_{\cF_f} \;
\hat\rho_f \hat A_f \; ,
\ee
where the statistical operator is
\be
\label{8.186}
\hat\rho_f =
\frac{\exp(-\bt\hat H_f)}{{\rm Tr}_{\cF_f}\exp(-\bt\hat H_f)}
\ee
and $\hat A_f$ is a representation of the operator $\hat A$ on the Fock
space $\cF_f$. Let us ascribe the index $f=1$ to the ordered phase, while
the index $f=2$, to the disordered phase. Then these phases are defined as
those for which
\be
\label{8.187}
< S_j^z >_1 \; \neq \; 0 \; , \qquad
< S_j^z >_2 \; \equiv \; 0 \; .
\ee

As a result of averaging over heterophase configurations, the renormalzied
Hamiltonian (\ref{8.184}) depends on the geometric probabilities of the
phases, $w_f$, which satisfy the conditions
\be
\label{8.188}
w_1 + w_2 = 1 \; , \qquad 0 \leq w_f \leq 1 \; .
\ee
Following the general procedure [54], for the considered case of the
double-well lattice, we have
\be
\label{8.189}
\hat H_f = w_f E_0 N +
\frac{w_f^2}{2} \; \sum_{i\neq j} A_{ij} \; - \;
w_f \Om \sum_j S_j^x \; + \;
w_f^2 \sum_{i\neq j} B_{ij} S_i^x S_j^x \; - \;
w_f^2 \sum_{i\neq j} I_{ij} S_i^z S_j^z \; .
\ee
Resorting again to the mean-field approximation (\ref{8.131}) and
introducing the notation
\be
\label{8.190}
\Om_f \equiv \Om - 2 w_f B < S_j^x >_f
\ee
yields
\be
\label{8.191}
\hat H_f = H_f^{non} - w_f \Om_f \sum_j S_j^x -
2 w_f^2 I \sum_j < S_i^z>_f S_j^z \; ,
\ee
with the first, nonoperator, term being
\be
\label{8.192}
H_f^{non} =  w_f E_0 N   +
\frac{w_f^2}{2} \left ( A - 2B < S_i^x>_f^2 \; + \;
2I < S_i^z>_f^2 \right ) N \; .
\ee
By defining the effective field
$$
{\bf H}_f^{eff} \equiv \{ w_f\Om_f, \; 0, \;
2w_f^2 I < S_j^z>_f \}
$$
reduces Hamiltonian (\ref{8.191}) to the form
\be
\label{8.193}
\hat H_f = H_f^{non} \; - \;
\sum_j {\bf H}_f^{eff} \cdot \bS_j \; .
\ee

The free energy of the whole system is
\be
\label{8.194}
F = F_1 + F_2 \; ,
\ee
with
\be
\label{8.195}
F_f = H_f^{non} - NT\ln \left ( 2 {\rm cosh}\;
\frac{H_f^{eff}}{2T} \right ) \; ,
\ee
where
\be
\label{8.196}
H_f^{eff} = w_f \; \sqrt{\Om_f^2 + 4w_f^2 I^2 < S_j^z>_f^2 } \; .
\ee

For the averages of the pseudospin operators, we get
$$
< S_j^x>_f \; = \; w_f \; \frac{\Om_f}{2H_f^{eff}} \; {\rm tanh}
\left ( \frac{H_f^{eff}}{2T} \right ) \; ,
$$
\be
\label{8.197}
< S_j^y>_f \; = \; 0 \; , \qquad
< S_j^z>_f \; = \; w_f^2 < S_j^z>_f \; \frac{I}{H_f^{eff}} \;
{\rm tanh} \left ( \frac{H_f^{eff}}{2T} \right ) \; .
\ee
It is convenient to employ the reduced variables
\be
\label{8.198}
x_f \equiv 2 < S_j^x>_f \; , \qquad
y_f \equiv 2 < S_j^y>_f \; , \qquad
z_f \equiv 2 < S_j^z>_f
\ee
and use the dimensionless quantities (\ref{8.144}). We define the
effective tunneling frequency
\be
\label{8.199}
\om_f \equiv \om - b w_f x_f
\ee
and introduce the notation
\be
\label{8.200}
h_f \equiv w_f \; \sqrt{\om_f^2 + (1 - b)^2 w_f^2 z_f^2 } \; .
\ee
Then averages (\ref{8.197}) transform into
\be
\label{8.201}
x_f = w_f \; \frac{\om_f}{h_f} \;
{\rm tanh} \left ( \frac{h_f}{2T} \right ) \; , \qquad y_f= 0 \; ,
\qquad z_f = w_f^2 z_f  \;
\frac{1-b}{h_f} \; {\rm tanh} \left ( \frac{h_f}{2T} \right ) \; ,
\ee
where the temperature $T$ is measured in units of $I+B$.

Condition (\ref{8.187}), distinguishing the ordered and disordered phases,
becomes
\be
\label{8.202}
z_1 \neq 0 \; , \qquad z_2 \equiv 0 \; .
\ee
For the ordered phase, $z_1$ is defined by the equation
\be
\label{8.203}
w_1^2 \; \frac{1-b}{h_1} \;
{\rm tanh} \left ( \frac{h_1}{2T} \right ) =  1 \; ,
\ee
while $x_1$, by the equation
$$
x_1 = \frac{\om_1}{(1-b)w_1} \; .
$$
The latter, in view of Eq. (\ref{8.199}), gives
\be
\label{8.204}
x_1 = \frac{\om}{w_1} \; .
\ee
From here, it follows that the ordered component can exist if
\be
\label{8.205}
0 \leq \om \leq w_1 \qquad ( 0 \leq x_1 \leq 1 ) \; .
\ee
Expressions (\ref{8.199}) and (\ref{8.200}) reduce to
\be
\label{8.206}
\om_1 = ( 1 - b) \om \; , \qquad
h_1 = ( 1  - b) w_1 \; \sqrt{\om^2 + w_1^2 z_1^2} \; .
\ee
When $z_1$ tends to zero, this can happen at the temperature
\be
\label{8.207}
T_c = \frac{w_1(1-b)\om}{2{\rm artanh}(\om/w_1)} \; ,
\ee
where $w_1=w_1(T_c)$.

For the disordered component, for which $z_2=0$, we have
\be
\label{8.208}
\om_2 = \om - b w_2 x_2 \; , \qquad h_2 = w_2 \om_2 \; .
\ee
And $x_2$ is defined by the equation
\be
\label{8.209}
x_2 = {\rm tanh} \left ( \frac{w_2\om_2}{2T} \right ) \; .
\ee

The proportions of the phases are prescribed by the system stability. The
equations for the phase probabilities $w_f$ can be found by minimizing the
free energy (\ref{8.194}) under the normalization condition (\ref{8.188}).
For that purpose, we define
\be
\label{8.210}
w_1 \equiv w \; , \qquad w_2 \equiv 1 - w \; .
\ee
Then, the free energy $F=F(w)$ is minimized with respect to $w$. From the
equation
\be
\label{8.211}
\frac{\prt F(w)}{\prt w} = 0 \; ,
\ee
we find
$$
w_1 = \frac{2u+\om_1x_1-\om_2x_2}{4u-(1-b)z_1^2} \; ,
$$
\be
\label{8.212}
w_2 = \frac{2u-\om_1x_1+\om_2x_2-(1-b)z_1^2}{4u-(1-b)z_1^2} \; ,
\ee
where the notation
\be
\label{8.213}
u \equiv \frac{A}{I+B}
\ee
is used.

Let us analyze the obtained equations for the case of zero temperature.
Then we have
$$
x_1  = \frac{\om}{w_1} \; , \qquad x_2 =  1 \; ,
$$
\be
\label{8.214}
z_1 = \sqrt{ 1  -\; \frac{\om^2}{w_1^2} } \; , \qquad z_2 = 0
\qquad ( T = 0 ) \; .
\ee
Also,
\be
\label{8.215}
\om_1 = (1-b) \om \; , \qquad \om_2 = \om - bx_2 \; , \qquad
h_1 = (1-b) w_1^2 \; , \qquad h_2 = w_2 (\om - bw_2 ) \; .
\ee
Probabilities (\ref{8.212}) reduce to
\be
\label{8.216}
w_1 = \frac{2u+ b - \om}{4u+2b-1} \; , \qquad
w_2 = \frac{2u+ b + \om - 1}{4u+2b-1} \; .
\ee
By definition, $0\leq w_f\leq 1$. This imposes the constraints under which
the heterophase mixture can exist,
\be
\label{8.217}
1 - b - 2u \; < \; \om \; < \; 2u + b \; .
\ee
In particular, one can notice that
\be
\label{8.218}
w_1 = w_2 = \frac{1}{2} \; , \qquad z_1 = 0 \qquad
\left ( \om = \frac{1}{2} \right ) \; .
\ee

To check whether the energy of the heterophase mixture is lower than that
of a pure phase, let us consider the internal energy
\be
\label{8.219}
E_{mix} \; \equiv \; \frac{1}{N} < \tilde H > \; = \; E_1 + E_2 \; ,
\ee
in which
\be
\label{8.220}
E_f \; \equiv \; \frac{1}{N} < \hat H_f>_f \; .
\ee
For the Hamiltonian (\ref{8.191}), we find
$$
E_f = \frac{H^{non}_f}{N} \; - \; \frac{I+B}{2} \;
w_f \left [ \om_f x_f + ( 1 -b ) w_f z_f^2 \right ] \; ,
$$
with
$$
\frac{H^{non}_f}{N} = w_f E_0 + \frac{I+B}{4} \;
w_f^2 \left [ 2u - bx_f^2 + (1 - b) z_f^2 \right ] \; .
$$
Combining the latter two expressions, we get
\be
\label{8.221}
E_f = w_f E_0 \; - \;  \frac{I+B}{2} \; \om w_f x_f +
\frac{I+B}{4} \; w_f^2
\left [ 2u + bx_f^2 - (1 - b) z_f^2 \right ] \; .
\ee
For the ordered and disordered components, Eq. (\ref{8.221}) gives,
respectively
$$
E_1 = w_1 E_0 + \frac{I+B}{4} \;
\left [ w_1^2 ( 2u + b -1) - \om^2 \right ] \; ,
$$
\be
\label{8.222}
E_2 = w_2 E_0 + \frac{I+B}{4} \;
\left [ w_2^2 ( 2u + b ) - 2\om w_2 \right ] \; .
\ee
The total sum (\ref{8.219}) becomes
\be
\label{8.223}
E_{mix} = E_0 + \frac{I+B}{4} \; \left [ 2u + b - 2\om
-\om^2 - 2w ( 2u + b - \om ) + w^2 ( 4u + 2b -1 ) \right ] \; ,
\ee
where notation (\ref{8.210}) is employed.

Minimizing Eq. (\ref{8.223}) with respect to $w$ implies that
\be
\label{8.224}
\frac{\prt E_{mix}}{\prt w} = 0 \; , \qquad
\frac{\prt^2 E_{mix}}{\prt w^2} > 0 \; .
\ee
From Eq. (\ref{8.223}), we have
$$
\frac{\prt E_{mix}}{\prt w} =
\frac{I+B}{2} \; [ w( 4u+ 2b-1) - 2u - b +\om ] \; ,
\qquad  \frac{\prt^2 E_{mix}}{\prt w^2} =
\frac{I+B}{2} \; (4u + 2b -1 ) \; .
$$
The first of Eqs. (\ref{8.224}) yields the expressions for the phase
probabilities (\ref{8.216}). And the second condition in Eqs.
(\ref{8.224})
requires that
\be
\label{8.225}
4u + 2b - 1 \; > \; 0 \; .
\ee
If one compares the energy (\ref{8.223}) of the heterophase mixture with
the energy of the pure ordered phase
\be
\label{8.226}
E_{ord} \equiv E_{mix} \qquad (w_1=1, \; w_2 = 0 ) \; ,
\ee
then one gets the difference
\be
\label{8.227} E_{mix} - E_{ord} = -\;
\frac{(2u+b+\om-1)^2}{4(4u+2b-1)} \; (I+B) \; ,
\ee
which shows that the energy of the mixture is lower than that of the pure
ordered phase under the same condition (\ref{8.225}).

The difference of the ordered-phase energy (\ref{8.226}) with the
disordered-phase energy
\be
\label{8.228}
E_{dis} \equiv E_{mix} \qquad (w_1=0, \; w_2 = 1 )
\ee
reads as
\be
\label{8.229}
E_{ord} - E_{dis} = - \; \frac{(\om-1)^2}{4} \; (I + B ) \; .
\ee
Hence, $E_{ord}\leq E_{dis}$ for all $\om$. Therefore the energy of the
heterophase mixture (\ref{8.223}) satisfies the inequality
\be
\label{8.230}
E_{mix} \; < \; E_{ord} \; \leq \; E_{dis} \; ,
\ee
provided that conditions (\ref{8.217}) and (\ref{8.225}) are valid. That
is, under these conditions, the mixed state is more profitable than the
pure phases.

\section{Tools for Quantum Computing}

\subsection{Entanglement Production}

Cold atoms in optical lattices are considered as a very promising
tool for realizing quantum information processing and quantum computation
[17,455]. General problems of quantum computation and information are
described in the books [456,457] and reviews [458,459]. Here we concentrate
our attention on the possibility of employing cold atoms in optical lattices
as a tool for this purpose.

Probably, the main advantage of quantum devices for information processing
and computation is the feasibility of creating entanglement. This purely
quantum property, which is absent in classical devices, should make quantum
processors much more powerful and miniature.

The notion of entanglement has two sides. The entanglement of a quantum
state describes the structure of this state. However, quantum states, as
such, are not measurable quantities, so that their entanglement could be
used only indirectly. Also, there is no uniquely defined measure of
entanglement for quantum states, especially when the latter are mixed
[456--459].

The other notion is the {\it entanglement production}, which shows how
much entanglement is generated by a quantum operation. There exists a
general measure of entanglement production, valid for arbitrary systems
[460,461]. This measure of entanglement production is introduced below,
with keeping in mind its application to lattices.

Let us have a lattice whose lattice sites are enumerated with the index
$j=1,2,\ldots, N_L$. For the purpose of information processing, it is
necessary to have a deep lattice potential, so that atoms be well
localized in the lattice sites. The appearance of BEC diminishes the
level of entanglement [461]. Therefore the insulating state is preferable.
Finite temperature reduces the feasibility of manipulating atoms. Hence
the system is to be deeply cooled down, so that atoms be almost at zero
temperature.

Suppose that atoms in a $j$-site can have different quantum numbers
labelled by the index $n_j$, such that these states $|n_j>$ form a basis
$\{|n_j>\}$. The closed linear envelope over this basis is the Hilbert
space
\be
\label{9.1}
{\cal H}_j \equiv \overline{\cal L}\{ | n_j>\} \; .
\ee
Denoting a given set $\{ n_j\}$ of the indices $n_j$ by ${\bf n}$, one
can define the states
\be
\label{9.2}
|{\bf n} > \; \equiv \; \bigotimes_j \; | n_j > \qquad
({\bf n}\equiv \{ n_j\} ) \; .
\ee
Using states (\ref{9.2}) as a basis $\{ |{\bf n_j} >\}$ makes it possible
to construct the closed linear envelope over this basis, which yields the
Hilbert space
\be
\label{9.3}
{\cal H} \equiv \overline{\cal L} \{ \; |{\bf n} > \} =
\bigotimes_j {\cal H}_j \; .
\ee

The states of space (\ref{9.3}) can be represented as
\be
\label{9.4}
\vp = \sum_\bn c_\bn | \bn > \qquad
(\vp \in \cH ) \; .
\ee
Generally, these states do not have the form of a tensor product
$\otimes_j\vp_j$, with $\vp_j\in\cH_j$. Let us separate out of the space
$\cH$ the {\it disentangled set}
\be
\label{9.5}
\cD \equiv \{ f = \bigotimes _j \vp_j |\; \vp_j \in \cH_j \} \; ,
\ee
whose members have the form of the tensor products. Then the compliment
$\cH\setminus\cD$ is the set of entangled states.

For any quantum operation, represented by an operator $\hat A$ on $\cH$,
we can introduce the norm over $\cD$,
\be
\label{9.6}
|| \hat A||_\cD \; \equiv \; \sup_{f,f'} | (f,\hat A f') | \; ,
\ee
where
$$
f \in \cD\; , \qquad f' \in \cD \; , \qquad
|| f|| = || f'|| = 1 \; ,
$$
with the norm $||f||\equiv\sqrt{{f,f}}$ generated by the scalar product.

It is worth noting that the restricted norm (\ref{9.6}) over the set $\cD$
can also be defined as a norm over a weighted Hilbert space [54,66]. With
this aim in view, we can introduce the weighted Hilbert space $\cH_\cD$ as
a projected space, in which the scalar product is defined as
$$
( f, f')_{\cH_\cD} \; \equiv \; ( P_f\vp, P_{f'}\vp')_\cH \; ,
$$
where $P_f$ is a projector, such that
$$
P_f\vp \equiv f \; \in \; \cD \; .
$$
The norm of $\hat A$ over the weighted Hilbert space $\cH_\cD$ is defined
as
$$
|| \hat A||_{\cH_\cD} \; \equiv \; || P_\cD \hat A P_\cD ||_\cH \; ,
$$
where
$$
P_\cD = \{ P_f | \; P_f\vp = f \; \in \; \cD \} \; .
$$
By this definition
$$
|| \hat A||_\cD = || \hat A||_{\cH_\cD} \; .
$$

Let us introduce the compliment space
\be
\label{9.7}
\cH \setminus \cH_j = \bigotimes_{i(\neq j)} \cH_i
\ee
and define the {\it partially traced operator}
\be
\label{9.8}
\hat A_j \equiv {\rm Tr}_{\cH \setminus \cH_j} \; \hat A \; .
\ee
Then we construct the {\it factor operator}
\be
\label{9.9}
\hat A^\otimes \equiv
\frac{{\rm Tr}_\cH\hat A}{{\rm Tr}_\cD\bigotimes_j\hat A_j} \;
\bigotimes_j \hat A_j \; ,
\ee
for which
$$
{\rm Tr}_\cD \hat A^\otimes =  {\rm Tr}_\cH \hat A \; .
$$

The measure of entanglement generated by the operator $\hat A$ is defined
[460,461] as
\be
\label{9.10}
\ep(\hat A) \equiv \log \;
\frac{||\hat A||_\cD}{||\hat A^\otimes||_\cD} \; ,
\ee
where $log$ is to the base two. This measure can be employed for any
operator possessing a trace. For example, one can consider how much
entanglement is produced by a Hamiltonian in a finite Hilbert space.
More often, one is interested in the level of entanglement produced by
a statistical operator.

\subsection{Topological Modes}

To realize any scheme of information processing it is necessary to
possess objects that could be transferred into different quantum states.
In the case of cold atoms, these could be internal atomic states [17].
Another possibility is to generate topological coherent modes in BEC,
as suggested in Ref. [174]. Various properties of these modes, representing
nonground-state Bose condensates, have been studied in several papers
[174--196,462-466]. The generation of such modes can be accomplished
in two ways. One method is the modulation of the trapping potential with
the frequency in resonance with the transition frequency between two
coherent modes [174--176]. The other way, as has been mentioned in Refs.
[189,190,192] and analysed in Ref. [464], is by the resonant modulation of
the atomic scattering length. Both these techniques are illustrated below.

Let us consider a deep lattice, in each site of which there are many Bose
atoms. Optical lattices with large filling factors, reaching $\nu\sim 10^4$,
are readily available in experiment [22,23]. All atoms inside a lattice site
can be made well localized, with the intersite tunneling almost completely
suppressed. Temperature can be kept low, so that practically all atoms
piling down to BEC.

Since lattice sites are very deep, we can start the consideration from
a single site, representing a kind of a trap. At low temperature and weak
interactions, the system inside the trap is described by the condensate
wave function satisfying the Gross-Pitaevskii equation
\be
\label{9.11}
i\; \frac{\prt}{\prt t} \; \eta(\br,t) = \left [ - \;
\frac{\nabla^2}{2m} + U(\br,t) - \mu_0 + \Phi_0(t) \;
|\eta(\br,t)|^2 \right ] \; \eta(\br,t) \; .
\ee
The condensate wave function is normalized to the number of atoms inside
the trap, that is, to the filling factor
\be
\label{9.12}
\int | \eta(\br,t) |^2 \; d\br  = \nu \; .
\ee
The external potential
\be
\label{9.13}
U(\br,t) = U(\br) + V(\br,t)
\ee
consists of a trapping potential $U(\br)$, characterizing the optical
potential at the considered lattice site, and of an additional modulating
potential $V(\br,t)$. The interaction strength can also be made
time-dependent by means of the Feshbach resonance techniques.

For convenience, one can use the relation
\be
\label{9.14}
\eta(\br,t) \equiv \sqrt{\nu} \; \vp(\br,t) \; ,
\ee
defining the function $\vp(\br,t)$ normalized to one,
\be
\label{9.15}
\int |\vp(\br,t)|^2 \; d\br = 1 \; .
\ee
Then Eq. (\ref{9.11}) reads as
\be
\label{9.16}
i\; \frac{\prt}{\prt t} \; \vp(\br,t) = \left [ - \;
\frac{\nabla^2}{2m} + U(\br,t) - \mu_0 + \nu\; \Phi_0(t) \;
|\vp(\br,t)|^2 \right ] \; \vp(\br,t) \; .
\ee

For a stationary case, when there is no modulating field, $V(\br,t)=0$,
and atomic interactions are constant, $\Phi_0(t)=\Phi_0$, Eq. (\ref{9.16})
becomes
\be
\label{9.17}
i\; \frac{\prt}{\prt t} \; \vp_n(\br,t) = \left [ - \;
\frac{\nabla^2}{2m} + U(\br) - \mu_0 + \nu\; \Phi_0 \;
|\vp_n(\br,t)|^2 \right ] \; \vp_n(\br,t) \; ,
\ee
where a multi-index $n$ enumerates quantum states. In stationary solutions,
the spatial and temporal variables can be separated as follows:
\be
\label{9.18}
\vp_n(\br,t) = \vp_n(\br) e^{-i\om_n t} \; .
\ee
As a result, Eq. (\ref{9.17}) reduces to the eigenproblem
\be
\label{9.19}
\left [ - \; \frac{\nabla^2}{2m} + U(\br)  + \nu\; \Phi_0 \;
|\vp_n(\br)|^2 \right ] \vp_n(\br) = E_n \; \vp_n(\br) \; ,
\ee
in which the eigenvalues
\be
\label{9.20}
E_n \equiv \om_n + \mu_0
\ee
have the property
$$
\min_n E_n = \mu_0 \; , \qquad \min_n \om_n = 0 \; .
$$
The eigenfunctions $\vp_n(\br)$ of Eq. (\ref{9.19}) are the {\it topological
coherent modes} [174--176]. Equally, the condensate functions
\be
\label{9.21}
\eta_n(\br) = \sqrt{\nu} \; \vp_n(\br)
\ee
can also be called the topological coherent modes. The condensate function
(\ref{9.21}), corresponding to the lowest energy $E_n=\mu_0$, characterizes
the standard BEC. The higher modes of $\eta_n(\br)$ describe the {\it
nonground-state condensates} [174--176]. The functions $\vp_n(\br)$ and
$\eta_n(\br)$ differ solely by their normalizations
$$
\int |\vp_n(\br)|^2 \; d\br = 1 \; , \qquad
\int |\eta_n(\br)|^2 \; d\br = \nu \; .
$$
In equilibrium, only the standard BEC is realized. To produce the
macroscopic occupation of the higher levels, it is necessary to apply
additional fields making the system nonequilibrium.

The modulating trapping potential can be taken in the form
\be
\label{9.22}
V(\br,t) = V_1(\br)\cos \om t + V_2(\br) \sin \om t \; .
\ee
Similarly, the interaction strength can be made time-dependent as
\be
\label{9.23}
\Phi_0(t) = \Phi_0 + \ep_1 \cos\om t + \ep_2 \sin\om t \; .
\ee

It is of principal importance to choose the frequency of the alternating
potentials to be tuned to a resonance with a transition frequency related
to the energy levels we wish to connect. Let us consider two energy levels,
$E_1$ and $E_2$, with the transition frequency being
\be
\label{9.24}
\om_{21} \equiv E_2 - E_1 \; .
\ee
Then the resonance condition is
\be
\label{9.25}
\left | \frac{\Dlt\om}{\om_{21} } \right | \; \ll \; 1 \qquad
(\Dlt \om \equiv \om - \om_{21} ) \; .
\ee
For instance, $E_1$ can correspond to the lowest energy level, equal
to $\mu_0$.

We can look for the solution to the temporal Eq. (\ref{9.16}) in the form
of the expansion over the coherent modes,
\be
\label{9.26}
\vp(\br,t) = \sum_n c_n(t) \vp_n(\br,t) \; .
\ee
The coefficient functions can be treated as slow in time, compared to the
exponential oscillations in Eq. (\ref{9.18}), such that
\be
\label{9.27}
\frac{1}{\om_n} \; \left |
\frac{dc_n}{dt} \right | \; \ll \; 1 \; .
\ee
The latter condition requires that atomic interactions and the pumping
alternating fields would not be too strong, which is easily realized in
experiment [174--176,182]. The normalization condition
\be
\label{9.28}
\sum_n | c_n(t) |^2 = 1
\ee
is assumed.

When there are two time scales, one can resort to the averaging techniques
[236,237] and to the scale separation approach [240--242]. To this end, we
substitute expansion (\ref{9.26}) into Eq. (\ref{9.16}), multiply the latter
by  $\vp_n^*(\br,t)$, integrate over $\br$, and accomplish the time averaging
according to the rule
\be
\label{9.29}
\{ f(t) \}_t \equiv \lim_{\tau\ra\infty} \;
\frac{1}{\tau} \; \int_0^\tau f(t) \; dt \; ,
\ee
where the slow variables are kept as quasi-integrals of motion [240--242].
For example, averaging (\ref{9.29}) gives
$$
\left \{ e^{i(\om_m-\om_n)t} \right \}_t = \dlt_{mn} \; ,
$$
from where
$$
\{ \vp_m^*(\br,t)\vp_n(\br,t) \}_t =
\dlt_{mn} |\vp_n(\br)|^2 \; .
$$
Therefore, the functions $\vp_n(\br,t)$ are orthogonal on average, though
the functions $\vp_n(\br)$ can be not orthogonal. Also, we have
$$
\left \{ e^{i(\om_m+\om_n-\om_k-\om_p)t} \right \}_t =
\dlt_{mk} \dlt_{np}  + \dlt_{mp} \dlt_{nk} -
\dlt_{mk}\dlt_{np} \dlt_{mn}\; .
$$

Let us introduce the notation for the matrix elements of the interaction
\be
\label{9.30}
\al_{mn} \equiv \nu \Phi_0 \int | \vp_m(\br)|^2 \left
[ 2 |\vp_n(\br)|^2 - |\vp_m(\br)|^2 \right ] \; d\br \; ,
\ee
of the pumping potential
\be
\label{9.31}
\bt_{mn} \equiv \int \vp_m^*(\br) [ V_1(\br) - i V_2(\br) ] \;
\vp_n(\br) \; d\br \; ,
\ee
and of the interaction modulation
\be
\label{9.32}
\gm_n \equiv \nu (\ep_1-i\ep_2) \int \vp_1^*(\br)
|\vp_n(\br)|^2 \vp_2(\br) \; d\br \; .
\ee
Then Eq. (\ref{9.16}) yields
$$
i\; \frac{dc_n}{dt} = \sum_{m(\neq n)} \al_{nm} | c_m|^2 c_n \; +
$$
$$
+ \frac{1}{2}\; \dlt_{n1} e^{i\Dlt\om t}
\left [ 2 \sum_{m(\neq 2)} \gm_{m} | c_m|^2 c_2 \; +
\; \gm_2|c_2|^2 c_2 + \bt_{12} c_2 \right ] \; +
\frac{1}{2}\; \dlt_{n1} e^{-i\Dlt\om t} \gm_1^* c_2^* c_1^2  \; +
$$
\be
\label{9.33}
+ \frac{1}{2} \dlt_{n2} e^{-i\Dlt\om t}
\left [ 2 \sum_{m(\neq 1)} \gm^*_{m} | c_m|^2 c_1 \; +
\; \gm_1^*|c_1|^2 c_1 + \bt^*_{12} c_1 \right ] \; + \;
\frac{1}{2}\; \dlt_{n2} e^{i\Dlt\om t} \gm_2 c_1^* c_2^2  \; .
\ee
It is not difficult to notice [182] that, if at the initial time $t=0$,
$c_n(0)=0$ for $n\neq 1,2$, then $c_n(t)=0$ for all $t\geq 0$ and $n\neq 1,2$.
Hence Eq. (\ref{9.33}) can be separated into two equations
$$
i\; \frac{dc_1}{dt} = \al_{12} | c_2|^2 c_1 +
\frac{1}{2}\; e^{i\Dlt\om t} \left ( 2\gm_1 |c_1|^2 c_2 +
\gm_2 |c_2|^2 c_2 + \bt_{12} c_2 \right ) +
\frac{1}{2}\; e^{-i\Dlt\om t} \gm_1^* c_2^* c_1^2 \; ,
$$
\be
\label{9.34}
i\; \frac{dc_2}{dt} = \al_{21} | c_1|^2 c_2 +
\frac{1}{2}\; e^{-i\Dlt\om t} \left ( 2\gm_2^* |c_2|^2 c_1 +
\gm_1^* |c_1|^2 c_1 + \bt^*_{12} c_1 \right ) +
\frac{1}{2}\; e^{i\Dlt\om t} \gm_2 c_1^* c_2^2 \; .
\ee
The solutions to these equations define the temporal behavior of the {\it
fractional mode populations}
\be
\label{9.35}
p_n(t) \equiv | c_n(t)|^2 \; .
\ee

The properties of Eqs. (\ref{9.34}) have been studied in detail for
the case of the mode generation by means of the trapping-potential
modulation, when $\bt_{12}\neq 0$ while $\gm_n=0$, in Refs.
[174--176,179,180,182,184--186,192--196]. The generalization for the
case of the multiple mode generation has been given [189,190]. It has
also been shown that the generation of the nonground-state condensate
is achievable at nonzero temperature [193,465]. The creation of the
topological coherent modes by the modulation of the interaction strength,
when $\bt_{12}=0$ but $\gm_n\neq 0$, is considered in Ref. [464].

\subsection{Coherent States}

Now let us turn to the situation, when there is a lattice with $N_L$ sites.
In each site a deep well is formed by an optical potential. The number of
atoms in a $j$-lattice site is $\nu_j\gg 1$. For an ideal lattice, the
filling factor $\nu_j$ does not depend on the site index. But, in general,
the lattice can be nonideal. Then $\nu_j$ can be different for different
sites.

By employing the resonant generation, described in the previous section,
one can excite in the $j$-site the topological coherent modes labelled by
a multi-index $n_j$. Suppose that $\eta_{n_j}$ are the coherent fields
 associated with the $j$-site and normalized to the corresponding occupation
number
\be
\label{9.36}
\int | \eta_{n_j}(\br )|^2 \; d\br = \nu_j \; .
\ee
Similarly to Eq. (\ref{9.21}), we can also define the functions
$\vp_{n_j}(\br)$ normalized to one, such that
\be
\label{9.37}
\eta_{n_j}(\br) \; = \; \sqrt{\nu_j}\; \vp_{n_j}(\br) \; .
\ee
Being the solutions to the nonlinear Schr\"odinger equation of type
(\ref{9.19}), the coherent fields $\eta_{n_i}(\br)$ and $\eta_{n_j}(\br)$,
with $i\neq j$, are not, generally, orthogonal, that is, the scalar product
\be
\label{9.38}
\int \eta^*_{n_i}(\br) \eta_{n_j}(\br) \; d\br \equiv \nu_{ij}
\ee
is not necessarily zero for $i\neq j$. The diagonal quantity
\be
\label{9.39}
\nu_{jj} = \nu_j
\ee
is the occupation number of the $j$-site.

In the Fock space, the coherent state, associated with the $n_j$-mode, is
given by the column
\be
\label{9.40}
| n_j > \; = \; \left [ \frac{\exp(-\nu_j/2)}{\sqrt{n!} }\;
\prod_{k=0}^n \eta_{n_j}(\br_k) \right ] \; ,
\ee
where $n=0,1,2,\ldots$. Expression (\ref{9.40}) is the short-hand notation
for the column state of the type
\begin{eqnarray}
\label{9.41}
|m>\; =\; e^{-\nu/2} \left [ \begin{array}{l}
1 \\
\eta_m(\br_1) \\
\frac{1}{\sqrt{2!}}\; \eta_m(\br_1)\eta_m(\br_2) \\
\cdot \\
\cdot \\
\cdot \\
\frac{1}{\sqrt{n!}}\; \eta_m(\br_1)\eta_m(\br_2)\ldots\eta_m(\br_n) \\
\cdot \\
\cdot \\
\cdot \\
\end{array} \right ] \; .
\end{eqnarray}
The coherent states (\ref{9.40}) are not necessarily orthogonal to each
other, so that the scalar product
\be
\label{9.42}
< n_i | n_j >\; = \; \exp\left ( -\; \frac{\nu_i+\nu_j}{2} +
\nu_{ij} \right )
\ee
is not, in general, zero for $i\neq j$. But the coherent states (\ref{9.40})
are normalized to one, since
\be
\label{9.43}
< n_j | n_j > \; = \; 1 \; .
\ee
However, as follows from Eqs. (\ref{9.38}) and (\ref{9.42}), the coherent
states are asymptotically orthogonal [467,468] in the sense that
\be
\label{9.44}
< n_i| n_j > \; \simeq \; \dlt_{ij} \qquad (\nu_i + \nu_j \gg 1) \; .
\ee
They also are asymptotically complete in the weak sense,
\be
\label{9.45}
\sum_{n_j} | n_j><n_j| \simeq 1 \qquad (\nu_i+\nu_j\gg 1) \; .
\ee
Therefore the states $|\bn>$, defined in Eq. (\ref{9.2}), form the
asymptotically orthogonal and complete basis $\{|\bn>\}$ in the Hilbert
space (\ref{9.3}).

\subsection{Coherent-Mode Register}

The topological coherent modes can be used as a tool for a quantum register
of information processing. These modes possess a rich variety of interesting
properties [174--196,462--466]. The most important, for the purpose of
quantum information processing, is the feasibility of producing entangled
states [192,467--469]. Entanglement production with topological coherent
modes and its temporal evolution can be regulated by external fields
[465,467--469].

The statistical state of a lattice with coherent modes generated in its
lattice sites, is characterized by the statistical operator $\hat\rho$
which can be expanded over the basis $\{ |\bn>\}$ of the coherent states,
\be
\label{9.46}
\hat\rho = \sum_\bn p_\bn | \bn > < \bn | \; .
\ee
The normalization condition
$$
{\rm Tr}_\cH \hat\rho =\sum_\bn p_\bn =  1
$$
is assumed. Following the procedure of Sec. 9.1, we construct the factor
operator
\be
\label{9.47}
\hat\rho^\otimes \equiv \bigotimes_j \hat\rho_j \; , \qquad
\hat\rho_j \; \equiv \; {\rm Tr}_{\cH\setminus\cH_j} \; \hat\rho \; .
\ee
The normalization conditions are valid:
$$
{\rm Tr}_{\cH_j} \hat\rho_j = 1 \; \qquad
{\rm Tr}_\cH \hat\rho^\otimes =
\prod_j {\rm Tr}_{\cH_j} \hat\rho_j = 1 \; ,
$$
where
$$
\hat\rho_j = \sum_\bn p_\bn\; | n_j > < n_j | \; .
$$

Using the measure of entanglement production (\ref{9.10}), we define the
level of entanglement produced by the statistical operator (\ref{9.46}),
\be
\label{9.48}
\ep(\hat\rho) = \log\;
\frac{||\hat\rho||_\cD}{||\hat\rho^\otimes||_\cD} \; .
\ee
Here
$$
||\hat\rho||_\cD = \sup_\bn p_\bn \; , \qquad
||\hat\rho_j||_{\cH_j} = \sup_{n_j} \;
\sum_{\bn(\neq n_j)} p_\bn \; , \qquad
||\hat\rho^\otimes||_\cD = \prod_j ||\hat\rho_j||_{\cH_j} \; .
$$
As a result, Eq. (\ref{9.48}) yields
\be
\label{9.49}
\ep(\hat\rho) = \log\;\frac{\sup_\bn p_\bn}
{\prod_j \sup_{n_j} \sum_{\bn(\neq n_j)} p_\bn} \; .
\ee

Entanglement in the lattice is generated when
$$
\sup_\bn p_\bn \neq
\prod_j \sup_{n_j} \sum_{\bn(\neq n_j)} p_\bn \; ,
$$
that is, when the lattice sites are somehow correlated. There are several
sources of their correlation. First, this is the common history of the
condensate preparation. Second, the lattice sites are never completely
independent, but there always exists at least a weak tunneling. Third, atoms
from different sites do interact, even though this interaction can be rather
weak. Finally, the  modulating resonant fields, producing the coherent modes,
can be common for all sites of the lattice. The maximal correlation between
the modes from different sites happens when all sites are identical and
modulated synchronously, so that
$$
p_\bn = p_n \prod_j \dlt_{n n_j} \; .
$$
Then the statistical operator (\ref{9.46}) is
$$
\hat\rho = \sum_n p_n \; | nn\ldots n > < nn\ldots n | \; .
$$
And we have
$$
\sup_\bn p_\bn = p_n \; , \qquad
\sum_{\bn(\neq n_j)} p_\bn = p_n \dlt_{n n_j} \; , \qquad
\sup_{n_j} p_n \dlt_{n n_j} = p_n \; .
$$
The entanglement-production measure (\ref{9.49}) reduces to
\be
\label{9.50}
\ep(\hat\rho) = - (N_L - 1) \log\; \sup_n p_n \; .
\ee
The quantity $p_n=p_n(t)$ is the same as in Eq. (\ref{9.35}), hence, is
defined by the evolution equations (\ref{9.34}). If the number of sites
$N_L$ is large, measure (\ref{9.50}) can be made very large. Since the
value $p_n(t)$ can be regulated, the evolution of measure (\ref{9.50}) can
also be regulated [465,467,469], thus, allowing for the realization of the
coherent-mode lattice register.

The specific features of the coherent-mode register are:

\vskip 2mm
(i) The working objects, multimode condensates, are mesoscopic. Entanglement
is accomplished for these mesoscopic objects, but not for separate particles.
\vskip 2mm
(ii) A very strong level of entanglement can be produced, when
$\ep(\hat\rho)\sim N_L\gg 1$.

\vskip 2mm
(iii) The computation dimension is very large. For $N_L$ lattice sites,
with $M$ modes in each, the computation dimension is $M^{N_L}$. Thus,
for two modes $(M=2)$ in a lattice of $N_L=100$ sites, the computation
dimension is $10^{30}$.

\vskip 2mm
(iv) The properties of the lattice, the strength of atomic interactions,
and the resonant modulating fields can be varied in a very wide range, thus,
making the mode register highly controllable.

\vskip 2mm
(v) It is feasible to organize parallel computation by producing operations
in different parts of the lattice.

\vskip 2mm
(vi) Erasing memory is a simple process that can be done by appropriately
varying the modulating fields.

\vskip 2mm
(vii) The decoherence time is sufficiently long. Estimates [182,465] give
it of the order of $10-100$ seconds.

\subsection{Double-Well Register}

Double-well lattices are considered as a very promising tool for quantum
information processing and quantum computing. Recently, such double-well
lattices have been realized experimentally in two-dimensional [420] and
three-dimensional [470--474] configurations. The lattices were loaded by
$^{87}$Rb atoms. The total number of lattice sites was around $3\times
10^3$. The filling factor could be varied between one and about 200 atoms.
The properties of the double wells, such as the barrier height, the distance
between the wells, and the relative energy offset, could be dynamically
controlled. The atoms could be transferred between the left and right wells
in a controllable way.

The possibility of dynamically varying the properties of the double-well
lattices allows for the regulation of their states and dynamics [475].
This controllable regulation is of high importance for realizing quantum
information processing and quantum computing with double-well lattices. For
the latter purpose, the lattices with the filling factor one seem to be the
most appropriate. The properties of such double-well lattices are described
in Chapter 8. Note that the double-well potentials can be made asymmetric,
which provides additional possibilities for regulating the system properties
[476].

Quantum information protocols hold the promise of technological applications
unattainable by purely classical means. In order to realize both, the storage
of quantum information and the faithful long-distance communication, combined
systems of atoms interacting with photons seem to be good candidates [477].
It would be interesting to consider the interaction of coherent electromagnetic
fields with atoms located at the cites of a double-well lattice.

\section{Brief Concluding Remarks}

The material, covered in the present review article, is so extensive that
it would take too much space for a more or less detailed concluding
discussion. And listing in short the considered topics would duplicate the
Contents. Therefore, instead of having a concluding summary, the reader is
advised to survey again the Contents.

At the present time, optical lattices is a fastly developing field of
research. There permanently appear new interesting results. For instance,
density modulations in an elongated BEC with a disorder potential were
observed [478]. The direct observation of the Anderson localization [358]
of boson matter waves in a one-dimensional non-interacting BEC with disorder
was announced [479,480]. The Anderson localization is a phenomenon typical
of the ideal gases, while rather weak interactions destroy this effect [481].

The phenomenon of the Anderson localization occurs in real space. There
exists an analogous effect, called the dynamical localization [482],
happening in momentum space. Such a dynamical localization can be realized
by means of the quasiperiodic kicked-rotator model [483].

Despite the variety of novel experimental observations, the basic theoretical
points remain the same. In this review, the emphasis was exactly on the main
theoretical ideas and methods. Therefore the material of this review should
remain useful in future for any researcher in the field of optical lattices.

Many techniques, related to periodic potentials, like those treated in the
review, are actually common for Bose as well as for Fermi systems. Although
the physics of the latter, in many respects, is different. The most detailed
description of the state of the art of ultracold Fermi gases has recently
been given by Ketterle and Zwierlein [484] (see also [15,485]).

In conclusion, it is worth mentioning that many properties of trapped atoms
are similar to those of particles in quantum dots, finite nuclei, and clusters.
The discussion of the latter systems can be found in the review articles
[486--490].

\vskip 5mm

{\bf Acknowledgements}

\vskip 2mm

I am very grateful for many useful discussions and collaboration to
V.S. Bagnato and E.P. Yukalova. Financial Support from the Russian Foundation
for Basic Research (Grant 08-02-00118) is appreciated.

\end{document}